\documentclass[twocolumn,aps,prd,amsmath,amssymb,preprintnumbers,longbibliography]{revtex4-1}
\usepackage{amsmath, etoolbox} 
\usepackage{amsfonts} 
\usepackage{amssymb}
\usepackage{bbm}
\usepackage{graphics}
\usepackage{graphicx}
\usepackage{titlesec}
\usepackage{mathtools}
\usepackage{environ}
\usepackage{dsfont}

\usepackage{mathrsfs}

\usepackage[colorlinks=true]{hyperref}
\hypersetup{urlcolor=black,linkcolor=black,citecolor=black}
\usepackage[toc,page]{appendix}

\usepackage{makecell,tabularx}
\setcellgapes{3pt}

\makeatletter
\renewcommand*\env@matrix[1][c]{\hskip -\arraycolsep
  \let\@ifnextchar\new@ifnextchar
  \array{*\c@MaxMatrixCols #1}}
\makeatother

\titleformat{\subsection}[block]{\normalfont\bfseries}{\thesubsection.}{1ex}{}
\titlespacing{\subsection}{0pt}{10pt}{1pt}[0pt]
\titleformat*{\section}{\large\bfseries}
\renewcommand{\thesubsection}{\arabic{subsection}}

\usepackage{natbib}

\usepackage{braket}

\usepackage{todonotes}

\definecolor{refkey}{rgb}{0,0,1}
\definecolor{labelkey}{rgb}{0,1,0}

\renewcommand{\thesection}{\arabic{section}}
\renewcommand{\thesubsection}{\thesection.\arabic{subsection}}

\makeatletter
\renewcommand{\p@subsection}{}
\renewcommand{\p@subsubsection}{}
\makeatother

\usepackage{enumitem}

\newcommand{\be}{\begin{equation}}
\newcommand{\bel}[1]{\be\label{#1}}
\newcommand{\ee}{\end{equation}}
\newcommand{\ba}{\begin{align}}
\newcommand{\ea}{\end{align}}
\newcommand{\bc}{\begin{center}}
\newcommand{\ec}{\end{center}}
\newcommand{\zwisch}[1]{\bc\textbf{#1}\ec}
\newcommand{\dub}{\ \ }
\newcommand{\trip}{\ \ \ }
\newcommand{\A}{A_{\mu mn}}
\newcommand{\e}{e_{\mu}^{\ m}}
\newcommand{\ei}{e_{m}^{\ \mu}}
\newcommand{\eps}{\varepsilon}

\newcommand{\deta}{\partial_{\eta}}
\newcommand{\dz}{\partial_{z}}
\newcommand{\hub}{\mathscr{H}}
\newcommand{\hubhat}{\widehat{\mathscr{H}}}
\newcommand{\Ccal}{\mathscr{C}}
\newcommand{\mub}{\mu_{b}^{2}}
\newcommand{\muc}{\mu_{c}^{2}}
\newcommand{\mubhat}{\hat{\mu}_{b}^{2}}
\newcommand{\muchat}{\hat{\mu}_{c}^{2}}
\newcommand{\mubtil}{\tilde{\mu}_{b}^{2}}
\newcommand{\muctil}{\tilde{\mu}_{c}^{2}}
\newcommand{\muh}{\mu_{H}^{2}}
\newcommand{\muvtil}{\tilde{\mu}_{v}^{2}}
\newcommand{\de}{\text{d}}
\newcommand{\dt}{\partial_{t}}
\newcommand{\infpast}{\eta\to -\infty}
\newcommand{\btil}{\tilde{b}}
\newcommand{\dtil}{\tilde{d}}
\newcommand{\ftil}{\tilde{f}}
\newcommand{\zetil}{\tilde{\zeta}}
\newcommand{\etil}{\tilde{e}}
\newcommand{\ytil}{\tilde{y}}
\newcommand{\ttil}{\tilde{t}}
\newcommand{\gtil}{\tilde{g}}
\newcommand{\wtil}{\tilde{w}}
\newcommand{\ctil}{\tilde{c}}
\newcommand{\chitil}{\tilde{\chi}}
\newcommand{\Rtil}{\tilde{R}}

\newcommand{\Util}{\tilde{U}}
\newcommand{\Ktil}{\tilde{K}}
\newcommand{\Htil}{\widetilde{H}}
\newcommand{\Ptil}{\tilde{P}}
\newcommand{\Ftil}{\tilde{F}}
\newcommand{\Ctil}{\tilde{C}}
\newcommand{\Etil}{\tilde{E}}
\newcommand{\Stil}{\tilde{S}}
\newcommand{\phitil}{\tilde{\varphi}}
\newcommand{\mtil}{\widetilde{m}^{2}}
\newcommand{\Ztil}{\widetilde{Z}}
\newcommand{\Hhat}{\widehat{H}}
\newcommand{\Vhat}{\widehat{V}}
\newcommand{\That}{\widehat{T}}
\newcommand{\Yhat}{\widehat{Y}}
\newcommand{\Khat}{\widehat{K}}
\newcommand{\fhat}{\hat{f}}
\newcommand{\rhohat}{\hat{\rho}}
\newcommand{\bhat}{\hat{b}}
\newcommand{\khat}{\hat{k}}
\newcommand{\that}{\hat{t}}
\newcommand{\chat}{\hat{c}}
\newcommand{\sbar}{\overline{s}}
\newcommand{\fbar}{\overline{f}}
\newcommand{\mubar}{\overline{\mu}}
\newcommand{\Vbar}{\overline{V}}
\newcommand{\Ybar}{\overline{Y}}
\newcommand{\Kbar}{\overline{K}}
\newcommand{\Mbar}{\overline{M}}
\newcommand{\Hbar}{\overline{H}}
\newcommand{\M}{M^{2}}
\newcommand{\m}{m^{2}}
\newcommand{\n}{n^{2}}
\newcommand{\abs}[1]{|#1|}
\newcommand{\half}{\frac{1}{2}}
\newcommand{\dchi}[1]{\frac{\partial#1}{\partial\chi}}
\newcommand{\nn}{\nonumber}
\newcommand{\Matrix}[4]{\begin{pmatrix}#1&#2\\#3&#4\end{pmatrix}}
\newcommand{\Vector}[2]{\begin{pmatrix}#1\\#2\end{pmatrix}}
\newcommand{\du}{\partial_{u}}

\newcommand{\gl}{\big(}
\newcommand{\gr}{\big)}

\begin{document}

\title[ ]{Cosmology from pregeometry}

\author{C. Wetterich}
\affiliation{Institut  f\"ur Theoretische Physik\\
Universit\"at Heidelberg\\
Philosophenweg 16, D-69120 Heidelberg}

\begin{abstract}

We discuss cosmological solutions for a diffeomorphism invariant gauge theory of the non-compact Lorentz group $SO(1,3)$. Besides the gauge bosons our model of pregeometry contains a vector field in the vector representation of $SO(1,3)$ and a scalar singlet. General relativity and variable gravity emerge as effective theories for large distances and times in Planck units. We propose an approximation to the effective action with up to two derivatives. For a suitable range of parameters the universe approaches for large times stable Minkowski space. For late cosmology the model predicts dynamical dark energy and provides for a candidate for dark matter. Early cosmology is characterized by an inflationary epoch. The beginning of the universe in the infinite past is great emptiness, corresponding to an ultraviolet fixed point with the associated quantum scale symmetry. The beginning universe is a vacuum state with vanishing expectation values and finite non-vanishing correlation functions for the fluctuations of all fields. There is no physical big bang singularity.

\end{abstract}

\maketitle

\onecolumngrid
\vspace{4cm}
\twocolumngrid

\tableofcontents

\newpage

\section{Introduction}
\label{section:I}

The geometric description of general relativity may only be an effective low energy or large distance theory. The metric could be a composite object or collective excitation, similar to pions and hadrons in strong interactions. The short distance quantum field theory for the gravitational interactions is then described by different degrees of freedom, similar to quarks and gluons for strong interactions. In this approach the dynamical geometry with its metric description emerges from more fundamental pregeometry.

A model for pregeometry based on a $SO(4)$-Yang Mills theory has been proposed recently \cite{CWPG} as a candidate for a consistent euclidean functional integral for the gravitational interactions. This is a promising starting point for the formulation of quantum gravity. The present paper investigates a similar model with $SO(1,3)$-gauge symmetry, related to the euclidean model by analytic continuation. It is our aim to demonstrate that the solutions of the field equations of this model lead to a realistic evolution history of our universe.

The main advantage of formulating a quantum field theory for pregeometry is the simple possible form of the action for short distances or high momenta, $q^{2}\to\infty$. This action may involve only up to two derivatives of the fields, admitting a starting point without ghost or tachyonic instabilities. In comparison, the issue in metric gravity seems to be more complex. The functional renormalization group \cite{Wetterich_1993}, \cite{RWEAEE} allows for an investigation of the flow of the effective action for metric gravity \cite{MR}. There is increasing evidence that metric gravity may be a non-perturbatively renormalizable quantum field theory according to the asymptotic safety paradigm \cite{WEIN, MR, SOUM, DPER, RSAU, LAUR}, for a recent review see ref. \cite{bonanno2020critical}. Still, no simple short distance behavior of gravity is visible in this approach so far. One may hope that a similar investigation in pregeometry may reveal a simpler structure, according to the possibility of a simple action involving no more than two derivatives of the fields.

The issue is visible in the momentum dependence of the graviton propagator. For very short distances it is not likely that the simple Einstein-Hilbert action offers a valid setting for quantum gravity. In next order in a momentum or derivative expansion the action involves up to four derivatives, as encoded for a diffeomorphism invariant setting in the presence of higher derivative invariants as $R_{\mu\nu\rho\sigma}R^{\mu\nu\rho\sigma}$, with $R_{\mu\nu\rho\sigma}$ the curvature tensor constructed from the metric. In this order the theory is perturbatively renormalizable \cite{STE, FRATSE, AB, AGRA1, AGRA2}, but plagued by ghost instabilities. Functional renormalization group investigations of the full momentum dependence of the graviton propagator in metric gravity have found encouraging results \cite{GRP1, GRP2, GRP3, bonanno2021reconstructing}, suggesting that instabilities could be avoided, and showing that some of the critics formulated in ref. \cite{Don} do not apply. A simple convincing picture for the short distance behavior has not emerged so far, however. This contrasts with the situation in pregeometry. For a simple action for pregeometry mixing effects lead indeed to a term $\sim q^{4}$ for the momentum expansion of the inverse graviton propagator. Nevertheless, the full momentum dependence of the graviton propagator takes in this case an acceptable form \cite{CWFSI, CWPG, wetterich2021pregeometry}, obeying the criteria of ref. \cite{PLCW}, and avoiding any tachyon or ghost instabilities. The high momentum behavior remains very simple.

Furthermore, the black hole solutions obtained from the Einstein-Hilbert action show a central singularity. It is speculated but not established that the asymptotic safety picture for quantum gravity could remove this singularity. With the simple short distance action of the proposed model of pregeometry it will be interesting to investigate the consequences for the black hole singularity. The simple short distance behavior of the action and propagators could be helpful in this respect.

The present paper addresses cosmology. Any realistic proposal for quantum gravity has to entail an acceptable evolution history of our universe. The evolution of the universe cannot be investigated in a euclidean theory. It needs the Minkowski signature of the metric. Such a model is thought to be obtained from a euclidean theory by analytic continuation. As a particular feature of pregeometry this analytic continuation also affects the gauge bosons. The gauge group $SO(4)$ in the euclidean formulation becomes the non-compact Lorentz group $SO(1,3)$. Simultaneous analytic continuation of the vierbein and the gauge fields at fixed coordinates realizes this requirement \cite{CWES, wetterich2021pregeometry}.

A non-compact gauge group $SO(1,3)$ comes with potential problems. Due to contractions with the indefinite $SO(1,3)$-invariant tensor $\eta_{mn}$ it is no longer guaranteed that the kinetic term for all gauge bosons has the correct sign needed for stability. In the presence of mass terms this could lead to tachyonic instabilities for flat space, which would obstruct any realistic cosmology. For a realistic cosmology the universe has to become close to flat space at late times. This would not be possible in the presence of a tachyonic instability, since neighboring solutions turn away from flat space. Four our model of pregeometry we find suitable ranges of parameters for which no such instabilities occur and cosmology indeed approaches flat space at late times. For the same parameters one also finds generically an approach of the early universe to an approximate de Sitter solution, realizing an inflationary epoch. Our finding of rather consistent overall cosmologies demonstrates that from this side there seems to be no obstruction to formulate quantum gravity as a model of pregeometry.

The ingredients of our model of pregeometry are the gauge bosons of the group $SO(1,3)$, a vector field in the vector representation of $SO(1,3)$, which can be associated with a generalized vierbein, and a scalar singlet field. The formulation of a model for gravity involving a generalized vierbein brings several aspects close to Cartan's geometry \cite{CAR, SCI, KIB, HH, RP, PS, HCM, DR2, HR, DR1, HEI, UTI, SHA, SSTZ, Z_o_nik_2018}. Nevertheless, in our approach the covariant derivative of the generalized vierbein does not vanish. The gauge fields are independent dynamical fields which are not fixed as functionals of the vierbein. In this respect our model is similar to refs. \cite{Nair_2009, Nikiforova_2009, HayashiGrav, SNgf}, see ref.~\cite{CWPG} for more detailed relations.

Our model involves as a geometric connection the Levi-Civita connection constructed from the vierbein, together with the associated spin connection. This part is related to the bundle structure of tangent space and reflects the same geometric structure as for Cartan's geometry. In addition, it also involves the gauge fields as an independent connection related to the bundle structure of a Yang Mills gauge theory. An important ingredient is the covariant derivative of the vierbein which involves both connections \cite{CWGG, CWFSI}, see also refs. \cite{Nair_2009, Nikiforova_2009, HayashiGrav, SNgf, CWGG, Percacci:1984ai, MMOY, DTQG, VDPT, KPG}. In the limit of a vanishing covariant derivative one recovers many features of Cartan's geometric approach. The only independent geometric degree of freedom is then given by the metric and cosmology is the same as for Einstein gravity. In contrast, for our approach the vierbein and the gauge fields contain propagating degrees of freedom that are not related to the metric. These degrees of freedom guarantee the simple short distance behavior.

On the other hand, the additional degrees of freedom are the potential source of instabilities. For an appropriate range of parameters we will find that instabilities are avoided. In this case the additional degrees of freedom become effectively frozen for late cosmology. For late cosmology the metric and a scalar field related to dynamical dark energy are the central ingredients for the overall evolution of the universe. They can be accompanied by rapidly oscillating massive gauge fields with zero mean value, that could constitute dark matter. In this limit one recovers general relativity or, more generally, variable gravity \cite{CWVG}. This limit guarantees compatibility with the many precise local measurements of gravity.

Beyond the gauge fields and the generalized vierbein discussed in ref. \cite{CWPG} we include here a scalar singlet field $\chi$. This is motivated by the possibility of a simple realization of quantum scale symmetry \cite{CWQS}. Quantum scale symmetry is a key ingredient for any complete quantum field theory. The possibility to extrapolate consistently to arbitrarily short distances requires the presence of an ultraviolet fixed point, either in the form of asymptotic freedom of asymptotic safety. At the fixed point, quantum scale symmetry is an exact symmetry. A complete candidate for a simple form of the microscopic action should therefore also permit a simple realization of quantum scale symmetry.

In the presence of a scalar field running dimensionless couplings become functions of $\chi/k$, with $k$ some renormalization scale, typically given by some effective infrared cutoff. As a consequence, key properties of the renormalization flow with $k$ are mapped to a field-dependence on $\chi$. For cosmological solutions with a time variation of $\chi$ this typically induces "crossover cosmologies". Due to an increase of $\chi$ the universe can evolve away from the vicinity of an ultraviolet fixed point, which only characterizes the infinite past. For a simple ansatz for the $\chi$-dependent couplings our model of pregeometry is indeed characterized by a crossover cosmology. This leads rather genuinely to an early inflationary epoch, which ends due to the crossover.

\begin{figure}[h]
\centering
\includegraphics[width=0.5\textwidth]{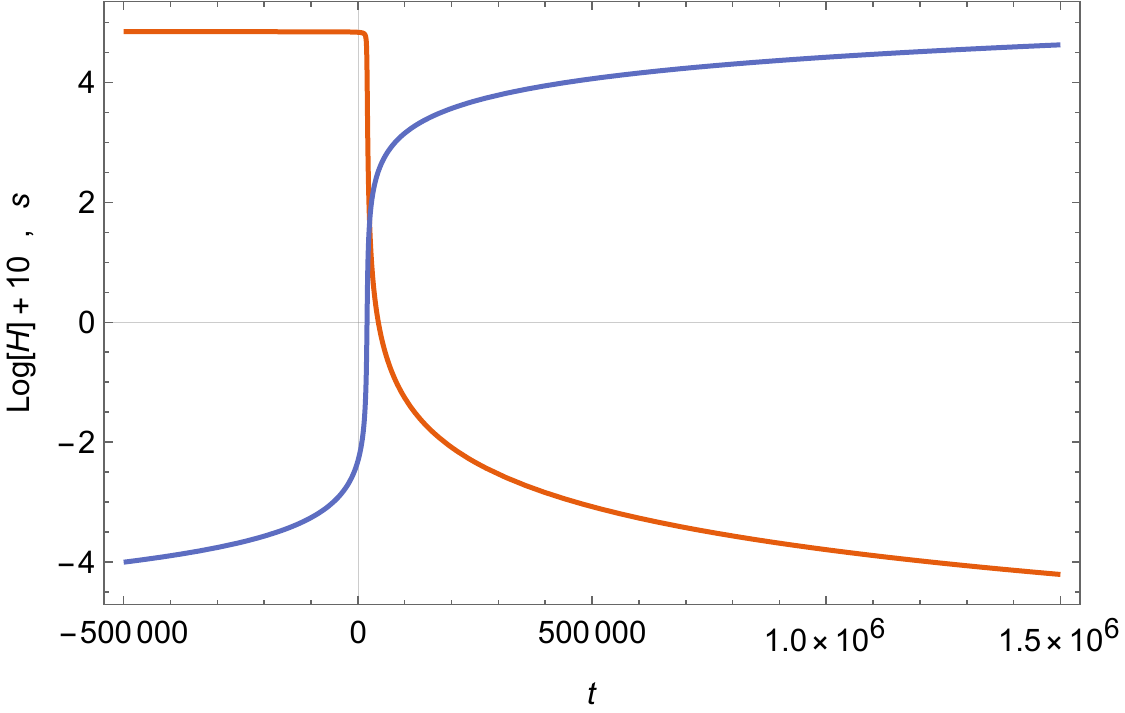}
\caption{\emph{Hubble parameter and scalar field}. We show the Hubble parameter $H$ and the scalar field $\chi$ as functions of cosmic time $t$. More precisely, we plot $\ln(H)+10$ (red) for better visibility, and $s=\ln(\chi)$ (blue). Units are set by the Planck mass. Parameters are $Z=-\Ztil=1$, $\m=-\mtil=5$, $Y=0$, $K=8$, $\Vbar=10^{-4}$. We observe an early inflationary epoch which can be continued to the infinite past. Once $s$ becomes positive inflation ends and the Hubble parameter decreases towards zero.}
\label{fig:A}
\end{figure}

As the main outcome of this investigation we find that realistic cosmology is possible for this type of models of pregeometry. This is demonstrated in Fig. \ref{fig:A} which shows the evolution of the Hubble parameter $H$ and the scalar field $s=\ln\chi$ with cosmic time $t$. An early inflationary epoch with almost constant $H$ and slow evolution of $s$ is followed by a transition to "late cosmology" for which $H$ decreases towards zero while $s$ continues growing. More details for this figure and a discussion of the properties of late cosmology will be provided later. Late cosmology is essentially general relativity with dynamical dark energy and dark matter.

We do not discuss in the main text the fermion fields which can be added in a standard way \cite{CWPG}. We also omit the additional gauge bosons and scalars needed for the standard model of particle physics or beyond. These fields have to be added for a fully realistic cosmology. In appendix~\ref{app:A*} we present a short description of the effective action for these additional fields, describing in particular their couplings to the fields of pregeometry. The additional fields are central for a realistic radiation dominated epoch and for the galaxies and stars in our present epoch. Nevertheless, for very early cosmology including the inflationary epoch these additional fields presumably play no important role. For the early epochs the fields included in the present paper may be sufficient for a realistic description.

For the present cosmological epoch and length or time scales much smaller than the present horizon or age of the universe all predictions of our model turn out to be identical to general relativity, up to tiny corrections. These corrections involve either processes mediated by very heavy fields, as the massive gauge bosons. Those are typically suppressed by factors $p^2/\m$, with $p$ the momentum, energy or inverse length or time of the considered process, and $\m$ typically larger than the effective Planck mass. For the physics of stars or galaxies these effects are unobservably small. A second source of new effects could arise from the time evolution of the scalar singlet field $\chi$. Neglecting the effects $\sim m^{-2}$ exact general relativity is recovered for constant values of the scalar singlet field $\chi$. Since the relative change of $\chi$ for late cosmology turns out to be proportional to the Hubble parameter all possible corrections for gravity on stars or galaxies are tiny. Furthermore, in the limit of quantum scale symmetry a time variation of couplings or apparent violations of the equivalence principle are absent. We discuss the emergence of general relativity for late cosmology in detail in sect.~\ref{section:EVG}.

First proposals for pregeometry have been based on fermionic constituents \cite{AK, AV, DS}. These models did not yet implement local Lorentz symmetry. In spinor gravity both diffeomorphism and local Lorentz symmetry can be realized \cite{CWTS, CWSG1, CWSG2}, see also refs. \cite{CWSGA, CWHEB} for models with global Lorentz symmetry. It is conceivable that the present model of pregeometry can be obtained as an effective theory from such an even more fundamental fermionic theory. This question remains beyond the present investigation.

The present paper is organized as follows: We introduce our model of pregeometry in sect. \ref{section:FEA}. In sect. \ref{section:FEHC} we derive the field equations that are relevant for homogeneous isotropic cosmological solutions. We establish the stability of flat space, which is discussed in more detail in appendix \ref{app:A}. Numerical solutions approach rather fast a constant Hubble parameter and a vanishing of the covariant derivative of the vierbein for a wide range of initial conditions. This early attractor solution dominates the inflationary epoch. In sect. \ref{section:CS} we discuss two types of de Sitter solutions. The first is the stable attractor solution towards which rather arbitrary initial conditions converge as long as the value of the scalar field $\chi$ remains small enough. The early attractor ends once the slowly evolving scalar field reaches a critical value. The second type of de Sitter solutions is unstable. It corresponds to the limit of the basin of attraction of the de Sitter solution of the first type. Both types of de Sitter solutions become exact in the limit of a vanishing scalar field.

In sect. \ref{section:EVG} we turn to the emergence of an effective low energy theory. This is variable gravity, an extension of general relativity which includes a scalar field, with a field dependent Planck mass, effective scalar potential and kinetic term. Also the masses of particles in the standard model will depend on this scalar field. The embedding of variable gravity in our model of pregeometry is not trivial since the covariant derivative of the vierbein does not vanish for a changing scalar field. Variable gravity becomes a very good approximation rather early in the cosmological evolution. Many results for rather realistic cosmologies in this context \cite{CWVG, CWIQM, RUCW, HMSS, HMSS2} can be taken over to our model. This simplifies the overall cosmological picture for our model of pregeometry. The additional degrees of freedom play a role in very early cosmology, which may be associated to the early stages of the inflationary epoch. Their average value is damped very rapidly to tiny values. Already for the epoch of inflation relevant for the observable primordial fluctuation spectrum, and for all later times, they only appear possibly in the form of rapidly oscillating fields that behave as non-relativistic fluids. If stable, these fluctuating fields could constitute dark matter.

Sects. \ref{section:GFE}-\ref{section:QS} expand the formal and conceptual setting. We first derive in sect. \ref{section:GFE} the general field equations. In sect. \ref{section:WS} we discuss field dependent conformal transformations of the metric or Weyl scalings. These transformations relate different choices of the metric or different "metric frames". The corresponding geometries are different, but all equivalent from the point of view of observation. We cast the effective action and the field equations into a frame invariant form. In sect. \ref{section:QS} we discuss quantum scale symmetry \cite{CWQS}. This symmetry plays a crucial role in the infinite past and infinite future of the universe, which are described by an ultraviolet and infrared fixed point, respectively. Quantum scale symmetry explains the almost scale invariant primordial fluctuation spectrum by the close vicinity to the ultraviolet fixed point. For late cosmology the vicinity of the infrared fixed point, where quantum scale symmetry is broken spontaneously, leads to a very light scalar field playing the role of a pseudo-Goldstone boson. This "cosmon"-field is the key ingredient for dynamical dark energy or quintessence \cite{Wetterich_1988}.

In sect. \ref{section:CM} we formulate the effective action for a family of models which are motivated by the functional renormalization flow from the ultraviolet to the infrared fixed point. We discuss in the following very simple representatives with frame- and scale-invariant couplings. While being an oversimplification for a quantitatively realistic universe, these simple models account for all main ingredients of a realistic cosmology. Sect. \ref{section:DMDE} turns to the consequences of our model of pregeometry for the late universe. We discuss rapidly oscillating gauge fields as dark matter candidates, show the natural emergence of dynamical dark energy associated to a very light scalar field with decreasing mass, and investigate the coupling between dark energy and dark matter \cite{CWCNC, CWCMAV, LACQ}. In sect. \ref{section:IC} we turn to inflationary cosmology. We establish that the epochs relevant for the observable properties of the primordial fluctuation spectrum are well approximated by variable gravity. We depict scenarios for which the small fluctuation amplitude as well as realistic spectral properties emerge in a natural way.

In sect. \ref{section:GF} we turn to the fate of inhomogeneous fluctuations in the vicinity of the homogeneous cosmic background solution. We concentrate on the graviton fluctuations. Our model contains two fields with the transformation properties of the graviton, one arising from the vierbein, the other from the gauge fields. The propagators of the two fields mix, resulting in flat space in a massless and a massive graviton. This mixing also leads to two different components for the primordial tensor fluctuations. For the early attractor solution the leading component behaves precisely as the graviton fluctuation in Einstein gravity.

Sect. \ref{section:BOU} addresses the beginning of the universe in the infinite past in physical time. For our model of pregeometry this turns out to be surprisingly simple. It is a vacuum state for which the expectation values of all fields vanish. The inhomogeneous fluctuations of the fields do not vanish, however. They are characterized by time-independent correlation functions. This vacuum state is unstable towards the growth of small vacuum expectation values of fields. Once the expectation values dominate, the universe becomes homogeneous. Relative inhomogeneities are strongly damped during the subsequent inflationary epoch. Finally, in sect. \ref{section:C} we summarize our conclusions.

\section{Fields and effective action}
\label{section:FEA}

Our formulation of pregeometry is based on a Yang-Mills gauge theory with non-compact gauge group $SO(1,3)$ and gauge fields $\A = -A_{\mu nm},\ m,n=0,1,2,3$. In addition to the six gauge fields we consider four vector fields $\e$ in the vector representation of $SO(1,3)$. We define the covariant field strength
\be
\label{eq:1}
F_{\mu\nu mn} = \partial_{\mu}A_{\nu mn} - \partial_{\nu}A_{\mu mn} + A_{\mu m}^{\trip p}A_{\nu pn} - A_{\nu m}^{\trip p}A_{\mu pn}\ ,
\ee
and the covariant derivative
\be
\label{eq:2}
U_{\mu\nu}^{\trip m} = D_{\mu}e_{\nu}^{\ m} = \partial_{\mu}e_{\nu}^{\ m} - \Gamma_{\mu\nu}^{\trip \sigma}e_{\sigma}^{\ m} + A_{\mu\dub n}^{\ m}e_{\nu}^{\ n}\ .
\ee
The vectors $\e$ play the role of a generalized vierbein for which the covariant derivative does not vanish, in contrast to most approaches based on Cartan's geometry. The vierbein is used to convert latin (gauge) indices to greek (world) indices,
\be
\label{eq:3}
F_{\mu\nu\rho\sigma} = e_{\rho}^{\ m}e_{\sigma}^{\ n}F_{\mu\nu mn}\ ,\ U_{\mu\nu\rho} = e_{\rho m}U_{\mu\nu}^{\trip m}\ .
\ee
The covariant vierbein derivative can be related to the torsion tensor
\bel{3A}
U_{\mu\nu\rho}=\frac{1}{2}\gl T_{\mu\nu\rho}-T_{\nu\rho\mu}+T_{\rho\nu\mu}\gr
\ee

We assume a positive determinant of the matrix $\e$ such that the inverse vierbein $\ei$ exists and can be used for the opposite index conversion
\be
\label{eq:4}
e = \text{det}(\e) > 0\ ,\ e_{m}^{\ \mu}e_{\mu}^{\ n} = \delta_{m}^{\ n}\ ,\ e_{\mu}^{\ m}e_{m}^{\ \nu} = \delta_{\mu}^{\ \nu}\ .
\ee
We define a composite metric $g_{\mu\nu}$ as a bilinear of the vierbein and its inverse
\be
\label{eq:5}
g_{\mu\nu} = \e e_{\nu}^{\ n}\eta_{mn}\ ,\ g^{\mu\nu} = \ei e_{n}^{\ \nu}\eta^{mn}\ .
\ee
It can be employed for raising and lowering greek indices in the usual way. Latin indices are raised and lowered with $\eta_{mn}$ and $\eta^{mn}$, $\eta = \text{diag}(-1,1,1,1)$. The connection $\Gamma_{\mu\nu}^{\trip \rho}$ is the Levi-Civita connection constructed from the composite metric. It is therefore a function of the vierbein and its first derivatives. At the present stage $g_{\mu\nu}$ is only a shorthand notation for a vierbein-bilinear. For the effective low energy theory this composite field will become the dominant degree of freedom, together with the scalar field $\chi$. It can then be used to define a useful, albeit not unique, geometry.

In addition to the "pregeometric fields" $\A$ and $\e$ we introduce a scalar field $\chi$. It will play the role of the inflaton or the cosmon for dynamical dark energy. The evolution of this scalar field will be the crucial ingredient which drives the transition between different cosmological epochs. It realizes the crossover from the ultraviolet fixed point at $\chi=0$ in the infinite past to the infrared fixed point for $\chi\to\infty$ in the infinite future.

\zwisch{Effective action}

A quantum effective action which is invariant under $SO(1,3)$-gauge transformations and diffeomorphisms takes in second order of a derivative expansion the form
\begin{align}
\label{eq:6}
\Gamma = \int_{x} e \bigg{\{} & \frac{Z}{8}F_{\mu\nu\rho\sigma}F^{\mu\nu\rho\sigma} +\frac{B}{2}F_{\mu\nu}F^{\mu\nu }+\frac{C}{2}F^{2}\nn\\
&- \frac{M^{2}}{2}F + \frac{m^{2}}{4}U_{\mu\nu\rho}U^{\mu\nu\rho} +\frac{\n}{2}U_{\mu\ \rho}^{\ \mu}U_{\nu}^{\ \nu\rho}\nonumber \\
&+ \frac{K}{2}\partial^{\mu}\chi\partial_{\mu}\chi + V + YU_{\mu}^{\ \mu\nu}\chi\partial_{\nu}\chi  \bigg{\}}\ ,
\end{align}
where
\be
\label{eq:7}
F_{\mu\rho}=F_{\mu\nu\rho\sigma}g^{\nu\sigma}\ ,\quad F = F_{\mu\nu\rho\sigma}g^{\mu\rho}g^{\nu\sigma}=F_{\mu\rho}g^{\mu\rho}\ .
\ee
The coupling functions $Z,\ B,\ C,\ m^{2},\ \n,\ M^{2},\ K,\ Y,\ \text{and}\ V$ are functions of the scalar field $\chi$.

The effective action includes all effects of quantum fluctuations. The field equations derived by variation of $\Gamma$ are exact. These are the field equations relevant for cosmology. We have not computed $\Gamma$ but rather made the ansatz of the validity of a derivative expansion. The action \eqref{eq:6} contains invariants in second oder in the derivative expansion which do not involve contractions with the totally antisymmetric tensor $\varepsilon^{\mu\nu\rho\sigma}$ or $\varepsilon^{mnpq}$. The omission of such terms can be justified by imposing discrete symmetries as parity and time reversal and by noting that $F_{\mu\nu mn}F_{\rho\sigma pq}\varepsilon^{\mu\nu\rho\sigma}\varepsilon^{mnpq}$ is a total derivative. There exist also other index contractions not involving the $\varepsilon$-tensor. Since these invariants do not seem to induce qualitative changes we have omitted them for the sake of simplicity. The effective action is therefore not in the most general form in second order in a derivative expansion. We also often omit the term $\sim B$. Again, this makes no qualitative difference. Our discussion of the invariant $\sim B$ can be considered as a proxy for other omitted invariants. The remaining invariants are all needed for an understanding of the structural properties of our model.

For $V=0$ Minkowski space with vanishing gauge fields and constant scalar field is a solution of the field equations derived from the effective action \eqref{eq:6}. The coupling functions should be chosen such that Minkowski space is stable. A detailed stability analysis is presented in appendix \ref{app:A}. We define the combinations
\bel{7A}
\Ztil = Z+4B+12C\ ,\quad \mtil=\m+3\n\ .
\ee
For constant coupling functions stability of Minkowski space can be realized for
\bel{7B}
Z>0\ ,\trip Z+2B>0\ ,\trip \Ztil<0\ ,\trip \m>0\ ,\trip \M>0\ ,
\ee
and either
\bel{9A}
\mtil>0\ ,\quad K>3Y^{2}\chi_{0}^{2}/\mtil\ ,
\ee
or for negative $\mtil$ obeying
\bel{9B}
\mtil<-2\M+\frac{3Y^{2}\chi_{0}^{2}}{K}\ .
\ee
The negative sign of $\Ztil$ is connected to the non-compact character of the gauge group $SO(1,3)$. The conditions \eqref{7B}-\eqref{9B} are extended later to $\chi$-dependent couplings.

\zwisch{Simple models}

For a first step beyond constant coupling functions we make an ansatz for the $\chi$-dependence which is motivated by quantum scale symmetry, fundamental scale invariance and asymptotic safety for gravity. Typical scaling solutions for functional flow equations for quantum gravity yield an approximate behavior \cite{CWQS, HPRW, HPW, PRWY, CWMY, CWESPA}
\be
\label{S7}
V = u_{0}k^{4}\ ,\ M^{2} = 2w_{0}k^{2} + \xi\chi^{2}\ .
\ee
We will asume that this extends to pregeometry, with a similar behavior for $\m$ and $\mtil$,
\be
\label{S8}
m^{2} = \m_{0}k^{2} + \zeta\chi^{2}\ ,\quad \mtil=\mtil_{0}k^{2}+\zetil\chi^{2}\ .
\ee
We take constant values of $Z$ and $\Ztil$. Furthermore, the dimensionless functions $K$ and $Y$ are assumed to depend only on $\chi^{2}/k^{2}$, in a form that will be discussed later. With dimensionless parameters $u_{0}$, $w_{0}$, $\xi$, $\m_{0}$, $\tilde{m}_{0}^{2}$, $\zeta$, $\zetil$ the renormalization scale $k$ is the only mass scale present. Quantum scale symmetry is realized in the limits $\chi^{2}\to\infty$ and $\chi\to 0$, as we will discuss in detail in section \ref{section:QS}. The $\chi$-dependence of the functions $Z$, $\Ztil$, $m^{2}$, $\mtil$, $M^{2}$ and $V$ according to the ansatz \eqref{S7}, \eqref{S8} is not exact, we only suppose here that it is a reasonable approximation.

We will assume the presence of a discrete symmetry $\chi\to -\chi$. For analytic $Z$ this implies for $\chi\to 0$ that the derivative vanishes, $\partial Z/\partial\chi\sim\chi$. We assume that analyticity holds for $Z$, $\Ztil$, $\m$, $\mtil$, $\M$ and $V$. For $\chi=0$ all terms in the field equations that are proportional to $\chi$-derivatives of these functions vanish. For $K$ and $Y$ we assume that they do not increase too fast for $\chi\to 0$, such that the scalar field equation is obeyed for $\chi=0$. We will discuss these functions more extensively in sect. \ref{section:CM}, where we embed our ansatz in a more complete crossover model.

\section{Field equations for homogeneous isotropic cosmology}
\label{section:FEHC}

In this section we derive the evolution equations for homogeneous isotropic cosmologies. We discuss the approach to flat space and the deviations of our model of pregeometry from general relativity. We also present numerical solutions for the early universe that show the natural emergence of an inflationary epoch.

\zwisch{Homogeneous isotropic cosmology}

We are first interested in homogeneous, isotropic and spatially flat cosmological solutions of the field equations derived from the effective action \eqref{eq:6}. The general ansatz is given for $\e$ and $\A$ by
\begin{align}
\label{eq:8}
\e &= a(\eta)\delta_{\mu}^{\ m}\ ,\ \chi=\chi(\eta)\ ,\nonumber \\
\ A_{kl0} &= -A_{k0l} = b(\eta)\delta_{kl}\ ,\ A_{klj} = c(\eta)\eps_{klj}\ ,
\end{align}
with $k,l,j = 1\dots3$. It involves four functions of conformal time $\eta$. We stress that the ansatz concerns the gauge fields $\A$. For $A_{\mu\nu\rho}$ an additional factor $a^{2}$ or $-a^{2}$ arises from the multiplication with vierbeins.

With this ansatz one finds for the non-vanishing components of $U_{\mu\nu\rho}$
\be
\label{eq:9}
U_{kl0} = -U_{k0l} = -a^{2}(b-\mathscr{H})\delta_{kl}\ ,\ U_{klj} = -a^{2}c\ \eps_{klj}\ ,
\ee
involving the conformal Hubble parameter
\be
\label{eq:10}
\mathscr{H} = \deta\text{ln}a = a^{-1}\frac{da}{d\eta}\ .
\ee
Similarly, the field strength $F_{\mu\nu mn}$ for the gauge bosons has the non-vanishing components
\begin{align}
\label{eq:11}
F_{klij} &= (b^{2}-c^{2})(\delta_{ki}\delta_{lj} - \delta_{kj}\delta_{li})\ ,\ F_{0klj} = \partial_{\eta}c\ \eps_{klj}\ , \nonumber \\
F_{klj0} &= -2cb\ \eps_{klj}\ ,\ F_{k0l0} = -\partial_{\eta}b\ \delta_{kl}\ ,
\end{align}
plus additional components obtained by the asymmetry in the first and last index pair,
\be
\label{eq:12}
F_{\mu\nu mn} = -F_{\mu\nu nm} = -F_{\nu\mu mn}\ .
\ee
This yields for the contracted scalar
\bel{16A}
F=\frac{6}{a^{2}}\big{(}\deta b+b^{2}-c^{2}\big{)}\ .
\ee

\zwisch{Field equations}

In order to derive the field equations for the functions $a(\eta),b(\eta),c(\eta)$ and $\chi(\eta)$ we insert the ansatz \eqref{eq:8} into the effective action \eqref{eq:6}
\begin{align}
\label{eq:13}
\Gamma = \Omega_{3}\int_{\eta} \bigg{\{} & \frac{3}{2}\big{[}\Ztil(\deta b)^{2}-Z(\deta c)^{2}\\
&+2(\Ztil -Z-2B)(b^{2}-c^{2})\deta b\nonumber \\
&+\Ztil(b^{4}+c^{4})-2(\Ztil+2Z)b^{2}c^{2}\big{]}\nonumber \\
&+\frac{3a^{2}}{2}\big{[}\m c^{2}-\mtil(b-\hub)^{2}\big{]}\nn\\
&- 3M^{2}a^{2}(\partial_{\eta}b + b^{2} - c^{2})\nn\\
&- \frac{Ka^{2}}{2}(\deta\chi)^{2} + a^{4}V +3Ya^{2}(b-\hub)\chi\deta\chi \bigg{\}}\nn\ .
\end{align}
Only even powers of $c$ appear. This is due to parity which changes the signs of all fields with an odd number of indices $\mu$ or $m$ taking the values $1,2,3$ and therefore transforms $c\to -c$.

The field equation for the "cosmic scale factor" $a(\eta)$ obtains by functional variation of the effective action \eqref{eq:13} with respect to the function $a(\eta)$,
\begin{align}
\label{eq:14}
\deta&\mathscr{H} + \mathscr{H}^{2} = \bigg{(}1 + \frac{2M^{2}}{\mtil}\bigg{)}(\partial_{\eta}b + b^{2}) -\bigg{(}\frac{\m+2\M}{\mtil}\bigg{)}c^{2}\nn\\
&-\frac{4a^{2}}{3\mtil}V+ \frac{K}{3\mtil}(\deta\chi)^{2} + \frac{\partial\text{ln}(\mtil)}{\partial\chi}\deta\chi(b-\mathscr{H}) \\
&-\frac{Y}{\mtil}\big{[}\chi\deta^{2}\chi+\bigg{(}1+\frac{\partial\ln Y}{\partial\ln\chi}\bigg{)}(\deta\chi)^{2}+2b\chi\deta\chi\big{]} \nn \ .
\end{align}
Similarly, the field equation for the scalar field $\chi(\eta)$ reads
\begin{align}
\label{eq:15}
K&\big{[}\deta^{2}\chi + 2\mathscr{H}\deta\chi + \frac{1}{2}\frac{\partial\text{ln}K}{\partial\chi}(\deta\chi)^{2}\big{]} = -a^{2}\frac{\partial V}{\partial\chi} \nonumber \\
&-\frac{3}{2a^{2}}\bigg{[}\frac{\partial \Ztil}{\partial\chi}\big{(}\deta b+b^{2}-c^{2}\big{)}^{2}-4\dchi{B}(b^{2}-c^{2})\deta b \nonumber \\
&-\dchi{Z}\big{[}(\deta c)^{2}+2(b^{2}-c^{2})\deta b+4b^{2}c^{2}\big{]}\bigg{]}\nn\\
&- \frac{3}{2}\frac{\partial m^{2}}{\partial\chi}c^{2} +\frac{3}{2}\dchi{\mtil}(b-\hub)^{2} + 3\frac{\partial M^{2}}{\partial\chi}(\partial_{\eta}b + b^{2} - c^{2})\nonumber \\
&+3Y\big{(}\deta b-\deta\hub+2\hub(b-\hub)\big{)}\chi \ .
\end{align}

Finally, we obtain the field equations for the two independent functions characterizing the gauge fields,
\begin{align}
\label{eq:16}
\deta^{2}b &+ \frac{\partial\text{ln}\Ztil}{\partial\chi}\deta\chi\partial_{\eta}b = - \frac{\mtil + 2M^{2}}{\Ztil}a^{2}(b-\mathscr{H}) \nonumber \\
&+2b(b^{2}-c^{2})-4\frac{Z}{\Ztil}bc^{2}+\frac{2}{\Ztil}\big{(}\Ztil-Z-2B\big{)}c\deta c\nn\\
&+ \frac{a^{2}}{\Ztil}\bigg{(}\frac{\partial M^{2}}{\partial\chi}+Y\chi\bigg{)}\deta\chi\nn \\
&-\frac{1}{\Ztil}\partial_{\chi}\big{(}\Ztil-Z-2B\big{)}\deta\chi(b^{2}-c^{2})\ ,
\end{align}
and
\begin{align}
\label{eq:17}
\deta^{2}c& + \frac{\partial\text{ln}Z}{\partial\chi}\deta\chi\partial_{\eta}c = - \frac{m^{2} + 2M^{2}}{Z}a^{2}c\\
&+\frac{2c}{Z}\big{[}\Ztil(b^{2}-c^{2}+\deta b)+Z(2b^{2}-\deta b)-2B\deta b\big{]}\nn\ .
\end{align}
Eqs. \eqref{eq:14}-\eqref{eq:17} constitute a system of four non-linear second order differential equations for $a,b,c\ \text{and}\ \chi$. In sect. \ref{section:GFE} we will obtain the same equations from the more general field equations for arbitrary fields.

\zwisch{Stability of flat space}

In the limit of a vanishing potential $V=0$, and for constant coupling functions as well as $Y=0$, the field equations \eqref{eq:14}-\eqref{eq:17} have solutions with Minkowski geometry, vanishing gauge fields, and constant scalar field
\bel{MS1}
\hub=b=c=0\ ,\quad \chi=\chi_{0}\ .
\ee
In the vicinity of this solution we can expand the field equations in $\hub$, $b$, $c$ and $\delta\chi=\chi-\chi_{0}$. Using cosmic time $t$ with $dt/d\eta=a$, $H=\hub/a$, $\btil=b/a$, $\ctil=c/a$, the linearized field equations read
\begin{align}
\label{MS2}
\dt H&=\Big{(}1+\frac{2\M}{\mtil}\Big{)}\dt\btil\ ,\quad \dt^{2}\delta\chi=0\ ,\\
\dt^{2}\btil&=-\frac{\mtil+2\M}{\Ztil}(\btil-H)\ ,\quad \dt^{2}\ctil=-\frac{\m+2\M}{Z}c\nn\ .
\end{align}
Inserting the linearized solution,
\bel{MS3}
H=\Big{(}1+\frac{2\M}{\mtil}\Big{)}\btil\ ,
\ee
into the evolution equation for $\btil$ yields
\bel{MS4}
\dt^{2}\btil=\frac{2\M}{\Ztil}\Big{(}1+\frac{2\M}{\mtil}\Big{)}\btil\ .
\ee
For $\M>0$, $1+2\M/\mtil>0$ one finds oscillations with a positive mass term $\mu^{2}$, $(\dt^{2}+\mu^{2})\btil=0$, provided that $\Ztil<0$. In contrast, for $\Ztil>0$ Minkowski space is unstable due to $\btil$ behaving as a tachyon. A second possibility for stable behavior is $\M>0$, $1+2\M/\mtil<0$, $\Ztil>0$. This agrees with the general stability discussion in appendix \ref{app:A}. For $\ctil$ we obtain a stable behavior for $\m+2\M>0$, $Z>0$.
The scalar fluctuation $\delta\chi$ behaves in this approximation as a massless field.

The stability properties can be understood in a simple way by replacing $H$ by a shifted variable $\dtil$,
\bel{MS5}
H=\dtil+(1+2\ytil)\btil\ ,\quad \ytil=\frac{\M}{\mtil}\ .
\ee
The field equations for $\btil$ and $\dtil$ become
\be
\label{MS6}
\dt^{2}\btil-\frac{2\ytil(1+2\ytil)\mtil}{\Ztil}\btil=\frac{\mtil}{\Ztil}(1+2\ytil)\dtil\ ,\quad
\dt\dtil=0\ .
\ee
For $\dtil=0$ we recover the stable oscillations of $\btil$ in the appropriate range of parameters.

Similar results will be found for $\chi$-dependent coupling functions. For the crossover model in sect. \ref{section:CM} the constant coupling functions will be replaced by constant scale invariant coupling functions. Only the frame invariant scalar potential will depend on $\chi$ in a way that vanishes for $\chi\to\infty$. Late cosmology will correspond to large $\chi$, such that the approximation $V=0$ applies. The stability discussion for flat space above therefore applies directly to late cosmology.

\zwisch{Evolution equations in cosmic time}

The use of cosmic time $t$ eliminates the explicit dependence on the scale factor $a$. We also replace $\mtil$ and $\m$ by the mass terms
\bel{MS7}
\mub=-\frac{2\M}{\Ztil}\Big{(}1+\frac{2\M}{\mtil}\Big{)}\ ,\quad \muc=\frac{\m+2\M}{Z}\ .
\ee
For early cosmology with $\chi\to 0$ the coupling functions can be taken constant, leading to constant mass terms $\mub$ and $\muc$. The same holds for late cosmology with $\chi\to\infty$ according to our ansatz \eqref{S7}, \eqref{S8} if we use frame invariant coupling functions. Thus the ansatz of constant coupling functions approximates well a wide range of models. In the following we only include a $\chi$-dependence of the scalar potential $V$.

For constant coupling functions (except $V(\chi)$) the field equations for the gauge fields \eqref{eq:16}, \eqref{eq:17} read
\begin{align}
\label{MS8}
(&\dt^{2}+3H\dt+\dt H+2H^{2})\btil = \frac{\mub}{2\ytil}(\btil-H)\nn\\
&+2\btil(\btil^{2}-\ctil^{2})-\frac{4Z}{\Ztil}\ctil^{2}\btil\nn\\
&+\frac{2\ctil}{\Ztil}(\Ztil-Z-2B)(\dt\ctil+H\ctil)+\frac{Y}{\Ztil}\chi\dt\chi\ ,
\end{align}
and
\begin{align}
\label{MS9}
(&\dt^{2}+3H\dt+\dt H+2H^{2}+\muc)\ctil\\
&=\frac{2\ctil}{Z}\big{[}\Ztil(\btil^{2}-\ctil^{2})+2Z\btil^{2}+(\Ztil-Z-2B)(\dt\btil+H\btil)\big{]}\nn\ .
\end{align}
Eq. \eqref{eq:14} becomes
\bel{MS9A}
\dt H+2H^{2}-(1+2\ytil)(\dt\btil+H\btil+\btil^{2})=-\frac{F_{H}}{\mtil}\ ,
\ee
with
\begin{align}
\label{MS10}
F_{H}&=Z\muc\ctil^{2}+\frac{4V}{3}-\frac{K}{3}(\dt\chi)^{2}\\
&+Y\big{[}\chi\dt^{2}\chi+(2\btil+H)\chi\dt\chi+(\dt\chi)^{2}\big{]}\ ,\nn
\end{align}
and the scalar field equation \eqref{eq:15} reads
\bel{MS11}
K(\dt^{2}\chi+3H\dt\chi)=-\dchi{V}+3Y(\dt\btil-\dt H+3\btil H-3H^{2})\chi\ .
\ee

For the potential we assume the form
\bel{33A}
V=V_{0}\Big{(}1+\frac{\chi^{2}}{\mu_{0}^{2}}\Big{)}^{-2}\ .
\ee
This form will be motivated in sects. \ref{section:WS}, \ref{section:CM} by the frame invariant potential or the potential in the Einstein frame. For $\chi^{2}\ll\mu_{0}^{2}$ the potential is almost constant. The decrease for $\chi^{2}\gg\mu_{0}^{2}$ allows for an exit of an inflationary epoch. For otherwise constant coupling functions the solution of the system of flow equations \eqref{MS8}-\eqref{MS11} already reveals several key aspects of the cosmology for our model of pregeometry. According to the ansatz \eqref{S7}, \eqref{S8} constant coupling functions are a valid approximation only for early cosmology when $\chi$ is small enough. We will see in sect. \ref{section:CM} that the region of validity of this approximation can be extended if one uses frame invariant coupling functions.

\vspace{3\baselineskip}
\zwisch{Initial approach to de Sitter solution}

\begin{figure}[h]
\centering
\includegraphics[width=0.5\textwidth]{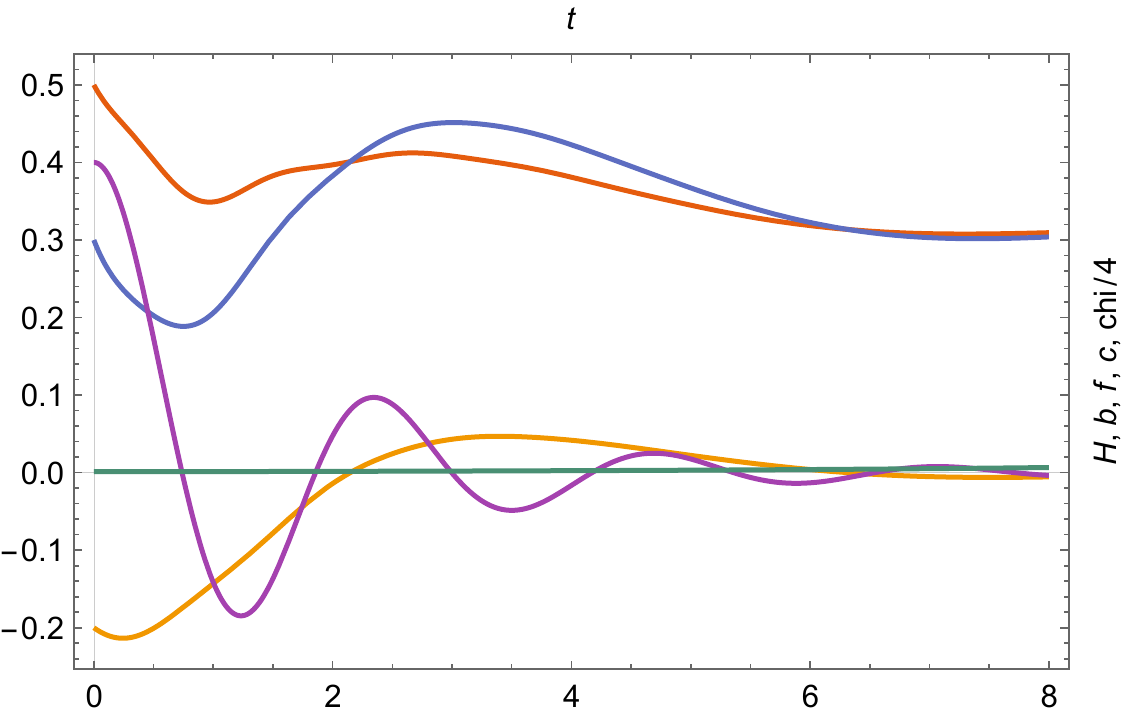}
\caption{\emph{Initial approach to de Sitter solution}. We plot $H$ (red), $\btil$ (blue), $\ftil$ (yellow), $\ctil$ (magenta) and $\chi/4$ (green) as a function of cosmic time $t$. The different curves can also be identified by the initial values $H=0.5$, $\btil=0.3$, $\ftil=-0.2$, $\ctil=0.4$, $\chi=0.005$. Initial derivatives are set to zero, except for $\dt\btil=-0.4$ chosen for better visibility. Parameters are $\M=1$, $\m=-\mtil=5$, $Z=-\Ztil=1$, $B=0$, $K=4$, $V_{0}=0.3$, $\mu_{0}^{2}=1$, $Y=0$.}
\label{fig:1}
\end{figure}

We show in Fig. \ref{fig:1} a typical numerical solution of the system of flow equations \eqref{MS8}-\eqref{MS11}. Parameters are chosen as $\M=1$, $\m=-\mtil=5$, $Z=-\Ztil=1$, $B=0$, $K=4$, $Y=0$, $V_{0}=0.3$, $\mu_{0}^{2}=1$. For the initial conditions at $t=0$ we have taken $\chi(0)=0.005$, $\dt\chi(0)=0$, $\btil(0)=0.3$, $\dt \btil(0)=-0.4$, $H(0)=0.5$, $\ctil(0)=0.4$, $\dt \ctil(0)=0$.
As a function of $t$ the figure displays $H(t)$, $\btil(t)$, $\ftil(t)=\btil(t)-H(t)$, $\ctil(t)$ and $\chi(t)/4$. An initial oscillatory behavior is damped. For the parameters chosen the solution approaches first a de Sitter solution with constant $H$. The difference $\ftil=\btil-H$ and the gauge field $\ctil$ approach zero rapidly already in an early stage and stay there. During the de Sitter solution the scalar field $\chi$ increases only slowly. Starting with a smaller initial value of $\chi$ the initial approach to the de Sitter solution is almost identical. Since the evolution of $\chi$ is even slower, the de Sitter solution extends to a much longer period. In the limit of the initial value of $\chi$ going to zero the de Sitter solution lasts for infinite time.

\zwisch{Exit from de Sitter solution}

\begin{figure}[h]
\centering
\includegraphics[width=0.5\textwidth]{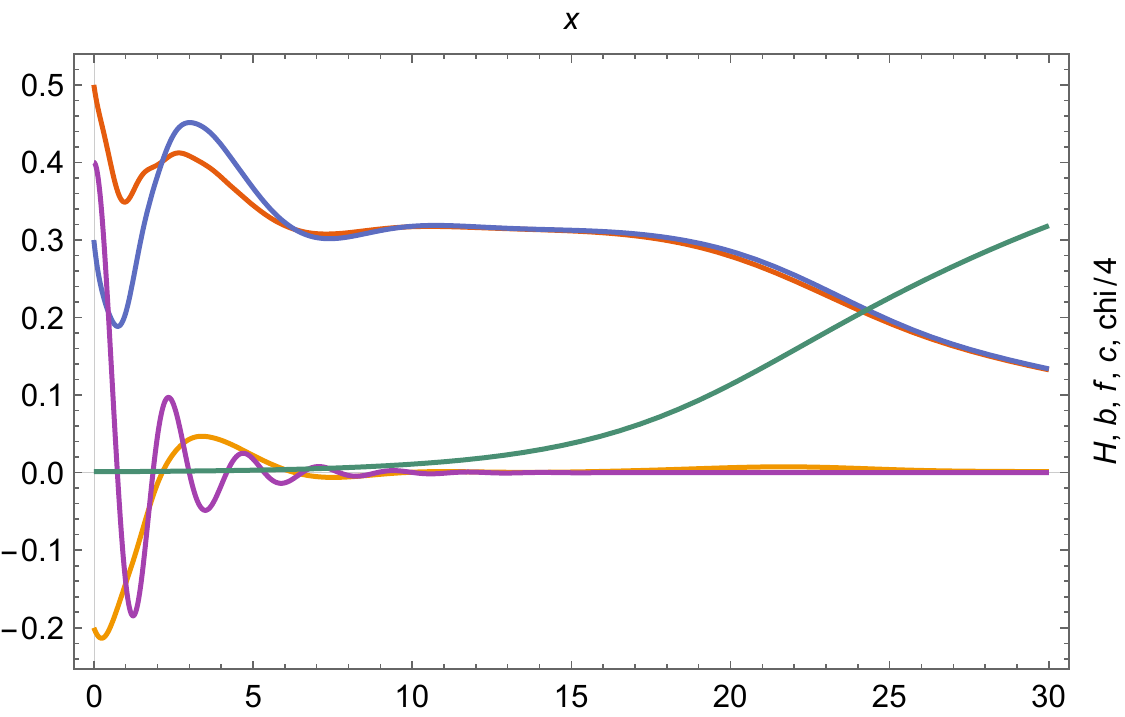}
\caption{\emph{Early cosmology}. The functions $H$ (red), $\btil$ (blue), $\ftil$ (yellow), $\ctil$ (magenta) and $\chi/4$ (green) are shown in dependence on $t$. Parameters, initial values and color coding are the same as for Fig. \ref{fig:1}. The de Sitter solution with almost constant $H$ and $b$ ends once $\chi$ is around $0.5$. Subsequently $H$ and $b$ show a powerlike decay.}
\label{fig:2}
\end{figure}

Even if we start with a tiny initial $\chi>0$ the value of $\chi$ will grow until it reaches a value of the order $0.5$. Subsequently, the derivative of the potential will become important in the scalar field equation \eqref{MS11} and the de Sitter solution ends. We have depicted this in Fig. \ref{fig:2} which uses the same parameters, initial conditions and color coding as in Fig. \ref{fig:1}. This figure only extends the range of $t$. After the end of the de Sitter epoch, which can be associated with inflationary cosmology, the functions $\Htil\approx\btil$ approach zero in a powerlike behavior. Small oscillations of $\Htil$, $\btil$, $\ctil$, $\ftil$ are not visible with the resolution of this plot.

The three stages - damped initial oscillations, de Sitter solution, approach to flat space are similar for a large range or parameters as long as some stability conditions are met, e.g. $Z>0$, $\Ztil<0$, $\m>0$. We have chosen initial conditions such that all three stages are easily visible in Fig. \ref{fig:2}.

The duration of the different stages depends on the parameters. For $V_{0}\to 0$ one observes many damped oscillations, with an oscillatory approach to flat space in the limit $V_{0}=0$. The de Sitter solution can last for many $e$-foldings if the evolution of the scalar field $\chi$ is slow, e.g. for very small initial values of $\chi$. A larger $K$ slows down the evolution of the scalar field.

\zwisch{Approach to flat space}

For $V=0$, $Y=0$ one has a solution with a constant scalar field $\chi=\chi_{0}$. The scalar field no longer influences the evolution of $H$, $\btil$ and $\ctil$ and we obtain an effective model without a scalar field. Eliminating $H$ 
in favor of $\dtil$ by eq. \eqref{MS5}, and concentrating on $\ctil=0$, the field equation for $\btil$ becomes in this case
\begin{align}
\label{MS12}
\dt^{2}&\btil+\big{[}4(1+2\ytil)\btil+3\dtil\ \big{]}\dt\btil+\mub\Big{(}\btil+\frac{\dtil}{2\ytil}\Big{)}+\btil\dt\dtil\nn\\
=&-8\ytil(1+\ytil)\btil^{3}-\big{[}4(1+2\ytil)\btil\dtil+2\dtil^{2}\big{]}\btil\ .
\end{align}
For the approach to flat space we can neglect the r.h.s. of eq. \eqref{MS12} in next to leading order of an expansion in small $\btil$, $\dtil$. We observe the appearance of a damping term $\sim\dt\btil$. For the evolution of $\dtil$ one finds
\be
\label{MS13}
\dt\dtil+2\dtil+3(1+2\ytil)\btil\dtil+2\ytil(1+2\ytil)\btil^{2}=0\ .
\ee
Eqs. \eqref{MS12}, \eqref{MS13} are a closed system of two coupled non-linear differential equations.

The late time behavior can be understood by setting in eq. \eqref{MS12}
\bel{MS14}
\ftil=0\ ,\quad \btil=H\ ,\quad \frac{\dtil}{2\ytil}=-\btil\ ,
\ee
such that the quadratic approximation becomes
\bel{MS15}
\dt^{2}\btil+4\btil\dt\btil=0\ .
\ee
From eq. \eqref{MS13} one infers
\bel{MS17}
\dt\btil=-2\btil^{2}\ .
\ee
One possible solution of eqs. \eqref{MS14}, \eqref{MS15} is a powerlike approach to flat space
\bel{MS18}
H=\btil=\frac{1}{2t}\ ,
\ee
as characteristic for radiation, the other one is $\btil=H=0$ corresponding to flat space. We will see later that for the full system of equations only the flat space solution is possible in the range $H^{2}\ll\M$ if $c=f=0$, $Y=0$, $\chi=\chi_{0}$.

Beyond the approximation \eqref{MS14} there are small oscillations of the gauge field with frequency given by $\mu_{b}$. They influence the evolution of the Hubble parameter as a form of cold dark matter. This will be discussed in sect. \ref{section:DMDE}. Furthermore, the scalar potential for large $\chi$ gets very small, but does not vanish. This will result in a dynamical dark energy component that also influences the evolution of the Hubble parameter. This will also be discussed in sect. \ref{section:DMDE}. The overall picture of Fig. \ref{fig:2} remains valid, however. After the end of the inflationary epoch the universe makes a transition to a cosmology approaching flat space asymptotically.

The approximation of constant coupling functions leads to an instability for late times if $Y\neq0$. This becomes visible if we insert the scalar field equation in eq. \eqref{MS9A}, 
\begin{align}
\label{MS19}
\Big{(}1&-\frac{3Y^{2}\chi^{2}}{K\mtil}\Big{)}(\dt H-\dt\btil)=2\ytil\big{[}\dt\btil+\btil(\btil+H)\big{]}\nn\\
&+(\btil+2H)(\btil-H)\nn\\
&-\frac{1}{\mtil}\bigg{\{}\frac{4V}{3}-\frac{Y\chi}{K}\dchi{V}+\Big{(}Y-\frac{K}{3}\Big{)}(\dt\chi)^{2}+Z\muc\ctil^{2}\nn\\
&\quad\quad+Y(\btil-H)\Big{(}2\chi\dt\chi+\frac{9HY\chi^{2}}{K}\Big{)}\bigg{\}}\ .
\end{align}
As  $\chi$ increases the differential equations become unstable as the prefactor of $\dt H$ approaches zero. For $\mtil$ increasing for large $\chi$ proportional to $\chi^{2}$ this instability can be avoided. Since this instability is absent in the concrete models discussed in sect. \ref{section:CM}, we will not discuss this issue further.

\zwisch{Evolution of covariant vierbein derivative}

The covariant derivative $U_{\mu\nu\rho}$ involves $c$ and the combination
\be
\label{eq:18}
f = b - \mathscr{H}\ .
\ee
We observe in Figs. \ref{fig:1}, \ref{fig:2} that both quantities approach zero already in early stages of the evolution. Since a vanishing covariant derivative of the vierbein leads to variable gravity we conclude that the variable gravity approximation becomes valid already for very early cosmology. It is instructive to understand the dynamical vanishing of the covariant vierbein derivative for $Y=0$ dynamically. This investigation also reveals that for $Y\neq 0$ the covariant vierbein derivative $U_{\mu\nu\rho}$ does not vanish but rather approaches a fixed value proportional to the time-derivative of the scalar field $\chi$. For the purpose of this discussion we take $B=0$ and consider constant $Z$ and $\Ztil$. We observe that $c=0$ solves eq. \eqref{eq:17} and we will in the following concentrate on this solution. 

In terms of $f$ the evolution of the scale factor follows from eq. \eqref{eq:14},
\begin{align}
\label{eq:19}
&6M^{2}(\deta\mathscr{H} + \mathscr{H}^{2}) = 4a^{2}V - K(\deta\chi)^{2} \\
&- 3\mtil(1 + \frac{2M^{2}}{\mtil})(\partial_{\eta}f + 2\mathscr{H}f + f^{2}) - 3\frac{\partial \mtil}{\partial\chi}\deta\chi f\nonumber \\
&+3Y\bigg{[}\chi\deta^{2}\chi+\bigg{(}1+\frac{\partial\ln Y}{\partial\ln\chi}\bigg{)}(\deta\chi)^{2}+2(\hub+f)\chi\deta\chi\bigg{]}\ .\nn
\end{align}
We observe that the curvature scalar formed from the composite metric \eqref{eq:5} is given by
\bel{43A}
R=\frac{6}{a^{2}}\big{(}\hub^{2}+\deta\hub\big{)}\ .
\ee
For $f=0$, $Y=0$ eq. \eqref{eq:19} corresponds to the contraction of the Einstein equation in general relativity coupled to a scalar field.
The scalar field equation simplifies for constant $Z$, $\Ztil$ and $B=0$, as well as $c=0$ to
\begin{align}
\label{eq:20}
K\big{[}\deta^{2}\chi &+ 2\mathscr{H}\deta\chi + \frac{1}{2}\frac{\partial\text{ln}K}{\partial\chi}(\deta\chi)^{2}\big{]} = -a^{2}\frac{\partial V}{\partial\chi} + \frac{3}{2}\frac{\partial \mtil}{\partial\chi}f^{2} \nonumber \\
&+ 3\frac{\partial M^{2}}{\partial\chi}\big{[}\deta\mathscr{H} + \mathscr{H}^{2} + \partial_{\eta}f + 2\mathscr{H}f + f^{2}\big{]} \nonumber \\
&+3Y(\deta f+2\hub f)\chi\ .
\end{align}
For $f=0$ this is the scalar field equation for variable gravity \cite{CWVG}.

The field equation for $f$ follows from the field equation \eqref{eq:16} for $b$. It can be written in the form
\begin{align}
\label{eq:21}
\deta^{2}f &= 2\mathscr{H}^{3} - \deta^{2}\mathscr{H} + \frac{a^{2}}{\Ztil}\bigg{(}\frac{\partial M^{2}}{\partial\chi}+Y\chi\bigg{)}\deta\chi \nonumber \\
&-\frac{\mtil+2\M}{\Ztil}a^{2}f+6\hub^{2}f+6\hub f^{2}+2f^{3}\ .
\end{align}
We observe a driving term that does not vanish for $f=0$. A solution with $f=0$ is possible only if the driving term vanishes. This happens for a de Sitter space where $\deta^{2}\hub=2\hub^{3}$ if $\deta\chi=0$. These conditions are obeyed for the plateaus in Figs. \ref{fig:1}, \ref{fig:2}. This is the reason why $f$ can relax rapidly to zero. For $Y\neq 0$ and constant $\M$ the driving term is of the order $\hub^{3}$ once the de Sitter solution with constant $H$ ends. The driving term, and therefore the size of $f$, is small and the deviation from zero is not visible with the resolution of the Figs. \ref{fig:1}, \ref{fig:2}. We will discuss the evolution of $f$ in more detail in sects. \ref{section:EVG} and \ref{section:DMDE}.

\section{Cosmological solutions with de Sitter geometry}
\label{section:CS}

In this section we discuss a few solutions of the field equations with a de Sitter geometry. With rather mild conditions on the coupling functions they are realized in the limit of a vanishing scalar field $\chi=0$. We will associate this type of solution with a possible beginning of the universe in the infinite past in conformal time, $\infpast$. As time increases, cosmology moves away from the exact de Sitter solutions. The approximate de Sitter solutions realized for large finite negative $\eta$ can be associated with inflation. Once the curvature becomes small as compared to $\m$ and $\mtil$ cosmology can be described by an effective theory. This is the "variable gravity" \cite{CWVG} extension of general relativity. We will establish this in the next section. The stages of the cosmic evolution which are relevant for observation are well described by this effective theory. In sect. \ref{section:IC} we will discuss in more detail the connection between the de Sitter solutions presented in this section and realistic scenarios for inflation. This can be done by showing how variable gravity emerges dynamically from our model of pregeometry. In the following we take for simplicity $B=0$.

\vspace{\baselineskip}
\vspace{\baselineskip}
\vspace{\baselineskip}
\zwisch{De Sitter solutions for constant coupling functions}

The behavior of the solutions of the field equations \eqref{eq:17}\eqref{eq:19}\eqref{eq:20}\eqref{eq:21} depends on the shape of the functions $V$, $M^{2}$, $m^{2}$, $\mtil$, $K$, $Y$, $Z$ and $\Ztil$. Let us first consider the case of constant $Z$, $\Ztil$, $V$ and $M^{2}$. One finds a simple solution with de Sitter geometry and constant scalar field
\be
\label{S1}
\hub = -\frac{1}{\eta}\ ,\ \deta\hub = \hub^{2}\ ,\ \deta^{2}\hub = 2\hub^{3}\ ,
\ee
with
\be
\label{S2}
c = f = 0\ ,\ b = \hub\ ,\ \chi = \chi_{0}\ .
\ee
Eq. \eqref{eq:19} yields
\be
\label{S3}
\frac{6M^{2}}{a^{2}}(\deta\hub + \hub^{2}) = 6M^{2}(\dot{H} + 2H^{2}) = 4V\ ,
\ee
where cosmic time $t$ is related to conformal time $\eta$ by $a\de \eta = \de t$, $H = \dt\ln a = \hub/a$, $\dot{H} = \dt H$. For de Sitter space $H$ is constant, $\dot{H} = 0$, and obeys
\be
\label{S4}
H^{2} = \frac{V}{3M^{2}}\ .
\ee
This is the same result as for Einstein gravity.

We observe that for this solution the gauge field only vanishes in the infinite past in conformal time $\eta\to -\infty$,
\be
\label{S5}
A_{kl0} = \hub\delta_{kl} = -\frac{1}{\eta}\delta_{kl}\ .
\ee
In this limit also the vierbein vanishes 
\be
\label{S6}
\e = -\sqrt{\frac{3M^{2}}{V}}\frac{1}{\eta} \ .
\ee
For this solution the universe starts from vanishing fields. We will discuss the beginning of the universe in more detail in sect. \ref{section:BOU}. In sect. \ref{section:BOU} we also discuss that the same physics can be discussed in a different metric frame for which geometry approaches flat space for $\infpast$.

The exact de Sitter solution remains no longer valid if $V$, $\M$, $Z$ or $\Ztil$ vary with $\chi$. If this variation is weak, we expect the existence of solutions that are close to de Sitter solutions. We will encounter in sect. \ref{section:IC} models for inflation that are described by such approximate de Sitter solutions. We will assume for this $\chi$-dependence the simple model \eqref{S7}, \eqref{S8}. In the present section we only consider a few simple cosmological solutions for our ansatz.

For $\chi\to 0 $ we recover the de Sitter solution \eqref{S1}-\eqref{S6} with $\chi_{0} = 0$. This remains an exact solution since the field dependence of the coupling functions vanishes for $\chi=0$ due to the discrete symmetry $\chi\to-\chi$. The de Sitter solution is unstable, however, for small scalar fluctuations $\chi$. This is most easily seen from the ratio
\be
\label{S9}
\widehat{V} = \frac{V}{M^{4}} = \frac{u_{0}}{4w_{0}^{2}}\big{(}1+\frac{\xi\chi^{2}}{2w_{0}k^{2}}\big{)}^{-2}\ ,
\ee
which is proportional to the scalar potential in the Einstein frame as we will see in section \ref{section:WS}. For positive $u_{0}$ the quantity $\widehat{V}(\chi)$ has a maximum at $\chi = 0$ and vanishes for $\chi\to\infty$.

\zwisch{Beginning with gauge fields in Minkowski space?}

The de Sitter solution $\chi = 0$, $c = 0$, $b = \hub = -\eta^{-1}$ is not the only possible solution for a beginning of the universe for $\infpast$. As an alternative we discuss next the possibility for a beginning with flat Minkowski space, $a = a_{0}$, $\hub = 0$, as well as $\chi = 0$, $c = 0$. Eqs. \eqref{eq:15} and \eqref{eq:17} are obeyed for this ansatz. Eq. \eqref{eq:14} becomes
\be
\label{FS1}
\deta b + b^{2} = \frac{4a^{2}V}{3(\mtil+2M^{2})}\ ,
\ee
while eq. \eqref{eq:16} reads
\be
\label{FS2}
\deta^{2} b = 2b^{3} - \frac{(\mtil+2M^{2})a^{2}}{\Ztil} b\ ,
\ee
where $a^{2}$, $\mtil$, $M^{2}$, $V$, $\Ztil$ are constants. A solution with constant $b$ requires
\be
\label{FS3}
b^{2} = \frac{4a^{2}V}{3(\mtil+2M^{2})} = \frac{(\mtil+2M^{2})a^{2}}{2\Ztil}\ .
\ee

As a consequence, the couplings have to obey the particular relation
\be
\label{FS4}
V\Ztil = \frac{3}{8}(\mtil+2M^{2})^{2}\ .
\ee
For given values of $\mtil$, $M^{2}$ and $\Ztil$ there is a particular value for $V$, given by eq. \eqref{FS4}, for which a very simple beginning of the universe can be realized by a flat space geometry, vanishing scalar field, and a constant gauge field with $b/a_{0}$ determined by eq. \eqref{FS3}. The relation \eqref{FS4} requires $V\Ztil>0$, such that it exist only if $V(\chi=0)$ and $\Ztil(\chi=0)$ have the same sign. We will concentrate in this paper on $V>0$, $\Ztil<0$, such that no flat space solution of this type exists.

\zwisch{General de Sitter solutions with vanishing scalar field}

More generally, we may investigate the subspace of solutions with $\chi = 0$, $c = 0$. We assume here that $Y$ does not increase too fast for $\chi\to 0$ such that $\chi=0$ is a solution of the scalar field equation. An alternative case with $Y\sim\chi^{-2}$ will be discussed in sects. \ref{section:QS}, \ref{section:CM}. With this assumption only the field equations \eqref{eq:14} and \eqref{eq:16} remain to be satisfied. We can use cosmic time $t$ in order to eliminate the scale factor $a$ in eq. \eqref{eq:14},
\be
\label{FS5}
\dt H + 2H^{2} = (1+2\ytil)(\dt \btil + H\btil + \btil^{2}) - \frac{4V}{3\mtil}\ ,
\ee
where we define
\be
\label{FS6}
\btil = \frac{b}{a}\ ,\ \ytil = \frac{\M}{\mtil}\ .
\ee
Similarly eq. \eqref{eq:16} becomes
\begin{align}
\label{FS7}
\dt^{2}\btil + 3H\dt\btil = &-\frac{(1+2\ytil) \mtil}{\Ztil}(\btil - H) \nonumber \\
&+ 2\btil^{3} - (\dt H + 2H^{2})\btil\ .
\end{align}

Solutions with time independent $\btil$ and $H$ correspond to solutions of two algebraic equations for $\btil$ and $H$. They are understood best in terms of
\be
\label{38A}
\ftil = \btil-H\ ,
\ee
for which they take the form
\begin{equation}
\label{62A}
4\ytil\btil^{2} + (3-2\ytil)\btil\ftil - 2\ftil^{2} = \frac{4V}{3\mtil}\ ,
\end{equation}
and
\be
\label{FS8}
\ftil\bigg{[}\frac{(1+2\ytil) \mtil}{\Ztil} - 4\btil^{2} + 2\btil\ftil\bigg{]} = 0\ .
\ee
One finds two types of solutions. The first one,
\be
\label{FS9}
\ftil = 0\ ,\ \btil^{2} = \frac{V}{3\ytil \mtil} = \frac{V}{3M^{2}}\ ,
\ee
corresponds to the de Sitter solution discussed previously according to eqs. \eqref{S2} and \eqref{S4}.

The second family of de Sitter solutions,
\begin{align}
\label{FS10}
&\ftil = 2\btil - \frac{(1+2\ytil)\mtil}{2\Ztil\btil}\ , \\
&\btil^{2} = \frac{1}{2}\bigg{\{}\frac{(5+2\ytil)(1+2\ytil)\mtil}{4\Ztil} - \frac{2V}{3\mtil} \nonumber \\
&\pm\sqrt{\frac{(9+2\ytil)(1+2\ytil)^{3}\tilde{m}^{4}}{16\Ztil^{2}} - \frac{(5+2\ytil)(1+2\ytil)V}{3\Ztil} + \frac{4V^{2}}{9\tilde{m}^{4}}}\ \bigg{\}} \nn\ ,
\end{align}
generalizes the solution \eqref{FS3}. Indeed, for the special case \eqref{FS4}, $V=V_{c}$,
\be
\label{FS11}
V_{c} = \frac{3(1+2\ytil)^{2}\tilde{m}^{4}}{8\Ztil}\ ,
\ee
one obtains $\btil^{2} = (1+2\ytil)\mtil/(2\Ztil)$ and therefore $\ftil=\btil$, $H=0$. As mentioned above, this condition cannot be obeyed for $V\Ztil<0$, but the de Sitter solutions of type 2 exist for arbitrary $V\Ztil$. If the condition \eqref{FS11} is not met, the square root in eq. \eqref{FS10} does no longer vanish. In this case one obtains two solutions according to the two signs of the root. As long as the expression under the root remains positive we are left with three different solutions with a de Sitter geometry, with type 1 given by eqs. \eqref{FS9} and type 2 for the two possibilities for eq. \eqref{FS10}. The corresponding Hubble parameter is extracted as $H = \btil - \ftil$. 

\zwisch{De Sitter solutions with non-vanishing covariant derivative of vierbein}

For the de Sitter solutions \eqref{FS10} of the second type with $\ftil\neq0$ the covariant derivative of the vierbein $U_{\mu\nu}^{\trip m}=D_{\mu}e_{\nu}^{\ m}$ does not vanish. For these solutions pregeometry can differ strongly from metric gravity. We will see that in contrast to the de Sitter solutions \eqref{FS9} of the first type the Hubble parameter is not small for small $V$.

The existence of the two de Sitter solutions \eqref{FS10} requires the expression under the root to be positive. For a given $y$ this corresponds to a condition for the combination
\be
\label{FS11A}
x=\frac{V\Ztil}{3\tilde{m}^{4}}\ ,
\ee
namely $x<x_{c}$ or $x>x_{+}$. Here $x_{c}$ corresponds to $V_{c}$ in eq. \eqref{FS11},
\be
\label{FS12A}
x_{c}=\frac{1}{8}(1+2\ytil)^{2}\ ,\quad x_{+}=\frac{1}{8}(9+2\ytil)(1+2\ytil)\ .
\ee
For $1+2\ytil>0$ one has $x_{+}>x_{c}$. One finds at the boundary the values
\be
\label{FS13A}
\btil^{2}=\frac{\mtil}{\Ztil}\big{[}\frac{1}{8}(5+2\ytil)(1+2\ytil)-x\big{]}\ ,
\ee
or
\bel{48A}
\btil^{2}(x_{c})=\frac{\mtil(1+2\ytil)}{2\Ztil}\ ,\quad \btil^{2}(x_{+})=-\frac{\mtil(1+2\ytil)}{2\Ztil}\ .
\ee

For positive $\mtil\Ztil$, which is the case we often consider in this paper, the negative value of $\btil^{2}(x_{+})$ indicates that no solution is possible for $x=x_{+}$. This extends to the whole range $x>x_{+}$. We infer the existence condition for the solution \eqref{FS10}  for positive $V$
\bel{48B}
x<x_{c}\ ,\quad \Ztil<\frac{3\tilde{m}^{4}(1+2\ytil)^{2}}{8V}\ .
\ee
This is obeyed automatically for $V>0$, $\Ztil<0$.

For the limit $V=0$ one has
\bel{48C}
\btil^{2}=\frac{(1+2\ytil)\mtil}{8\Ztil}\Big{(}5+2\ytil\pm\sqrt{(9+2\ytil)(1+2\ytil)}\Big{)}\ .
\ee
This is of the order $\mtil/\Ztil$. The squared Hubble parameter is of the same order of magnitude. This demonstrates that the second type of de Sitter solutions is far away from general relativity. The Hubble parameter is no longer determined by the scalar potential. It is rather induced by non-vanishing values of the gauge fields and an associated covariant derivative of the vierbein different from zero.

We finally observe that no solution with vanishing gauge field $\btil=0$ is possible for $V\neq 0$. For $\btil=0$ eq. \eqref{FS5} requires $H\neq 0$, such that eq. \eqref{FS7} cannot be obeyed.

\zwisch{Early attractor solution}

\begin{figure}[h]
\centering
\includegraphics[width=0.5\textwidth]{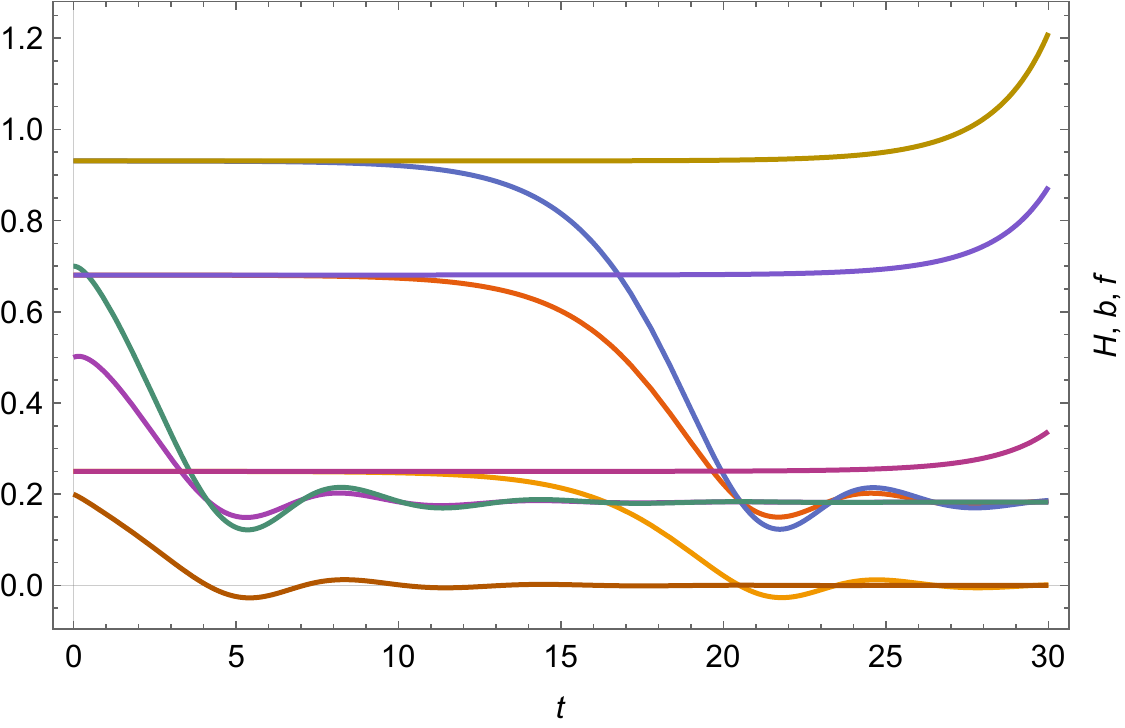}
\caption{\emph{Attractor solution}. The evolution of $H$, $\btil$ and $\ftil$ is shown as function of $t$. We employ three different initial conditions (with $\ctil=0$) explained in the text. The two pairs of upper curves show the instability of $\btil$ (upper pair) and $H$ (lower pair) for initial conditions very close to the de Sitter solution of type 2. The two middle curves on the left correspond to $\btil$ and $H$ with initial conditions $\btil=0.7$, $\Htil=0.5$. Finally the pair of curves that is almost horizontal for small $t$ displays $\ftil$ for initial conditions given almost by the de Sitter solution of type 2, while the lowest curve shows $\ftil$ for the third set of initial conditions. Parameters are $Z=-\Ztil=1$, $\M=1$, $\m=-\mtil=5$, $B=0$, $Y=0$, $K=4$. We take an almost constant potential $V=0.1$.}
\label{fig:3}
\end{figure}

The de Sitter solutions of type 1 with $\ftil=0$ are attractor solutions, while the de Sitter solutions of type 2 with $\ftil\neq0$ are unstable. This can be seen in Fig. \ref{fig:3} where we plot $H$, $\btil$ and $\ftil$ as a function of $t$ for three different intital conditions. The two upper curves starting horizontally correspond to the values of $\btil_{2}$ and $H_{2}$ given by the de Sitter solution of type 2 according to eq. \eqref{FS10}, with a minus sign of the root. A slight shift in the initial value for $\btil$, namely $\btil=\btil_{2}+10^{-7}$ for the two curves growing for large $t$, and $\btil=\btil_{2}-10^{-4}$ for the two curves dercreasing for large $\that$, demonstrates that the de Sitter solution of type 2 is unstable. The two corresponding curves for $\ftil$ are the ones with a constant behavior for small $t$. The unstable behavior of $H$, $\btil$ (top curve) and $\ftil$ is manifest, with initial values only slightly different from the exact solution leading for large $t$ to a very different behavior. 

The curves with initial value $\btil=\btil_{2}-10^{-4}$ approach for large $t$ the de Sitter solution of type 1. This solution is approached already for smaller $t$ by our third set of initial conditions with $\btil(0)=0.7$, $H(0)=0.5$. The de Sitter solution of type 1 is a stable attractor. It is approached by neighboring solutions within its range of attraction, as visible by $\ftil$ going to zero. The range of attraction is bounded by the de Sitter solution of type 2. The initial condition $\btil=\btil_{2}+10^{-7}$ is no longer attracted by the de Sitter solution of type 1. It leads to unstable behavior.

These considerations hold as long as $\chi$ is small enough such that the potential can be taken as constant. Once the gradient of the potential becomes important the evolution of the scalar field destabilizes the de Sitter solution of type 1. (For Fig. \ref{fig:3} we have employed a potential $V=V_{0}(1+\exp(2\chi/M))^{-2}$ that will be motivated in sect. \ref{section:CM}. For the small initial value of $\chi/M=-4$ the potential is constant to a very good approximation for the range shown.)

We discuss in appendix \ref{app:B} more details of the conditions of existence and the stability of the de Sitter solutions with vanishing scalar field. For the time being we are satisfied with the observation that simple solutions exist that are candidates for a beginning of the universe and a following inflationary epoch. We will later turn back to this issue.

\vspace{\baselineskip}
\zwisch{Hubble parameter}

For $\deta\chi=0$, $c=0$ an integral of the field equations \eqref{eq:14},\eqref{eq:16} determines the Hubble parameter by
\begin{align}
\label{FS12}
H^{2}&=(1+2\ytil)\btil^{2}-\frac{2V}{3\mtil}\nn \\
&+\frac{\Ztil}{\mtil}\Big{[}(\dt\btil)^{2}+2H\btil\dt\btil+H^{2}\btil^{2}-\btil^{4}\Big{]}\ .
\end{align}
Indeed, the $t$-derivative of eq. \eqref{FS12} yields a linear combination of eqs. \eqref{eq:14},\eqref{eq:16} and \eqref{FS12}. One can verify that the de Sitter solutions with $\dt\btil=0$, as given by the two types of solutions of eq. \eqref{FS8}, indeed obey eq. \eqref{FS12}. It is also straightforward to see that the explicit de Sitter solutions \eqref{FS9}, \eqref{FS10} are consistent with eq. \eqref{FS12}. We will discuss in sect. \ref{section:GFE} the origin of this relation in more detail.

We can use eq. \eqref{FS12} for establishing the general solutions with $\ctil=0$, $\ftil=0$, $\chi=\chi_{0}$ for constant $V$, $\M$, $\mtil$ $\Ztil$ and $Y=0$. Inserting $\btil=H$ in eq. \eqref{FS12} yields
\bel{70A*}
H^{2}=\frac{V}{3\M}-\frac{\Ztil}{2\M}\dt H\big{(}2H^{2}+\dt H\big{)}\ .
\ee
In particular, for $V=0$ this yields
\bel{70B*}
(\dt H)^{2}+2H^{2}\dt H-\muh H^{2}=0\ ,\quad \muh=-\frac{2\M}{\Ztil}\ .
\ee
Combined with eq. \eqref{MS9A},
\bel{70C*}
\dt H+2H^{2}=0\ ,
\ee
the only solution is $H=0$. The candidate solution \eqref{MS18} solves eq. \eqref{70C*} but not eq. \eqref{70A*}.

\zwisch{Instability of non-geometric state}

For the de Sitter solutions the scale factor $a(\eta)$ reaches zero for $\infpast$. The limit $a\to 0$ is actually well defined despite the vanishing determinant $e$ of the vierbein. This is visible in the effective action \eqref{eq:13} for which only the terms $\sim Z$, $\Ztil$ and $B$ remain for $a=0$. The limiting state $a\to 0$ may be called the "non-geometric state" since the metric vanishes in this limit.

The homogeneous isotropic field equations precisely on the non-geometric state involve the gauge fields. For constant $Z$, $\Ztil$, $B$ they read
\begin{align}
\label{50A}
-\Ztil\deta^{2}b&=-\frac{\partial W}{\partial b}-2(\Ztil-Z-2B)c\deta c\ ,\nn\\
Z\deta^{2}c&=-\frac{\partial W}{\partial c}+2(\Ztil-Z-2B)c\deta b\ ,
\end{align}
with effective gauge field potential
\bel{50B}
W(b,c)=\frac{\Ztil}{2}(\btil^{2}-c^{2})^{2}-2Zb^{2}c^{2}\ .
\ee
For $Z>0$, $\Ztil<0$ the potential is not bounded from below, with both terms in eq. \eqref{50B} being negative. Changing the sign of the $Z$ or $\Ztil$ factors induces the "wrong" sign of the derivative term. For all choices of parameters we find a solution with vanishing gauge fields, $b=c=0$. This non-geometric solution is unstable due to the unbounded potential.

The instability is a characteristic feature of the non-compact character or the gauge group. For small gauge fields this instability is rather weak since the gauge field potential is quartic in the gauge fields. As we have seen, also fluctuations with small non-zero $a$ are unstable. They are leading for the de Sitter solutions discussed in this section. The dynamics of $a$ can stabilize the evolution of the gauge field fluctuations in the sense that they get correlated with the increasing $a$.

\section{Emergence of variable gravity}
\label{section:EVG}

Due to the instability towards non-zero values of the scalar fields the de Sitter solutions may be considered as candidates for the beginning of the universe and an early inflationary epoch. In later stages the cosmological evolution departs from these solutions. We will next see how for late times the solutions approach the solutions for general relativity coupled to a scalar field. For late times an effective low energy theory becomes valid. It is based on a derivative expansion for the metric and a scalar field. This effective theory is variable gravity \cite{CWVG}, for which interesting cosmological solutions have been discussed previously \cite{CWVG,CWIQM, RUCW, HMSS, HMSS2}. Even earlier, variable gravity provides for an accurate description of the early attractor solution and the associated inflationary epoch. For a description of the evolution of the universe during inflation and after the end of inflation it is crucial to understand how variable gravity is embedded in, and dynamically emerges from, our model of pregeometry.

\zwisch{Variable Gravity}

Variable gravity couples a scalar field $\chi$ to the metric. In second order in a derivative expansion the effective action is given by
\be
\label{23}
\Gamma_{VG} = \int_{x}\sqrt{g}\big{\{}-\frac{M^{2}}{2}R + \frac{\tilde{K}}{2}\partial^{\mu}\chi\partial_{\mu}\chi + V\big{\}}\ .
\ee
Here $R$ is the curvature scalar formed from the metric $g_{\mu\nu}$ and $\sqrt{g}=\sqrt{-\text{det}(g_{\mu\nu})}=e$. This effective action involves three functions of $\chi$, i.e. the field dependent squared effective Planck mass $M^{2}(\chi)$, the kinetial $\tilde{K}(\chi)$ and the potential $V(\chi)$. Furthermore, also particle masses depend on $\chi$. This is crucial for the implementation of quantum scale symmetry \cite{CWQS} and the associated absence of apparent violations of the equivalence principle and time-varying fundamental constants. It is also a key ingredient for an acceptable matter dominated cosmological epoch despite the variation of the effective Planck mass $M(\chi)$ \cite{CWCNC}, \cite{Wetterich_1988}. For example, a field dependent Planck mass $M^{2} = \chi^{2}$ realizes quantum scale symmetry provided that all particle masses are proportional to $\chi$. For cosmologies with a continuous increase of $\chi(\eta)$ both the Planck mass and particle masses change without conflict with observation.
The $\chi$-dependence of the particle masses distinguishes variable gravity from many other generalizations of Brans-Dicke theories. This version of quantum scale symmetry also requires constant $\tilde{K}$ and $V/\chi^{4}$ . For our ansatz \eqref{S7},\eqref{S8} this behavior is reached for $k^{2}/\chi^{2}\to 0$. A different version of quantum scale symmetry with constant $V$ and $\M$ and $\Ktil\sim\chi^{-2}$ will be discussed in sect. \ref{section:QS}. This will become relevant for $k^{2}/\chi^{2}\to\infty$.

Variable gravity can describe interesting cosmologies for which the same scalar field accounts for inflation and dynamical dark energy \cite{CWVG, CWIQM, RUCW, HMSS, HMSS2}. In this section we establish how variable gravity arises from our formulation of pregeometry as an "effective low energy theory", once squared (covariant) momenta $q^{2}$ or geometric invariants as $R$ are sufficiently small as compared to $m^{2}$ and $\mtil$. In turn, Einstein's general relativity emerges from variable gravity in the limit where the effects of a time evolution or space-dependence of $\chi$ become negligible. This is realized for the physics on earth or on galactic scales.

For our model of pregeometry most observable features in cosmology can be described within the effective low energy model of variable gravity. This concerns the late stages of the inflationary epoch which are relevant for the observable properties of the primordial cosmic fluctuations. It also applies to the appearance of dynamical dark energy, typically both in the form of early dark energy (EDE) during the radiation and matter dominated epochs, as well as the present epoch of dark energy domination. Since variable gravity reduces to general relativity in the limit of a constant scalar field, the understanding of the emergence of variable gravity is also crucial for an understanding of how our model of pregeometry predicts all properties of general relativity for the effects of gravity on the scales of galaxies, stars or planets - in other words how geometry emerges from pregeometry.

The differences between pregeometry and variable gravity concern mainly the "beginning epoch" of our universe. In these very early stages the dynamics of the gauge fields is not yet tightly coupled to the dynamics of the vierbein. For the later epochs we mainly have to understand how the "parameters" of variable gravity, namely the coupling functions $M^{2}(\chi)$, $\tilde{K}(\chi)$ and $V(\chi)$, are connected to the "parameters" of our model of pregeometry. We will find that $M^{2}(\chi)$ and $V(\chi)$ are identical for pregeometry and variable gravity. For the coefficient of the scalar kinetic term the kinetial $\tilde{K}$ of variable gravity will be connected to the kinetial $K$ of pregeometry by a shift
\be
\label{23A}
\tilde{K}(\chi) = K(\chi) - \frac{3}{\mtil+2M^{2}}\bigg{(}\frac{\partial M^{2}}{\partial \chi}+Y\chi\bigg{)}^{2}\ .
\ee
The two kinetials are identical only in the limit where the $\chi$-dependence of the effective Planck mass $M(\chi)$ and $Y$ can be neglected.

\vspace{\baselineskip}
\vspace{\baselineskip}
\zwisch{Variable gravity from pregeometry}

For an understanding of this shift we write the effective action \eqref{eq:6} of pregeometry as a sum
\be
\label{24}
\Gamma = \Gamma_{1} + \Gamma_{2}\ ,
\ee
with
\be
\label{24A}
\Gamma_{1} = \int_{x}e\bigg{\{}-\frac{M^{2}}{2}R + \frac{K}{2}\partial^{\mu}\chi\partial_{\mu}\chi + V\bigg{\}}
\ee
a first form of variable gravity. The "completion" of the effective action in pregeometry reads
\begin{align}
\label{25}
\Gamma_{2} &= \int_{x}e\bigg{\{}\frac{M^{2}}{2}\Delta + \frac{Z}{8}F_{\mu\nu\rho\sigma}F^{\mu\nu\rho\sigma} +\frac{B}{2}F_{\mu\nu}F^{\mu\nu}+\frac{C}{2}F^{2}\nn\\
&+ \frac{m^{2}}{4}U_{\mu\nu\rho}U^{\mu\nu\rho} +\frac{\n}{2}U_{\mu\ \rho}^{\ \mu}U_{\nu}^{\ \nu\rho}+Y\chi\partial_{\nu}\chi U_{\mu}^{\ \mu\nu}\bigg{\}}\ ,
\end{align}
with
\be
\label{26}
\Delta = R - F\ .
\ee
One possibility for the emergence of variable gravity is a situation where $\Gamma_{2}$ can be neglected. We will see that this is indeed realized for $\partial M^{2}/\partial \chi = 0$, $Y=0$. More generally, $\Gamma_{2}$ will also contribute to the effective action of variable gravity. This is the origin of the difference between $\tilde{K}$ and $K$ in eq. \eqref{23A}.

We can relate the field strength $F_{\mu\nu\rho\sigma}$ to the Riemann curvature tensor $R_{\mu\nu\rho\sigma}$. This is done \cite{CWGG} by use of the commutator of two covariant derivatives of the vierbein
\begin{align}
\label{VG1}
[D_{\mu},D_{\nu}]e_{\rho}^{\ m} &= F_{\mu\nu\ n}^{\dub m}e_{\rho}^{\ n} - R_{\mu\nu\ \rho}^{\dub\sigma}e_{\sigma}^{\ m} \nonumber \\
&= D_{\mu}U_{\nu\rho}^{\dub m} - D_{\nu}U_{\mu\rho}^{\dub m} = V_{\mu\nu\rho}^{\trip m}\ .
\end{align}
We employ the relation
\be
\label{VG2}
F_{\mu\nu\rho\sigma} = R_{\mu\nu\rho\sigma} - V_{\mu\nu\rho\sigma}
\ee
in order to write the effective action \eqref{eq:6} in terms of the Riemann tensor $R_{\mu\nu\rho\sigma}$ and the tensor $U_{\mu\nu}^{\dub m}$,
\begin{align}
\label{VG3}
\Gamma &= \int_{x}e\bigg{\{}-\frac{\M}{2}R+\frac{K}{2}\partial^{\mu}\chi\partial_{\mu}\chi+V \\
&+\frac{Z}{8}R^{\mu\nu\rho\sigma}R_{\mu\nu\rho\sigma} +\frac{B}{2}R_{\mu\nu}R^{\mu\nu}+\frac{C}{2}R^{2}\bigg{\}}+\Delta\Gamma_{1}+\Delta\Gamma_{2}\ .\nn
\end{align}
The first terms involve only the metric and the scalar field. They extend the two-derivative approximation \eqref{23} by inclusion of terms with four derivatives. These terms reproduce Stelle's gravity in the limit where $Z$, $B$, and $C$ are independent of $\chi$. The additional terms $\Delta\Gamma_{1}+\Delta\Gamma_{2}$ vanish for $U_{\mu\nu\rho}=0$. The leading "correction term" is given by
\begin{align}
\label{70A}
\Delta\Gamma_{1}&=\int_{x}e\bigg{\{}\frac{\m}{4}U^{\mu\nu\rho}U_{\mu\nu\rho}+\frac{\n}{2}U_{\mu\ \rho}^{\ \mu}U_{\nu}^{\ \nu\rho} \\
&-\frac{\M}{2}\big{(}U^{\nu\mu\rho}U_{\mu\nu\rho}-U_{\nu}^{\ \nu\rho}U_{\mu\dub\rho}^{\ \mu}\big{)}\nn \\
&+\bigg{(}\frac{\partial\M}{\partial\chi}+Y\chi\bigg{)}\partial_{\mu}\chi U_{\nu}^{\ \nu\mu} \nonumber \\
&+\frac{Z}{2}\big{(}D_{\mu}R^{\mu\nu\rho\sigma}U_{\nu\rho\sigma}+R^{\mu\nu\rho\sigma}U_{\nu\rho}^{\dub\tau}U_{\mu\sigma\tau}\big{)} \nonumber \\
&-B\Big{[}D_{\mu}R^{\mu\nu}U_{\rho\ \nu}^{\ \rho}+D_{\rho}R^{\mu\nu}U_{\mu\nu}^{\dub\rho}\nn\\
&\quad\quad +R^{\mu\nu}\big{(}U_{\mu\nu\sigma}U_{\rho}^{\ \rho\sigma}-U_{\mu}^{\ \rho\sigma}U_{\rho\nu\sigma}\big{)}\Big{]}\nn\\
&-2C\partial_{\mu}RU_{\nu}^{\ \nu\mu}+CR\big{(}U^{\rho\mu\nu}U_{\mu\rho\nu}-U_{\mu\ \rho}^{\ \mu}U_{\nu}^{\ \nu\rho}\big{)}\bigg{\}}\ .\nn
\end{align}
Further corrections arise from covariant derivatives of $U_{\mu\nu}^{\dub m}$ and from a $\chi$-dependence of the coupling functions,
\begin{align}
\label{84*}
\Delta\Gamma_{2}&=\int_{x}e\bigg{\{}\half\frac{\partial Z}{\partial\chi}\partial_{\mu}\chi R^{\mu\nu\rho\sigma}U_{\nu\rho\sigma}\\
&-\dchi{B}\partial_{\mu}\chi\big{(}R^{\mu\nu}U_{\rho\ \nu}^{\ \rho}+R^{\rho\nu}U_{\rho\nu}^{\dub\mu}\big{)}-2\dchi{C}\partial_{\mu}\chi RU_{\nu}^{\ \nu\mu}\nn\\
&+\frac{Z}{4}\big{(}D^{\mu}U_{\dub m}^{\nu\rho}D_{\mu}U_{\nu\rho}^{\dub m} - D^{\nu}U_{\dub m}^{\mu\rho}D_{\mu}U_{\nu\rho}^{\dub m}\big{)}\nn\\
&+\frac{B}{2}\big{(}D_{\mu}U_{\rho\nu}^{\dub m}-D_{\rho}U_{\mu\nu}^{\dub m}\big{)}e_{m}^{\ \rho}\big{(}D^{\mu}U_{\sigma}^{\ \nu n}-D_{\sigma}U^{\mu\nu n}\big{)}e_{n}^{\ \sigma}\nn\\
&+\frac{C}{2}\Big{[}\big{(}D_{\mu}U_{\rho}^{\ \mu m}-D_{\rho}U_{\mu}^{\ \mu m}\big{)}e_{m}^{\ \rho}\Big{]}^{2}\bigg{\}}\ .\nn
\end{align}

An effective theory for the vierbein or the associated metric is obtained by "integrating out" the gauge fields $A_{\mu mn}$. For this purpose one has to solve the field equations for $A_{\mu mn}$ as a functional of $\e$. Subsequently, one reinserts the solution into the effective action \eqref{VG3}. The reduced effective action obtained in this way only depends on the vierbein and its derivatives. The field equations for the reduced effective action remain exact. They are, however, rather complicated and useful only in the presence of suitable approximations.

\zwisch{Field equations for $\boldsymbol{U_{\mu\nu\rho}}$}

Instead of solving the field equations for $A_{\mu mn}$ at fixed $\e$ and reinserting the result, we can also solve the field equations for $U_{\mu\nu\rho}$ at fixed vierbein $\e$ and reinsert the result. Taking into account the antisymmetry $U_{\mu\nu\rho} = -U_{\mu\rho\nu}$ the field equation for $U_{\mu\nu\rho}$ reads
\begin{align}
\label{VG4}
&\m U^{\mu\nu\rho}+\n\big{(}U_{\sigma}^{\ \sigma\rho}g^{\mu\nu}-U_{\sigma}^{\ \sigma\nu}g^{\mu\rho}\big{)}\nn\\
&-\M\big{(}U^{\nu\mu\rho}-U^{\rho\mu\nu}-g^{\mu\nu}U_{\tau}^{\ \tau\rho}+g^{\mu\rho}U_{\tau}^{\ \tau\nu}\big{\}}\nn \\
&=-\bigg{[}\bigg{(}\frac{\partial\M}{\partial\chi}+Y\chi\bigg{)}\big{(}g^{\mu\nu}\partial^{\rho}\chi-g^{\mu\rho}\partial^{\nu}\chi\big{)} + ZD_{\sigma}R^{\sigma\mu\nu\rho} \nonumber \\
&-B\big{(}D_{\sigma}R^{\sigma\rho}g^{\mu\nu}-D_{\sigma}R^{\sigma\nu}g^{\mu\rho}+D^{\rho}R^{\mu\nu}-D^{\nu}R^{\mu\rho}\big{)}\nn\\
&-2C\big{(}\partial^{\rho}Rg^{\mu\nu}-\partial^{\nu}Rg^{\mu\rho}\big{)}+\frac{\partial Z}{\partial\chi}\partial_{\sigma}\chi R^{\sigma\mu\nu\rho}\nn\\
&-\dchi{B}\Big{[}\big{(}R^{\sigma\rho}g^{\mu\nu}-R^{\sigma\nu}g^{\mu\rho}\big{)}\partial_{\sigma}\chi+R^{\mu\nu}\partial^{\rho}\chi-R^{\mu\rho}\partial^{\nu}\chi\Big{]}\nn\\
&-2\dchi{C}R\big{(}\partial^{\rho}\chi g^{\mu\nu}-\partial^{\nu}\chi g^{\mu\rho}\big{)}\bigg{]} +\dots
\end{align}
The dots denote contributions from eq. \eqref{84*} which involve covariant derivatives of $U_{\mu\nu}^{\dub m}$, and from terms $\sim R^{\mu\nu\rho\sigma}U_{\nu\rho}^{\dub\tau}U_{\mu\sigma\tau}$, $RU_{\mu\ \rho}^{\ \mu}U_{\nu}^{\ \nu\rho}$ etc., which are of the order $RU^{2}$. These contributions can be neglected for (squared covariant) momenta or curvature tensor sufficiently small as compared to $\mtil$.

We decompose
\begin{align}
\label{VG5}
U^{\mu\nu\rho} &= \frac{1}{3}\big{(}g^{\mu\nu}U_{\tau}^{\ \tau\rho}-g^{\mu\rho}U_{\tau}^{\ \tau\nu}\big{)} + \tilde{U}^{\mu\nu\rho}\ , \nonumber \\
g_{\mu\nu}&\tilde{U}^{\mu\nu\rho}= 0\trip ,\trip g_{\mu\rho}\tilde{U}^{\mu\nu\rho} = 0\ .
\end{align}
Contracting eq. \eqref{VG4} with $g_{\mu\nu}$ yields
\begin{align}
\label{VG6}
U_{\tau}^{\ \tau\rho} &= -\frac{1}{\mtil+2\M}\bigg{[}3\bigg{(}\frac{\partial\M}{\partial\chi}+Y\chi\bigg{)}\partial^{\rho}\chi - ZD_{\nu}R^{\nu\rho}\nn \\
&-B\big{(}2D_{\mu}R^{\mu\rho}+\partial^{\rho}R\big{)}-6C\partial^{\rho}R-\frac{\partial Z}{\partial\chi}\partial_{\nu}\chi R^{\nu\rho}\nn\\
&-\dchi{B}\big{(}2R^{\mu\rho}\partial_{\mu}\chi+R\partial^{\rho}\chi\big{)}-6\dchi{C}R\partial^{\rho}\chi\bigg{]}\ ,
\end{align}
while $\tilde{U}^{\mu\nu\rho}$ obeys
\begin{align}
\label{VG7}
&\m \Util^{\mu\nu\rho} + \frac{\M}{2}\big{(}\Util^{\rho\mu\nu}-\Util^{\nu\mu\rho}+\Util^{\rho\nu\mu}-\Util^{\nu\rho\mu}\big{)} \nonumber \\
&=-Z\Big{(}D_{\sigma}R^{\sigma\mu\nu\rho} + \frac{1}{3}\big{(}D_{\sigma}R^{\sigma\rho}g^{\mu\nu}-D_{\sigma}R^{\sigma\nu}g^{\mu\rho}\big{)}\Big{)} \nonumber \\
&-\frac{\partial Z}{\partial\chi}\partial_{\sigma}\chi\Big{(}R^{\sigma\mu\nu\rho} + \frac{1}{3}\big{(}R^{\sigma\rho}g^{\mu\nu}-R^{\sigma\nu}g^{\mu\rho}\big{)}\Big{)}+\dots\ ,
\end{align}
with dots involving similar terms $\sim B$, $C$ or $\partial B/\partial\chi$, $\partial C/\partial\chi$.
The two equations can be solved separately and the terms quadratic in $U$ in the effective action \eqref{VG3} do not mix $U_{\tau}^{\ \tau\rho}$ and $\Util^{\mu\nu\rho}$.
One could further decompose $\Util$, but we will not need this here in practice.

\vspace{\baselineskip}
\zwisch{Pregeometry corrections in variable gravity}

The leading term in the low energy effective theory arises from the term $\sim(\partial\M/\partial\chi+Y\chi)\partial_{\mu}\chi$ in eq. \eqref{VG6}. Insertion into the effective action contributes to the scalar kinetic term, resulting in a shift \eqref{23A} between $\Ktil$ and $K$. This explains in a simple way why for $\partial\M/\partial\chi+Y\chi \neq 0$ the kinetial $K$ in pregeometry and the kinetial $\Ktil$ in the effective variable gravity model differ.

In addition to the shift between $\Ktil$ and $K$ the insertion of \eqref{VG6} into the effective action leads to a series of "non-renormalizable operators", corresponding to invariants with higher dimension, involving couplings or prefactors suppressed by increasing powers of $\widetilde{m}^{-1}$,
\begin{align}
\label{VG9}
\Delta\Gamma_{1} = -\frac{3}{2}\int_{x}e\frac{1}{\mtil+2\M}\bigg{(}\frac{\partial\M}{\partial\chi}+Y\chi\bigg{)}^{2}\partial^{\mu}\chi\partial_{\mu}\chi+\Delta\Gamma_{3}\ ,
\end{align}
where
\begin{align}
\label{75A}
\Delta\Gamma_{3}=\int_{x}e\bigg{\{}&Z\bigg{(}\frac{\partial\M}{\partial\chi}+Y\chi\bigg{)}\bigg{[}Z\partial_{\mu}\chi D_{\nu}R^{\nu\mu}\nn\\
&+ \frac{\partial Z}{\partial\chi}\partial_{\mu}\chi\partial_{\nu}\chi R^{\mu\nu}\bigg{]} \nonumber \\
&- \frac{Z^{2}}{6}D_{\mu}R^{\mu\nu}D_{\rho}R_{\ \nu}^{\rho} - \frac{Z}{3}\frac{\partial Z}{\partial\chi}D_{\mu}R^{\mu\nu}R_{\ \nu}^{\rho}\partial_{\rho}\chi \nonumber \\
&- \frac{1}{6}\bigg{(}\frac{\partial Z}{\partial\chi}\bigg{)}^{2}\partial_{\mu}\chi\partial_{\nu}\chi R^{\mu\rho}R_{\rho}^{\ \nu}+\dots\bigg{\}}\ ,
\end{align}
with dots denoting again similar contributions from $B$ and $C$.
For the counting we recall that by dimensional analysis $\partial Z/\partial\chi\sim\chi/\m$, $\partial\M/\partial\chi+Y\chi\sim\chi$. The insertion of the solution for $\Util^{\mu\nu\rho}$ yields additional terms of the same structure as the last three terms in eq. \eqref{75A}. In leading order for an effective low energy theory only the first term in eq. \eqref{VG9} matters. We observe that the dimensionally next to leading term vanishes for all geometries with $D_{\mu}R^{\mu\nu}=0$.

The variable gravity approximation becomes valid if $\Delta\Gamma_{2}$ can be neglected. For the homogeneous field equations this holds if both $f$ and $c$ are small. We will discuss this in detail in the appendix \ref{app:C}. If furthermore the higher powers and derivatives of the curvature tensor in $\Delta\Gamma_{3}$ become small, one can effectively work with a derivative expansion of variable gravity to fourth order. For a curvature tensor much smaller than $\M$ the second order derivative expansion \eqref{23} of variable gravity becomes valid. The variable gravity approximation is an important simplification since only the metric degree of freedom and the scalar remain. 

\zwisch{Homogeneous field equations for variable gravity}

We can compare the homogeneous field equations of pregeometry with the homogeneous field equations for variable gravity with effective action \eqref{23}. For a Robertson-Walker metric, with cosmic time $t$ related to conformal time $\eta$ by $a\de\eta = \de t$, $\deta = a\dt$, $H = \dt\ln a = \hub/a$, $\dot{\chi} = \dt\chi = \deta\chi/a$, the field equations for the scale factor derived from eq. \eqref{23} for homogeneous cosmology can be written in the form \cite{CWVG}
\be
\label{31}
\M R = 4V - \tilde{K}\dot{\chi}^{2} - C_{R}\ ,
\ee
with
\begin{align}
\label{32}
C_{R} &= 6\frac{\partial\M}{\partial\chi^{2}}\big{(}\dot{\chi}^{2}+(\ddot{\chi}+3H\dot{\chi})\chi\big{)} + 12\frac{\partial^{2}\M}{(\partial\chi^{2})^{2}}\dot{\chi}^{2}\chi^{2} \nonumber \\
&= \frac{3}{a^{3}}\dt\Big{(}\frac{\partial\M}{\partial\chi}\dot\chi a^{3}\Big{)} = 3(\dt+3H)\dt\M\ .
\end{align}
The scalar field equation is given by
\be
\label{33}
\tilde{K}(\ddot{\chi}+3H\dot{\chi})+\half \frac{\partial\tilde{K}}{\partial{\chi}}\dot{\chi}^{2} = -\frac{\partial V}{\partial\chi} + \half\frac{\partial\M}{\partial\chi}R\ .
\ee
Here the curvature scalar obeys
\be
\label{34}
R = 12H^{2}+6\dot{H}=\frac{6}{a^{2}}(\hub^{2}+\deta\hub)\ .
\ee
These equations are indeed much simpler than the full field equations of pregeometry. In appendix \ref{app:C} we discuss in detail how the field equations of variable gravity follow from the field equations of pregeometry. This gives a more precise intuition about the neglected terms for the case of homogeneous cosmologies.

In the presence of additional radiation or matter the field equations of variable gravity will involve an additional energy momentum tensor, as well as a possible source term for the evolution of the scalar field. Fluctuations of the fields of pregeometry beyond the metric can contribute to this energy momentum tensor. We will discuss this in sect. \ref{section:DMDE}.

\zwisch{Overall cosmology in pregeometry}

At this stage the main features for any realistic cosmology in pregeometry have already become clear. For cosmic epochs for which the characteristic length or time scale is much larger than $m^{-1}$ and $\tilde{m}^{-1}$ the effective theory of variable gravity becomes valid. With inverse characteristic length scale given by the Hubble parameter $H$, and in view of the stability conditions $\M<\m$, $\M<\abs{\mtil}/2$, this applies to all epochs for which $H^{2}\ll\M$. The homogeneous isotropic solutions of variable gravity are known to be stable attractors for the inflationary epoch, provided the coupling functions $\M$, $\Ktil$ and $V$ obey some stability conditions as the positivity of $\M$, the boundedness from below for $V$ and an inequality for $\Ktil$. For realistic cosmological models variable gravity can be used for the epoch of inflation, the end of inflation, and the radiation dominated epoch which is expected after inclusion of the particles of the standard model.
In a later matter dominated epoch small fluctuations in the energy density of (additional) matter increase. This rather mild instability is the origin of the emergence of the structures in the universe from the small primordial fluctuations.

Even in the context of our restricted model without the standard model particles a semi-realistic "late" cosmology can be obtained provided that small fluctuations of the gauge fields beyond variable gravity, as $f$ or $c$ , are stable, and the functions $V(\chi)$, $\Ktil(\chi)$, $\M(\chi)$ take a form compatible with observation. This will be discussed in sects. \ref{section:DMDE}, \ref{section:IC}. The residual effects of small fluctuations of $f$ or $c$ typically contribute to the energy momentum tensor in variable gravity.

It is possible that the validity of variable gravity actually extends to the beginning phase of the universe. For example, this can be the case for the de Sitter solution \eqref{S1}-\eqref{S3} for $\chi_{0}=0$, corresponding to eq. \eqref{FS9}. For $V\ll M^{4}$ the conditions for an approximation by variable gravity are met. If this de Sitter solution for $\chi\to 0$ is realized for $\infpast$, and connected smoothly to late cosmology, variable gravity can describe the whole history of the universe.

It is also possible that variable gravity with only a few derivatives in the effective action is not a good approximation for the beginning phase. In this case one has to understand the transition from the beginning solution to the later epoch of validity of variable gravity. We have discussed this in the two preceding sections.

\section{General field equations}
\label{section:GFE}

Before a further investigation of cosmology we proceed in the next two sections to a more detailed discussion of the field equations. In the present section we derive the general field equations for the quantum effective action \eqref{eq:6}. They will be needed for the discussion of inhomogeneous cosmologies, or for the consequences of pregeometry for black holes. The general field equations also define key objects as the energy momentum tensor. Specializing to homogeneous isotropic configurations we will identify the energy density as a useful partial integral of these equations.

\zwisch{Field equations for gauge fields}

The field equations obtain by variation of the effective action \eqref{eq:6}. For the gauge fields they read
\begin{align}
\label{FE1}
&ZD_{\nu}F^{\mu\nu mn}+2C\partial_{\nu}F(e^{m\mu}e^{n\nu}-e^{m\nu}e^{n\mu})\\
&+B\Big{[}D_{\nu}F^{\mu m}e^{n\nu}-D_{\nu}F^{\mu n}e^{m\nu}-D_{\nu}F^{\nu m}e^{n\mu}+D_{\nu}F^{\nu n}e^{m\mu}\Big{]}\nn\\
&= J^{\mu mn}\ ,\nn
\end{align}
where the current is given by
\begin{align}
\label{FE2}
&J^{\mu mn} = m^{2}U^{\mu mn} \\
&+(2CF-\M)\big{[}U^{m\mu n} -U^{n\mu m} + U_{\rho}^{\dub\rho m}e^{n\mu}-U_{\rho}^{\dub\rho n}e^{m\mu}\big{]} \nonumber \\
&-\n\big{(}U_{\rho}^{\ \rho m}e^{n\mu}-U_{\rho}^{\ \rho n}e^{m\mu}\big{)}\nn\\
&+B\Big{(}F^{\nu m}U_{\nu}^{\ \mu n}-F^{\nu n}U_{\nu}^{\ \mu m}-F^{\mu m}U_{\nu}^{\ \nu n}+F^{\mu n}U_{\nu}^{\ \nu m}\Big{)}\nn\\
&-\frac{\partial Z}{\partial \chi}\partial_{\nu}\chi F^{\mu\nu mn}\nn\\
&-\dchi{B}\partial_{\nu}\chi\Big{(}F^{\mu m}e^{n \nu}-F^{\mu n}e^{m \nu}-F^{\nu m}e^{n \mu}+F^{\nu n}e^{m \mu}\Big{)} \nonumber \\
&+ \bigg{(}\frac{\partial M^{2}}{\partial \chi}+Y\chi-2\dchi{C}F\bigg{)}\partial_{\nu}\chi\big{(}e^{m\mu}e^{n\nu} - e^{n\mu}e^{m\nu}\big{)}\nn\ .
\end{align}
The covariant derivative of the field strength
\begin{align}
\label{FE2A}
D_{\nu}F^{\mu\nu mn} &= \partial_{\nu}F^{\mu\nu mn} + \Gamma_{\nu\rho}^{\trip\nu}F^{\mu\rho mn} \nonumber \\
&+ A_{\nu\ p}^{\ m}F^{\mu\nu p n} + A_{\nu\ p}^{\ n}F^{\mu\nu m p}
\end{align}
involves the contracted Levi-Civita connection
\be
\label{7FE2B}
\Gamma_{\nu\rho}^{\trip\nu} = \frac{1}{e}\partial_{\rho}e\ .
\ee

\vspace{\baselineskip}
\zwisch{Field equations for vierbein and energy momentum tensor}

For the vierbein one obtains two equations, one from the symmetric and the other from the antisymmetric fluctuations. The symmetric part is composed of terms that can be identified with contributions to the energy momentum tensor,
\be
\label{FE3}
T_{\mu\nu}^{(U)} = -\bigg{(} T_{\mu\nu}^{(F)} + T_{\mu\nu}^{(\chi)} + T_{\mu\nu}^{(R)} + T_{\mu\nu}^{(Y)}\bigg{)}\ .
\ee
Indeed, we can write eq. \eqref{FE3} in the form of the Einstein equation
\be
\label{FE18}
M^{2}(R_{\mu\nu} - \frac{1}{2}Rg_{\mu\nu}) ) = T_{\mu\nu}\ ,
\ee
with energy momentum tensor
\be
\label{FE19}
T_{\mu\nu} = T_{\mu\nu}^{(U)} + T_{\mu\nu}^{(F)} + T_{\mu\nu}^{(\chi)} + T_{\mu\nu}^{(Y)} + T_{\mu\nu}^{(\Delta)}\ ,
\ee
where
\be
\label{FE20}
T_{\mu\nu}^{(\Delta)} = \frac{M^{2}}{2}\big{[}2R_{\mu\nu} - F_{\mu\nu} - F_{\nu\mu} - (R-F)g_{\mu\nu}\big{]}\ .
\ee
The tensor $T_{\mu\nu}^{(\Delta)}$ vanishes in the limit for which the symmetric part of $F_{\mu\nu}$ agrees with $R_{\mu\nu}$.

The left hand side of eq. \eqref{FE3} involves covariant derivatives of the vierbein,
\begin{align}
\label{FE4}
T_{\mu\nu}^{(U)} &= \frac{m^{2}}{2}\big{(}D_{\rho}U_{\mu\nu}^{\trip\rho}+D_{\rho}U_{\nu\mu}^{\trip\rho}\nn\\
&\quad\quad+U_{\mu}^{\ \tau\rho}U_{\nu\tau\rho} -\half U^{\sigma\tau\rho}U_{\sigma\tau\rho}g_{\mu\nu}\big{)}\nn\\
&+\frac{\n}{2}\Big{[}2D_{\rho}U_{\tau}^{\ \tau\rho}g_{\mu\nu}-D_{\mu}U_{\rho\ \nu}^{\ \rho}-D_{\nu}U_{\rho\ \mu}^{\ \rho}\nn\\
&\quad\quad+(U_{\mu\nu\rho}+U_{\nu\mu\rho})U_{\tau}^{\ \tau\rho}-U_{\rho\ \sigma}^{\ \rho}U_{\tau}^{\ \tau\sigma}g_{\mu\nu}\Big{]}\ .
\end{align}
With up to second derivatives of the vierbein this renders the symmetric fluctuations of the vierbein dynamical. The term $T_{\mu\nu}^{(U)}$ vanishes for a vanishing covariant derivative of the vierbein, $U_{\mu\nu}^{\dub m}=0$.

On the r.h.s. of eq. \eqref{FE3} one finds various "source terms" that can drive the evolution of $U_{\mu\nu\rho}$. The contribution to the energy momentum tensor involving the gauge fields is given by
\begin{align}
\label{FE5}
T&_{\mu\nu}^{(F)} = \frac{Z}{2}\big{(}F_{\mu}^{\ \rho mn}F_{\nu\rho mn} - \frac{1}{4}F^{\sigma\rho mn}F_{\sigma\rho mn}g_{\mu\nu}\big{)}\nn\\
&+B\Big{[}F_{\mu\rho}F_{\nu}^{\ \rho}+\half F^{\rho\sigma}(F_{\rho\mu\sigma\nu}+F_{\rho\nu\sigma\mu})-\half F^{\sigma\tau}F_{\sigma\tau}g_{\mu\nu}\Big{]}\nn\\
&+C\Big{[}F(F_{\mu\nu}+F_{\nu\mu})-\half F^{2}g_{\mu\nu}\Big{]}\ .
\end{align}
It is traceless,
\bel{109*}
g^{\mu\nu}T_{\mu\nu}^{(F)}=0\ .
\ee
For $B=C=0$ this is the standard energy momentum tensor for Yang-Mills theories. We observe, however, that the contractions with $\eta_{mn}$, corresponding to the non-compact character of the gauge group $SO(1,3)$, render some contributions to $T_{00}^{(F)}$ negative. This is another aspect of the need of negative $C$ for a stable theory.

One further has the scalar contribution to the energy momentum tensor,
\begin{align}
\label{FE6}
T_{\mu\nu}^{(\chi)} &= K\big{(}\partial_{\mu}\chi\partial_{\nu}\chi - \frac{1}{2}\partial^{\rho}\chi\partial_{\rho}\chi g_{\mu\nu}\big{)} - Vg_{\mu\nu} \nn\\
&+\half\frac{\partial m^{2}}{\partial \chi}\partial_{\rho}\chi\big{(}U_{\mu\nu}^{\trip\rho}+U_{\nu\mu}^{\trip\rho}\big{)}\nn\\
&+\half\dchi{\n}\big{[}2\partial_{\rho}\chi U_{\tau}^{\ \tau\rho}g_{\mu\nu}-\partial_{\mu}\chi U_{\tau\ \nu}^{\ \tau}-\partial_{\nu}\chi U_{\tau\ \mu}^{\ \tau}\big{]}\ .
\end{align}
The first terms are the standard energy momentum tensor in general relativity. We have added here mixed terms from the $\chi$-derivatives of $\m$ and $\n$. They vanish for $U_{\mu\nu\rho}=0$. Another type of mixed terms arises from the invariant $\sim Y$,
\begin{align}
\label{82B}
T_{\mu\nu}^{(Y)} &= \half Y\chi\big{[}\partial_{\mu}\chi U_{\rho\ \nu}^{\ \rho}+\partial_{\nu}\chi U_{\rho\ \mu}^{\ \rho}+\partial_{\rho}\chi(U_{\mu\nu}^{\dub\rho}+U_{\nu\mu}^{\dub\rho}) \nonumber \\
&\trip\trip\trip\trip -2\partial_{\rho}\chi U_{\tau}^{\ \tau\rho}g_{\mu\nu}\big{]} \nonumber \\
&+g_{\mu\nu}D^{\rho}(Y\chi\partial_{\rho}\chi)-\half D_{\mu}(Y\chi\partial_{\nu}\chi)-\half D_{\nu}(Y\chi\partial_{\mu}\chi)\ .
\end{align}

Finally, we have a term linear in the field strength for the gauge bosons
\be
\label{82A}
T_{\mu\nu}^{(R)} = -\frac{\M}{2}\big{(}F_{\mu\nu}+F_{\nu\mu}-Fg_{\mu\nu}\big{)}\ ,
\ee
For $F_{\mu\nu}=R_{\mu\nu}$ this term involves second derivatives of the metric. This limit yields exactly the expression appearing in Einstein's field equation.

The field equations for the vierbein have also an antisymmetric part. This antisymmetric part is given by
\begin{align}
\label{FE7}
&m^{2}D_{\rho}U_{\ \mu\nu}^{\rho} + (\M-2CF)\big{(}F_{\mu\nu}-F_{\nu\mu}\big{)}\\
&-BF^{\rho\sigma}\big{(}F_{\rho\mu\sigma\nu}-F_{\rho\nu\sigma\mu}\big{)}+ \frac{\partial m^{2}}{\partial \chi}\partial_{\rho}\chi U_{\ \mu\nu}^{\rho} \nonumber \\
&-\n\big{[}(U_{\mu\nu\rho}-U_{\nu\mu\rho})U_{\tau}^{\ \tau\rho}-D_{\mu}U_{\rho\ \nu}^{\ \rho}+D_{\nu}U_{\rho\ \mu}^{\ \rho}\big{]}\nn\\
&+\dchi{\n}\big{[}\partial_{\mu}\chi U_{\rho\ \nu}^{\ \rho}-\partial_{\nu}\chi U_{\rho\ \mu}^{\ \rho}\big{]}\nn\\
&+Y\chi\big{[}\partial_{\rho}\chi(U_{\mu\nu}^{\dub \rho} -U_{\nu\mu}^{\dub \rho})-\partial_{\mu}\chi U_{\rho\ \nu}^{\ \rho} + \partial_{\nu}\chi U_{\rho\ \mu}^{\ \rho}\big{]} = 0\nn\ .
\end{align}
This equation is obeyed identically for homogeneous isotropic solutions. In this case it it does not yield additional information.

\vspace{8\baselineskip}
\zwisch{Scalar field equation}

The scalar field equation reads
\begin{align}
\label{FE8}
-D_{\mu}&\big{(}K\partial^{\mu}\chi\big{)} + \frac{1}{2}\frac{\partial K}{\partial \chi}\partial^{\mu}\chi\partial_{\mu}\chi\nn\\
= &-\frac{\partial V}{\partial \chi}-\frac{1}{8}\frac{\partial Z}{\partial \chi}F_{\mu\nu\rho\sigma}F^{\mu\nu\rho\sigma} \nonumber \\
&-\half\dchi{B}F_{\mu\nu}F^{\mu\nu}-\half\dchi{C}F^{2}\nn\\
&- \frac{1}{4}\frac{\partial m^{2}}{\partial \chi}U_{\mu\nu\rho}U^{\mu\nu\rho} -\half\dchi{\n}U_{\mu\ \rho}^{\ \mu}U_{\nu}^{\ \nu\rho}\nn\\
&+ \frac{1}{2}\frac{\partial M^{2}}{\partial \chi}F +Y\chi D_{\nu}U_{\mu}^{\ \mu\nu}\ .
\end{align}
The $\chi$-dependence of coupling functions generates various sources beyond the usual potential gradient $\partial V/\partial\chi$. In particular, the term $\sim\partial\M/\partial\chi$ becomes important for large $\chi$ if $\M\sim\chi^{2}$. With $F$ approximated by $R$ this additional driving force explains why the relevant potential for the evolution of $\chi$ is the frame invariant potential which we will discuss in the next section, rather than $V(\chi)$. Furthermore, the term $\sim Y$ is connected to the difference \eqref{23A} between the kinetial $\Ktil$ of variable gravity and $K$.

\zwisch{Homogeneous field equations}

We may first recover the homogeneous field equations by inserting the ansatz \eqref{eq:8}. For the non-vanishing components of $D_{\nu}F^{\mu\nu mn}$ one finds
\begin{align}
\label{FE9}
D_{\nu}F^{k\nu l0} &= -a^{-4}\big{[}\deta^{2}b - 2b(b^{2}-3c^{2})\big{]}\delta^{kl}\ , \nonumber \\
D_{\nu}F^{k\nu l j} &= a^{-4}\big{[}\deta^{2}c + 2c(c^{2}-3b^{2})\big{]}\varepsilon^{klj}\ ,
\end{align}
and the non-vanishing components of $J^{\mu mn}$ obtain as
\begin{align}
\label{FE10}
J^{kl0} &= \frac{1}{a^{2}}\delta^{kl}\Big{[}(\mtil+2M^{2})(b-\hub)-\bigg{(}\frac{\partial M^{2}}{\partial \chi}+Y\chi\bigg{)}\deta\chi\nn\\
&-\frac{6}{a^{2}}(b-\hub)\big{[}(B+4C)\deta b+(2B+4C)(b^{2}-c^{2})\big{]}\nn\\
&+\frac{1}{a^{2}}\frac{\partial \Ztil}{\partial \chi}\deta\chi\deta b+\frac{2}{a^{2}}\bigg{(}\dchi{B}+6\dchi{C}\bigg{)}\deta\chi(b^{2}-c^{2})\Big{]}\ , \nonumber \\
J^{klj} &= -\frac{1}{a^{2}}\varepsilon^{klj}\Big{[}(m^{2}+2M^{2})c + \frac{1}{a^{2}}\frac{\partial Z}{\partial \chi}\deta\chi\deta c\nn\\
&+\frac{2c}{a^{2}}\big{[}(B-12C)\deta b+(2B-12C)(b^{2}-c^{2})\big{]}\Big{]}\ .
\end{align}

The two independent parts of the field equations \eqref{FE9}, \eqref{FE10} are identical to eq. \eqref{eq:16} and eq. \eqref{eq:17}. For the ansatz \eqref{eq:8} eq. \eqref{FE7} is obeyed identically. The scalar field equation \eqref{FE8} coincides with eq. \eqref{eq:15}. For the remaining field equations we need the energy momentum tensor for homogeneous isotropic field configurations.

\vspace{3\baselineskip}
\zwisch{Homogeneous energy momentum tensor}

With the ansatz \eqref{eq:8} the non-zero components of $T_{\mu\nu}^{(U)}$ are
\begin{align}
\label{96A}
T_{00}^{(U)} &= \frac{3m^{2}}{2}\big{[}c^{2} - (b+\hub)(b-\hub)\big{]}\nn\\
&-\frac{9\n}{2}(b+\hub)(b-\hub) \nonumber \\
&=-\frac{3\mtil}{2}(b+\hub)(b-\hub)+\frac{3\m}{2}c^{2}\ ,
\end{align}
and
\begin{align}
\label{FE11}
T_{kl}^{(U)} &= \frac{\mtil}{2}\big{[}2\deta(b-\hub) + (b+\hub)(b-\hub)\big{]}\delta_{kl}\nn\\
&-\frac{\m}{2}c^{2}\delta_{kl}\ ,
\end{align}
The trace,
\be
\label{FE12}
g^{\mu\nu}T_{\mu\nu}^{(U)} = \frac{3\mtil}{a^{2}}\big{[}\deta(b-\hub) + (b+\hub)(b-\hub)\big{]}-\frac{3\m}{a^{2}}c^{2}\ ,
\ee
vanishes for $f=0$, $c=0$. We observe a positive contribution to the energy density $T_{00}^{(U)}$ from $c$ if $\m>0$. On the other hand, the contribution from $b$ and $\hub$ has no definitive sign.

Similarly, we obtain for the gauge field contributions
\begin{align}
\label{98A}
T_{00}^{(F)} &= \frac{3Z}{2a^{2}}\big{[}(\deta c)^{2}-(\deta b)^{2} + b^{4}+c^{4}-6b^{2}c^{2}\big{]}\nn\\
&-\frac{6}{a^{2}}(B+3C)\big{[}(\deta b)^{2}-(b^{2}-c^{2})^{2}\big{]}\nn\\
&=-\frac{3\Ztil}{2a^{2}}\big{[}(\deta b)^{2}-(b^{2}-c^{2})^{2}\big{]}+\frac{3Z}{2a^{2}}\big{[}(\deta c)^{2}-4b^{2}c^{2}\big{]}\ ,
\end{align}
and
\begin{align}
\label{FE13}
T_{kl}^{(F)} &= \frac{Z}{2a^{2}}\big{[}(\deta c)^{2}-(\deta b)^{2} + b^{4}+c^{4}-6b^{2}c^{2}\big{]}\delta_{kl}\nn\\
&-\frac{2}{a^{2}}(B+3C)\big{[}(\deta b)^{2}-(b^{2}-c^{2})^{2}\big{]}\delta_{kl}\ ,
\end{align}
with
\bel{119A}
g^{\mu\nu}T_{\mu\nu}^{(F)} = 0\ .
\ee
For $Z>0$ the kinetic contribution to the energy density $T_{00}^{(F)}$ from $(\deta c)^{2}$ is positive, while the one $\sim (\deta b)^{2}$ would be negative for $B=C=0$. For $\Ztil<0$ the kinetic energy density $\sim (\deta b)^{2}$ is positive as well. This demonstrates the need of a parameter choice in an appropriate range. For positive kinetic energy densities of the gauge fields the quartic terms in $T_{00}^{(F)}$ are negative. Stability needs a domination of the quadratic terms $\sim b^{2}$, $c^{2}$ from $T_{00}^{(U)}$. Nevertheless, the negative quartic terms suggest that stability does not hold for very large values of $b^{2}$ or $c^{2}$. The discussion of $T_{00}^{(U)}+T_{00}^{(F)}$ underlines that stability is not a trivial issue.

The scalar contribution is given by
\begin{align}
\label{FE14}
T_{00}^{(\chi)} &= \frac{K}{2}(\deta\chi)^{2} + a^{2}V\ , \nonumber \\
T_{kl}^{(\chi)} &= \bigg{[}\frac{K}{2}(\deta\chi)^{2} - a^{2}V + (b- \hub)\frac{\partial \mtil}{\partial \chi}\deta\chi\bigg{]}\delta_{kl}\ .
\end{align}
The scalar energy density $T_{00}^{(\chi)}$ is positive for positive $K$ and $V$. Further, one finds
\begin{align}
\label{90A}
T_{00}^{(Y)} &= 3Y\hub\chi\deta\chi \nonumber\ , \\
T_{kl}^{(Y)} &= -Y\bigg{[}\bigg{(}1+\frac{\partial\ln Y}{\partial\ln\chi}\bigg{)}(\deta\chi)^{2} \\
&\quad\quad\quad+\chi\deta^{2}\chi+(2b-\hub)\chi\deta\chi\bigg{]}\delta_{kl}\nn\ .
\end{align}

Finally, for the homogeneous isotropic ansatz \eqref{eq:8} the non-vanishing components of $T_{\mu\nu}^{(\Delta)}$ are given by
\begin{align}
\label{FE21}
T_{00}^{(\Delta)} &= 3\M\big{[}c^{2}-(b+\hub)(b-\hub)\big{]}\ , \\
T_{kl}^{(\Delta)} &= \M\big{[}2\deta (b-\hub)+(b+\hub)(b-\hub)-c^{2}\big{]}\delta_{kl}\ .\nn
\end{align}
For $\M>0$ the contribution $T_{00}^{(\Delta)}$ to the energy density is positive for $c$ and $\hub$ and negative for $b$. Opposite signs for $\hub$ and $b$ obtain from eq. \eqref{FE12} if $\mtil$ is negative. This results in a positive energy density $\sim b^{2}$ if $-\mtil>2\M$. We note, however, that the part involving $b$ and $\hub$ vanishes for $b=\hub$. At this stage both signs of $\mtil$ seem possible.

\zwisch{Homogeneous vierbein field equations}

We can write the non-vanishing components of the field equation \eqref{FE3} in the form
\begin{align}
\label{FE15}
M^{2}(F_{00} -& \frac{1}{2}Fg_{00}) = 3M^{2}(b^{2}-c^{2}) \nonumber \\
&= T_{00}^{(U)} + T_{00}^{(F)} + T_{00}^{(\chi)}+T_{00}^{(Y)}\ ,
\end{align}
and
\begin{align}
\label{FE16}
\frac{M^{2}}{2}(F_{kl} + F_{lk} &- Fg_{kl}) = -M^{2}(2\deta b + b^{2} - c^{2})\delta_{kl} \nonumber \\
&= T_{kl}^{(U)} + T_{kl}^{(F)} + T_{kl}^{(\chi)} + T_{kl}^{(Y)}\ .
\end{align}
Combining eqs. \eqref{FE15}, \eqref{FE16} one finds
\begin{align}
\label{FE17}
-6&M^{2}(\deta b + b^{2} - c^{2}) \nonumber \\
&= a^{2}g^{\mu\nu}(T_{\mu\nu}^{(U)} + T_{\mu\nu}^{(F)} + T_{\mu\nu}^{(\chi)} + T_{\mu\nu}^{(Y)}) \nonumber \\
&= 3\mtil\big{[}\deta (b-\hub) + b^{2} -\hub^{2}\big{]}-3\m c^{2} \\
&+ K(\deta\chi)^{2} - 4a^{2}V +3(b-\hub)\frac{\partial \mtil}{\partial \chi}\deta\chi \nonumber \\
&-3Y\bigg{[}\bigg{(}1+\frac{\partial\ln Y}{\partial\ln\chi}\bigg{)}(\deta\chi)^{2}+\chi\deta^{2}\chi+2b\chi\deta\chi\bigg{]}\nn\ .
\end{align}
This reproduces eq. \eqref{eq:14}. Besides the four equations \eqref{eq:14}-\eqref{eq:17} we have an additional equation that we may take as eq. \eqref{FE15}. This equation is not independent, however, as we will show next.

\zwisch{Energy-Momentum conservation}

Due to the Bianchi-identity the effective energy momentum tensor $T_{\mu\nu}$ obeys the relation
\bel{103A}
D^{\mu}\Big{(}\frac{1}{\M} T_{\mu\nu}\Big{)} = 0\ .
\ee
Similar to general relativity the energy momentum tensor is covariantly conserved if $\M$ does not depend on $\chi$. For $\M$ depending on $\chi$ one can define a modified energy momentum tensor by multiplication with $M^{-2}$. This corresponds to the covariant conservation of the frame invariant energy momentum tensor.

With the relation \eqref{103A} the derivative of eq. \eqref{FE15} can be expressed as a linear combination of the other field equations. In more detail, $T_{\mu\nu}$ obeys the relation
\be
\label{FE23}
g^{\mu\rho}\partial_{\rho}T_{\mu\nu} - \Gamma_{\mu}^{\ \mu\rho}T_{\rho\nu}-\Gamma_{\ \nu}^{\mu\ \rho}T_{\mu\rho} = \frac{\partial\ln(\M)}{\partial\chi}\partial_{\rho}\chi T_{\ \nu}^{\rho}\ .
\ee
For homogeneous isotropic configurations this reads
\be
\label{FE24}
(\deta +2\hub)T_{00} + a^{2}\hub T_{\mu}^{\ \mu} = \dchi{\ln(\M)}\deta\chi T_{00}\ .
\ee
We can write the two field equations involving $T_{00}$ and $T_{kl}$ as
\be
\label{FE25}
T_{\mu}^{\ \mu}+\M R = 0\ ,
\ee
and
\be
\label{FE26}
T_{00}-3\M\hub^{2} = 0\ .
\ee
Eq. \eqref{FE24} can be written in the form
\be
\label{FE27}
(\deta +2\hub)(T_{00}-3\M\hub^{2})+a^{2}\hub(T_{\mu}^{\ \mu}+\M R)=0\ ,
\ee
demonstrating that the two equations \eqref{FE25} and \eqref{FE26} are not independent.

One may evaluate the different terms in eq. \eqref{FE27} explicitly. From eqs. \eqref{FE15}, \eqref{FE21} one infers
\begin{align}
\label{FE28}
a^{2}T_{\mu}^{\ \mu} &= 3\mtil(1+2\ytil)\big{[}\deta (b-\hub)+b^{2}-\hub^{2}\big{]}\\
&-3\m(1+2y)c^{2} \nonumber \\
&+3\dchi{\mtil}\deta\chi (b-\hub)+K(\deta\chi)^{2}-4a^{2}V\nn \\
&-3Y\bigg{[}\bigg{(}1+\frac{\partial\ln Y}{\partial\ln\chi}\bigg{)}(\deta\chi)^{2}+\chi\deta^{2}\chi+2b\chi\deta\chi\bigg{]}\ .\nn
\end{align}
On the other hand, one has
\begin{align}
\label{FE29}
T_{00}&-3\M\hub^{2} = -\frac{3\mtil}{2}(b^{2}-\hub^{2})+\frac{3\m}{2}c^{2} \\
&-3\M (b^{2}-c^{2})+\frac{K}{2}(\deta\chi)^{2}+a^{2}V+3Y\hub\chi\deta\chi \nonumber \\
&-\frac{3\Ztil}{2a^{2}}\big{[}(\deta b)^{2}-(b^{2}-c^{2})^{2}\big{]}+\frac{3Z}{2a^{2}}\big{[}(\deta c)^{2}-4b^{2}c^{2}\big{]}\ .\nn
\end{align}
It is straightforward to verify that eq. \eqref{FE24} holds identically by virtue of the scalar field equation \eqref{eq:15} and the gauge boson field equations \eqref{eq:16},\eqref{eq:17}. We conclude that the field equation \eqref{FE26} does not impose additional constraints. We may equivalently use eq. \eqref{FE26} or eq. \eqref{FE25} as independent geometric field equation.

It is instructive to evaluate the energy density for $Y=0$, $f=0$,
\begin{align}
\label{136A}
T_{00}&=\frac{3}{2}\big{(}\m+2\M\big{)}c^{2}+\frac{3Z}{2a^{2}}\big{[}(\deta c)^{2}-4\hub^{2} c^{2}\big{]}\\
&-\frac{3\Ztil}{2a^{2}}\big{[}(\deta\hub)^{2}-\hub^{4}-c^{4}+2\hub^{2}c^{2}\big{]}\nn\\
&+\frac{K}{2}(\deta\chi)^{2}+a^{2}V\ .\nn
\end{align}
Even for flat space, $\hub=0$, the energy density can become negative for $\Ztil<0$ and large values of $c$. Only for small values of $\hub$ and $c$, for which we can neglect the term $\sim\Ztil$ and the term $\sim Z\hub^{2}c^{2}$, the energy density is positive. This tells us that stability of flat space cannot simply be based on a positive energy theorem as for general relativity. A more detailed discussion is necessary. This restricts the range of stability in field space.

\zwisch{Hubble parameter}

Eq. \eqref{FE26} expresses the Hubble parameter as a function of the other fields,
\begin{align}
\label{FE30}
\hub^{2} &= (1+2\ytil)b^{2}-\bigg{(}\frac{\m}{\mtil}+2\ytil\bigg{)}c^{2}\\
&-\frac{K}{3\mtil}(\deta\chi)^{2}-\frac{2a^{2}V}{3\mtil}-\frac{2Y\hub\chi\deta\chi}{\mtil}\nn\\
&+\frac{1}{a^{2}\mtil}\Big{\{}\Ztil\big{[}(\deta b)^{2}-(b^{2}-c^{2})^{2}\big{]}-Z\big{[}(\deta c)^{2}-4b^{2}c^{2}\big{]}\Big{\}}\ .\nn
\end{align}
We can consider this relation as a partial solution or partial integral of the differential equations \eqref{eq:14}-\eqref{eq:17}. If we interpret $T_{\mu\nu}$ as a generalized version of the energy momentum tensor, eq. \eqref{FE26} is the same relation as for general relativity. In the low energy limit only the "light" fields contribute to $T_{00}$, which will take a familiar form. The limit $c=0$, $\deta\chi=0$ of eq. \eqref{FE30} yields eq. \eqref{FS12}.

Despite the formal analogies there are also important differences as compared to general relativity. The energy density $T_{00}$, as given by eq. \eqref{FE29}, depends on $\hub$. For $Y\neq0$ eq. \eqref{FE30} is a quadratic equation of $\hub$ whose solution will be somewhat lengthy. We note the negative signs of the scalar potential and kinetic term on the r.h.s. of eq. \eqref{FE30}. For configurations with $b=\hub$, $f=0$ the insertion of $b(\hub)$ switches this sign effectively.

\section{Weyl scaling and frame invariant field equations}
\label{section:WS}

In this section we discuss scalar-field-dependent conformal transformations of the vierbein and associated metric. Such Weyl transformations \cite{HWGE, RDMPI} are changes of field variables. For all quantities that can be derived from the quantum effective action the choice of field variables has no influence on observables provided they are transformed accordingly. Field relativity \cite{CWUWE, CWIQM} states that all models related by Weyl transformations describe the same physical content - they are equivalent. We can therefore discuss a whole class of seemingly different models at once. This is done by the use of frame invariant field equations \cite{CWPFVG}. The frame invariant field equations offer an important technical simplification since the number of coupling functions gets reduced. The equivalence of cosmological models related by Weyl scaling for the quantum effective action has been advocated in ref. \cite{CWCNC}. Detailed work has mapped explicitly many quantities relevant for cosmology between different metric frames \cite{FCDDGE, DTTG, FCFFG, CEJFC, DCECG, CCCO, CWEU, PEEJF, JIQG, JFICIM, KFCMI}.

\zwisch{Weyl scaling}

A conformal transformation or Weyl scaling multiplies the vierbein by a scalar function $w$, while the gauge fields $A_{\mu m n}$ are left invariant
\be\label{Eq: W01}
e_{\mu}{}^{m}=we'_{\mu}{}^{m}\;,\quad A_{\mu mn}=A'_{\mu mn}\;.
\ee
In our case $w$ is a function of the scalar field~$\chi$.
Accordingly, one has
\be\label{Eq: W02}
e=w^{4}e'\;,\quad g_{\mu\nu}=w^{2}g'_{\mu\nu}\;,\quad g^{\mu\nu}=w^{-2}g'^{\mu\nu}\;.
\ee
With invariant $F_{\mu\nu mn}$ the gauge boson kinetic term $\sim Z$ is invariant under Weyl scaling. On the other hand, one has
\be\label{Eq: W03}
F_{\mu\nu}=F'_{\mu\nu}\;,\quad F=w^{-2}F'\;,\quad eF=w^{2}e'F'\;,
\ee
such that also the terms $\sim B$, $C$ are invariant.

Observing for the Levi-Civita connection
\be\label{Eq: W04}
\Gamma_{\mu\nu}{}^{\rho}=\Gamma'_{\mu\nu}{}^{\rho}+\partial_{\mu} \ln w \,\delta_{\nu}^{\rho}+\partial_{\nu}\ln\! w \, \delta_{\mu}^{\rho}-g'^{\rho\lambda}g'_{\mu\nu}\partial_{\lambda}\ln w\;,
\ee
one obtains for the covariant derivative of the vierbein
\be\label{Eq: W05}
U_{\mu\nu}{}^{m}=w\Big{\lbrace} U'_{\mu\nu}{}^{m}-\partial_{\nu}\ln w\ e'_{\mu}{}^{m}+e'^{m\rho}\partial_{\rho}\ln w\ g'_{\mu\nu}\Big{\rbrace}\;.
\ee
For the transformed quantities covariant derivatives involve ${\Gamma'}_{\mu\nu}{}^{\rho}$ and indices are transformed or contracted with ${e'}_{\mu}{}^{m}$ or $g'_{\mu\nu}$, such that
\be\label{Eq: W06}
U_{\mu\nu\rho}=w^{2}\Big{\lbrace} U'_{\mu\nu\rho}-\partial_{\nu}\ln w\ g'_{\mu\rho}+\partial_{\rho}\ln w\ g'_{\mu\nu}\Big{\rbrace}\;.
\ee
Using $\partial_{\mu}\ln w=\big( \partial \ln w /\partial\chi\big)\partial_{\mu}\chi$ one finds
\begin{align}\label{Eq: W07}
eU^{\mu\nu\rho}&U_{\mu\nu\rho}=w^{2}e'\biggl{\lbrace} U'^{\mu\nu\rho}U'_{\mu\nu\rho}\nn\\
+&4\dfrac{\partial \ln w}{\partial\chi}{U'}_{\mu}{}^{\mu\nu}\partial_{\nu}\chi+6\Big(\dfrac{\partial \ln w}{\partial\chi}\Big)^{2}\partial^{\mu}\chi\partial_{\mu}\chi\biggr{\rbrace}\;.
\end{align}
For the scalar field we observe
\be\label{Eq: W08}
e\Bigl{\lbrace} \dfrac{K}{2}\partial^{\mu}\chi\partial_{\mu}\chi+V\Bigr{\rbrace}=e'\Bigl{\lbrace}\dfrac{w^{2}K}{2}\partial^{\mu}\chi\partial_{\mu}\chi+w^{4}V\Bigr{\rbrace}\;.
\ee

Let us consider the effective action \eqref{eq:6}. After Weyl scaling this keeps the same form in terms of ${e'}_{\mu}{}^{m}$, with coefficient functions changed according to
\begin{align}\label{Eq: W10}
m'^{2}=w^{2}m^{2}\;,\quad n'^{2}&=w^{2}n^{2}\ ,\quad \tilde{m}'^{2}=w^{2}\mtil\ ,\nn\\
M'^{2}=w^{2}M^{2}\ ,&\quad V'=w^{4}V\;, \nn\\
K'=w^{2}\bigg[K+3\mtil\Big(&\dfrac{\partial \ln w}{\partial\chi}\Big)^{2}+6Y\chi\dfrac{\partial\ln w}{\partial\chi}\bigg]\;,\nn\\
Y'=w^{2}\bigg[Y&+\dfrac{\mtil}{\chi}\dfrac{\partial\ln w}{\partial\chi}\bigg]\ ,
\end{align}
and invariant $Z$, $B$, $C$.
The coupling $Y$ needs to be kept for a closed set of invariants. 

\zwisch{Frame invariant coupling functions}

We can construct frame invariant combinations of coupling functions that have the same functional dependence on $\chi$ in all frames related by Weyl scaling. For this purpose we choose (besides $Z$, $B$, $C$)
\begin{align}\label{Eq: W11}
&\widehat{V}=\dfrac{V}{M^{4}}\;,\quad y=\dfrac{M^{2}}{m^{2}}\;,\quad \ytil=\frac{\M}{\mtil}\ ,\nn\\
\widehat{K}&=\dfrac{1}{M^{2}}\biggl{\lbrace}K+\dfrac{3\mtil}{4}\Bigl(\dfrac{\partial \ln M^{2}}{\partial\chi}\Bigr)^{2}-3Y\chi\dfrac{\partial\ln M^{2}}{\partial\chi}\biggr{\rbrace}\;,\nn\\
&\widehat{Y}=\dfrac{1}{M^{2}}\biggl{\lbrace}Y-\dfrac{\mtil}{2\chi}\, \dfrac{\partial \ln M^{2}}{\partial\chi}\biggr{\rbrace}\;.
\end{align}
They are the equivalent of similar frame invariant combinations in variable gravity \cite{CWPFVG}. In particular, $\widehat{V}\overline{M}^{4}$ corresponds to the scalar potential in the Einstein frame with fixed Planck mass $\overline{M}$ .

We can also define a frame invariant vierbein variable,
\be\label{Eq: W12}
\tilde{e}_{\mu}{}^{m}=M(\chi)e_{\mu}{}^{m}\;.
\ee
In terms of this variable the effective action reads
\begin{align}\label{Eq: W13}
\Gamma=\int_{x} \tilde{e}\,&\bigg{\lbrace}\dfrac{Z}{8}\tilde{F}_{\mu\nu\rho\sigma}\tilde{F}^{\mu\nu\rho\sigma}+\frac{B}{2}\tilde{F}_{\mu\nu}\tilde{F}^{\mu\nu}+\frac{C}{2}\tilde{F}^{2}\nn\\
&+\dfrac{1}{4y}\tilde{U}_{\mu\nu\rho}\tilde{U}^{\mu\nu\rho}+\frac{1}{6}\bigg{(}\frac{1}{\ytil}-\frac{1}{y}\bigg{)}\tilde{U}_{\mu\ \rho}^{\ \mu}\tilde{U}_{\nu}^{\ \nu\rho}-\dfrac{1}{2}\tilde{F}\nn\\
&+\dfrac{\widehat{K}}{2}\partial^{\mu}\chi\partial_{\mu}\chi+\widehat{V}+\widehat{Y}\chi\partial_{\nu}\chi\tilde{U}_{\mu}{}^{\mu\nu}\biggr{\rbrace}\;.
\end{align}
Here $\tilde{U}_{\mu\nu}{}^{m}=\tilde{D}_{\mu}\tilde{e}_{\nu}{}^{m}$ and indices are transformed and contracted with $\tilde{e}_{\mu}{}^{m}$ and $\tilde{g}_{\mu\nu}=M^{2}(\chi)g_{\mu\nu}$, which are also used for the Levi-Civita connection in the covariant derivative $\tilde{D}_{\mu}$. The effective action~\eqref{Eq: W13} involves only the frame invariant coupling functions. It is manifestly invariant under the Weyl scaling~\eqref{Eq: W01},\eqref{Eq: W10} since it only involves invariant quantities. The identity of the effective actions \eqref{Eq: W13} and \eqref{eq:6} can be verified by explicit computation.
The field equations derived by a variation of $\Gamma$ with respect to $\tilde{e}_{\mu}{}^{m}$ and $A_{\mu mn}$ are valid in all metric frames related by a conformal transformation. They follow from the field equations \eqref{FE1}, \eqref{FE3}, \eqref{FE7}, \eqref{FE8} by the replacement $e_{\mu}{}^{m}\rightarrow\tilde{e}_{\mu}{}^{m}$, $m^{2}\rightarrow 1/y$, $\mtil\to 1/\ytil$, $M^{2}\rightarrow 1$, $K\rightarrow\widehat{K}$, $V\rightarrow \widehat{V}$, $Y\rightarrow\widehat{Y}$.
Besides the advantage of validity in all frames they offer also a substantial technical simplification. 
The function $M^{2}(\chi)$ no longer appears, and all terms involving $\partial M^{2}/\partial\chi$ are omitted.
 
\zwisch{Frame invariant homogeneous field equations}

In the following we take again the simplified setting with $B=0$. For the homogeneous isotropic field equations one replaces the scale factor $a$ by the frame invariant scale factor
\be\label{Eq: W14}
A=M(\chi)a\;,
\ee
while $b, c$ and $\chi$ remain unchanged. With the corresponding shift to frame invariant coupling functions eq. \eqref{eq:13} becomes
\begin{align}\label{Eq: W15}
\Gamma=&\Omega_{3}\int_{\eta}\biggl{\lbrace}\frac{3}{2}\Big{[}\Ztil(\deta b)^{2}-Z(\deta c)^{2}+2(\Ztil-Z)(b^{2}-c^{2})\deta b\nn\\
&\quad\quad\quad\quad+\Ztil(b^{4}+c^{4})-2(\Ztil+2Z)b^{2}c^{2}\Big{]}\nn\\
&+\dfrac{3A^{2}}{2y}c^{2}-\frac{3A^{2}}{2\ytil}(b-\hub)^{2}-3A^{2}\bigl[\partial_{\eta}b+b^{2}-c^{2}\bigr] \nn\\
&-\dfrac{\widehat{K}A^{2}}{2}(\partial_{\eta}\chi)^{2}+A^{4}\widehat{V}+3\widehat{Y}A^{2}(b- \widehat{\hub})\chi\partial_{\eta}\chi\biggr{\rbrace}\;.
\end{align}

The Hubble parameter is replaced by the frame - invariant combination \cite{CWEU}
\be\label{Eq: W16}
\widehat{\hub}=\dfrac{\partial \ln A}{\partial\eta}=\hub+\dfrac{1}{2}\, \dfrac{\partial\ln M^{2}}{\partial\chi}\, \partial_{\eta}\chi\;.
\ee
It obeys according to eq. \eqref{FE30}
\begin{align}\label{Eq: W17}
\widehat{\hub}^{2}&=(1+2\ytil)b^{2}-\bigg{(}\frac{\ytil}{y}+2\ytil\bigg{)}c^{2}-\dfrac{1}{3}\ytil\widehat{K}(\partial_{\eta}\chi)^{2}-\dfrac{2}{3}\ytil A^{2}\widehat{V} \nn\\
&+\frac{\ytil}{A^{2}}\Big{\{}\Ztil\Big{[}(\deta b)^{2}-(b^{2}-c^{2})^{2}\Big{]}-Z\Big{[}(\deta c)^{2}-4b^{2}c^{2}\Big{]}\Big{\}}\nn\\
&-2\ytil\widehat{Y}\widehat{\hub}\chi\deta\chi\ .
\end{align}

We will often directly use the frame invariant field equations derived from the effective action \eqref{Eq: W13} or \eqref{Eq: W15}. Occasionally, we will also employ field equations in a given frame, say the Einstein frame. Replacing all quantities subsequently by frame invariant quantities such results are easily translated to other frames. For the Einstein frame one takes in eq. \eqref{Eq: W01} $w^{2}=\Mbar^{2}/\M(\chi)$, such that in eq. \eqref{Eq: W10} one has $M'^{2}=\Mbar^{2}$, with $\Mbar=2.44\cdot 10^{18}$ GeV the fixed observed Planck mass.

\zwisch{Frame invariant variable gravity}

Starting from the effective action \eqref{Eq: W15} we can compute the coupling functions of the effective low energy theory similar to sect. \ref{section:EVG}. For the corresponding frame invariant formulation of the variable gravity model one obtains the frame invariant kinetial
\bel{Eq: W18}
\tilde{\widehat{K}}=\widehat{K}-\frac{3\ytil}{1+2\ytil}\widehat{Y}^{2}\chi^{2}\ .
\ee
Equivalently, we could start with the effective action \eqref{23} for the variable gravity model and define the frame invariant couplings according to ref \cite{CWPFVG}, with $\widehat{V}=V/M^{4}$ and
\bel{Eq: W19}
\widehat{\tilde{K}}=\frac{\tilde{K}}{\M}+\frac{3}{2}\bigg{(}\dchi{\ln\M}\bigg{)}^{2}\ .
\ee
Inserting the relation \eqref{23A} the two procedures are equivalent
\bel{Eq: W20}
\tilde{\widehat{K}}=\widehat{\tilde{K}}\ .
\ee
Not surprisingly, a frame invariant formulation of pregeometry leads to a frame invariant effective theory of variable gravity.

\section{Quantum scale symmetry}
\label{section:QS}

In this section we introduce quantum scale symmetry \cite{CWQS} as a central ingredient for the understanding of the beginning of the universe in the infinite past, and its ending in the infinite future. We describe two different versions of scale transformations, relevant for an ultraviolet and an infrared fixed point, respectively. We will later understand the evolution of the universe as a crossover between the two fixed points.

Quantum scale symmetry will be one of the guides for specifying a concrete family of models by restricting the properties of the coupling functions. Indeed, for a more detailed cosmological model we need to specify the functions $Z(\chi), \Ztil(\chi), m^2(\chi), \mtil(\chi), M^2(\chi), V(\chi), K(\chi), Y(\chi),$ or the corresponding frame invariant combinations $Z, \Ztil, y, \ytil, \widehat{V}, \widehat{K}, \widehat{Y}$. We take the dimensionless functions $Z, \Ztil, m^2/k^2, \mtil/k^{2}, M^2/k^2, V/k^4$ to depend only on the ratio $\chi^2/k^2$. In case of fundamental scale invariance \cite{CWFSI} the scale $k$ is a ''renormalization scale`` rather than an intrinsic mass scale. In this case the coupling functions correspond to scaling solutions of functional flow equations for a scale dependent effective action $\Gamma_{k}$. This point is not crucial here, we also can consider the case where $k$ is an intrinsic mass scale. The frame invariant combinations $Z, \Ztil, y, \ytil, \widehat{V}$ are dimensionless and therefore only depend on the dimensionless ratio $\chi^2/k^2$. The same holds for $\chi^2 \widehat{K}$ and $\chi^2 \widehat{Y}$. 

\zwisch{Quantum scale symmetry at UV- and IR-fixed points}

Quantum scale symmetry \cite{CWQS} becomes exact for points or regions in field space for which the effective action does not involve any mass scale. In our formulation, $\Gamma_{k}$ has to become independent of $k$ \cite{CWFSI}. This statement needs to specify a choice of fields since some $k$-dependence may be absorbed by a suitable choice of fields. As a consequence, there exist different versions of quantum scale symmetry, with different transformation properties of the fields. Running couplings correspond to a dependence of dimensionless functions on the ratio $\chi^2/k^2$. Such a running violates quantum scale symmetry. At a fixed point the running stops and quantum scale symmetry is realized. 

We will consider two fixed points -- the ultraviolet (UV)-fixed point for $k^2/\chi^2 \to \infty$, and the infrared (IR)-fixed point for $k^2/\chi^2 \to 0$. For a fixed $k$ they are realized by the limits $\chi \to 0$ (UV) and $\chi \to \infty$ (IR). We will see that the quantum scale symmetry for the UV-fixed point differs from the one for the IR-fixed point. Interesting ''crossover cosmologies`` will be characterized by an evolution of the scalar field from $\chi \to 0$ for the infinite past to $\chi \to \infty$ for the infinite future. Such cosmologies describe a crossover from the UV-fixed point in the past to the IR-fixed point in the future. The approximate scale invariance of the primordial fluctuation spectrum will originate from the vicinity of the UV-fixed point, reflecting directly the approximate quantum scale symmetry.   

The quantum scale symmetry relevant for the IR-fixed point is a global symmetry corresponding to a multiplicative scaling of fields
\begin{equation}
e_{\mu}^{\ m} \to \alpha^{-1} e_{\mu}^{\ m}, \quad \chi \to \alpha \chi, \quad A_{\mu mn} \to A_{\mu mn}.
\label{eq:QS1} 
\end{equation}
Coordinates remain fixed. The effective action \eqref{eq:6} 
is invariant under this transformation provided $Z, \Ztil, Y$ and $K$ are constant, while 
\begin{equation}
M^2 =\tfrac{1}{2}\xi \chi^2, \dub m^2=\tfrac{1}{2} \zeta \chi^2, \dub \mtil=\tfrac{1}{2}\zetil\chi^{2}\ ,\dub V=\lambda \chi^4,
\label{eq:QS2} 
\end{equation} 
with constant dimensionless couplings $\lambda$, $\xi$, $\zeta$ and $\zetil$. Correspondingly, the frame invariant combinations behave as 
\begin{align}
\widehat{V}=\frac{\lambda}{\xi^2}, \quad &y=\frac{\xi}{\zeta}, \quad \ytil=\frac{\xi}{\zetil}\ ,\nn\\
\widehat{K}=\frac{1}{\xi \chi^2}(K+3 \tilde{\zeta}-6Y)\ ,&\quad \widehat{Y}=\frac{1}{\xi \chi^2}(Y-\tilde{\zeta})\ .
\label{eq:QS3} 
\end{align}
With our ansatz \eqref{S7},\eqref{S8} 
quantum scale symmetry is indeed realized in the IR-limit $\chi \to \infty$, provided $K$ and $Y$ approach constants. With 
\begin{equation}
\widehat{V}=\frac{u_0 k^4}{\xi^2 \chi^4} \to 0
\label{eq:QS4} 
\end{equation}
the effective coupling $\lambda=u_0 k^4/\chi^4$ goes to zero.

The quantum scale symmetry for the UV-fixed point $\chi \to 0$ is different. Now only the scalar field is transformed 
\begin{equation}
\chi \to \alpha \chi, \quad e_{\mu}^{\ m} \to e_{\mu}^{\ m}, \quad A_{\mu mn} \to A_{\mu mn}.	
\label{eq:QS5} 
\end{equation}
This symmetry requires a divergence of $K$ and $Y$ for $\chi \to 0$
\begin{equation}
\lim_{\chi \to 0} K(\chi)=\frac{\kappa k^2}{\chi^2}, \quad \lim_{\chi \to 0} Y(\chi)=\frac{Y_0 k^2}{\chi^2},
\label{eq:QS6} 
\end{equation}
with constant $\kappa$ and $Y_0$. The functions $Z$, $\Ztil$, $m^2$, $\mtil$, $M^2$, $V$ have to approach constants for $\chi \to 0$. This is indeed realized for our ansatz \eqref{S7},\eqref{S8}. The limiting behavior of the frame invariant quantities obeys for $\chi \to 0$
\begin{align}
\widehat{V}=\frac{u_0}{4 w_0^2}, \quad y&=\frac{2w_{0}}{\m_{0}}\ ,\quad \ytil=\frac{2 w_0}{\tilde{m}_0^2},\nn\\
\widehat{K}=\frac{\kappa}{2 w_0 \chi^2},& \quad \widehat{Y}=\frac{Y_0}{2 w_0 \chi^2}.
\label{eq:QS7} 
\end{align} 

We observe that in terms of frame invariant quantities the consequences of quantum scale symmetry are very similar for the UV- and IR-fixed points. In both cases $Z, \Ztil, \widehat{V}$, $y$ and $\ytil$ approach constants, while $\widehat{K}$ and $\widehat{Y}$ scale $\sim \chi^{-2}$. Indeed, in the frame invariant formulation \eqref{Eq: W13} 
quantum scale symmetry is realized for fixed $\tilde{e}_\mu^{\;\;m}$ and $A_{\mu mn}$ by the scaling \eqref{eq:QS5} of $\chi$. The different versions of quantum scale symmetry at the UV- and IR-fixed points are only connected to a different choice of fields. At the UV-fixed point we can choose a scaling frame with $M'^2(\chi \to 0)=\chi^2$. From the ansatz \eqref{S7}, \eqref{S8} this obtains by a Weyl scaling \eqref{Eq: W01}
with $w=\chi/M(\chi)$. For $\chi \to 0$ one has constant $\M(\chi)=2w_{0}k^{2}$ such that $w$ scales $\sim \alpha$. In consequence, the scale transformation reads in this frame
\begin{equation}
e^{\prime\ m}_{\mu} \to \alpha^{-1} e^{\prime\ m}_{\mu}, \quad \chi \to \alpha \chi,
\label{eq:QS8} 
\end{equation}
which is the same as for the IR-scaling \eqref{eq:QS1}. In this scaling frame the dilatation transformation takes its usual form and quantum scale symmetry can be defined by the absence of any mass scale once the fields $e^{\prime\ m}_{\mu}$ or $g^\prime_{\mu \nu}$ are used for the geometry. 

\section{Crossover models}
\label{section:CM}

In this section we propose a family of more concrete models of pregeometry. They are based on the relations \eqref{S7}, \eqref{S8}, motivated by asymptotic safety for quantum gravity, fundamental scale invariance and quantum scale symmetry. For simplicity we approximate $\M(\chi)$, $\m(\chi)$ and $\mtil(\chi)$ by eqs. \eqref{S7}, \eqref{S8} for the whole range of $\chi$. These functions entail a crossover between a qualitatively different behavior for $\chi\to 0$, where they are approximately constant, and for $\chi\to\infty$ where they scale $\sim\chi^{2}$. This change may be associated with a crossover from the vicinity of the UV-fixed point to the vicinity of the IR-fixed point. The characteristic field for the crossover in $\M$ is given by $\chi^{2}\approx 2w_{0}k^{2}/\xi$, while for $\m$ it occurs for $\chi^{2}\approx \m_{0}k^{2}/\zeta$. A crossover can occur in different steps if the two crossover scales are of different magnitude. This extends to the crossover in $\mtil(\chi)$.

For simplicity we consider here constant $Z$ and $\Ztil$. More generally, one expects that these quantities assume different values at the UV- and IR-fixed points. This would induce an additional scale of crossover that we neglect here. For the kinetial we assume for $\chi\to 0$ the form \eqref{eq:QS6} required by quantum scale symmetry. For $\chi\to\infty$ quantum scale symmetry requires that $K$ approaches a constant and we assume here
\bel{145A}
K(\chi)=\frac{\kappa k^{2}}{\chi^{2}}+K_{\infty}\ .
\ee
If the IR-fixed point exhibits conformal symmetry the value of $K_{\infty}$ is fixed. It is typically approached logarithmically \cite{CWIQM}. This will play an important role for the detailed evolution of dynamical dark energy in the late universe. We will not focus on this issue here and work with the simple form \eqref{145A}. Again, the effective scale of crossover in the kinetial may be different from the crossover in $\M$ or $\m$. Finally, for $Y$ we may take
\bel{145B}
Y(\chi)=\frac{Y_{0}k^{2}}{\chi^{2}}+Y_{\infty}.
\ee
We could actually consider $Y_{0}=0$. The frame invariant function $\Yhat$ would then be given by
\bel{145C}
\Yhat=-\frac{\xi(\mtil_{0}+\zetil\chi^{2})}{(2w_{0}+\xi\chi^{2})^{2}}\ .
\ee
We conclude that the frame invariant function $\Yhat$ differs from zero rather generically if $\xi\neq 0$.

In the frame with variables $e_{\mu}^{\ m}$ the divergence of $K$ for $\chi \to 0$ can be associated with an ''anomalous dimension``. Models of this type have been discussed in ref. \cite{CWIQM}
, where the behavior \eqref{eq:QS6} corresponds to $\sigma=2$. This type of models leads to a successful implementation of inflation. The increase of the kinetial $K$ can ensure the ''slow roll`` behavior for the evolution of the scalar field. We will come back to this issue in sect. \ref{section:IC}. It is, of course, possible to choose a different definition of the scalar field which changes the power of $K \sim \chi^{-\sigma}$. We will stick here to a normalization for which $\sigma=2$.  

\zwisch{Frame invariant formulation}

For a first investigation we will simplify further by assuming a single crossover, as given by the crossover in $\M$. This is an over-simplification, but many characteristic features of this family of models can be understood already at this level. We define our simple model directly in the frame invariant formulation by parametrizing  
\begin{equation}
\widehat{V}=\overline{V}(1+\frac{\chi^2}{\mu^2})^{-2}, \quad \widehat{K}=\frac{\overline{K}}{ \chi^2}, \quad \widehat{Y}=\frac{\overline{Y}}{\chi^2},
\label{eq:QS9} 
\end{equation}
and take constant $Z$, $\Ztil$, $y$ and $\ytil$. In this case the only violation of quantum scale symmetry arises from the scale $\mu$. In terms of our parameters it is given by the crossover scale,
\bel{146A}
\mu^{2}=\frac{2w_{0}k^{2}}{\xi}\ ,\quad \Vbar=\frac{u_{0}}{4w_{0}^{2}}\ .
\ee

The effective action of this rather simple model takes the frame invariant form
\begin{align}
\label{146B}
\Gamma=\int_{x}&\etil\bigg{\{}\frac{Z}{8}\Ftil_{\mu\nu\rho\sigma}\Ftil^{\mu\nu\rho\sigma}+\frac{\Ztil-Z}{24}\Ftil^{2}-\half\Ftil\nn\\
&+\frac{1}{4y}\Util_{\mu\nu\rho}\Util^{\mu\nu\rho}+\frac{1}{6}\bigg{(}\frac{1}{\ytil}-\frac{1}{y}\bigg{)}\Util_{\mu\ \rho}^{\ \mu}\Util_{\nu}^{\ \nu\rho}\nn\\
&+\frac{\Kbar}{2}\partial^{\mu}s\partial_{\mu}s+\Vhat+\Ybar\Util_{\mu}^{\ \mu\nu}\partial_{\nu}s\bigg{\}}\ ,
\end{align}
with
\bel{146C}
\Vhat=\Vbar(1+e^{2s})^{-2}\ ,\quad 2s=\ln\bigg{(}\frac{\chi^{2}}{\mu^{2}}\bigg{)}\ .
\ee
Up to the different variable for the scalar field and the different form of the potential this action has almost the same form as the effective action with constant coupling functions and $\chi\to s$. The only difference concerns the factor $\chi$ in the term $\sim Y$ which has no corresponding factor $s$ in eq. \eqref{146C}.

Besides $\mu$ the model has seven dimensionless parameters $Z, \Ztil, y, \ytil, \overline{V}, \overline{K}, \overline{Y}$. We will in the following concentrate on the effective action \eqref{146B} with constant frame invariant couplings except for $\Vhat$. In this formulation the crossover is completely encoded in the function $\Vhat(\chi)$. As we have argued before the true crossover in a model of quantum gravity may provide for a substantial richer picture of the crossover with possibly several rather distinct crossover scales. Exploiting this richness can give rise to a rather rich and realistic picture of the cosmological evolution. In the present paper we want to concentrate on the most characteristic features of cosmology and therefore concentrate on a single crossover scale as encoded in the simple effective action \eqref{146B}. A few remarks beyond this oversimplified picture, in particular concerning an additional crossover scale in the kinetial $\Kbar$, will be presented in the discussion of realistic inflationary scenarios in sect. \ref{section:IC}.

For $Y=0$ we can directly take over all our results of sect. \ref{section:FEHC}, \ref{section:CS} if we replace $\chi\to s$ and use frame invariant couplings. The only difference concerns the potential $V(s)$ in eq. \eqref{146C}. This does not matter for small $\chi$ or large negative $s$ for which $\Vhat$ is almost constant. For Fig. \ref{fig:3} we have already employed the potential for the frame invariant formulation with $\chi\to s$. The figures \ref{fig:A}, \ref{fig:1}, \ref{fig:3} are valid for the frame invariant formulation of the crossover model. In particular, for Fig. \ref{fig:A} we have set initial conditions already by the early attractor solution. The corresponding frame invariant couplings are $y=-\ytil=0.2$, $\Ybar=0.2$, $\Kbar=8$. Fig. \ref{fig:A} covers a much larger time interval than Figs. \ref{fig:1}-\ref{fig:3} which concentrate more on the approach to the early scaling solution. We can extend Fig. \ref{fig:A} to $t\to-\infty$ by starting with more and more negative $s$. We also can extend this figure to much larger values of $t$.

The homogeneous isotropic field equations take for $c=0$ the form
\begin{align}
\begin{split}
& \partial_\eta \widehat{\hub}+\widehat{\hub}^2=(1+2\ytil)(\partial_\eta b+ b^2)-\frac{4A^2 \overline{V}\ytil}{3(1+\chi^2/\mu^2)^2}\\
& \quad \quad +\frac{\overline{K}\ytil}{3}(\partial_\eta \ln \chi)^2-\overline{Y}\ytil(\partial_\eta^2 \ln \chi+2 b \partial_\eta \ln \chi),
\end{split}
\label{eq:QS10} 
\end{align}   
\begin{align}
\overline{K}(\partial_\eta^2 &\ln \chi+2\widehat{\hub}\partial_\eta \ln \chi)=4\overline{V}A^2 \tfrac{\chi^2}{\mu^2}(1+\tfrac{\chi^2}{\mu^2})^{-3} \nn \\
&+3\overline{Y}(\partial_\eta +2\widehat{\hub})(b-\widehat{\hub}),
\label{eq:QS11} 
\end{align}  
\begin{equation}
\partial_\eta^2 b=-\tfrac{A^2}{\Ztil}[\tfrac{1+2\ytil}{\ytil}(b-\widehat{\hub})-\overline{Y}\partial_\eta \ln \chi]+2b^3.
\label{eq:QS12} 
\end{equation}

Up to replacements of time-derivatives $\deta\chi$ by suitable time-derivatives of $\ln\chi$ the field equations \eqref{eq:QS10}-\eqref{eq:QS12} are identical to eqs. \eqref{eq:14}-\eqref{eq:16} for constant couplings. This holds provided we replace these constant couplings by the corresponding constant frame-invariant coupling functions.

\zwisch{Scale-free formulation}

The scale $\mu$ is the only mass scale appearing in the homogeneous field equations \eqref{eq:QS10},\eqref{eq:QS11}. It can be absorbed into the definition of a dimensionless scalar field,
\begin{equation}
s=\ln(\tfrac{\chi}{\mu}), \quad \tfrac{\chi^2}{\mu^2}=e^{2s}.
\label{eq:QS13} 
\end{equation}
Quantum scale transformations act as a constant additive shift in $s$. Quantum scale symmetry is realized for $s\to-\infty$ ($\chi\to 0$) where $\widehat{V}$ becomes a constant $\overline{V}$, and for $s\to\infty$ ($\chi\to\infty$) where $\widehat{V}$ vanishes identically.
By use of $s$ any explicit dependence on $\mu$ is eliminated, as visible in eq. \eqref{146B}. For $s\to\infty$ the potential is flat and $s$ corresponds to the massless Goldstone boson of spontaneously broken quantum scale symmetry.

For a theory with fundamental scale invariance \cite{CWFSI} the fields $\chitil=\chi/\mu$ and $\etil_{\mu}^{\ m}=Me_{\mu}^{\ m}$ are scale invariant fields. If the effective action is given by a scaling solution of functional renormalization equations (flow equations) the dimensionless coupling functions only depend on $\chitil$ or $s$. Couplings with dimension are multiplied by appropriate powers of the renormalization scale $k$, as $\M=2w(\chitil)k^{2}$. In consequence, $\mu^{2}$ is proportional to $k^{2}$, as seen for our explicit model. The use of the scale invariant variables $\chitil$, $\etil_{\mu}^{\ m}$, $A_{\mu mn}$ completely eliminates from the effective action any dependence on the renormalization scale $k$. This would be different in the presence of relevant parameters in the renormalization flow. In this case coupling functions do not only depend on $\chitil$, but in addition on $k/\mubar$ with $\mubar$ some intrinsic scale generated by the running of relevant parameters away from the scaling solution.

\zwisch{Scale free and frame invariant homogeneous field equations}

For the homogeneous field equations we can make a further step by eliminating the dependence on the frame invariant scale factor $A$. Indeed, the explicit dependence of the field equations \eqref{eq:QS10}-\eqref{eq:QS12} on $A$ can be absorbed by the definition of a new time coordinate $\hat{t}$, 
\begin{equation}
d\hat{t}= A \hspace{0.1cm}d\eta, \quad \widehat{H}=\partial_{\hat{t}} \ln A=\tfrac{\widehat{\hub}}{A}\ ,
\label{eq:QS14} 
\end{equation} 
and use of the variables
\bel{151A}
\bhat=\frac{b}{A}\ ,\quad \chat=\frac{c}{A}\ .
\ee
The new time variable is a type of frame invariant dimensionless cosmic time,
\bel{172A}
\text{d}\that=Ma\text{d}\eta=M\text{d}t\ .
\ee

In terms of these variables the evolution equations for the gauge fields read for $B=0$
\begin{align}
\label{151B}
\bhat''+3\Hhat\bhat'&=-(\Hhat'+2\Hhat^{2})\bhat+\frac{\hat{\mu}_{b}^{2}}{2\ytil}(\bhat-\Hhat)\nn\\
&+2(\bhat^{2}-\chat^{2})\bhat-\frac{4Z}{\Ztil}\chat^{2}\bhat\nn\\
&+2\chat\Big{(}1-\frac{Z}{\Ztil}\Big{)}(\chat'+\Hhat\chat)+\frac{\Ybar}{\Ztil}s'\ ,
\end{align}
and
\begin{align}
\label{151C}
\chat''&+3\Hhat\chat'=-\big{(}\hat{\mu}_{c}^{2}+2\Hhat^{2}+\Hhat'\big{)}\chat\nn\\
&+\frac{2\chat}{Z}\Big{[}\Ztil(\bhat^{2}-\chat^{2})+2Z\bhat^{2}+(\Ztil-Z)(\bhat'+\Hhat\bhat)\Big{]}\ .
\end{align}
Here primes denote derivatives with respect to $\that$. The mass terms are given by
\bel{151D}
\hat{\mu}_{b}^{2}=-\frac{2(1+2\ytil)}{\Ztil}\ ,\quad \hat{\mu}_{c}^{2}=\frac{1+2y}{Zy}\ .
\ee
The geometric field equation reads
\begin{align}
\label{151E}
\Hhat'&+2\Hhat^{2}=(1+2\ytil)(\bhat'+\Hhat\bhat+\bhat^{2})\nn\\
&-\ytil\bigg{\{}\frac{4\Vbar}{3}-\frac{\Kbar}{3}s'^{2}+Z\hat{\mu}_{c}^{2}\chat^{2}+\Ybar\big{[}s''+(2\bhat+\Hhat)s'\big{]}\bigg{\}}\ ,
\end{align}
and for the scalar field equation one obtains
\bel{151F}
\Kbar(s''+3\Hhat s')=-\frac{\partial\Vhat}{\partial s}+3\Ybar(\bhat'-\Hhat'+3\bhat\Hhat-3\Hhat^{2})\ .
\ee
Comparison with eqs. \eqref{MS8}-\eqref{MS11} reveals a very similar structure if one replaces $\chi\to s$, $Y\chi\to\Ybar$. Only the term $\sim Y(\dt\chi)^{2}$ in eq. \eqref{MS10} is now absent. Eqs. \eqref{151B}-\eqref{151F} can be considered as the central homogeneous field equations for our crossover model of pregeometry. They are used for the numerical solutions.

\begin{figure}[h]
\centering
\includegraphics[width=0.5\textwidth]{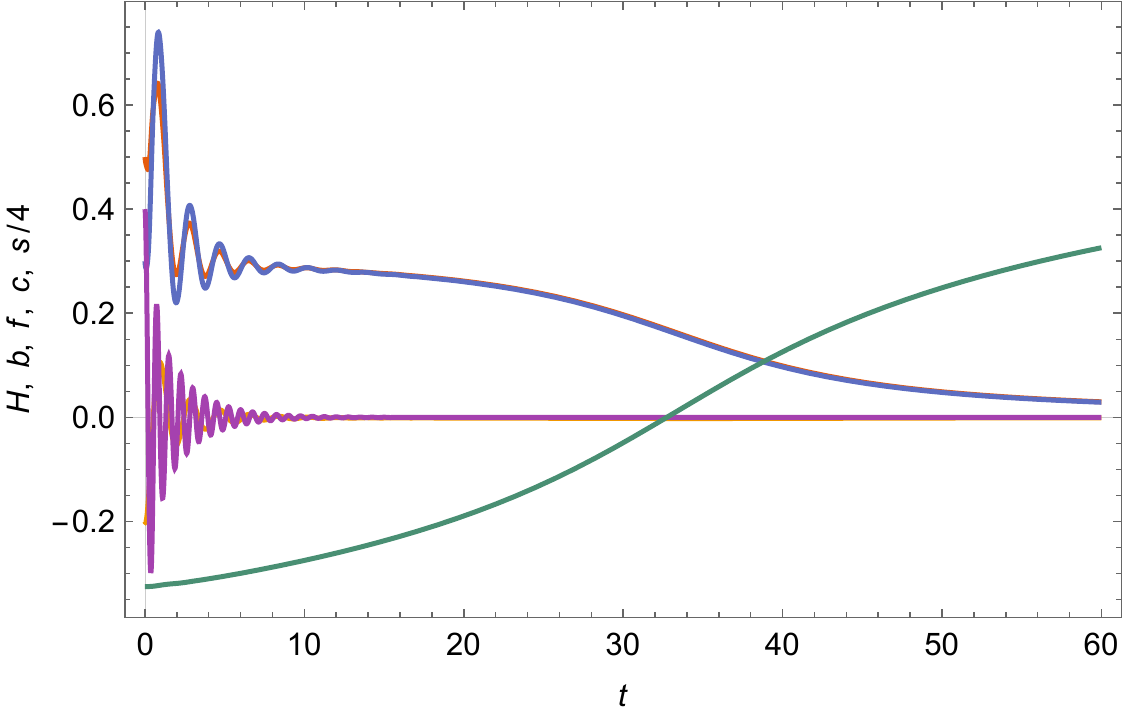}
\caption{\emph{Evolution in the crossover model}. We show the frame invariant fields $\Hhat$ (red), $\bhat$ (blue), $\fhat$ (yellow), $\chat$ (magenta) and $s/4$ (green) as function of $\that$. The functions $H$ and $b$ are almost indistinguishable on this scale, and the same holds for the pair $c$ and $f$. The parameters $Z=0.1$, $\Ztil=-0.1$, $y=0.2$, $\ytil=-0.2$, $\Kbar=4$, $\Vbar=0.3$, $\Ybar=0.1$ correspond to the ones if Figs. \ref{fig:1}, \ref{fig:2} except for $Z$, $\Ztil$, $\Ybar$. We employ here constant frame invariant coupling functions instead of constant coupling functions. The resulting sequence of initial damping, de Sitter solution and transition to flat space is very similar to Figs. \ref{fig:1}, \ref{fig:2}.}
\label{fig:4}
\end{figure}

For the crossover model the late time instability for $Y\neq 0$ is no longer present. We show a numerical solution of the system of differential equations \eqref{151B}-\eqref{151F} in Fig. \ref{fig:4}. Parameters are $y=0.2$, $\ytil=-0.2$, $Z=0.1$, $\Ztil=-0.1$, $\Vbar=0.3$, $\Kbar=4$, $\Ybar=0.1$. This plot shows the frame invariant functions $\Hhat$, $\bhat$, $\fhat$, $\chat$ and $s$ as functions of $\that$. The result is very similar to Figs. \ref{fig:1}, \ref{fig:2} if we take the replacements given above. In contrast to Figs. \ref{fig:1}, \ref{fig:2} we take here $\Ybar=0.1$ since there is no longer a late-time instability for non-vanishing $\Ybar$.
Another difference as compared to Fig. \ref{fig:2} concerns the smaller values of the couplings $Z$, $\Ztil$. According to eq. \eqref{151D} this leads to larger mass terms $\mubhat$, $\muchat$, and therefore to shorter oscillation periods.

For more negative initial $s$ the solution approaches the de Sitter solution \eqref{S1}-\eqref{S4} after a short initial oscillatory evolution. Due to the slow increase of $s$ it remains for many $e$-foldings very close to a de Sitter solution. Only once $s$ increases to positive values the de Sitter solutions ends and the solution makes a transition towards flat space. This demonstrates the natural occurrence of an inflationary epoch.

\zwisch{Evolution of the covariant vierbein derivative}

For the parameters used in Figs. \ref{fig:A}, \ref{fig:1}, \ref{fig:2}, \ref{fig:4} variable gravity becomes a valid approximation already in very early stages of cosmology. For $\Ybar=0$ the covariant derivative of the vierbein vanishes rather rapidly, while for $\Ybar\neq 0$ it approaches rapidly a fixed function. Indeed, in Fig. \ref{fig:4} we also show the combination
\begin{equation}
\fhat=\frac{b-\widehat{\hub}}{A}.
\label{eq:QS15} 
\end{equation} 
It approaches rapidly a very small value. Also $\chat$ goes rapidly to zero. As a result, the cosmological evolution leads rather rapidly to a very small value of the covariant derivative of the vierbein and the variable gravity approximation becomes valid. The frequencies of the initial oscillations of $\fhat$ and $\chat$ are actually different (barely visible in Fig. \ref{fig:4}), but the main message is that for $\that\gtrsim10$ the difference of both fields from zero becomes too small to remain visible.

It is instructive to understand the evolution of $\fhat$ analytically. For this purpose we consider solutions with $\chat=0$.
The field equation for $\fhat$ follows from eq. \eqref{eq:QS12},
\begin{align}
\label{eq:QS18}
\fhat^{\prime\prime}&+3\Hhat\fhat^{\prime}=-4\Hhat\Hhat'-\Hhat''+\frac{\overline{Y}}{\Ztil}s'\\
&-\frac{1+2\ytil}{\ytil\Ztil}\fhat+(4\Hhat^{2}-\Hhat')\fhat+6\Hhat\fhat^{2}+2\fhat^{3}\nn\ .
\end{align}
We observe a "source term" for $\fhat$ which does not vanish for $\fhat=0$. For a non-vanishing source term the field $\fhat$ does not vanish even for late time. As a consequence, the covariant derivative of the vierbein does not vanish. Cartan's geometry does not become exact for asymptotically large times.

We also express the other field equations in terms of $\fhat$ by eliminating $\bhat$ in favor of $\Hhat$ and $\fhat$.
In terms of $\fhat$ eq. \eqref{eq:QS10} becomes
\begin{align}
\label{eq:QS16}
6(\widehat{H}^{\prime}&+2\widehat{H}^2)=-\frac{3(1+2\ytil)}{\ytil}(\fhat^\prime + 3\widehat{H}\fhat+\fhat^2)\\
&+4\overline{V}(1+e^{2s})^{-2}-\overline{K}s^{\prime 2}+3\overline{Y}(s^{\prime \prime}+3\widehat{H}s^{\prime}+2\fhat s^{\prime})\ .\nn
\end{align} 
An equivalent relation obtains from eq. \eqref{Eq: W17},
\begin{align}
\label{eq:QS17C}
\widehat{H}^{2}&=\frac{\overline{V}}{3}(1+e^{2s})^{-2}+\frac{\overline{K}}{6}s^{\prime 2}+\overline{Y}\widehat{H}s^{\prime}\\
&-\Ztil(\widehat{H}^{2}\widehat{H}^{\prime}+\half\widehat{H}^{\prime 2})+\Delta_{1}+\Delta_{2}\nn\ ,
\end{align}
with
\bel{eq:QS17D}
\Delta_{1}=-\frac{1+2\ytil}{2\ytil}(2\widehat{H}\fhat+\fhat^{2})+\Ztil(\widehat{H}^{3}-\widehat{H}\widehat{H}^{\prime})\fhat
\ee
and
\begin{align}
\label{eq:QS17E}
\Delta_{2}=\frac{\Ztil}{2}\Big{[}&5\widehat{H}^{2}\fhat^{2}+4\widehat{H}\fhat^{3}+\fhat^{4}\nn\\
&-2(\Hhat^{2}+\Hhat^{\prime}+\Hhat\fhat)\fhat^{\prime}-\fhat^{\prime 2}\Big{]}\ .
\end{align}

The scalar field equation \eqref{eq:QS11} reads
\begin{equation}
\overline{K}(s^{\prime \prime} +3 \widehat{H}s^\prime)= 4\overline{V}e^{2s}(1+e^{2s})^{-3}+3\overline{Y}(\fhat^\prime+3\widehat{H}\fhat)\ .
\label{eq:QS17}
\end{equation}
For $Y\neq 0$, which is generic, we observe an additional source term for $\fhat\neq 0$, which is of the type discussed in ref. \cite{CWCNC}. As a consequence, the frame invariant version of the energy momentum tensor for the scalar field is not conserved separately. We will interpret this term in the context of an interaction between dark energy and dark matter in sect. \ref{section:DMDE}.

For our ansatz \eqref{eq:QS9} we may directly express the effective action \eqref{Eq: W15} in terms of the variables \eqref{eq:QS13}-\eqref{151A}, \eqref{eq:QS15}, which reads for $c=0$
\begin{align}
\label{eq:QS19A}
\Gamma=\Omega_{3}\int_{\that}A^{3}\bigg{\{}&\frac{3\Ztil}{2}\Big{[}2\Hhat^{4}+2\Hhat^{2}\Hhat'+\Hhat'^{2}-2(\Hhat''+4\Hhat\Hhat')\fhat\nn\\
&+(4\Hhat^{2}-\Hhat')\fhat^{2}+\fhat'^{2}+4\Hhat\fhat^{3}+\fhat^{4}\Big{]}\nn\\
&-\frac{3}{2\ytil}\fhat^{2}-3(2\Hhat^{2}+\Hhat'+\fhat^{2})-\frac{\overline{K}}{2}s'^{2}\nn\\
&+3\overline{Y}\fhat s'+\Vbar(1+e^{2s})^{-2}\bigg{\}}\ .
\end{align}
We have employed here partial integration in order to eliminate terms linear in $\fhat'$. The field equations \eqref{eq:QS17} and \eqref{eq:QS18} follow directly from the variation of $\Gamma$ with respect to $s$ and $\fhat$. The derivation of the field equations for $A$ is more cumbersome in this formulation due to the appearance of up to three derivatives of $A$ in $\Gamma$.

For $\overline{Y}=0$ and $\Hhat'=\fhat'=s'=0$, $s\to-\infty$ we recover the algebraic equations \eqref{FS8} for the de Sitter solutions in sect. \ref{section:CS}. In this case one has $A=Ma$, $\that=Mt$, $\Hhat=H/M$, $\fhat=\ftil/M$, with constant $M=M(\chi=0)$. For $\overline{Y}\neq 0$ the de Sitter solutions of the second family are no longer consistent with the scalar field equation if $H\ftil\neq 0$. Only the first family of solutions with $\ftil=0$ persists in this case for $\chi=0$ or $s\to-\infty$. For $\overline{Y}\Hhat\fhat\neq 0$ one can still have solutions with constant $\Hhat$, $\fhat$ and $s$, now for finite $s$ as given by
\bel{eq:QS19}
4\overline{V}e^{2s}(1+e^{2s})^{-3}+9\overline{Y}\Hhat\fhat=0\ .
\ee

The general solution for $\fhat$ is characterized by two different regimes. For the early regime one can neglect the evolution of $s$, such that one approximates the derivative of eq. \eqref{eq:QS16} by
\bel{159A}
\Hhat''+4\Hhat\Hhat'=-\frac{1+2\ytil}{2\ytil}\big{(}\fhat''+3\Hhat\fhat'+3\Hhat'\fhat+2\fhat\fhat'\big{)}\ .
\ee
Insertion into eq. \eqref{eq:QS18} yields
\begin{align}
\label{159B}
\fhat''+3\Hhat\fhat'&+\big{[}\hat{\mu}_{b}^{2}+8\ytil\Hhat^{2}+(3+4\ytil)\Hhat'\big{]}\fhat+2(1+\ytil)\fhat\fhat'\nn\\
&+12\ytil\Hhat\fhat^{2}+4\ytil\fhat^{3}=0\ .
\end{align}
This describes damped oscillations, for which the Hubble damping drives $\fhat$ rapidly towards zero. For the late regime we can linearize in $\fhat$. The evolution of $\fhat$ is typically slow and given in leading order by
\bel{159C}
\fhat=\Big{(}\frac{\hat{\mu}_{b}^{2}}{2\ytil}+4\Hhat^{2}-\Hhat'\Big{)}^{-1}\Big{(}4\Hhat\Hhat'+\Hhat''-\frac{\Ybar}{\Ztil}s'\Big{)}\ .
\ee
The first stage is visible in Figs. \ref{fig:1}, \ref{fig:2}, \ref{fig:4} while a better resolution would be needed in order to see the small deviation from zero in the late stage. On top of this slow evolution there may be very small damped oscillations that we discuss in sect. \ref{section:DMDE}.

\zwisch{Variable gravity}

For small $\fhat$ and small curvature one finds for the effective model of variable gravity an effective action
\bel{eq:QS20}
\Gamma=\int_{x}\etil\bigg{\{}-\half \Rtil+\half\khat^{2}\ \partial^{\mu}s\partial_{\mu}s+\Vhat\bigg{\}}\ ,
\ee
with
\bel{eq:QS21}
\Vhat=\Vbar(1+e^{2s})^{-2}\ ,
\ee
and
\bel{eq:QS22}
\khat^{2}=\Kbar-\frac{3\ytil}{1+2\ytil}\Ybar^{2}=\widehat{\tilde{K}}\chi^{2}\ .
\ee
Here $\Rtil$ is the curvature scalar formed from $\gtil_{\mu\nu}$. In this version it is obvious that the criterion for stability is a positive scalar kinetic term $\khat^{2}>0$, together with some boundedness properties of $\Vhat$ which are obeyed.

For constant $\khat^{2}$ the scalar field equation reads
\bel{eq:QS23}
\khat^{2}D^{\mu}D_{\mu}s=\frac{\partial\Vhat}{\partial s}\ .
\ee
For the independent geometrical field equation we can take
\bel{eq:QS24}
\Rtil_{00}-\half\Rtil\gtil_{00}=\khat^{2}(\partial_{0}s\partial_{0}s-\half\partial^{\rho}s\partial_{\rho}s\ g_{00})-\Vhat\gtil_{00}\ .
\ee
The homogeneous isotropic equations read in this frame invariant form
\bel{eq:QS25}
\khat^{2}(s''+3\Hhat s')=-\frac{\partial\Vhat}{\partial s}\ ,
\ee
with frame invariant Hubble parameter given by
\bel{eq:QS26}
\Hhat^{2}=\frac{1}{3}(\Vhat+\half\khat^{2}s'^{2})\ .
\ee
The two equations \eqref{eq:QS25}, \eqref{eq:QS26} are closed. They constitute an enormous simplification as compared to the system of field equations \eqref{151B}-\eqref{151F}. In case of additional fields for matter and radiation there will be additional contributions to $\Hhat^{2}$ in the usual way. This includes small fluctuations of $\chat$ or $\fhat$ acting as cold dark matter. In case of masses depending on $s$ also the r.h.s. of eq. \eqref{eq:QS23} contains an additional source term.

\zwisch{Einstein frame}

The frame invariant formulation allows for a very simple transition to the familiar Einstein frame, which is characterized by a constant Planck mass, $\M=\overline{M}^{2}$. The transition simply provides for canonical mass dimensions by inserting appropriate powers of $\overline{M}=2.44\cdot 10^{18}$ GeV according to
\begin{align}
\label{eq:QS27}
\etil_{\mu}^{\ m}&=\Mbar\e\ ,\quad \gtil_{\mu\nu}=\Mbar^{2}g_{\mu\nu}\ ,\quad A=\Mbar a\ ,\nn\\
\Hhat&=\frac{H}{\Mbar}\ ,\quad \that=\Mbar t\ ,\quad s=\frac{\phitil}{\Mbar}\ ,
\end{align}
where $t$ is cosmic time. All solutions of the frame invariant field equations can be translated directly to the Einstein frame. We observe that conformal time is frame invariant, while $\that$ is related to cosmic time by eq. \eqref{eq:QS27}.

For the limit of variable gravity the field equations take the familiar form for a scalar field coupled to gravity, with dots denoting derivatives with respect to $t$, namely
\bel{eq:QS28}
\khat^{2}(\ddot{\phitil}+3H\dot{\phitil})=-\frac{\partial V}{\partial\phitil}\ ,\quad V=\Mbar^{4}\Vhat\ ,
\ee
and
\bel{eq:QS29}
H^{2}=\frac{1}{3\Mbar^{2}}(V+\half\khat^{2}\dot{\phitil}^{2})\ .
\ee

The Einstein frame is particularly useful after the end of inflation. For late cosmology the approximate quantum scale symmetry near the IR-fixed point results in constant ratios between particle masses and the Planck mass \cite{CWQS}. Keeping a fixed value of $\M$ one therefore also has fixed particle masses. This facilitates many practical computations since varying masses need not to be taken into account. The corresponding picture is the hot big bang. We recall, nevertheless, that the hot big bang is a particular picture, not physical reality. The frame invariant decrease of the ratio between temperature and particle masses can be described as well in a frame with increasing particle masses.

\section{Dark matter and dynamical dark \\ energy}
\label{section:DMDE}

At the end of inflation starts a transition epoch which can be rather complex. Entropy can be produced effectively through rapidly oscillating modes. This process can be associated with a "heating" of the universe after inflation. In a more complete model additional particles will play a role, as fermions or the gauge bosons of the standard model of particle physics or some grand unified extension. For a realistic cosmology one expects a transition to a radiation dominated epoch, followed later by matter domination, and finally the onset of dark energy domination in the present epoch. We do not aim here for the description of cosmology with all the additional degrees of freedom from particle physics. We only show here that the present model of pregeometry has the ingredients to provide both for dark matter and dynamical dark energy.

Dynamical dark energy can be associated with the evolution of the scalar field $s$. It consists of the potential energy $\Vhat$ and the kinetic energy $\sim\khat^{2}s'^{2}/2$ of a homogeneous field. The crossover model leads to a potential that vanishes for large $s$ exponentially. Together with suitable properties of the kinetial $\khat^{2}$ this can induce a dynamical dark energy that will relax to zero in the infinite future \cite{Wetterich_1988}. Dark matter can arise from the oscillations of the gauge fields $\chat$ or $\bhat$. For the range of parameters needed for the stability of flat space the rapid fluctuations can be associated to particles with masses of the order of the Planck mass or larger. If these masses depend on the scalar field $s$ one will find versions of "coupled dark energy", with a non-zero interaction between dark energy and dark matter.

We will investigate these phenomena by an exploration of the solutions of the field equations for the present model of pregeometry without additional particles. The main effects of these neglected particles are twofold. Once produced after the end of inflation they are expected to dominate the energy density in the radiation dominated epoch. In turn, this will modify the evolution of the Hubble parameter and of the scalar field $s$. Second, it is possible that the dark matter candidates arising from the fluctuations of the gauge bosons $\chat$ and $\bhat$, or from other fluctuations of the gauge bosons and vierbein not considered here, can decay into the particles of the standard model. In this case they will play no role for the late epochs of cosmology and the presently observed dark matter. If one of the heavy excitations is stable, however, it could constitute the observed dark matter. In this case the dynamics has to be such that the energy density stored in the rapid oscillations is small enough in order not to dominate the universe too early.

Within the solutions of the field equations for our model we will investigate the late epochs where $\Hhat^{2}$ is already a small quantity. One expects the validity of the variable gravity approximation. In the presence of the rapidly oscillating gauge fields this is supplemented, however, by an effective energy momentum tensor for non-relativistic dark matter. Nevertheless, the field equations can be simplified considerably by neglecting higher powers of $\Hhat$, $\bhat$, and $\chat$.

\zwisch{Dark matter from $\mathbf{\chat}$-oscillations}

For sufficiently small $\chat$ we can linearize the field equation for this gauge field. Since the effective action contains no linear terms in $\chat$ there will be no source term. The evolution of $\chat$ can be considered independently of other small oscillations for $\Hhat$ and $\bhat$ or $\fhat$. For these fields we can use a "background solution" that can be influenced by other ingredients, including additional degrees of freedom in a more realistic model of particle physics.

Neglecting terms $\sim\chat^{3}$ we can write the field equation \eqref{151C} in the form
\bel{CC1}
\chat''+3\Hhat\chat'+(\muchat+\frac{9\Hhat^{2}}{4}+\frac{3\Hhat'}{2})\chat=-\delta\muchat\chat\ ,
\ee
with
\begin{align}
\label{CC2}
\delta\muchat&=-\Big{(}\frac{9}{4}+\frac{4\Ztil}{Z}\Big{)}\Hhat^{2}+\Big{(}\frac{3}{2}-2\frac{\Ztil}{Z}\Big{)}\Hhat'-6\Big{(}1+\frac{\Ztil}{Z}\Big{)}\Hhat\fhat\nn\\
&-2\Big{(}2+\frac{\Ztil}{Z}\Big{)}\fhat^{2}+2\Big{(}1-\frac{\Ztil}{Z}\Big{)}\fhat'\ .
\end{align}
In terms of the variable
\bel{CC3}
\Ccal=A^{\frac{3}{2}}\chat
\ee
this reads
\bel{CC4}
\Ccal''+\muchat\Ccal=-\delta\muchat\Ccal\ .
\ee
A late time $\delta\muchat$ is much smaller than $\muchat$ and we may neglect it in a first approximation. The solution
\bel{CC5}
\Ccal=\Ccal_{0}\cos(\muchat\that+\alpha_{c})
\ee
describes oscillations with constant
\bel{CC6}
\eps_{c}=\Ccal'^{2}+\muchat\Ccal^{2}=\muchat\Ccal_{0}^{2}\ .
\ee

\zwisch{Energy density for non-relativistic dark matter}

The oscillating field contributes to the energy density $\That_{00}$,
\begin{align}
\label{CC6}
\rhohat^{(c)}&=\frac{1}{A^{2}}\big{(}\That_{00}^{(U,c)}+\That_{00}^{(F,c)}+\That_{00}^{(\Delta,c)}\big{)}\\
&=\frac{3}{2}\bigg{\{}\frac{1+2y}{y}\chat^{2}+Z\big{[}(\chat'+\Hhat\chat)^{2}-4\bhat^{2}\chat^{2}\big{]}-2\Ztil\bhat^{2}\chat^{2}\bigg{\}}\nn\\
&=\frac{3Z}{2A^{3}}\bigg{\{}\muchat\Ccal^{2}+\Ccal'^{2}-\Hhat\Ccal\Ccal'\nn\\
&\quad\quad\quad+\Big{(}\frac{\Hhat^{2}}{4}-4\bhat^{2}-\frac{2\Ztil}{Z}\bhat^{2}\Big{)}\Ccal^{2}\bigg{\}}\ .\nn
\end{align}
Eq. \eqref{FE26} or \eqref{eq:QS29} for the Hubble parameter,
\bel{CC7}
\Hhat^{2}=\frac{1}{3A^{2}}\That_{00}=\frac{\rho^{2}}{3}+\dots
\ee
receives from the oscillating $c$-field a contribution
\bel{CC8}
\Hhat_{c}^{2}=\frac{Z}{2A^{3}}(\eps_{c}+\delta_{c})
\ee
with
\bel{CC9}
\delta_{c}=\Big{(}\frac{\Hhat^{2}}{4}-4\bhat^{2}-\frac{2\Ztil}{Z}\bhat^{2}\Big{)}\Ccal^{2}-\Hhat\Ccal\Ccal'\ .
\ee

The leading contribution is given by the constant $\eps_{c}$. Neglecting $\delta_{c}$ the oscillating $c$-field gives a contribution to cold dark matter. In the Einstein frame one has
\bel{CC10}
H^{2}=\frac{1}{3\Mbar^{2}}\rho
\ee
where the $c$-oscillations contribute to the energy density $\rho$,
\bel{CC11}
\rho_{c}=\frac{3Z\eps_{c}\Mbar}{2a^{3}}=\frac{3(1+2y)}{2y}\bigg{(}\frac{a_{c}}{a}\bigg{)}^{3}\chat_{c}^{2}\Mbar^{4}\ .
\ee
This contribution decays $\sim a^{-3}$, as characteristic for cold dark matter.

For a given model and given cosmological solution the amount of dark matter from the $c$-fluctuations is a computable quantity. In eq. \eqref{CC11} the constant $a_{c}$ is the scale factor at some time $t_{c}$ at which $\chat$ takes the value $\chat_{c}$. The time $t_{c}$ should be chosen such that the leading approximation becomes valid. For a realistic cosmology one needs at a time somewhat after the end of inflation a very small value $\chat_{c}$. Otherwise the dark matter component of the oscillating $c$-field would dominate the evolution too early, in contrast to the observed extended radiation dominated epoch. Very small values of $\chat_{c}$ are expected due to the Hubble damping in the inflationary epoch.

\zwisch{Entropy production}

\begin{figure}[h]
\centering
\includegraphics[width=0.5\textwidth]{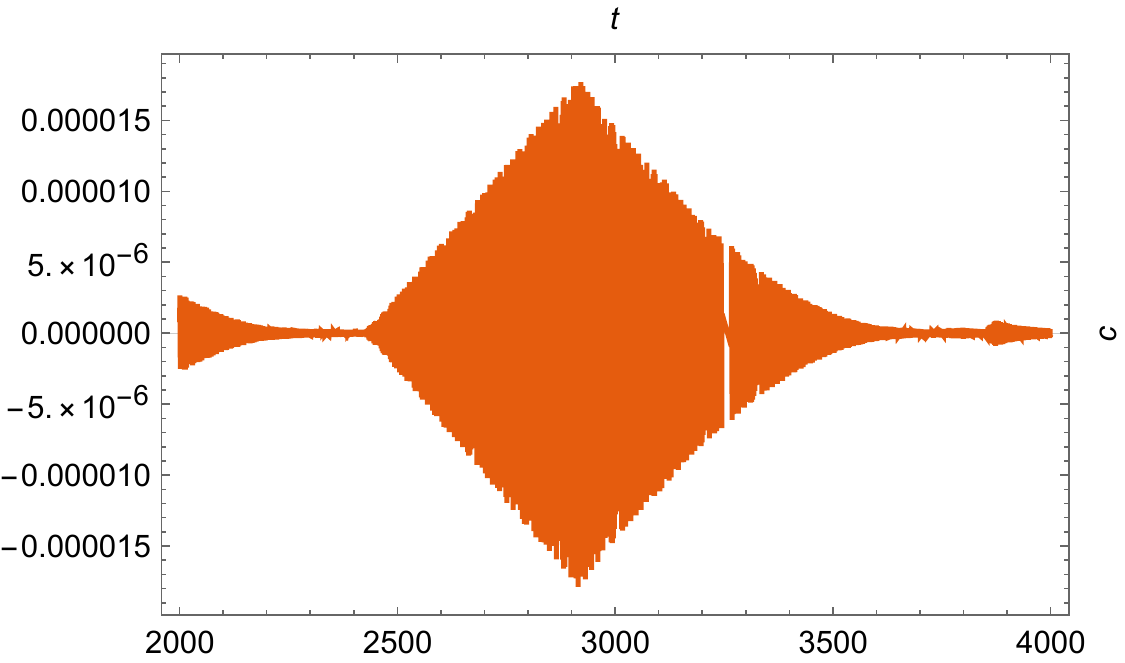}
\caption{\emph{Excursion of small gauge field fluctuations}. We show $\chat$ as function of $\that$ for an appropriate time window. The oscillations on short time scales are not resolved in this figure. Parameters are $Z=-\Ztil=1$, $y=-\ytil=0.2$, $\Kbar=4$, $\Vbar=0.3$, $\Ybar=0$, corresponding to Figs. \ref{fig:1}, \ref{fig:2}. Initial values are the same as for Figs. \ref{fig:1}, \ref{fig:2}, except for a smaller value of the initial $\chi$ or $s$. The very small fluctuation amplitude of $\chat$ increases for short periods and decays again.}
\label{fig:5A}
\end{figure}

Beyond the leading approximation the behavior is more complex. This concerns, in particular, the transition epoch after the end of inflation. The term $\delta_{c}$ in eqs. \eqref{CC8}, \eqref{CC9} is rapidly oscillating on a time scale given by $\hat{\mu}_{c}^{-1}$. This induces a corresponding oscillatory components in the Hubble parameter. In turn, the contribution $\delta\muchat$ is eqs. \eqref{CC1}, \eqref{CC2} induces an effective oscillation of the mass term. This can induce effects similar to parametric resonance \cite{TBRA, KLS, STB, GKLS}. The resulting evolution of the oscillations is rather irregular, with occasional "excursions" for which the amplitude gets enhanced for certain time intervals.
We show a typical "excursion" of $\chat$ in Fig. \ref{fig:5A}. These excursions have rather small effects on the overall cosmology. They are not visible in figures showing also quantities much larger than $\chat$. Excursions of this type typically repeat with irregular shapes. This is one of the reasons why a numerical solution for very large intervals in $\that$ becomes very hard. Not enough resolution is available to follow the irregular oscillations in detail. For $\chat$ this problem could be avoided by starting with $\chat=0$. Similar phenomena occur, however, for the oscillations of $\fhat$ for which a zero amplitude cannot be maintained.

Due to the highly irregular behavior induced by the non-linear terms the detailed initial information concerning the fluctuations is practically lost. The only remaining information is the overall size of the energy density averaged over a sufficiently large time interval. We may associate this effective loss of information with an effective production of entropy. In a more complete picture the entropy associated to this particular oscillation mode will only be a small part of the overall entropy. Nevertheless, it demonstrates the principle of entropy production.

For late stages of the evolution $\Hhat$ gets very small. In this case one can neglect $\delta\muchat$ such that $\eps_{c}$ becomes a constant. In this range the contribution of the $\chat$-oscillations to the energy density becomes simple
\bel{217A}
\rho_{c}=\frac{3Z\eps_{c}}{2A^{3}}\ .
\ee
This is the standard contribution of cold dark matter. One may therefore switch from a description which resolves the oscillations of $\chat$ to an averaged description which accounts for the effects of the $\chat$-oscillations only in a time-averaged sense. For this purpose one omits the explicit $\chat$-field and uses instead for the evolution of the other fields $\Hhat$ and $s$ a source term according to eq. \eqref{217A}. This procedure seems numerically feasible since the resolution of oscillations is needed only to determine the matching conditions, i.e. to compute $\eps_{c}$ from the evolution of $\rho_{c}$ at some suitable time when eq. \eqref{217A} is already valid.

\zwisch{Dark matter from $\mathbf{\fhat}$-oscillations}

A second dark matter candidate arises from the oscillations of the gauge field $\bhat$ or $\fhat$. Due to mixing effects the correct treatment is somewhat more complicated than for $\chat$. For small $\fhat$ and $\Hhat$, as appropriate for the late universe, we can approximate the effective action \eqref{eq:QS19A} by expanding in $\Hhat$, $\fhat$ and $\Hhat'$ up to quadratic order
\begin{align}
\label{FF1}
\Gamma&=\Omega_{3}\int_{\that}A^{3}\bigg{\{}\frac{3\Ztil}{2}\big{(}\fhat'^{2}+\Hhat'^{2}-2\Hhat''\fhat\big{)}\\
&-\bigg{(}\frac{3}{2\ytil}+3\bigg{)}\fhat^{2}-3\big{(}2\Hhat^{2}+\Hhat'\big{)}-\frac{\Kbar}{2}s'^{2}+3\Ybar\fhat s'+\Vhat\bigg{\}}\nn\ .
\end{align}
Let us first discuss the case $\Ybar=0$ for which $\fhat$ and $s$ can be treated separately. For $\Ztil<0$, $-1/2<\ytil<0$, and neglecting $\Hhat$ one would have oscillations of $\fhat$ corresponding to a massive particle with $\mu^{2}=(1+2\ytil)/(\ytil\Ztil)$. While this yields a qualitatively correct picture, the quantitative value of the mass term is modified by the coupling to $\Hhat$ such that the oscillation frequency is given by $\mubhat$.

The reason of this shift in the mass term is the term linear in $\fhat'$ in the evolution equation \eqref{Eq: W17} for $\Hhat$. This differs from the influence of $\chat$ on $\Hhat$ which occurs only quadratically in eq. \eqref{151E}. The coupling of $\Hhat$ to $\fhat$ in linear order corresponds to the off-diagonal term in the propagator matrix \eqref{M30} in appendix \ref{app:A}. We can split $\Hhat$ into a slowly evolving part $\Hbar$ and a part $H_{f}$ that oscillates together with $\fhat$, $\Hhat=\Hbar+H_{f}$. Here $\Hbar$ obeys eq. \eqref{Eq: W17} with $\fhat$ set to zero. For the evolution of $H_{f}$ we linearize eq. \eqref{Eq: W17} in $\fhat$ and $H_{f}$,
\bel{FF2}
H_{f}'+4\Hbar H_{f}=-\frac{1+2\ytil}{2\ytil}(\ftil'+3\Hbar\fhat)\ .
\ee
With the ansatz
\bel{FF3}
H_{f}=-\frac{1+2\ytil}{2\ytil}\fhat+\delta_{f}\ ,
\ee
one finds
\bel{FF4}
\delta_{f}'=\Hbar\bigg{(}\frac{1+2\ytil}{2\ytil}\fhat-4\delta_{f}\bigg{)}\ .
\ee
In the limit $\Hbar=0$ we can identify in eq. \eqref{MS5} $\dtil=-2\ytil\delta_{f}$. Insertion of $H_{f}$ into eq. \eqref{eq:QS18} yields in linear order in $\fhat$ an evolution equation similar to eq. \eqref{CC1},
\bel{FF5}
\fhat''+3\Hbar\fhat'+\big{(}\mubhat+\frac{9\Hbar^{2}}{4}+\frac{3\Hbar'}{2}\big{)}\fhat=-\delta\mubhat\fhat\ ,
\ee
with
\bel{FF6}
\delta\mubhat=-\Big{(}\frac{9}{4}+8\ytil\Big{)}\Hbar^{2}-\Big{(}\frac{9}{2}+4\ytil\Big{)}\Hbar'\ .
\ee
The frequency of the fast oscillations of $\fhat$ is governed by $\mubhat$ as expected. For $\Hbar\to 0$ we recover the results of the stability analysis in flat space. In summary, the oscillations of $\fhat$ constitute a second dark matter component, in complete analogy to the fluctuations of $\chat$.

For $\Ybar\neq 0$ the evolution of $\fhat$ and $s$ are more closely coupled. Similar to the treatment of $\Hhat$ above, we decompose $\fhat$ and $s$ into slowly evolving parts $\fbar$ and $\sbar$, and rapidly oscillating parts $\Delta$ and $s_{f}$
\bel{FF7}
\fhat=\fbar+\Delta\ ,\quad s=\sbar+s_{f}\ ,\quad \Hhat=\Hbar+H_{f}\ .
\ee
For $\fbar$ we take the leading order expression
\bel{FF8}
\fbar=\frac{\ytil\Ybar}{1+2\ytil}s'\ .
\ee
Insertion into the other field equations yields the shift \eqref{23} for the scalar kinetic term in variable gravity. Expressing $s_{f}$ and $H_{f}$ in terms of $\Delta$ by use of their linearized field equations, and insertion into the field equations for $\fhat$ yields the linearized field equation for $\Delta$. For $\Vhat=0$ the oscillation frequency is now given by a mass term similar to eq. \eqref{M50} in the appendix \ref{app:A}. In the presence of non-vanishing derivatives of $\Vhat$ the mass term for $\Delta$ will depend on $s$. This leads to a coupling between dark matter and dark energy that we will discuss below.

\begin{figure}[h]
\centering
\includegraphics[width=0.5\textwidth]{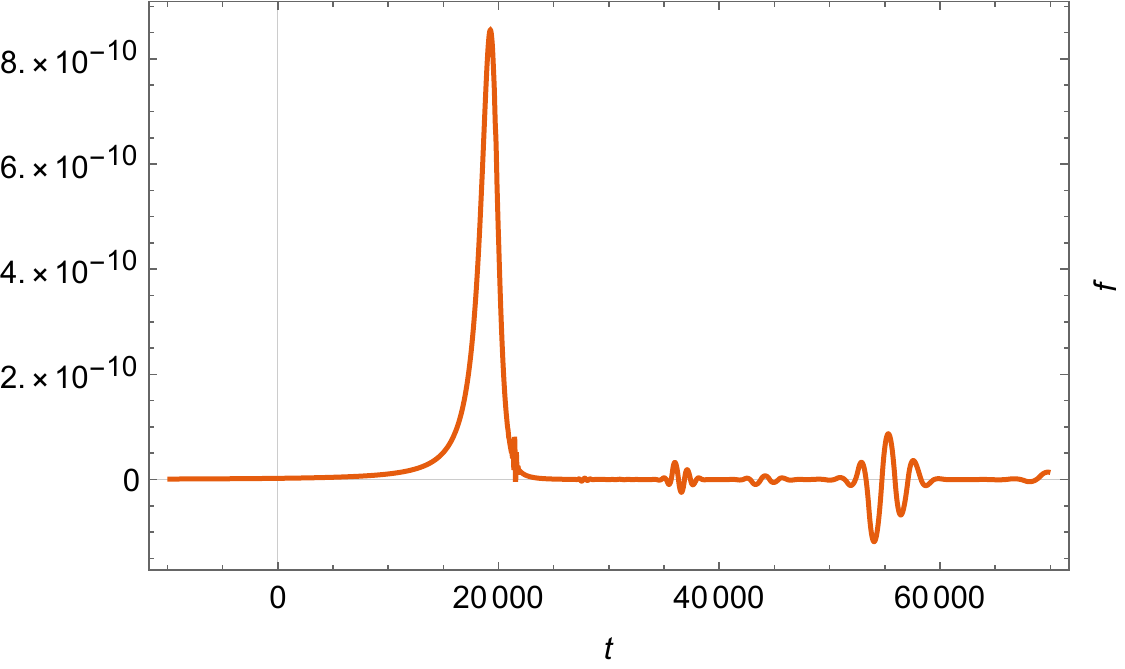}
\caption{\emph{Small oscillations of $\mathit{\fhat}$ as dark matter candidate}. We show $\fhat$ as a function of $\that$ for initial conditions corresponding to a de Sitter solution of type 1. For the inflationary epoch the initial value $\fhat=0$ remains stable. The tiny oscillations are triggered by the end of inflation. They continue to be damped with $\fhat\approx 10^{-18}$ for $\that\approx 10^{6}$. Parameters are $Z=-\Ztil=0.1$, $y=-\ytil=0.2$, $\Kbar=8$, $\Ybar=0$, $V_{0}=10^{-4}$.}
\label{fig:B}
\end{figure}

An important qualitative difference as compared to the $\chat$-oscillation concerns the energy density stored in the fluctuations after the end of the inflationary epoch. If the $\chat$-fluctuations equal zero at the end of inflation this will remain so afterwards. The reason is that due to conserved parity the $\chat$-field appears at least quadratically in the effective action, and there is therefore no source term in its field equation. If inflation lasts long enough, in particular if the inflationary epoch extends to the infinite past, the $\chat$-field vanishes for all practical purposes due to the strong damping during inflation. Also the $\fhat$-field may be very close to zero towards the end of inflation. In contrast to the $\chat$-field the field equation for $\fhat$ has a source term. This source typically differs from zero at the end of inflation. It induces a non-zero amplitude for the $\fhat$-oscillations even in case that $\fhat=0$ during inflation.
Fig. \ref{fig:B} demonstrates that very small oscillations of $\fhat$ are generated by the crossover behavior at the end of the inflationary epoch. This figure shows the result for $\Ybar=0$, for which the oscillations are around $\fhat=0$. For $\Ybar\neq 0$ the oscillations occur around the solution \eqref{FF8}. These fluctuations are further damped as $\that$ progresses. For $\that\approx 10^{6}$ one finds $\fhat\approx 10^{-18}$.
The irregular behavior seen in this figure reflects the non-linearity of the field equation for $\fhat$ that is similar to the one for $\chat$ discussed below.

Once radiation domination sets in the energy density in the $\fhat$-fluctuations increases relative to the dominant energy density in radiation. If the $\fhat$-fluctuations do not decay into other particles, their energy density will equal the radiation energy density at some time $\that_{\text{eq}}$ and associated $\Hhat_{\text{eq}}=\Hhat(\that_{\text{eq}})$ for "equality". The value of $\Hhat_{\text{eq}}$ is fixed by observation of the cosmic microwave background. In order to have $\Hhat_{\text{eq}}$ sufficiently small the value of the energy density in the $\fhat$-fluctuations has to be sufficiently small at the beginning of the radiation dominated era. This may turn to a restriction for realistic models.

\vspace{2\baselineskip}
\zwisch{Dynamical dark energy}

The key ingredient for the prediction of dynamical dark energy or quintessence in our crossover model is the flatness of the potential $V(\chi)$ for large values of $\chi$, as encoded in the ansatz \eqref{S7}, $V=u_{0}k^{4}$. This flatness seems to be a generic feature of scaling solutions for functional flow equations \cite{HPRW, HPW, CWQS, CWESPA}. The precise form of $V(\chi)$ is not important in this respect. For example, the constant value of $V$ for $\chi\to\infty$ could be different from the value for $\chi\to 0$, e.g.
\bel{DE1}
V(\chi\to\infty)=u_{\infty}k^{4}\ .
\ee
One only needs a positive value $u_{\infty}>0$. Only in order to keep the number of parameters small we choose here $u_{\infty}=u_{0}$. The constant value of the potential combined with an increase of $\M\sim\chi^{2}$ leads to a frame invariant potential $\Vhat$ that vanishes for large $\chi$ or $s=\ln(\chi/k)$
\bel{DE2}
\Vhat=\Vbar e^{-4s}\ ,\quad \Vbar=\frac{u_{\infty}}{\xi^{2}}\ .
\ee
The results are cosmic runaway solutions for which $s$ diverges in the infinite future. The dynamical dark energy stored in $\Vhat$ vanishes asymptotically.

The detailed dynamics depends on the kinetial $\khat^{2}(s)$ and the dynamics at the end of inflation. After the end of inflation one typically finds a "kination" epoch for which the energy density is dominated by the scalar kinetic energy. During this epoch $s$ can grow substantially. At the end of the kination epoch $s$ is frozen at $s_{f}$, or it changes much slower for $s>s_{f}$. There are two possibilities for dynamical dark energy. If $s_{f}$ obeys
\bel{DE3}
\Mbar^{4}\Vbar e^{-4s_{f}}=\big{(}2\cdot 10^{-3}\text{eV}\big{)}^{4}\ ,\quad s_{f}=s_{cr}\approx 70+\frac{1}{4}\ln\Vbar\ ,
\ee
the potential energy in the Einstein frame, $\Mbar^{4}\Vhat(s_{f})$, can be identified with the present dark energy density. The value of $s$ does no longer change until the total energy density reaches a similar value. This constitutes "thawing quintessence". If $s_{f}$ is smaller than the critical value $s_{cr}$ the universe may consecutively follow a scaling solution ("cosmological attractor solution", "asymptotic solution", "tracker solution") for which dynamical dark energy is an almost constant fraction of the dominant radiation or matter density. Both possibilities are first discussed in ref \cite{Wetterich_1988}.

In the second case some event has to "kick out" the evolution from the scaling solution around redshift $z=5$, realizing a scenario of "freezing quintessence". One can achieve realistic models by a suitable choice of the kinetial $\khat^{2}(s)$. A cosmic trigger event ending the scaling solution could be the growth of the ratio between neutrino and electron mass, induced by a new step of crossover in the beyond standard model sector \cite{AQCGM}, \cite{CWGNCS}. It is not the purpose of the present paper to develop a detailed scenario for realistic quintessence. Short summaries of realistic variable gravity models can be found in refs. \cite{CWIQM, CWQS}.

In the absence of matter and radiation the scaling solution is dominated by the potential and kinetic energy of the scalar field \cite{Wetterich_1988}. Adapting the results of ref. \cite{Wetterich_1988} for our notation, the field equations in the Einstein frame \eqref{eq:QS28}, \eqref{eq:QS29} have for large $\phitil/\Mbar$ the asymptotic solution
\begin{align}
\label{DE3A}
H=\frac{\khat^{2}}{8t}\ &,\quad \phitil=\frac{\Mbar}{2}\big{[}\ln(\Mbar t)-\half\ln c_{0}^{2}\big{]}\ ,\nn\\
\dot{\phitil}=\frac{\Mbar}{2t}\ &,\quad A=A_{0}\Big{(}\frac{t}{t_{0}}\Big{)}^{\khat^{2}/8}\ ,
\end{align}
with
\bel{DE3B}
c_{0}^{2}=\frac{\khat^{2}}{8\Vbar}\Big{(}\frac{3\khat^{2}}{8}-1\Big{)}\ .
\ee

\begin{figure}[h]
\centering
\includegraphics[width=0.5\textwidth]{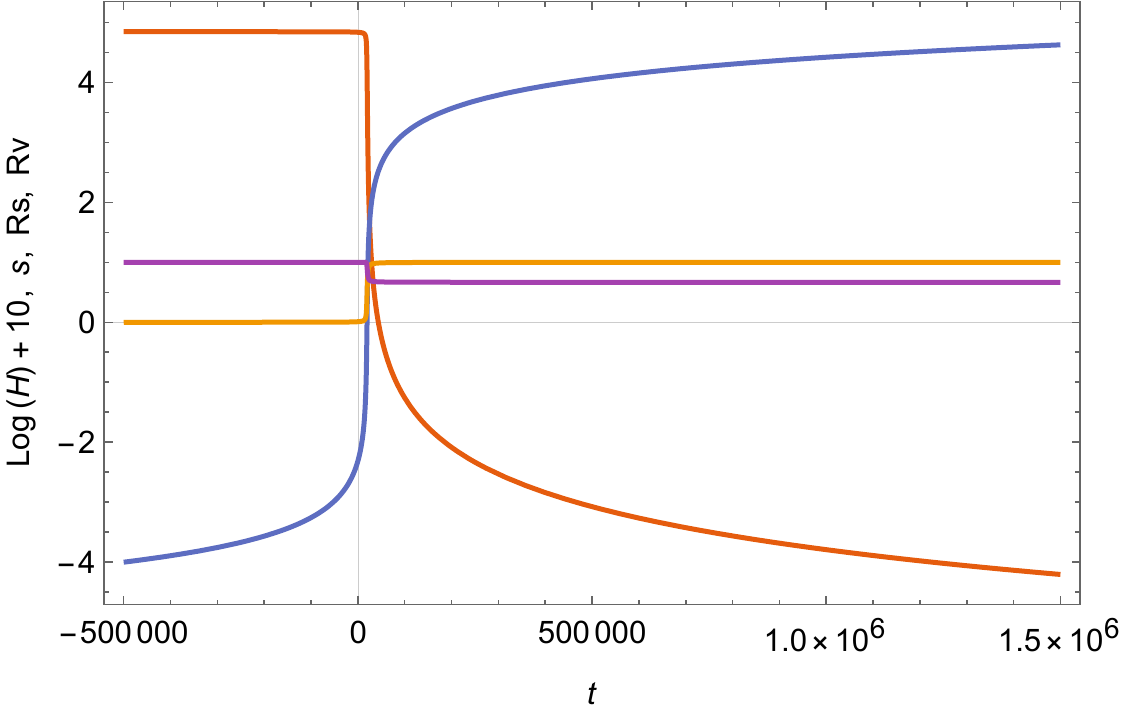}
\caption{\emph{Transition to scalar dominated dynamical dark energy}. As a function of cosmic time $\that$ we plot the Hubble parameter logarithmically as $\ln H+10$ (red) and the scalar field $s$ (blue). In addition, we display $R_{s}$ (orange) and $R_{V}$ (magenta). Initial conditions are the de Sitter solution of type 1, and parameters are the same as for Fig. \ref{fig:B}, $Z=-\Ztil=0.1$, $y=-\ytil=0.2$, $\Kbar=8$, $\Ybar=0$, $V_{0}=10^{-4}$.}
\label{fig:C}
\end{figure}

We illustrate the transition from inflation to a scalar field dominated scaling solution for dynamical dark energy in Fig. \ref{fig:C}. The quantities
\bel{B27A}
R_{s}=\frac{2\dot{s}}{\Hhat}\ ,\quad R_{V}=\frac{\Vhat}{3\Hhat^{2}}
\ee
allow us to follow the crossover at the end of inflation quantitatively. For the inflationary period one has $R_{s}\ll 1$, $R_{V}=1$, while for the scaling solution both quantities reach rapidly the corresponding values.

The scaling solution \eqref{DE3A} exists for $\khat^{2}>8/3$. For $\khat^{2}<8/3$ the potential becomes negligible as compared to the scalar kinetic term, with
\bel{DE3C}
H=\frac{1}{3t}\ ,\quad \dot{\phitil}=\sqrt{\frac{2}{3}}\ \frac{\Mbar}{\khat t}\ ,\quad A=A_{0}\Big{(}\frac{t}{t_{0}}\Big{)}^{\frac{1}{3}}\ .
\ee
In the presence of an additional radiation component the latter will finally dominate if $\khat^{2}<4$. From there on the dark energy density follows the dominant radiation or matter energy density, typically with an almost constant fraction of "early dark energy". Rather realistic cosmologies obtain if $\khat^{2}$ depends on $s$ or $\phitil$, decreasing from large values to small values as $s$ increases \cite{CWIQM}, \cite{RUCW}.

\zwisch{Coupled quintessence}

Dynamical dark energy and matter may interact leading to coupled quintessence \cite{CWCMAV}, \cite{LACQ}. This happens if the mass of a dark matter particle depends on the value of the cosmon field. One defines the coupling $\beta$ by the dependence of the dark matter particle mass $\mu$ on the normalized scalar field in the Einstein frame
\bel{DE4}
\beta=-\Mbar\frac{\partial}{\partial\sigma}\ln\mu\ .
\ee
Here $\sigma$ is related to $s$ by
\bel{DE5}
\frac{\text{d}\sigma}{\text{d}s}=\Mbar\khat(s)\ .
\ee
In the presence of a coupling the scalar field equation becomes in the Einstein frame, $V_{\text{E}}=\Mbar^{4}\Vhat$,
\bel{DE6}
\dt^{2}\sigma+3H\dt\sigma=-\frac{\partial V_{\text{E}}}{\partial\sigma}+\frac{\beta}{\Mbar}\rho_{\text{dm}}\ ,
\ee
and the conservation equation for the dark matter energy density $\rho_{\text{dm}}$ gets modified
\bel{DE7}
\dt\rho_{\text{dm}}+3H\rho_{\text{dm}}+\frac{\beta}{\Mbar}\rho_{\text{dm}}\dt\sigma=0\ .
\ee
The energy momentum tensor for dark matter and dark energy are no longer conserved separately since energy can be exchanged between the two components.

If dark matter arises from $\fhat$-fluctuations the relevant mass in the Einstein frame follows from eq. \eqref{M50},
\bel{DE8}
\mu^{2}=-\frac{2\Mbar^{2}}{\Ztil}\bigg{(}1-\frac{2\ytil\ \Kbar}{(1+2\ytil)\khat^{2}}\bigg{)}^{-1}\ .
\ee
This yields
\bel{DE9}
\beta=-\frac{1}{2\khat}\frac{\partial\ln\mu^{2}}{\partial s}=\frac{1}{2\khat}\frac{\partial}{\partial s}\bigg{\{}\ln\big{(}-\Ztil\big{)}+\ln\Big{(}1-\frac{2\ytil\ \Kbar}{(1+2\ytil)\khat^{2}}\Big{)}\bigg{\}}\ .
\ee
In the limit of exact quantum scale symmetry the coupling $\beta$ vanishes since all couplings are independent of $s$. On the other hand the kinetials $\khat^{2}$ and $\Kbar$ may still vary with $s$, reaching fixed point values only for $s\to\infty$. If dark matter arises from $\fhat$-fluctuations one expects a non-vanishing coupling to dark energy.

\section{Inflationary cosmology}
\label{section:IC}

The field equations for the model presented in section \ref{section:CM} lead to inflationary cosmology in a very natural way. 
For $s \to -\infty$, corresponding to $\chi\to 0$, the potential \eqref{eq:QS21} becomes flat. This flat tail will lead to the "slow roll" behavior of the scalar field characteristic for inflation, whenever a "beginning epoch" with very small $\chi$ is realized. We have discussed in sect.~\ref{section:CS} possible beginnings with de Sitter solutions for $\chi=0$. A slow evolution away from this exact de Sitter solution results in an early inflationary epoch.

\begin{figure}[h]
\centering
\includegraphics[width=0.5\textwidth]{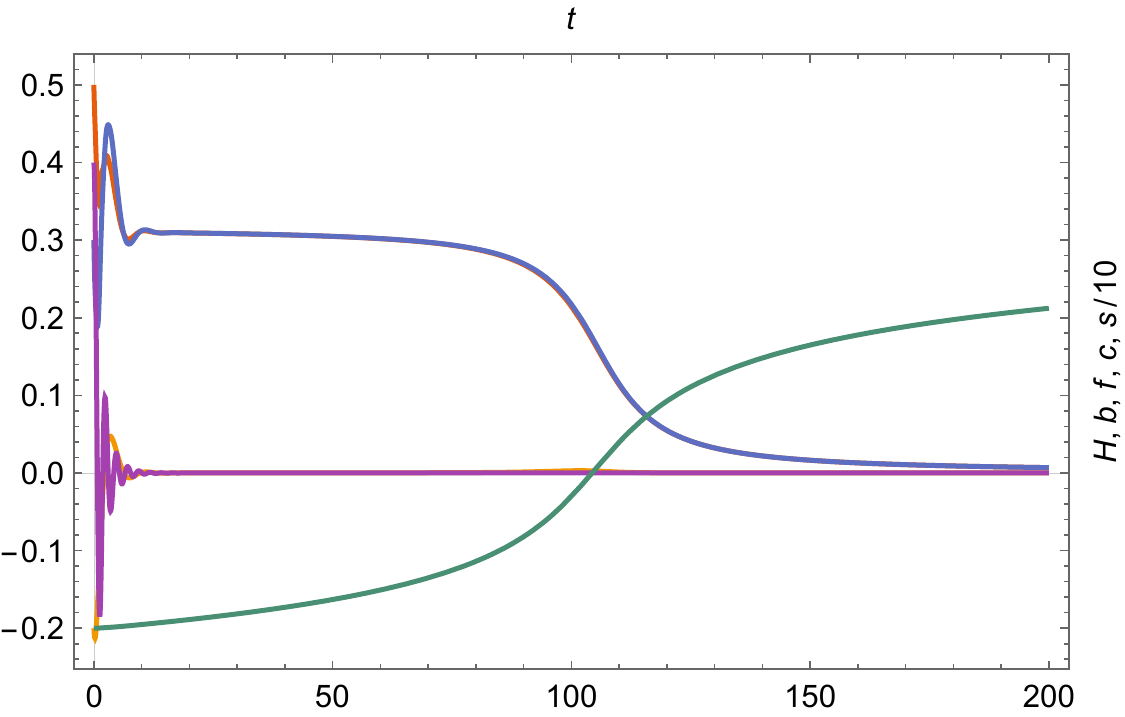}
\caption{\emph{Inflationary cosmology}. We plot the frame invariant functions $\Hhat$, $\bhat$, $\fhat$, $\chat$ and $s/10$ as function of $\that$. Parameters are $Z=-\Ztil=1$, $y=-\ytil=0.2$, $\Kbar=4$, $\Vbar=0.3$, $\Ybar=0$ corresponding to Figs. \ref{fig:1}, \ref{fig:2}. Initial values and color coding of the cuves are the same as for Figs. \ref{fig:1}, \ref{fig:2}, except a smaller value of the initial $\chi$ or $s$ which extends the epoch of inflation. All parameters and initial conditions are identical to fig. \ref{fig:5A}. Bot $\chat$ and $\fhat$ are almost zero after a short initial epoch, while $\bhat$ and $\Hhat$ reach the same plateau characteristic for inflation, and decay towards zero after the end of inflation.}
\label{fig:5}
\end{figure}

Since the de Sitter solution is an attractor for small enough $\chi$ or large enough negative $s$ there is no need for a beginning with a de Sitter solution. A very large family of "initial states" with small $\chi$ will approach quickly the de Sitter solution. This solution is depicted in Fig. \ref{fig:5} where we show the frame invariant functions $\Hhat$, $\bhat$, $\fhat$, $\chat$ and $s$ for parameters $Z=-\Ztil=1$, $y=-\ytil=0.2$, $\Vbar=0.3$, $\Ybar=0$ corresponding to Figs. \ref{fig:1}, \ref{fig:2}. The frame- and scale invariant kinetial with $\Kbar=4$ corresponds for constant $\M$ to $K=4k^{2}/\chi^{2}$. In the Einstein frame all quantities are in units of the Planck mass $\Mbar$. The time $\that$ corresponds in this case to cosmic time in units of the Planck time $\Mbar^{-1}$.

By choosing smaller initial values of $\chi$ the end of the inflationary epoch moves to large $\that$ and inflation extends over a larger time interval. (The position of the zero of $\that$ is, of course, arbitrary.) Smaller values of the Hubble parameter during inflation obtain for smaller values of $\Vbar$. The end of inflation and the value of the Hubble parameter at the end of inflation can also be modified by a $\chi$-dependent function $\Khat$ as we will discuss below. We observe that $\fhat$ and $\chat$ reach very small values rapidly. One therefore expects that variable gravity provides for a rather accurate approximation.

The potential~\eqref{eq:QS21} corresponds precisely to the potential of our ansatz \eqref{S7}, with
\be
\label{IA}
\mu^{2}=\frac{2w_{0}k^{2}}{\xi}\;,\quad \overline{V}=\frac{u_{0}}{4w_{0}^{2}} \ .
\ee
This potential has its maximum for $\chi=0$, $\tilde{\varphi}=0$.
Any small nonzero $\chi$ or $\tilde{\varphi}$ will move slowly away from this maximum. We employ this potential here without modification, while we will discuss below also the effect of kinetials different form eq. \eqref{eq:QS9}.

We will discuss the inflationary epoch first in terms of the associated variable gravity model. Subsequently, we start from the full field equations for pregeometry and investigate the precise circumstances under which variable gravity becomes a valid approximation. This will be the case for the cosmological epoch responsible for the properties of the observable primordial fluctuations.

\zwisch{Inflation in variable gravity}

We start from the effective action \eqref{eq:QS20} in the Einstein frame $\big{(}\sqrt{g}=e\big{)}$, 
\be
\label{I1}
\Gamma=\int\sqrt{g}\bigg{\lbrace}-\frac{\overline{M}^{2}}{2}R+\half\overline{M}^{2}\widehat{k}^{2}\partial^{\mu}s\partial_{\mu}s+\overline{M}^{4}\widehat{V}\bigg{\rbrace}\ .
\ee
A convenient normalization of the scalar field relates it directly to the potential~\cite{CWIQM}
\be
\label{I2}
\widehat{V}=\overline{V}\exp\bigg{(}-\frac{\varphi}{\overline{M}}\bigg{)}\ .
\ee
This choice is always possible if the potential is monotonic for the range of fields of interest, as realized in our case. For this normalization one has
\be
\label{I3}
\Gamma=\int\sqrt{g}\bigg{\lbrace}-\frac{\overline{M}^{2}}{2}R+\half k^{2}(\varphi)\partial^{\mu}\varphi\partial_{\mu}\varphi+\overline{M}^{4}\exp\bigg{(}-\frac{\varphi}{\overline{M}}\bigg{)}\bigg{\rbrace}\ ,
\ee
with kinetial
\be
\label{I4}
k^{2}(\varphi)=\widehat{k}^{2}\overline{M}^{2}\bigg{(}\frac{\partial s}{\partial\varphi}\bigg{)}^{2}\ .
\ee
All properties of the spectrum of primordial cosmic fluctuations can be read off directly from the kinetial $k^{2}(\varphi)$~\cite{CWIQM}.

For the potential~\eqref{eq:QS21} the identification
\be
\label{I5}
(1+e^{2s})^{-2}=e^{-\tfrac{\varphi}{\overline{M}}}
\ee
relates $\varphi$ and $s$
\be
\label{I6}
s=\half \ln\bigg{[}\exp\bigg{(}\frac{\varphi}{2\overline{M}}\bigg{)}-1\bigg{]}\ .
\ee
The kinetial shows an exponential dependence on $\varphi$,
\be\label{I7}
k^{2}(\varphi)=\frac{\widehat{k}^{2}}{16}\exp\big{(}\frac{\varphi}{\overline{M}}\big{)}\bigg{[}\exp\bigg{(}\frac{\varphi}{2\overline{M}}\bigg{)}-1\bigg{]}^{-2}\ .
\ee
For $s\to -\infty$, corresponding to $\chi\to 0$ and $\varphi\to 0$ according to
\be
\label{I8}
\chi^{2}=\mu^{2}\bigg{(}\exp\bigg{(}\frac{\varphi}{2\overline{M}}\bigg{)}-1\bigg{)}\approx\frac{\mu^{2}\varphi}{2\overline{M}}\ ,
\ee
the kinetial diverges
\be
\label{I9}
k^{2}(\varphi\to 0)=\frac{M^{2}\widehat{k}^{2}}{4\varphi^{2}}\ .
\ee
This reflects the flat tail of the potential $V(s\to -\infty)$. As $\varphi$ increases the kinetial $k^{2}(\varphi)$ decreases. Inflation ends once $k^{2}(\varphi)$ decays below one.

\zwisch{Primordial fluctuation spectrum}

For a standard form of the potential the details of the inflationary model are encoded in the kinetial~\cite{CWCI, CWIQM, CWVG}. (See also ref.~\cite{GKLR} for a different form of the standard potential.) The advantage of the form~\eqref{I2} is that the characteristic properties of inflation can be read off directly~\cite{CWVG, CWIQM, CWCI} from the kinetial $k^{2}(\varphi)$. This is easily seen  by relating $\varphi$ to a field $\sigma$ with canonical kinetic term by 
\be
\label{I10}
\frac{\mathrm{d}\sigma}{\mathrm{d}\varphi}= k(\varphi)\;. 
\ee
The slow roll parameters are given by
\begin{align}
\label{I11}
\varepsilon=&\frac{\overline{M}^{2}}{2}\bigg{(}\frac{\partial\ln V}{\partial\sigma}\bigg{)}^{2}=\frac{1}{2k^{2}(\varphi )}\ , \nn\\
\eta=&\frac{\overline{M}^{2}}{V}\frac{\partial^{2}V}{\partial\sigma^{2}}=2\varepsilon-\overline{M}\frac{\partial\varepsilon}{\partial\varphi}=\frac{1}{k^{2}(\varphi)}\bigg{(}1+\frac{\overline{M}}{2}\frac{\partial\ln k^{2}(\varphi)}{\partial\varphi}\bigg{)}\ , 
\end{align}
and can therefore be read off directly from $k^{2}(\varphi)$.
The slow roll behavior is realized for small $\varepsilon$, corresponding to large $k^{2}(\varphi)$ as realized for $\varphi\to 0$. 
As $\varphi$ increases due to the gradient of the potential~\eqref{I2}, the kinetial decreases and $\varepsilon(\varphi)$ increases. The inflationary epoch ends once $\varphi$ has reached the value $\varphi_{f}$ defined by
\be
\label{I12}
k^{2}(\varphi_{f})=1\ .
\ee

The kinetial can also be used for relating the number of $e$-foldings before the end of inflation $N$ to the value of $\varphi$. One obtains
\be
\label{I13}
N(\varphi)=\frac{1}{\overline{M}}\int_{\varphi}^{\varphi_{f}} \mathrm{d}\varphi' k^{2}(\varphi')\ .
\ee
In turn, one can use the inverse function $\varphi(N)$ in order to express the slow roll parameters $\varepsilon(N)$, $\eta(N)$ as functions of $N$. 
The properties of the spectrum of primordial fluctuations are directly related to $\varepsilon(N)$ and $\eta(N)$, with $N\approx 50-60$ depending on the heating after inflation. (We do not discuss heating in this paper since it involves the couplings to other fields in the standard model or extensions. For a detailed discussion for this type of inflationary models motivated by quantum scale symmetry see ref.~\cite{RUCW}.) In particular, the spectral index for the primordial scalar fluctuations is given by
\be
\label{I14}
n=1-6\varepsilon(N)+2\eta(N)\ ,
\ee
and the ratio of tensor fluctuations over scalar fluctuations obeys
\be
\label{I15}
r=16\varepsilon(N)=\frac{8}{k^{2}(N)}\ .
\ee

The amplitude of the primordial scalar fluctuations is proportional to the value of the potential for~$\varphi(N)$ . Since we employ a standard form of the potential our formulation provides for a direct relation between $\varphi(N)$ and the fluctuation amplitude. More quantitatively the scalar fluctuation amplitude $\Delta$ obeys
\be
\label{I16}
24\pi^{2}\Delta^{2}=\frac{V\big{(}\varphi(N)\big{)}}{\varepsilon(N)\overline{M}^{4}}\approx 5\cdot10^{-7} \ ,
\ee
where the last relation indicates the observed value. Compatibility with observation therefore requires
\be
\label{I17}
\exp\bigg{(}-\frac{\varphi(N)}{\overline{M}}\bigg{)}=\frac{2.5\cdot 10^{-7}}{\overline{V}k^{2}\big{(}\varphi(N)\big{)}}\ .
\ee
For a given model the function $k^{2}(\varphi)$ is fixed and the value of $\varphi(N)$ required for the observed amplitude can be computed.

\zwisch{Realistic inflationary scenarios}

There are two classes of realistic inflationary models. For the first $\varphi(N)/\overline{M	}$ is small as compared to one. In this case the parameter combination $\overline{V}k^{2}(N)$ is fixed by observation
\be
\label{I18}
\overline{V}k^{2}(N)\approx 2.5\cdot 10^{-7} \ .
\ee
This is the range where the approximation~\eqref{I9} is valid, such that the parameter combination~$\overline{V}\widehat{k}^{2}$ is determined by
\be
\label{I19}
\overline{V}\widehat{k}^{2}\approx 10^{-6}\frac{\varphi^{2}(N)}{\overline{M}^{2}}\ .
\ee
If the approximation~\eqref{I9} remains roughly valid until the end of inflation the relations~\eqref{I12},~\eqref{I13} yield
\be
\label{I20}
\frac{\varphi_{f}^{2}}{\overline{M}^{2}}\approx\frac{\widehat{k}^{2}}{4}
\ee
and
\be
\label{I21}
N=\frac{\widehat{k}}{2}\bigg{(}\frac{\widehat{k}\overline{M}}{2\varphi(N)}-1\bigg{)}\ ,
\ee
or
\be
\label{I22}
\frac{\varphi(N)}{\overline{M}}=\frac{\widehat{k}^{2}}{4N}\bigg{(}1+\frac{\widehat{k}}{2N}\bigg{)}^{-1} \ .
\ee
From eq.~\eqref{I19} we conclude the approximate relation
\be
\label{I23}
\overline{V}\approx\frac{10^{-6}}{16N^{2}}\widehat{k}^{2}\approx 2\cdot 10^{-11}\widehat{k}^{2}\ .
\ee
Unless $\khat^{2}$ is huge a small fluctuation amplitude requires for this scenario a rather small value of $\overline{V}$ .

For this type of models we can compute $k^{2}(N)$ as
\be
\label{I23A}
k^{2}(N)=\frac{4N^{2}}{\widehat{k}^{2}}\bigg{(}1+\frac{\widehat{k}}{2N}\bigg{)}^{2}\ .
\ee
This yields
\be
\label{I23B}
\varepsilon=\frac{\widehat{k}^{2}}{8N^{2}}\bigg{(}1+\frac{\widehat{k}}{2N}\bigg{)}^{-2}=\frac{2\varphi^{2}(N)}{\widehat{k}^{2}\overline{M}^{2}}
\ee
and
\be
\label{I23C}
\eta=-\frac{1}{N}\bigg{(}1+\frac{\widehat{k}}{2N}\bigg{)}^{-1}+2\varepsilon=-\dfrac{4\varphi(N)}{\widehat{k}^{2}M}+2\varepsilon\ .
\ee
For $\widehat{k}^{2}\ll N^{2}$ the resulting tensor to scalar ratio is very small,
\be
\label{I23D}
r=\frac{2\widehat{k}^{2}}{N^{2}}\bigg{(}1+\frac{\widehat{k}}{2N}\bigg{)}^{-2}\ ,
\ee
and the special index obtains as
\be
\label{I23E}
n=1-\frac{2}{N}\bigg{(}1+\frac{\widehat{k}}{2N}\bigg{)}^{-1}-\frac{\widehat{k}^{2}}{4N^{2}}\bigg{(}1+\frac{\widehat{k}}{2N}\bigg{)}^{-2}\ .
\ee
This range of values is compatible with observation.

For the second class of possible realistic inflationary scenarios the tiny value of the primordial fluctuation amplitude is explained by the exponential suppression of $V(\varphi)$ for $\varphi / \overline{M}\gg 1 $. This is the type of models investigated in refs.~\cite{CWCI, CWVG, CWIQM, RUCW}. This scenario can be realized if we extend the simple ansatz of sect.~\ref{section:CM} to a kinetial $\widehat{k}^{2}(s)$ that is no longer constant. For $\varphi/\Mbar\gg 1$ the kinetial $k^{2}(\varphi)$ in eq. \eqref{I7} is well approximated by $\khat^{2}/16$. As long as $\khat^{2}$ remains larger than $16$ inflation cannot end. The end of inflation is now triggered by a crossover from $\widehat{k}^{2}(s)>16$ for $s$ smaller than $s_{f}$ to $\widehat{k}^{2}(s)<16$ for $s$ larger than $s_{f}$.

A variation of $\khat^{2}$ with $\chi$ or $s$ is well motivated within our family of crossover models. It extends the setting with a single crossover due to the $\chi$-dependence of $\M$ to the possibility of a second crossover scale related to the kinetial.
Indeed, $\widehat{k}^{2}$ is determined by different parameter combinations for $\chi\to 0$ and $\chi\to\infty$. For $\chi\to 0$ one infers from eq.~\eqref{eq:QS22},~\eqref{eq:QS7}
\be
\label{I24}
\widehat{k}_{0}^{2}=\overline{K}-\frac{3\ytil}{1+2\ytil}\Ybar^{2}=\frac{\kappa}{2w_{0}}-\frac{3\ytil Y_{0}^{2}}{4(1+2\ytil)w_{0}^{2}}\ ,
\ee
with $\ytil=2w_{0}/\tilde{m}_{0}^{2}$. On the other hand, for $\chi^{2}\to\infty$ eqs.~\eqref{eq:QS22},~\eqref{eq:QS3} imply, with $\ytil=\xi/\tilde{\zeta}$,
\be
\label{I25}
\widehat{k}_{\infty}^{2}=\frac{K_{\infty}}{\xi}-\frac{3\ytil}{1+2\ytil}\bigg{\lbrace}\bigg{(}\frac{Y_{\infty}}{\xi}+2\bigg{)}^{2}-6\bigg{\rbrace}\  ,
\ee
with $K_{\infty}=K(\chi\to\infty)$, $Y_{\infty}=Y(\chi\to\infty)$. One therefore expects a crossover between the two limiting values. Values $\widehat{k}_{0}\gg 16$ and $\widehat{k}_{\infty}\ll 16$ seem not to be unnatural.

In this second scenario the end of inflation occurs in a range of large $\varphi/\overline{M}$ for which
\be
\label{I26}
s\approx\frac{\varphi}{4\overline{M}}\quad , \quad k^{2}(\varphi)\approx\frac{\widehat{k}^{2}(\varphi)}{16} \ .
\ee
The change from large to small $k^{2}(\varphi)$ ending inflation is then almost entirely due to the dependence of $\widehat{k}^{2}$ on $s$ or $\varphi$. If this behavior is already relevant around $50-60$ $e$-foldings before the end of inflation a rather large $\widehat{k}^{2}(N)$ is needed for a small tensor to scalar ratio, $r\approx 128/\widehat{k}^{2}(N)$. On the other hand, the parameter~$\overline{V}$ no longer has to be tiny as if $\varphi_{f}/\overline{M}$ is sufficiently large.

Realistic inflationary models are also possible in between the two scenarios depicted above. Only some calculation of the parameters in our ansatz can decide between the many possible inflationary models. For the time being we just impose observational constraints which limit the allowed parameter ranges. Fixing them in detail is not the purpose of this paper.

\zwisch{Validity of the variable gravity approximation.}

We have seen the possibility of realistic inflationary models within the variable gravity approximation for pregeometry. We next want to investigate the conditions for this approximation to be valid. Variable gravity is valid for small enough $f$ and $c$. We take here $c=0$ and concentrate on $f$. Our aim are quantitative relations for the inflationary scenarios discussed above.

We start from the frame invariant field equations~\eqref{eq:QS17}-\eqref{eq:QS18} for our model of pregeometry. For a solution of eq.~\eqref{eq:QS18} we consider the approximations
\bel{276A}
\frac{4\ytil\Ztil\widehat{H}^{2}}{1+2\ytil}\ll 1\quad,\quad\frac{\fhat^{2}}{\widehat{H}^{2}}\ll 1\ ,
\ee
and
\bel{276B}
\frac{3\ytil\Ztil |\widehat{H}\fhat'|}{1+2\ytil}\ll |\fhat|\ .
\ee
We further assume that the order of magnitude of $\widehat{H}'$ does not exceed $\widehat{H}^{2}$, and that $\fhat''$ does not exceed by much $3\widehat{H}\fhat'$. With these approximations the solution of eq.~\eqref{eq:QS18} reads
\be
\label{I27}
\fhat=\frac{\ytil\overline{Y}s'}{1+2\ytil}-\frac{\ytil\Ztil}{1+2\ytil}(4\widehat{H}\widehat{H}'+\widehat{H}'')\ .
\ee
Insertion into the scalar field equation~\eqref{eq:QS17} yields
\begin{align}
\label{I28}
&\bigg{(}\overline{K}-\frac{3\ytil\overline{Y}^{2}}{1+2\ytil}\bigg{)}(s''+3\widehat{H}s')=-\frac{\partial\widehat{V}}{\partial s} \nn \\
&-\frac{3\ytil\overline{Y}\Ztil}{1+2\ytil}(12\widehat{H}^{2}\widehat{H}'+7\widehat{H}\widehat{H}''+4\widehat{H}'^{2}+\widehat{H}''')\   .
\end{align}
We observe the appearance of the combination $\widehat{k}^{2}$ given by eq.~\eqref{eq:QS22}. Omitting the term $\sim \Ztil$ yields the scalar field equation~\eqref{eq:QS25} of variable gravity that we employ in the first part of this section.

Similarly, inserting eq.~\eqref{I27} into eq.~\eqref{eq:QS16} one finds
\begin{align}
\label{I29}
&\widehat{H}^{2}+\half\widehat{H}'=\frac{\widehat{V}}{3}-\frac{\widehat{k}^{2}s'^{2}}{12}\\
\!\!+\frac{\Ztil}{4}&\bigg{[}12\widehat{H}^{2}\widehat{H}'\!+4\widehat{H}'^{2}\!+7\widehat{H}\widehat{H}''\!\!+\widehat{H}'''
\!\!-\frac{\ytil\Ztil}{1+2\ytil}(4\widehat{H}\widehat{H}'\!+\widehat{H}'')^{2}\bigg{]} \, ,\nn
\end{align}
while eq.~\eqref{eq:QS17C} becomes
\begin{align}
\label{I30}
\widehat{H}^{2}=&\frac{\widehat{V}}{3}+\frac{\widehat{k}^{2}}{6}s'^{2}+\Ztil\bigg{\lbrace}3\widehat{H}^{2}\widehat{H}'-\half \widehat{H}'^{2}+\widehat{H}\widehat{H}'' \nn\\
&+\frac{\ytil\overline{Y}s'}{1+2\ytil}(\widehat{H}^{3}+3\widehat{H}\widehat{H}'+\widehat{H}'')\bigg{\rbrace} \\
&-\frac{\ytil\Ztil^{2}}{2(1+2\ytil)}(4\widehat{H}\widehat{H}'+\widehat{H}'')(2\widehat{H}^{3}+2\widehat{H}\widehat{H}'+\widehat{H}'')+\Delta_{2}\ ,\nn
\end{align}
where we use in eq.~\eqref{eq:QS17D}
\begin{align}
\label{I31}
\Delta_{1}=\overline{Y}\widehat{H}s'-\frac{\ytil}{2(1+2\ytil)}\overline{Y}^{2}s'^{2}+\Ztil(4\widehat{H}^{2}\widehat{H}'+\widehat{H}\widehat{H}''+\dots )
\end{align}
and $\Delta_{2}$ contains subleading terms $\sim \Ztil$ as given by eq.~\eqref{eq:QS17E}. Omitting the terms~$\sim \Ztil$ we find the geometrical field equations for variable gravity~\eqref{eq:QS26} from which we have started our investigation of inflationary cosmology above. We conclude that the approximation of variable gravity becomes valid if the terms~$\sim \Ztil$ in eqs.~(\ref{I28}~-~\ref{I30}) can be neglected. These terms all involve higher powers of the Hubble parameter and its derivatives. For any model of realistic inflation the Hubble parameter in units of the Planck mass, as given by $\Hhat$, is a very small quantity for the epoch relevant for the properties of the primordial fluctuation spectrum. In this range the variable gravity approximation holds with high precision.

\vspace{3\baselineskip}
\zwisch{Slow roll approximation}

The corrections beyond the variable gravity approximation vanish if both $\Ztil\widehat{H}'=0$ and $\ytil\overline{Y}s'=0$. This is obeyed for the de Sitter solution~\eqref{FS9} where $\Hhat'=0$, $s'=0$. A beginning within the range of validity of the variable gravity approximation is therefore possible. Variable gravity remains a good approximation for the evolution away from the de Sitter solution as long as the "slow roll approximation" holds, as specified by
\be
\label{I32}
\widehat{k}^{2}s'^{2}\ll 6\widehat{H}^{2}\ .
\ee
In this approximation one has
\be
\label{I33}
\widehat{H}^{2}\approx\frac{\widehat{V}}{3}\approx\frac{\overline{V}}{3}(1+e^{2s})^{-2}
\ ,
\ee
while the scalar field equation yields for $|s''|\ll |3\widehat{H}s'|$
\be
\label{I33A}
s'\approx\frac{4e^{2s}\widehat{H}}{(1+e^{2s})\widehat{k}^{2}}\ .
\ee
The slow roll approximation~\eqref{I32} remains valid for the region of large negative $s$ and small $\chi$. It ends once the slow roll condition,
\be
\label{I33B}
\frac{e^{2s}}{1+e^{2s}}\ll \sqrt{\frac{3\widehat{k}^{2}}{8}}\ ,
\ee
gets violated. For large $\widehat{k}^{2}\gg 8/3$ the condition~\eqref{I33B} or~\eqref{I32} holdes for all $s$.

In the slow roll epoch the Hubble parameter changes only slowly. With
\be
\label{I34}
\frac{\widehat{H}'}{\widehat{H}^{2}}\approx\frac{s'}{6\widehat{H}^{3}}\frac{\partial\widehat{V}}{\partial s}=-\frac{2e^{2s}\; \widehat{V}s'}{3(1+e^{2s})\widehat{H}^{3}}\approx-\frac{2e^{2s}\;   s'}{(1+e^{2s})\widehat{H}}\ ,
\ee
the condition~\eqref{I32} implies for the region of large negative $s$ or small $\chi$ a slow relative change of $\widehat{H}$,
\be
\label{I35}
\bigg{|}\frac{\widehat{H}'}{\widehat{H}^{2}}\bigg{|}\ll\frac{2\sqrt{6}\; e^{2s}}{|\widehat{k}|(1+e^{2s})}\ .
\ee
For $\widehat{k}^{2}\gg 6 $ the approximation $|\widehat{H}'|\ll 2\widehat{H}^{2}$ extends to all $s$ and the slow roll epoch does not end. For the two realistic inflationary scenarios discussed above the slow roll condition is obeyed during the whole inflationary epoch. These solutions can connect smoothly to a beginning with the de Sitter solution~\eqref{S1} -~\eqref{S4} or \eqref{FS9}.

So far we have discussed the slow roll approximation within variable gravity. We next establish that this approximation is self-consistent. From eq.~\eqref{I30} one infers that the variable gravity approximation is valid for 
\be
\label{I36}
3\Ztil|\widehat{H}'|\ll1\quad ,\quad \frac{\ytil\Ztil\overline{Y}\widehat{H}s'}{1+2\ytil}\ll 1 \ .
\ee
The slow roll conditions~\eqref{I32} and 
\bel{288A}
\frac{|\widehat{H}'|}{2\widehat{H}^{2}}\ll 1
\ee
guarantee these conditions provided
\bel{288B}
\abs{\Ztil\widehat{H}^{2}}\lesssim 1\ , \quad \Big{|}\frac{\ytil\overline{Y}}{1+2\ytil}\Big{|}\lesssim 1\ .
\ee
With these condition also the terms~$\sim  \Ztil$ in eqs.~\eqref{I28},~\eqref{I29} remain small.

In the Einstein frame one has $\widehat{H}=H/\overline{M}$ and $\widehat{V}(s)=V(s)/\overline{M}^{4}$. A small amplitude of the primordial fluctuations requires according to eq.~\eqref{I16}
\be 
\nn
\widehat{V}(s)=3\widehat{H}^{2}=5\varepsilon\cdot 10^{-7}\ .
\ee
For any realistic model of inflation the product~$\Ztil\widehat{H}^{2}$ is tiny unless $\Ztil$ is huge. For $\Ztil$ and $\overline{Y}$ not exceeding unity by many orders of magnitude all corrections beyond the variable gravity approximation are suppressed by the very small value of $\widehat{H}^{2}$ and correspondingly $s'^{2}$.

From the point of view of pregeometry this brings us to an interesting conclusion. For homogeneous and isotropic solutions it is well possible that the whole history of our universe can be described within the variable gravity approximation. Higher derivative terms are present for the effective metric theory. They do not play a role for homogeneous isotropic cosmology, however. In particular, for the small values of $\overline{V}$ according to our first family of realistic inflationary models, the value of $\widehat{H}^{2}$ is already tiny for the de Sitter solution \eqref{S1}-\eqref{S4} that may describe the beginning of the Universe.

\section{Graviton fluctuations}
\label{section:GF}

The homogeneous and isotropic universe is an idealization. The real universe is inhomogeneous, as manifest by the observed structures from galaxy clusters to stars. Homogeneity and isotropy may be a good approximation for averages over large enough volumes. This property may hold for the present universe and more generally for the late universe after inflation. We have no guarantee, however, that it holds also towards the beginning, and we will argue in the next section that the beginning is actually highly inhomogeneous. For this reason we address in this section the issue of inhomogeneous solutions of the general field equations derived in sect. \ref{section:GFE}.

We will concentrate in this section on small inhomogeneities for which a linearization of the field equations is a good approximation. In linear order the various inhomogeneities can be treated by separate field equations for the different representations of the rotation group. They do not mix in linear order. We will not perform the most general discussion here but rather concentrate on one particular mode - the graviton fluctuations. Many general features can be found from the evolution of the inhomogeneous cosmologies that can be described by a superposition of graviton modes. We can consider the detailed discussion of the graviton modes as a case study, from which we will draw more general conclusions in the next section. Since the graviton fluctuations mix degrees of freedom from the vierbein and the gauge fields, the new features of pregeometry become visible for this investigation. We perform the computation in the fixed metric frame used in sect. \ref{section:CS}. The generalization to a frame-invariant formulation \cite{CWPFVG} is straightforward and will be employed in the next section.

\zwisch{Traceless transverse tensor fluctuations}

For the fluctuations in the graviton sector we consider the traceless transverse symmetric tensor fluctuations $t_{\mu\nu}$ and $E_{\mu\nu}$. They are encoded in the ansatz
\begin{align}
\label{G1}
e_{\mu}^{\prime\ m} &= \e+\half t_{\mu\nu}e^{m\nu}\ , \nn \\
A_{\mu mn}^{\prime} &= A_{\mu mn}+\half e_{m}^{\ \nu} e_{n}^{\ \rho}\big{(} D_{\nu}E_{\mu\rho}-D_{\rho}E_{\mu\nu}\big{)}\ .
\end{align}
Here $\e$ and $A_{\mu mn}$ are the vierbein and gauge field of the homogeneous isotropic background solution, and the covariant derivative $D_{\nu}$ is evaluated for these background fields. For an isotropic background we decompose the fluctuations into $SO(3)$-representations. In the graviton sector we concentrate on the transversal traceless space components $t_{ij}$ and $E_{ij}$, $i,j=1,2,3$. In Fourier space with comoving momenta $k_{i}$ they obey the constraints
\begin{align}
\label{G2}
\delta^{ij}t_{ij} &= 0\ ,\quad\ k^{i}t_{ij} = 0\ ,\quad\ t_{ji} = t_{ij}\ ,\nn \\
\delta^{ij}E_{ij} &= 0\ ,\quad k^{i}E_{ij} = 0\ ,\quad E_{ji} = E_{ij}\ .
\end{align}

The non-vanishing components of $D_{\nu} E_{\mu\rho}$ are, with $i,j,l=1,2,3$,
\begin{align}
\label{G3}
D_{0} E_{ij}&=(\deta-2\hub)E_{ij}\ ,\quad D_{l} E_{ij}=\partial_{l} E_{ij}=ik_{l} E_{ij}\ , \nn \\
D_{l} E_{i0} &=-\hub E_{il}\ ,\quad D_{l} E_{0j}=-\hub E_{lj}\ .
\end{align}
Correspondingly, one finds for the non-zero components of $a_{\mu mn}=A_{\mu mn}^{\prime}-A_{\mu mn}$,
\begin{align}
\label{G4}
a_{li0}&=-a_{l0i}=-\frac{1}{2a^{2}}(\deta-\hub)E_{li}\ , \nn \\
a_{lij}&=\frac{i}{2a^{2}}(k_{i}E_{lj}-k_{j}E_{li})\ .
\end{align}
The fields $t_{ij}$ and $E_{ij}$ do not mix with other representations of the rotation group in the linear approximation. Also modes with different $\vec{k}$ do not mix. For the evolution of an inhomogeneous universe which can be described by a superposition of graviton modes $t_{ij}(\eta,\vec{k})$ and $E_{ij}(\eta,\vec{k})$ we can therefore investigate the linearized field equations for each $\vec{k}$-mode separately.

\vspace{4\baselineskip}
\zwisch{Linear expansion}

For the derivation of the linear field equations for the graviton fluctuations we first compute their contribution to the field strength $F_{\mu\nu mn}$ and the covariant vierbein derivative $U_{\mu\nu}^{\dub m}$ in linear order. We denote the linear term in an expansion of $F_{\mu\nu mn}$ by $F_{\mu\nu mn}^{(1)}$,
\begin{align}
\label{G5}
F&_{\mu\nu mn}^{(1)}=\partial_{\mu}a_{\nu mn}-\partial_{\nu}a_{\mu mn} \nn \\
&+A_{\mu m}^{\trip p}a_{\nu pn}-A_{\mu n}^{\trip p}a_{\nu pm}-A_{\nu m}^{\trip p}a_{\mu pn}+A_{\nu n}^{\trip p}a_{\mu pm}\ .
\end{align}
For the components that differ from zero for graviton fluctuations one obtains
\begin{align}
\label{G6}
F_{0i0j}^{(1)}&=\frac{1}{2a^{2}}(\deta-2\hub)(\deta-\hub)E_{ij}\ , \nn \\
F_{0ilj}^{(1)}&=\frac{i}{2a^{2}}\big{[}k_{l}(\deta-2\hub)E_{ij}-k_{j}(\deta-2\hub)E_{il}\big{]}\ , \nn \\
F_{li0j}^{(1)}&=\frac{i}{2a^{2}}\big{[}k_{l}(\deta-\hub-b)E_{ij}-k_{i}(\deta-\hub-b)E_{lj}\big{]} \nn \\
&+\frac{c}{2a^{2}}\big{[}\varepsilon_{lj}^{\dub s}(\deta-\hub)E_{is}-\varepsilon_{ij}^{\dub s}(\deta-\hub)E_{ls}\big{]}\ , \nn \\
F_{lisj}^{(1)}&=-\frac{b}{2a^{2}}(\deta-\hub)\big{[}\delta_{ls}E_{ij}+\delta_{ij}E_{ls}-\delta_{lj}E_{is}-\delta_{is}E_{lj}\big{]} \nn \\
&-\frac{1}{2a^{2}}\big{[}k_{l}k_{s}E_{ij}+k_{i}k_{j}E_{ls}-k_{l}k_{j}E_{is}-k_{i}k_{s}E_{lj}\big{]} \nn \\
&+\frac{ic}{2a^{2}}\big{[}\varepsilon_{ls}^{\dub q}(k_{q}E_{ij}-k_{j}E_{iq})-\varepsilon_{lj}^{\dub q}(k_{q}E_{is}-k_{s}E_{iq}) \nn \\
&\quad -\varepsilon_{is}^{\dub q}(k_{q}E_{lj}-k_{j}E_{lq})+\varepsilon_{ij}^{\dub q}(k_{q}E_{ls}-k_{s}E_{lq})\big{]}\ .
\end{align}
In addition, we have the components from the antisymmetry in the first and second pair of indices of $F_{\mu\nu mn}$.

We next turn to the covariant derivative of the vierbein. For a general linear expansion
\be
\label{G7}
e_{\mu}^{\prime\ m}=\e +\half H_{\mu\nu}e^{m\nu}\ ,
\ee
the linear expansion of $U_{\mu\nu\rho}$ reads
\begin{align}
\label{G8}
U_{\mu\nu\rho}^{(1)}&=\omega_{\mu\nu\rho}^{(1)}-a_{\mu mn}e_{\nu}^{\ m}e_{\rho}^{\ n}\nn\\
&-\half A_{\mu mn}e^{n\sigma}(e_{\nu}^{\ m}H_{\rho\sigma}-e_{\rho}^{\ m}H_{\nu\sigma})\ .
\end{align}
Here $\omega_{\mu\nu\rho}^{(1)}$ is given by
\begin{align}
\label{G9}
\omega_{\mu\nu\rho}^{(1)}=\half\Big{[}&\partial_{\rho}H_{\mu\nu}^{(S)}-\partial_{\nu}H_{\mu\rho}^{(S)}+\partial_{\mu}H_{\nu\rho}^{(A)}\nn\\
&-B_{\rho\mu}^{\dub\sigma}H_{\nu\sigma}+B_{\nu\mu}^{\dub\sigma}H_{\rho\sigma}\Big{]}\ ,
\end{align}
where
\be
\label{G10}
H_{\mu\nu}^{(S)}=\half (H_{\mu\nu}+H_{\nu\mu})\ ,\quad H_{\mu\nu}^{(A)}=\half (H_{\mu\nu}-H_{\nu\mu})
\ee
and
\be
\label{G11}
B_{\mu\rho}^{\dub\sigma}=e_{m}^{\dub \sigma}\partial_{\rho}\e\ .
\ee
Insertion of the homogeneous isotropic background and the graviton fluctuations yields for $U_{\mu\nu\rho}^{(1)}=-U_{\mu\rho\nu}^{(1)}$ the following non-zero components
\begin{align}
\label{G12}
U_{ij0}^{(1)}&=-U_{i0j}^{(1)}=\half (\deta-b)t_{ij}+\half (\deta-\hub)E_{ij}\ , \nn \\
U_{ijl}^{(1)}&=\frac{i}{2}\big{[}k_{l}(t_{ij}+E_{ij})-k_{j}(t_{il}+E_{il})\big{]} \nn \\
&-\frac{c}{2}\big{(}\varepsilon_{ij}^{\dub s}t_{ls}-\varepsilon_{il}^{\dub s}t_{js}\big{)}\ .
\end{align}

For the contraction $F_{\mu\nu}$ one obtains in linear order the non-vanishing components
\begin{align}
\label{248A}
F_{ij}^{(1)}&=-\frac{1}{2a^{2}}\big{[}(\deta-2\hub+b)(\deta-\hub)+k^{2}\big{]}E_{ij}\nn\\
&+\frac{ic}{2a^{2}}k_{q}\big{(}2\eps_{j}^{\ lq}E_{il}-\eps_{i}^{\ lq}E_{jl}\big{)}\ ,
\end{align}
and therefore
\bel{248B}
F^{(1)}=0\ .
\ee
We conclude that the invariant $CF^{2}$ does not contribute to the linear dynamics in the graviton sector. We take here $B=0$ as often in the main text. Including $B\neq 0$ essentially replaces $Z\to Z+B$. Furthermore, the contraction $U_{\mu\ \rho}^{\ \mu}$ does not contribute in linear order
\bel{248C}
U_{\dub\mu\ \rho}^{(1) \mu}=0\ .
\ee
We can therefore omit the invariant $\sim\n U_{\mu\ \rho}^{\ \mu}U_{\nu}^{\ \nu\rho}$.

\zwisch{Linearized field equations}

We next compute the field equation \eqref{FE3} in linear order in the graviton fluctuation. We concentrate on background solutions with $c=0$. For $T_{\mu\nu}^{(U)}$ we need the non-zero components of $D_{\rho}U_{\mu\nu}^{\trip\rho}$ in linear order,
\begin{align}
\label{G13}
(D_{\rho}U_{ij}^{\dub\rho})^{(1)}=&-\frac{1}{2a^{2}}\Big{\{}\big{[}\deta^{2}-2\hub\deta+k^{2}\big{]}(t_{ij}+E_{ij}) \nn \\
&+2(b-\hub)\deta t_{ij}+(2\hub^{2}-\hub b -\deta b)t_{ij} \nn \\
&+(\hub^{2}-\deta\hub)E_{ij}\Big{\}}\ .
\end{align}
One obtains the non-zero components of $T^{(U)}$,
\begin{align}
\label{G14}
T_{ij}^{(U,1)}&=-\frac{\m}{2a^{2}}\Big{\{}\Big{[}\deta^{2}-2\hub\deta+k^{2}+\hub^{2} \\
&-\deta\hub\Big{]}(t_{ij}+E_{ij})-(\deta+3b)(b-\hub)t_{ij}\Big{\}}\nn\ .
\end{align}
Similarly, the energy momentum tensor of the gauge bosons receives a contribution
\begin{align}
\label{G15}
T_{ij}^{(F,1)}&=-\frac{Z}{2a^{4}}\Big{\{}2\Big{[}\deta b(\deta-2\hub)(\deta-\hub) \nn \\
&\dub+b^{3}(\deta-\hub)+b^{2}k^{2}\Big{]}E_{ij}+\big{[}b^{4}+3(\deta b)^{2}\big{]}t_{ij}\Big{\}}\ .
\end{align}
Finally, we need the scalar contribution
\begin{align}
\label{G16}
T_{ij}^{(\chi,1)}&=\Big{[}\frac{K}{2a^{2}}(\deta\chi)^{2}-V\Big{]}t_{ij} \\
&-\frac{1}{2a^{2}}\dchi{\m}\deta\chi\big{[}(\deta-b)t_{ij}+(\deta-\hub)E_{ij}\big{]}\nn\ ,
\end{align}
and
\begin{align}
\label{G17}
T_{ij}^{(R,1)}=\frac{\M}{2a^{2}}\Big{\{}&\Big{[}(\deta-2\hub+b)(\deta-\hub)+k^{2}\Big{]}E_{ij} \nn \\
&+(3b^{2}+5\deta b)t_{ij}\Big{\}}\ .
\end{align}
Grouping things together the linearized field equation \eqref{FE3},
\be
\label{G19}
2a^{2}\Big{(}T_{ij}^{(U,1)}+T_{ij}^{(F,1)}+T_{ij}^{(R,1)}+T_{ij}^{(\chi,1)}\Big{)}=0\ ,
\ee
yields a linear differential equation which involves up to two $\eta$-derivatives of $t_{ij}$ and $E_{ij}$.

For a second field equation of a similar type we employ the linear approximation to the pure space components of the gauge field equation \eqref{FE1}
\be
\label{G20}
Z(D_{\nu}F_{\ ijl}^{\nu})^{(1)}=-J_{ijl}^{(1)}\ ,
\ee
where $i$ is a world index and $j,l$ are Lorentz indices. For the linear expansions one obtains
\begin{align}
\label{G21}
(D_{\nu}F_{\ ijl}^{\nu})^{(1)}&=\frac{i}{2a^{4}}\Big{\{}2b^{2}(k_{l}t_{ij}-k_{j}t_{il}) \\
&+\Big{[}\deta^{2}+(2b-4\hub)\deta+4\hub^{2} \nn \\
&-2\deta\hub-2\hub b+k^{2}\Big{]}(k_{l}E_{ij}-k_{j}E_{il})\Big{\}}\nn
\end{align}
and
\begin{align}
\label{G22}
J_{ijl}^{(1)}=\frac{i}{2a^{4}}\Big{\{}&a^{2}(\m-\M)\big{[}k_{l}(E_{ij}+t_{ij})-k_{j}(E_{il}+t_{il})\big{]} \nn \\
&+\dchi{Z}\deta\chi(\deta-2\hub)\big{[}k_{l}E_{ij}-k_{j}E_{il}\big{]}\Big{\}}\ .
\end{align}
The combination of eqs. \eqref{G19}, \eqref{G20} yields a system of two closed linearized differential equations for the two functions $E_{ij}$ and $t_{ij}$. It can be solved by standard linear analysis.

\zwisch{Mode mixing}

The modes $t_{ij}$ and $E_{ij}$ are mixed. As a next task we want to find the linear combinations (eigenmodes) whose time evolution can be solved separately. Multiplication of eq. \eqref{G20} with $k_{l}/k^{2}$ yields the second oder differential equation
\be
\label{G23}
AE_{ij}+Bt_{ij} = 0\ ,
\ee
with operators
\begin{align}
\label{G24}
A&=\deta^{2}+(2b-4\hub)\deta+2(2\hub^{2}-\deta\hub-\hub b)+k^{2} \nn \\
&+a^2\frac{\m-\M}{Z}+\dchi{\ln Z}\deta\chi(\deta-2\hub)\ , \nn \\
B&=2b^{2}+a^{2}\frac{\m-\M}{Z}\ .
\end{align}
Eq. \eqref{G19} can be written in a similar form 
\be
\label{G25}
CE_{ij}+Dt_{ij} = 0\ ,
\ee
with
\begin{align}
\label{G26}
C&=\m\Big{[}\deta^{2}-2\hub\deta+\hub^{2}-\deta\hub+k^{2} \nn \\
&\quad\quad\quad+\dchi{\ln\m}\deta\chi(\deta-\hub)\Big{]} \nn \\
&-\M\Big{[}(\deta-2\hub+b)(\deta-\hub)+k^{2}\Big{]} \nn \\
&+\frac{2Z}{a^{2}}\Big{[}\deta b(\deta-2\hub)(\deta-\hub)+b^{3}(\deta-\hub)+b^{2}k^{2}\Big{]}\ ,
\end{align}
and
\begin{align}
\label{G27}
D&=\m\Big{[}\deta^{2}-(\hub+b)\deta+\hub^{2}-\deta b-3b^{2}+3\hub b+k^{2} \nn \\
&\quad\quad\quad+\dchi{\ln\m}\deta\chi(\deta-b)\Big{]} \nn \\
&-\M\big{[}3b^{2}+5\deta b\big{]}+\frac{Z}{a^{2}}\big{[}b^{4}+3(\deta b)^{2}\big{]} \nn \\
&-K(\deta\chi)^{2}+2a^{2}V\ .
\end{align}

We can organize the two equations \eqref{G23}, \eqref{G25} as a matrix equation
\be
\label{G28}
\Matrix{A}{B}{C}{D}\Vector{E_{ij}}{t_{ij}}=\tilde{P}\Vector{E_{ij}}{t_{ij}}=\tilde{P}\psi=0\ .
\ee
Non-trivial solutions correspond to vanishing eigenvalues of the operator $\Ptil$. They are the corresponding eigenfunctions. These eigenfunctions depend on the background configuration $\chi$, $\e$ and $\A$ or the functions $\chi(\eta)$, $b(\eta)$ and $\hub(\eta)$. We will discuss separately a few important cases.

\zwisch{Fluctuations for Minkowski geometry and vanishing homogeneous gauge fields}

Let us first concentrate for $V=0$ on a flat background geometry with vanishing gauge fields, $\hub=b=0$, and $\deta\chi=0$. We take $a=1$ for simplicity and define the squared four-momentum in Fourier space
\be
\label{G29}
q^{2}=\deta^{2}+k^{2}=q^{\mu}q_{\mu}=-q_{0}^{2}+k^{2}\ .
\ee
The matrix $\Ptil$ reads
\be
\label{G30}
\Ptil=\renewcommand*{\arraystretch}{1.2}\Matrix{q^{2}+\frac{\m-\M}{Z}}{\frac{\m-\M}{Z}}{(\m-\M)q^{2}}{\m q^{2}}\ ,
\ee
and the zero eigenvalues obey $\det\Ptil=0$, or
\be
\label{G31}
q^{2}\big{[}Zq^{2}+\m y(1-y)\big{]}=0\ ,
\ee
where $y=\M/\m$.

The eigenvalue $q^{2}=0$ corresponds to the massless graviton. The corresponding eigenvector is a plane wave, $k=\abs{\vec{k}}$,
\be
\label{G33}
\psi(\vec{k})=\psi_{+}(\vec{k})e^{ik\eta}+\psi_{-}e^{-ik\eta}\ .
\ee
The second eigenvalue,
\be
\label{G34}
q^{2}=-\mu^{2}\ ,\quad \mu=\frac{\m y(1-y)}{Z}\ ,
\ee
corresponds to the solutions
\be
\label{G35}
\psi(\vec{k})=\psi_{+}^{\prime}(\vec{k})e^{i\sqrt{k^{2}+\mu^{2}}\eta}+\psi_{-}^{\prime}(\vec{k})e^{-i\sqrt{k^{2}+\mu^{2}}\eta}\ .
\ee
This describes a massive spin-two excitation if $\mu^{2}>0$. For $\mu^{2}<0$ one obtains a tachyonic behavior with exponentially growing modes. For stability we require the range
\be
\label{G36}
0<\M<\m\ .
\ee
As it should be, these results coincide with the general stability analysis in appendix \ref{app:A}.

\zwisch{Fluctuations for de Sitter solutions of type 1}

We next turn to the de Sitter solution \eqref{S1}-\eqref{S4}. This early attractor solution describes the inflationary epoch. It is therefore the relevant background configuration for the investigation of primordial tensor fluctuations. We find for the transversal traceless fluctuations the same behavior for the massless graviton as for general relativity. This is supplemented by tensor fluctuations arising from an additional massive graviton. The latter shows the same behavior as for a massive particle in general relativity.

Defining
\be
\label{G37}
\beta=\frac{2ZV}{3\M\m}
\ee
one finds with $2Z\hub^{2}/a^{2}=\beta\m$ the matrix elements
\begin{align}
\label{G38}
A&=\deta^{2}-2\hub\deta+k^{2}+\frac{2(1-y)}{\beta}\hub^{2} \nn\ , \\
B&=2\Big{(}\frac{1-y}{\beta}+1\Big{)}\hub^{2} \nn\ , \\
C&=\m\big{(}1+\beta-y\big{)}\big{(}\deta^{2}-2\hub\deta+k^{2}\big{)} \nn\ , \\
D&=\m\big{[}\deta^{2}-2\hub\deta+k^{2}+2(\beta-y)\hub^{2}\big{]}\ .
\end{align}
The general solution of the system of equations \eqref{G28},\eqref{G38} depends on the two parameters $\beta$ and $y$, while $\m$ can be absorbed into the normalization of the eigenmodes.

For the eigenmodes we make the general ansatz
\begin{align}
\label{G39}
E_{ij}&=c_{Et}\ t_{ij}\ , \nn \\
\big{[}\deta^{2}-2\hub\deta&+k^{2}+\gamma\hub^{2}\big{]}t_{ij}=0\ .
\end{align}
Insertion into eq. \eqref{G28} yields two algebraic equations,
\begin{align}
\label{G40}
\bigg{(}\frac{2(1-y)}{\beta}-\gamma\bigg{)}c_{Et}+\frac{2(1-y)}{\beta}+2&=0\ , \nn \\
-\big{(}1+\beta-y\big{)}\gamma c_{Et}+2(\beta-y)-\gamma&=0\ .
\end{align}
The two solutions for $\gamma$ are given by
\be
\label{G41}
\gamma_{+}=-\frac{2}{\beta}(\beta-y)(1-y)\ ,\quad c_{Et}^{+}=-\frac{1}{1-y}\ ,
\ee
and 
\be
\label{G42}
\gamma_{-}=-2\ ,\quad c_{Et}^{-}=-1\ .
\ee
For given constant $\gamma$ we can solve the evolution equation \eqref{G39} and establish the behavior of the corresponding eigenmodes.

A particularly simple solution occurs for $\beta=y$, where $\gamma_{+}=0$, and eq. \eqref{G39} is solved with $\hub=-1/\eta$ by 
\begin{align}
\label{G43}
t_{ij}(\eta,\vec{k})&=a(\eta)\tilde{t}_{ij}(\eta,\vec{k})\ , \nn \\
\tilde{t}_{ij}(\eta,\vec{k})&=\tilde{t}_{ij}^{+}(\vec{k})e^{ik\eta}+\tilde{t}_{ij}^{-}(\vec{k})e^{-ik\eta}\ .
\end{align}
The corresponding perturbation of the vierbein \eqref{G1} is a simple oscillating function
\be
\label{G44}
e_{i}^{\ j}(\eta,\vec{k})=a(\eta)\delta_{ij}+\half\tilde{t}_{ij}(\eta,\vec{k})\ .
\ee
The metric fluctuation is given by
\begin{align}
\label{G45}
g_{ij}(\eta,\vec{k})&=a^{2}(\eta)\delta_{ij}+a(\eta)\tilde{t}_{ij}(\eta,\vec{k})+\frac{1}{4}\tilde{t}_{i}^{\ l}(\eta,\vec{k})\tilde{t}_{lj}(\eta,\vec{k})\nn\\
&\approx a^{2}(\eta)\big{(}\delta_{ij}+\frac{\tilde{t}_{ij}(\eta,\vec{k})}{a(\eta)}\big{)}\ .
\end{align}
The second line holds for the linear approximation, such that linear relative metric fluctuations diverge $\sim a^{-1}(\eta)$ for $\infpast$. We will see that the asymptotic behavior of the solution for $\infpast$ is independent of the precise value of $\gamma$. The asymptotic behavior \eqref{G43}-\eqref{G45} is general and does not require $\beta=y$.

\zwisch{Graviton mode}

For general $\gamma$ we may write eq. \eqref{G39} in terms of the relative metric fluctuation
\be
\label{G46}
\gamma_{ij}=a^{-2}t_{ij}\ .
\ee
The relative graviton fluctuations $\gamma_{ij}$ are frame-invariant quantities \cite{CWPFVG}. The solutions $\gamma_{ij}(\eta,\vec{k})$ are the same in all metric frames related by Weyl scalings.
Using $\deta\hub=\hub^{2}$ one obtains from eq. \eqref{G39}
\be
\label{G47}
\big{(}\deta^{2}+2\hub\deta+(2+\gamma)\hub^{2}+k^{2}\big{)}\gamma_{ij}=0\ .
\ee
For the class of solutions \eqref{G42} with $\gamma=-2$ eq. \eqref{G47} is the standard equation for graviton fluctuations in de Sitter space for general relativity. We conclude that for the de Sitter solution \eqref{S1}-\eqref{S2} of type 1 the graviton fluctuations in pregeometry and general relativity obey the same evolution equation. The solution for $\gamma=-2$ takes the well known form
\be
\label{G48}
\gamma_{ij}(\eta,\vec{k})=\gamma_{ij}^{+}(\vec{k})w_{k}^{+}(\eta)+\gamma_{ij}^{-}(\vec{k})w_{k}^{-}(\eta)\ ,
\ee
with mode functions $w_{k}^{\pm}$ given by
\be
\label{G49}
w_{k}^{-}(\eta)=\big{(}w_{k}^{+}(\eta)\big{)}^{\ast}=\frac{1}{a(\eta)\sqrt{2k}}\Big{(}1-\frac{i}{u}\Big{)}e^{-iu}\ ,
\ee
where
\be
\label{G50}
u=k\eta=-\frac{k}{\hub(\eta)}=-\frac{k}{a(\eta)H}\ .
\ee
The asymptotic behavior for $k\eta\to -\infty$ is given by the solution \eqref{G43}-\eqref{G45}.

In the limit $k\eta\to -\infty$ we can neglect in eq. \eqref{G39} the term $\gamma\hub^{2}$ as compared to $k^{2}$. The solution becomes therefore independent of $\gamma$ or $\beta$. For fixed $k$ the asymptotic behavior in the infinite past $\infpast$ is the same for both modes with $\gamma_{+}$ or $\gamma_{-}$. It is universally given by the plane waves \eqref{G43} for $\ttil_{ij}$. This universal property holds for the de Sitter solutions of type 1. We will see that it is not realized for the de Sitter solutions of type 2.

We have normalized the mode functions $w_{k}^{\pm}(\eta)$ such that they are directly related to the graviton propagator $G_{\text{grav}}(k,\eta)$ and the primordial tensor spectrum $\Delta_{T}^{2}(k,\eta)$,
\begin{align}
\label{292A}
G_{\text{grav}}(k,\eta)&=\frac{4}{\M}\abs{w_{k}^{-}(\eta)}^{2}\ ,\nn\\
\Delta_{T}^{2}(k,\eta_{\text{hc}})&=\frac{k^{3}G_{\text{grav}}(k,\eta_{\text{hc}}(k))}{\pi^{2}}\ ,
\end{align}
where $\eta_{\text{hc}}(k)$ denotes the time when a given $k$-mode crosses the horizon. This happens for $u=-1$, or 
\bel{292B}
-k\eta_{\text{hc}}(k)=\frac{k}{\hub_{\text{hc}}}=\frac{k}{a_{\text{hc}}H_{\text{hc}}}=1\ .
\ee
In other words "horizon crossing" occurs once the physical momentum $k/a$ gets smaller than $H$. The graviton propagator in eq. \eqref{292A} refers directly to the relative graviton fluctuations.

The graviton propagator is given by the equal time two-point correlation function for (infinitesimal) graviton fluctuations,
\begin{align}
\label{292C}
\langle t_{ij}(\eta,\vec{k})&t_{lm}^{*}(\eta',\vec{k}')\rangle=\\
&a^{2}(\eta)a^{2}(\eta')G_{\text{grav}}(k,\eta,\eta')P_{ijlm}^{(\gamma)}(\vec{k})\delta(\vec{k}-\vec{k}')\ ,\nn
\end{align}
with $P_{ijlm}^{(\gamma)}(k)$ a projector on the graviton fluctuations and $G_{\text{grav}}(k,\eta)=G_{\text{grav}}(k,\eta,\eta)$. The tensor fluctuations at horizon crossing obey
\bel{292D}
\frac{\Delta_{T}^{2}(k,\eta_{\text{hc}})}{\Delta^{2}(k)}=2r\ ,
\ee
where the factor $2$ relates $\Delta_{T}^{2}(k,\eta_{\text{hc}})$ to the observable tensor spectrum for $k\eta\to 0$. Eq. \eqref{292D} expresses $\Delta_{T}^{2}(k,\eta_{\text{hc}})$ in terms of the observed amplitude of the scalar fluctuation spectrum
\bel{292E}
\Delta_{T}^{2}(k,\eta_{\text{hc}})\approx 4\cdot 10^{-9}r\ .
\ee

Inserting $u=-1$ for $\abs{w_{k}^{-}(u)}^{2}$ one finds the familiar relation to the Hubble parameter at horizon crossing
\bel{292F}
\Delta_{T}^{2}(k,\eta_{\text{hc}})=\frac{4H_{\text{hc}}^{2}}{\pi^{2}\M}\ ,
\ee
determining
\bel{292G}
\frac{H_{\text{hc}}^{2}}{\M}\approx 10^{-8}r .
\ee

This brief summary reveals that for the early attractor solution (de Sitter solution of type 1) the behavior of one of the modes corresponds precisely to the graviton in general relativity. This finding should not be too surprising, given the previous observation that variable gravity is a very good approximation for the de Sitter solutions of type 1.

\zwisch{Massive tensor particle mode}

Besides the standard graviton fluctuation our model of pregeometry has an additional spin-two mode corresponding to the solution \eqref{G41}. For this mode the asymptotic behavior for $k\eta\to -\infty$ is given by eqs. \eqref{G43}-\eqref{G45}, independently of the precise value of $\gamma_{+}$ or $\beta$. In this limit the relative influence of the term $\sim\gamma\hub^{2}$ in eq. \eqref{G39} is suppressed by a factor $\gamma/(k\eta)^{2}$. We observe that $2+\gamma$ can be associated to a mass term $m_{p}^{2}$
\be
\label{G51}
2+\gamma=\frac{m_{p}^{2}}{H^{2}}=\frac{a^{2}m_{p}^{2}}{\hub^{2}}=\tilde{m}_{p}^{2}\ .
\ee
For our model of pregeometry the primordial tensor spectrum will receive an additional contribution from a massive graviton. This corresponds to the stable massive graviton found in the stability analysis for flat space.

The general solution of eq. \eqref{G39} involves again mode functions $w_{k}^{\pm}(\eta)$ as for eq. \eqref{G48}. Their asymptotic behavior for $u\to -\infty$ is the same as in eq. \eqref{G49}, while the behavior for smaller $\abs{u}$ depends on $\gamma$. Eq. \eqref{G47} for $w_{\pm}$ can be rewritten as an equation for $v_{\pm}=aw_{\pm}$,
\begin{align}
\label{G52}
\big{(}\deta^{2}+\gamma\hub^{2}+k^{2}\big{)}v_{\pm}&=0\ ,\quad v_{\pm}=aw_{\pm}\ , \nn \\
\big{(}\du^{2}+\frac{\gamma}{u^{2}}+1\big{)}v_{\pm}&=0\ .
\end{align}
With the ansatz
\be
\label{G53}
v_{k}^{-}=\frac{1}{\sqrt{2k}}b(u)e^{-iu}\ ,\quad \lim_{u\to -\infty}b(u)=1\ ,
\ee
the function $b(u)$ obeys the differential equation
\be
\label{G54}
\big{(}\du^{2}-2i\du+\frac{\gamma}{u^{2}}\big{)}b=0\ .
\ee
For $u\to 0$ one finds the approximate solution by neglecting the term $-2i\du$,
\be
\label{G55}
b=c_{b}(-u)^{d_{b}}\ ,\quad d_{b}=\half\big{(}1-\sqrt{1-4\gamma}\big{)}\ .
\ee
For $\gamma<0$ one has $d_{b}<0$ and $b$ diverges for $u\to 0$. The particular value $\gamma=-2$, $d_{b}=-1$ reproduces the mode function \eqref{G49}, which corresponds to
\be
\label{G56}
b=1-\frac{i}{u}\ .
\ee
Positive $\beta$ implies $\gamma_{+}>-2$. In turn, this corresponds to a positive mass term $m_{p}^{2}$ in eq. \eqref{G51}. In this case one finds $d_{b}>-1$, such that the leading increase for $u\to 0$ corresponds to the "graviton mode" with $\gamma_{-}=-2$. The solution $b(u)$ interpolates smoothly between $b(u\to -\infty)=1$ and the increase \eqref{G55} for $u\to 0$. Horizon crossing for a given $k$-mode occurs for $u=-1$. Solutions for more general geometries close to de Sitter space with $\deta\hub=(1+\nu)\hub^{2}$ can be found in ref. \cite{Wetterich_2015}. They show a qualitatively similar behavior as for de Sitter space with $\nu=0$.

Since we always consider $V\geq0$ the stability conditions $Z>0$, $\m>0$, $\M>0$ imply $\beta>0$. Our model of pregeometry therefore describes a massless graviton and an additional massive spin-two particle. For both the evolution of the primordial fluctuations is the same as for massless and massive particles in general relativity. This is one more facet of the validity of variable gravity for the inflationary epoch. If the amplitude of both modes is of similar size at early times with $-u\gg 1$, the contribution of the massive excitation is suppressed for $-u\ll 1$.

\zwisch{Fluctuations for de Sitter solutions of type 2}

We have already seen that the de Sitter solutions of type 2 describe the limit of the basin of attraction towards the early attractor solution in the space of homogeneous fields. We find that an instability can occur as well in the sector of transversal traceless tensor modes. This investigation also reveals that the behavior of fluctuations similar to general relativity, that we have found for the de Sitter solution of type 1, is not general. In general, the evolution of fluctuations in pregeometry differs from general relativity or variable gravity. We will display a detailed discussion in appendix \ref{app:D}.

\section{The beginning of the universe}
\label{section:BOU}

Our model of pregeometry is a proposal for a completion of the gravitational interactions at the shortest distances. It is natural to ask what are the consequences for the beginning of the universe. We have advocated that this beginning can be associated with an ultraviolet fixed point in the flow of couplings, or more generally the effective action. The quantitative properties will depend on the behavior of $Z(\chi^{2}/k^{2})$ for $\chi^{2}/k^{2}\to 0$, and similar for other couplings. It is not guaranteed that our approximation of constant coupling functions for $\chi^{2}/k^{2}\to 0$ is valid. Nevertheless, several interesting results follow in this approximation. Important general features may remain valid beyond the approximation of constant couplings.

The main outcome of our investigation of cosmological solutions for our simple model of pregeometry is a beginning with a "non-geometric state". For this state the expectation value of the vierbein $\etil_{\mu}^{\ m}$ vanishes. Therefore also the metric $\gtil_{\mu\nu}$ vanishes and can no longer be used for the definition of a geometry. The gauge fields $\A$ and the scalar field $\chi$ vanish as well. The beginning is a vacuum state with zero expectation values of all fields. The two-point correlation functions for these fields do not vanish, however. They take constant values in the beginning state. The beginning is similar to many probabilistic systems in the disordered phase. Vacuum expectation values of fields associated to order parameters vanish, while correlations differ from zero.

The vacuum state characterizing the beginning is unstable with respect to the increase of small expectation values of $\etil_{\mu}^{\ m}$, $\A$ and $\chi$. These expectation values define the homogeneous isotropic background configuration. As long as the correlation functions dominate the universe is highly inhomogeneous. As $\etil_{\mu}^{\ m}$ increases, and the size of the inhomogeneities reflected by the correlation functions remains constant, the relative size of the inhomogeneities decreases. Once the expectation values dominate, the universe is dominantly homogeneous, with remaining inhomogeneities reflected by the small primordial density fluctuations.

Once the effects of quantum fluctuations are included in the computation or assumption of the quantum effective action, the evolution of both expectation values and correlations follow exact "classical" field equations derived from the first and second functional derivative of the effective action. On the level of the quantum effective action, no further quantum effects should be included - they would amount to double counting. The solutions of field equations discussed in the present paper are supposed to reflect the full quantum field theory already from the beginning. The universe becoming "classical" is in this picture nothing else than the domination of expectation values of fields over the (connected) correlation functions. Before turning to the physical properties of the beginning we will first have to clarify a few conceptual issues about the meaning of a "beginning state".

\zwisch{Infinite past}

The homogeneous solutions discussed in this paper can be extrapolated backwards to the infinite past in conformal time, $\infpast$. It has been argued that conformal time is a good proxy for physical time, which may be defined by counting the number of oscillations of the photon wave function or the graviton fluctuations \cite{CWEU, CWGE}. Indeed, the graviton fluctuations discussed in the preceding section oscillate $\sim e^{ik\eta}$, such that for any given $k$ the number of oscillations if infinite for $\infpast$. A discussion of the beginning of the universe therefore concerns its state in the infinite past for $\infpast$. If one wants to avoid strict infinities one can replace $\infpast$ by finite very large negative $\eta$.

For the de Sitter solutions one has in a frame invariant formulation $\hubhat=-1/(\eta-\eta_{0})$ and we choose for simplicity the additive constant $\eta_{0}=0$. From
\bel{B1}
\hubhat=\deta\ln A=-\frac{1}{\eta}\ ,
\ee
one infers that $A(\infpast)$ vanishes as
\bel{B2}
A=M(\chi)a=-\frac{1}{\Hhat\eta}\ .
\ee
with constant $\Hhat$.
The vanishing of $A$ can be accounted for by $a\to 0$ or $M(\chi)\to 0$. This issue depends on the choice of the metric frame, as we will discuss below. In turn, the time $\that$ obeys
\bel{B3}
\frac{\text{d}\that}{\text{d}\eta}=A=-\frac{1}{\Hhat\eta}\ ,\quad \that=\that_{0}+\frac{1}{\Hhat}\ln\Big{(}-\frac{1}{\Hhat\eta}\Big{)}\ .
\ee
The infinite past in conformal time, $\infpast$, is also the infinite past in the frame invariant cosmic time, $\that\to -\infty$. The constant frame invariant Hubble parameter $\Hhat$ is fixed for a given de Sitter solution by the parameters of the model.

\zwisch{Cosmic attractors and uncertainty of beginning}

We have seen that the de Sitter solution of type 1, with $b=\hub$, or $\bhat=\Hhat$ in a frame invariant formulation, is a cosmic attractor. This holds if the evolution of the scalar field $\chi$ and the corresponding change of $\Vhat(\chi)$ can be neglected. This situation is realized for the beginning epoch for all solutions discussed so far, for which the scalar field vanishes in the infinite past, $\chi(\infpast)=0$. As can be seen from Figs. \ref{fig:1}, \ref{fig:3}, for a large family of initial conditions the homogeneous solution rapidly approaches the de Sitter solution of type 1, with typical time scales given by $m^{-1}$, $\widetilde{m}^{-1}$ or $M^{-1}$.
The de Sitter solutions of type 1 are the "early attractor solutions".

Relative inhomogeneities have been found to decrease and the homogeneous early attractor solution is approached in this sense. We may take the graviton solutions as a proxy for more general inhomogeneous solutions. The results of the previous section can be taken over to the frame invariant formulation by replacing $a\to A$, $\hub\to\hubhat$ etc. Conformal time $\eta$, the comoving momenta $\vec{k}$ and the relative tensor fluctuations,
\bel{B4}
\gamma_{ij}=A^{-2}t_{ij}\ ,
\ee
are frame invariant quantities. Le us start at some time $\eta_{0}$ far in the past with sufficiently small
\bel{B5}
\gamma_{ij}(\eta_{0},\vec{k})=\gamma_{ij}^{-}(\vec{k})w_{k}^{-}(\eta_{0})\approx\frac{\gamma_{ij}^{-}(\vec{k})}{A(\eta_{0})\sqrt{2k}}e^{-ik\eta_{0}}\ ,
\ee
such that the linear approximation applies. These fluctuations are damped at some later time $\eta>\eta_{0}$ due to the increase of $A$,
\bel{B6}
\gamma_{ij}(\eta,\vec{k})=\frac{A(\eta_{0})}{A(\eta)}e^{-ik(\eta-\eta_{0})}\gamma_{ij}(\eta_{0},\vec{k})\ .
\ee
For $\abs{\eta}\ll\abs{\eta_{0}}$ the suppression factor is tiny,
\bel{B7}
\frac{A(\eta_{0})}{A(\eta)}=\frac{\abs{\eta}}{\abs{\eta_{0}}}\ .
\ee
The relative inhomogeneity is erased effectively, and the homogeneous solution becomes a better and better approximation.

If we move for a fixed $\eta$ the time $\eta_{0}$, where initial conditions are set, to the infinite past, $\eta_{0}\to -\infty$, all such inhomogeneities are erased completely, if the initial inhomogeneity $\gamma_{ij}(\eta_{0},\vec{k})$ is not too large such that the linear approximation is valid. For small enough $\gamma_{ij}(\eta_{0},\vec{k})$ in the infinite past the model predicts at any finite $\eta$ a homogeneous universe, $\gamma_{ij}(\eta,\vec{k})=0$. The situation remains the same if we generalize the initial conditions to include $\gamma_{ij}^{+}(\vec{k})\neq 0$ or if we consider more general initial conditions for $t_{ij}$ and $E_{ij}$ that include contributions from both modes according to eqs. \eqref{G41}, \eqref{G42}. For a suitable range of parameters this behavior also holds for the other inhomogeneous fluctuations.

Let us define as the "basin of attraction" of the de Sitter solution of type 1 all initial conditions for which the solution is attracted to this de Sitter solution. The rate of attraction is given by $\Hhat$. Already after the first 20 $e$-foldings the memory of the precise initial conditions is erased (almost) completely if one starts well within the basin of attraction (not very close to the boundary). There is no way to distinguish by observation the different initial conditions if they are set in the infinite past. An attractor solution necessarily leads to an ambiguity or uncertainty of the precise beginning since different beginnings well withing the basin of attraction cannot be distinguished. The question of the precise beginning becomes then a metaphysical rather than a physical question. On the other hand, all initial conditions well within the basin of attraction can be considered as equivalent. They cannot be distinguished by observation. Within this equivalence class we may choose the precise attractor solution as a representative. This is what we will do in the following. We will often call the de Sitter solution of type 1 the "early attractor solution". Our notion of a "beginning state" should be viewed in this context. It is a representative of a family of states that cannot be distinguished by observation at finite $\eta$.

\zwisch{Crossover at boundary of basin of attraction}

The situation changes for initial conditions at the boundary of the basin of attraction. Such boundaries are often a type of "watershed" between different basins of attraction, as visible in Fig. \ref{fig:3}. The de Sitter solutions of type 2 are precisely on the boundary of the basin of attraction for the early attractor solution. Since they are exact solutions, the universe can stay for infinite time on the boundary if the initial conditions are chosen precisely on the boundary. Any small deviation from the exact de Sitter solution of type 2 will lead to an evolution away from the watershed into one or the other basin of attraction.

The two types of de Sitter solutions can be considered as "scaling solutions" for the time evolution since their properties do not change in time. For initial conditions very close to the de Sitter solution of type 2 one observes a "crossover" from this scaling solution to the other scaling solution, namely the de Sitter solution of type 1. This crossover in time is very similar to the crossover between two fixed points in the renormalization flow of couplings. The evolution from one scaling solution to another defines a "crossover trajectory". These crossover trajectories are displayed in Fig. \ref{fig:3}. The crossover trajectories are not arbitrary. Often they can be characterized by a single parameter, namely the time at which a given point on the trajectory is reached. It can also happen that a family of crossover trajectories exists, characterized by more than one parameter. The number of parameters corresponds to the number of unstable directions (the correspondence to relevant parameters at fixed points of the renormalization flow) of the first scaling solution (de Sitter solution of type 2). This is generically a small number.

\zwisch{Initial conditions and fine tuning}

For any point on the crossover trajectory one can follow the time evolution forwards and backwards. For the forward evolution the solution approaches the stable early attractor solution, while the backwards evolution leads towards the unstable de Sitter solution of type 2. Let us now consider general "initial conditions" for the solution of the differential field equations at some finite time $\eta$. We first restrict these initial conditions to points on the crossover trajectory. Extrapolating backwards to the infinite past one always ends at the unstable de Sitter solution of type 2. The only exception is an initial condition precisely on the de Sitter solution of type 1. Moving forwards in time one always ends in the infinite future on the de Sitter solution of type 1. This can only be avoided by tuning the initial conditions such that they correspond precisely to the de Sitter solution of type 2. Looking at the universe as a whole for all times one can set initial conditions on the crossover trajectory for arbitrary $\eta$. They only fix the value of $\eta$ at which a given point on the crossover trajectory is reached. While initial conditions can be set, in principle, for arbitrary $\eta_{0}$, one has to accept tuning if one wants to describe a universe that for a given $\eta$ is at some particular point on the crossover trajectory and $\eta-\eta_{0}$ is large.
This is manifest for Fig. \ref{fig:3} where we have employed a high degree of tuning in order to have a crossover at the value of $\that$ shown in the figure. More and more tuning would be needed in order to postpone the crossover to later $\that$.

One can choose at some given $\eta$ arbitrary initial conditions on the crossover trajectory and extrapolate both forwards and backwards. This does not imply that arbitrary general initial conditions away from the crossover trajectory can be set at $\eta$ if a given model is valid since the infinite past. The reason is the predictivity related to the restricted number of parameters characterizing a crossover trajectory. A given model for a universe existing since the infinite past predicts certain properties at $\eta$. Any "observer" at $\eta$ can only find states that are compatible with these predictions. Setting his initial conditions at $\eta$ these "allowed states" can be extrapolated backwards to the infinite past and forwards to the infinite future. In contrast, setting initial conditions violating the predictions should lead to inconsistencies. If initial conditions violating the predictions would lead to consistent solutions up to the infinite past, this would contradict the existence of a prediction. In summary, allowed initial conditions at finite $\eta$ have to be consistent with the properties of the crossover solutions and the associated predictions.

There is a simple way how to achieve at finite $\eta$ consistency with the predictions of a given model. One sets initial conditions at $\eta_{0}$ in the far distant past and considers the limit $\eta_{0}\to -\infty$. The predictivity of a model is connected to the restricted set of "relevant parameters" or "non-decaying modes". "Decaying modes", which are the equivalent to irrelevant parameters for the renormalization flow, approach fixed values for $\eta-\eta_{0}\to\infty$. They correspond to partial fixed points that are necessarily reached if the evolution lasts for infinite time. The predictivity of a model is precisely related to those partial fixed points. Since for arbitrary initial conditions at $\eta_{0}$ the evolution reaches precisely the partial fixed point if $\eta-\eta_{0}\to\infty$, the consistency with the prediction at $\eta$ is guaranteed if initial conditions are set at $\eta_{0}$.

Predictions for some quantities can be extremely precise. In this case a very high degree of fine-tuning is necessary if one wants to set initial conditions at finite $\eta$ and extrapolate backwards to the infinite past. This fine tuning is not needed for an evolution towards the future. Since the decaying modes anyhow approach their partial fixed points with increasing $\eta$, a small deviation from the partial fixed point for the initial condition does not matter. By the same reasoning it is sufficient to set initial conditions in some finite far distant past. A strict limit $\eta_{0}\to -\infty$ is not necessary.

\zwisch{Arrow of time and backward singularities}

The reason why initial conditions are better chosen in the distant past and not at arbitrary finite $\eta$ is the presence of an arrow of time \cite{CWVG}. While the field equations are time-reversal invariant, the presence of scaling solutions singles out an arrow of time. The positive time direction is the one in which the stable attractor solution is reached. A large region of initial conditions well within the basin of attraction leads to solutions of the field equations that can be continued to arbitrary large $\eta$. They all approach the attractor solution. Only very special initial conditions, namely those on the crossover trajectory with the sign of all time derivatives switched, evolve with increasing $\eta$ towards the unstable de Sitter solution of type 2. For tiny deviations from these initial conditions the solution will only come close to the de Sitter solution of type 2, but finally deviate from it and approach for large enough $\eta$ the stable scaling solution. The situation is similar to the evolution of correlation functions in the process of thermalization, where the presence of thermal equilibrium as a stable attractor solution induces an arrow of time \cite{ABW1, ABW2}.

The backward extrapolation of arbitrary "initial conditions" at finite $\eta$ will typically lead to divergent solutions or a singularity if one does not start on a point of the crossover trajectory. Seen from initial conditions set at $\eta_{0}$ the explanation is simple. The corresponding "initial condition" at $\eta$ can simply not be reached by any solution if $\eta-\eta_{0}\to\infty$. Only a certain range of field values at $\eta$ is possible, namely those corresponding to a point on the crossover trajectory.

Starting in the far distant past with finite relative inhomogeneities within the range of validity of a linear expansion this argument predicts a completely homogeneous universe. This contrasts with structures observed in our universe, or the observed anisotropies in the cosmic microwave background. We will resolve this puzzle below. The solution is related to the fact that the attractive character of the homogeneous solutions concerns only the relative size of inhomogeneities. The absolute size of inhomogenous fluctuations does not decrease. This will be discussed in more detail below.

\vspace{\baselineskip}
\zwisch{Instability of early attractor solution}

The early attractor solution attracts the neighboring homogeneous solutions only as long as $\chi$ can be approximated by zero or a small constant. In a more complete space of field configurations it has an unstable mode which is given by a nonzero homogeneous scalar field $\chi(\eta)$. More precisely, for $\chi=0$ the de Sitter solution of type 1 is an attractor solution for the gauge fields and vierbein. In the absence of a scalar field the universe would approach the early attractor solution and stay there forever. It is the presence of the scalar field that induces an instability of the early attractor solution and permits a further crossover to a different scaling solution for the infinite future. It is this crossover away from the early scaling solution that leads to an interesting and realistic cosmology.

For our crossover models the scalar field equation \eqref{151F} is given by
\bel{B8}
\Kbar(s''+3\Hhat s')=-\frac{\partial\Vhat}{\partial s}+3\Ybar (\fhat'+3\Hhat\fhat)\ .
\ee
For $s'\neq 0$ and $\Ybar\neq 0$ the relation $\fhat=0$ for the early scaling solution holds no longer exactly. Inserting in eq. \eqref{I28} $\Hhat'=0$, as appropriate for a de Sitter solution, the scalar field equation is given by eq. \eqref{eq:QS25},
\bel{B9}
s''+3\Hhat s'=-\khat^{-2}\frac{\partial\Vhat}{\partial s}=\frac{4\Vbar e^{2s}}{\khat^{2}(1+e^{2s})^{3}}\ .
\ee
The beginning with $\chi=0$ corresponds to $s\to -\infty$. In this limit the slow roll approximation becomes very precise, with
\bel{B10}
s'=\frac{4\Vbar e^{2s}}{3\Hhat\khat^{2}(1+e^{2s})^{3}}\approx\frac{4\Vbar e^{2s}}{3\Hhat\khat^{2}}=\frac{4\Hhat e^{2s}}{\khat^{2}}\ .
\ee
This implies
\bel{B11}
s''=2s'^{2}\ ,\quad \frac{s''}{3\Hhat s'}=\frac{2s'}{3\Hhat}=\frac{8\Vbar e^{2s}}{9\Hhat^{2}\khat^{2}}=\frac{8e^{2s}}{3\khat^{2}}\ ,
\ee
such that the term $s''$ can indeed be neglected for $s\to -\infty$.

The general solution of eq. \eqref{B10} is given by ($\that<\that_{s}$)
\bel{B12}
s=\half\ln\bigg{(}-\frac{\khat^{2}}{8\Hhat(\that-\that_{s})}\bigg{)}\ ,
\ee
with free integration constant $\that_{s}$. With eq. \eqref{B3} this yields
\bel{B12A}
s=\half\ln\bigg{(}-\frac{\khat^{2}}{8\ln\big{(}-c_{s}/(\Hhat\eta)\big{)}}\bigg{)}\ ,
\ee
or
\bel{B13}
\frac{\chi}{\mu}=\bigg{(}-\frac{\khat^{2}}{8\ln\big{(}-c_{s}/(\Hhat\eta)\big{)}}\bigg{)}^{\half}=\bigg{(}-\frac{\khat^{2}}{8\ln(c_{s}A)}\bigg{)}\ .
\ee
The integration constant $c_{s}$ reads
\bel{B13A}
c_{s}=\exp\big{[}\Hhat(\that_{0}-\that_{s})\big{]}\ .
\ee
This solution is part of the crossover trajectory away from the early scaling solution. Extrapolating the crossover solution backwards in time the universe starts for $\eta\to -\infty$ with $\chi=0$.

For $\infpast$ or $\that\to -\infty$ the change of $\chi$ is very slow, as visible from
\bel{B14}
s'=-\frac{1}{2(\that-\that_{s})}\ .
\ee
This very slow motion continues until $\that$ comes close to $\that_{s}$ or $\eta$ approaches $-c_{s}/\Hhat$. During one $e$-folding in the evolution of $A$ the ratio $\mu^{2}/\chi^{2}$ decreases by
\bel{B15}
\Delta\bigg{(}\frac{\mu^{2}}{\chi^{2}}\bigg{)}=-\frac{8}{\khat^{2}}\ .
\ee
Starting with very large values of $\mu^{2}/\chi^{2}$ it takes a huge number of $e$-foldings until $\mu^{2}/\chi^{2}$ reaches a value close to one. For a start at $\chi_{0}\to 0$ the number of $e$-foldings before $\chi$ reaches $\mu$ grows to infinity.

During this long period of slow evolution of $s$ the de Sitter solution of type 1 remains a very good approximate solution. For the early scaling solution one only has to replace $\Vbar$ by $\Vhat(s)$. This solution remains an attractor in the space of gauge fields and vierbeins. It is also an attractor with respect to small relative inhomogeneous scalar fluctuations.

Nevertheless, at some time $\that$ close to $\that_{s}$ the solution \eqref{B12}-\eqref{B14} ceases to be a valid approximation. More precisely, this end of the early scaling solution occurs for $s$ near zero. From there on the evolution of the universe changes qualitatively. The motion away from the early scaling solution can be viewed as a crossover from the de Sitter solution to a different scaling solution for the infinite future $\eta\to\infty$. This crossover can take place in several steps. For a given choice of an additive constant in $\eta$ the different crossover trajectories are characterized by a single further free parameter $c_{s}$. In the early stages these crossover trajectories are given by eq. \eqref{B12A}, with $\Vhat(s)$ replacing $\Vbar$ and therefore modifying $\Hhat$ for the associated de Sitter solution. From eq. \eqref{B13} we infer that the memory of initial conditions only concerns observables that depend on the size of $A$. Typical observables of interest do not depend on the absolute size of $A$.

\zwisch{Predictivity}

For a given set of model parameters our setting of pregeometry is very predictive. Up to a multiplicative factor for $A$ no memory of initial conditions remains if we start in the infinite past well within the basin of attraction of the early scaling solution, and with $\chi$ very close to zero. All quantities are, in principle, computable in terms of the model parameters. In particular, $\chat(\eta)$ is predicted to be equal to zero for any finite $\eta$. Since $\chat$ follows damped oscillations for the approach to the early scaling solution, no memory of any finite initial value $\chat(\eta_{0})$ is left at finite $\eta$ if $\eta_{0}\to -\infty$. After the end of validity of the early scaling solution $\chat(\eta)=0$ remains a solution. The potential dark matter contribution due to $\chat$-fluctuations vanishes if the inflationary epoch lasts for an infinite duration $\eta-\eta_{0}$.

The situation for the early evolution of $\fhat(\eta)$ is similar. Memory of initial conditions is lost, and $\fhat$ is given by eq. \eqref{I27} to a good approximation. Even though the oscillations of $\fhat$ have almost died out at $\that$ for a finite value $\Hhat(\that_{s}-\that)$ sufficiently large, such oscillations can be generated again by the more rapid evolution when $s$ approaches zero. One expects a small amount of dark matter due to the $\fhat$-fluctuations. This amount is calculable for given model parameters. The quantitative value of the dark matter contribution depends on the details of the behavior of the coupling functions. We have not attempted here to compute it. It is an interesting question if for a reasonable choice of model parameters the $\fhat$-fluctuations could constitute the observed dark matter in our universe. This also requires that the associated particles are stable or have a large enough lifetime once the couplings to other particles of the standard model or beyond are included.

\zwisch{Homogeneous and inhomogeneous Universe}

Let us next look closer at the issue of inhomogeneities. For the homogeneous early attractor solution both the frame-invariant vierbein and the gauge fields $A_{\mu m n }$ vanish for $\eta\to -\infty$ , 
\ba
\label{B16}
\tilde{e}_{\mu}{}^{m}&=A(\eta ) \delta_{\mu}^{m}=-\frac{1}{\widehat{H}\eta}  \delta_{\mu}^{m} \ ,
\nn\\
A_{i j 0}&= \hubhat(\eta )  \delta_{ij}=A(\eta )\widehat{H} \delta_{ij}=-\frac{1}{\eta}\delta_{ij} \ .
\end{align}
The Universe originates from the non-geometric state $\tilde{e}_{\mu}{}^{m}=0$ , $A_{\mu mn}=0$ in the infinite past.
In contrast, if we add the graviton fluctuations the vierbein no longer vanishes. Translating eq.~\eqref{G44} to the frame invariant formulation yields for our crossover model the asymptotic behavior for $\eta\to -\infty$
\be
\label{B17}
\tilde{e}_{i}{}^{\ j}(\eta , \vec{x})=A(\eta )\delta_{i}^{j}+\frac{1}{2}\int_{k} e^{i\vec{k}\vec{x}}\Big{(}\tilde{t}_{i}^{+j}(\vec{k}) e^{ik\eta}+\tilde{t}_{i}^{-j}(\vec{k}) e^{-ik\eta}\Big{)} \ .
\ee
For any non-zero amplitude $\ttil^{\ \pm  j}_{i}$ the inhomogeneous part dominates over the homogeneous part for $\eta\to -\infty$,  $ A\to 0$. The beginning of the Universe becomes inhomogeneous. 

We have derived the evolution equation for the graviton fluctuations under the assumption that a linear approximation for small fluctuations is valid. This is obviously no longer the case far in the past when the inhomogeneous part dominates. The breakdown of the linear approximation for a given $k$-mode occurs for some critical frame-invariant scale factor $A_{\text{nl}}(k)$ which depends on $k$, as the associated critical conformal time $\eta_{\text{nl}}(k)$.

We may estimate the value of $A_{\text{nl}}(k)$ for models with a realistic inflationary epoch. In this case we choose for $\tilde{t}_{i}^{\pm j}(\vec{k})$ a value which is comparable to the amplitude of the primordial tensor fluctuations. The tensor fluctuations with momentum $\vec{k}$ are frozen at some scale factor $A_{\text{hc}}$, which is related to horizon crossing in the Einstein frame. At this moment the linear approximation is valid. For graviton fluctuations with an amplitude corresponding to the graviton propagator~\eqref{292C} one has 
\ba 
\label{B18}
&|\tilde{t}_{i}^{j}(k)|^{2}=A_{\text{hc}}^{2}G_{\text{grav}}(k,\eta_{\text{hc}})k^{-3}\phantom{\bigg{|}}
\nn\\
=&\frac{\pi^{2}A_{\text{hc}}^{2}}{k^{6}}\Delta_{T}^{2}(k, \eta_{\text{hc}})=\frac{\pi^{2}}{k^{4} \widehat{H}_{\text{hc}}^{2}}\Delta_{T}^{2}(k, \eta_{\text{hc}})=\frac{4}{k ^{4}} \ .
\end{align}
The two terms in eq.~\eqref{B17} are of similar size at $A_{\text{nl}}(k)$ given by
\be 
\label{B19}
A_{\text{nl}}(k)=k^{3}\frac{|\tilde{t}_{i}^{j}(\vec{k})|}{2}=k\quad , \quad \frac{k}{a_{\text{nl}}}=M \ .
\ee
In the Einstein frame the linear approximation breaks down when the physical momentum $k/a$ reaches the Planck mass $\overline{M}$.

As compared to the scale factor $A_{\text{hc}}(k)$ at "horizon crossing" one finds
\be
\label{B20}
\frac{A_{\text{nl}}(k)}{A_{\text{hc}}(k)}=\widehat{H}_{\text{hc}}\ .
\ee
We denote the time when fluctuations become non-linear by the corresponding number of e-foldings before the end of inflation.
From eq.\eqref{B20} one infers
\be
\label{B21}
N_{\text{nl}}=N+\ln\big{(}\widehat{H}_{\text{hc}}^{-1}\big{)}\ .
\ee
In view of eq.~\eqref{292G}, $\widehat{H}_{\text{hc}}^{-1}=10^{4}/\sqrt{r}$, one concludes that $N_{\text{nl}}$ is larger than $N$, but not by a huge factor, $(N_{\text{nl}}-N)/N\ll 1$ . This estimate is not particular to pregeometry - it is the same for standard inflationary  models in general relativity. We conclude that in its early stages for $A<A_{\text{nl}}$ the Universe is inhomogeneous. The linear approximation for small inhomogeneities no longer applies.

Let us now make the assumption that the relation~\eqref{B17} continues to hold in the non-linear regime. While this cannot be inferred from the linear computation of the preceding section, we will argue below in favor of this assumption on the basis of a discussion of the graviton propagator. We further assume that for the other components of the vierbein, in particular the diagonal components $\tilde{e}_{0}{}^{0}$ and $\tilde{e_{i}}{}^{i}$ , a relation similar to eq.~\eqref{B17} holds. These assumptions entail a rather simple beginning of the Universe. For $\eta\to\ -\infty$ , $A\to 0$ the Universe is characterized by inhomogeneous fluctuations. The homogeneous expectation value of the vierbein vanishes. The same holds for the expectation value of the scalar field, $\chi=0$, and of the gauge fields, $A_{\mu mn}=0$. The "non-geometric state" discussed above is actually a state for which the inhomogeneous fluctuations dominate. 

Besides the inhomogeneous vierbein field one also has inhomogeneous gauge fields. According to eqs. \eqref{G1},~\eqref{G39} the graviton contribution to $A_{\mu mn}$ is given in the linear approximation by
\ba 
\label{B22}
A_{ij0}&=\frac{1}{2A^{2}}\big{(}D_{0}t_{ij}-D_{j}t_{i0}\big{)}_{,}=\frac{1}{2A^{2}}\big{(}\partial_{\eta}-\widehat{\hub}\big{)}t_{ij}\nn\\
&=\frac{1}{2A}\partial_{\eta}\tilde{t}_{ij}=\frac{1}{2}\tilde{t}_{ij}' \ .
\end{align}
In close analogy to eq.~\eqref{B17} the sum of the homogeneous and inhomogeneous contributions reads
\be
\label{B23}
A_{ij0}(\eta , \vec{x})=A\widehat{H}\delta_{ij}+\frac{i}{2}\int_{k}\frac{k}{A}\bigg{(}\tilde{t}_{ij}^{+}(\vec{k}) e^{ik\eta}-\tilde{t}_{ij}^{-}(\vec{k}) e^{-ik\eta}\bigg{)}\ .
\ee
The characteristic values for gauge fields are typically derivatives of the vierbein. For the inhomogeneous contribution this results in multiplication with the "dimensionless physical momentum" $k/A$ , while for the homogeneous part the dimensionless time-derivative $\sim\widehat{H}$ matters. 


While in the very early stages the homogeneous expectation values~\eqref{B16} are tiny as compared to the inhomogeneous fluctuations, they grow as $A$ increases, while the inhomogeneous fluctuations of $\tilde{e}_{\mu}{}^{m}$ are constant and the ones for $A_{\mu mn}$ decrease. For the vierbein the homogeneous and inhomogeneous parts are equal at $A_{\text{nl}}$, and subsequently for $A>A_{\text{nl}}$ the homogeneous contribution dominates. The process that the Universe becomes more and more homogeneous is not due to a decrease of the vierbein fluctuations. It is rather a consequence of the increase of the homogeneous expectation value. In consequence, the \textit{relative} size of the inhomogeneities decreases. In this picture it is natural that the relatives size of the inhomogeneities diverges in the infinite past for $A\to 0$ . This is simply due to the vanishing of the homogeneous expectation value, which appears for the relative size in the denominator.

\zwisch{The fluctuating beginning}

A quantum field theory makes statements about the probability distribution for field configurations at a given time. These probability distributions obey quantum constraints ~\cite{CWPW} which ensure the positivity of the quantum density matrix and the associated uncertainly relations. A description of the history of the Universe in quantum gravity has to be probabilistic. It is not about the history of a single field configuration, but rather about the evolution of the density matrix and the associated probability distribution for all field configurations.
The dominance of a particular field configuration is only a particular case, often associated to the "classical limit".
There is, in principle, no distinction between classical and quantum fluctuations. The concept of the time evolution of a probability distribution for field configurations is general, covering all sorts of fluctuations, classical, quantum or thermal. The particularity of quantum systems is only the quantum constraint that the density matrix and probability distribution have to obey, and the particular unitary evolution law for the time dependence of the density matrix.

Instead of following the time evolution of the density matrix or probability distribution, one can also follow the time evolution of the associated correlations functions or $n$-point functions. This much more economical approach is the one relevant in practice. Only correlation functions with low $n$ are observable in practice, and the correlation functions with high $n$ have typically little influence on the properties of correlation functions with low $n$. The dominant correlation functions are the expectation values of fields $(n=1)$ and the field-correlators or two-point functions $(n=2)$ . The latter are directly related to the propagator. The evolution of the expectation values follows the field equations obtained from the first functional derivative of the effective action. This is what we have investigated so far. Similarly, the evolution law for the two point correlation functions or propagators follows from inverting the second functional derivative of the effective action~\cite{Wetterich_2015, CWMF}. The effective action therefore specifies a system of evolution equations both for the expectation values and the propagators. 

The homogeneous field equations describe the time evolution of vacuum expectation values of fields, in our case $\tilde{e}_{\mu}{}^{m}$ , $A_{\mu mn}$ and $\chi$. For a given solution of the field equations for the expectation values one can compute the evolution equation for the propagators~\cite{Wetterich_2015}. This constitutes again a system of differential equations with initial conditions. The solution for the propagator is therefore not unique. A particular scaling solution or partial fixed point is given by the propagator which is associated to the Bunch-Davies vacuum~\cite{Bunch:1978yq}. The graviton propagator~\eqref{292C} corresponds precisely to this particular scaling solution.

We conclude that the beginning state is characterized by a probability distribution for field configurations for which the correlation functions differ from zero, while the expectation values vanish. Such a situation in rather common in statistical physics and not a sign of any singular behavior. For example, if the expectation value corresponds to the order parameter for a possible spontaneous breaking of some symmetry, the state with zero order parameter corresponds simply to the symmetric phase. In the symmetric phase the correlation function does not vanish. The only thing that is no longer possible in this "non-geometric state" is the definition of a geometry by use of the metric~\eqref{eq:5}. This is, a priori, not a problem since for models of pregeometry the metric and associated geometry can be considered as emergent quantities.

The beginning of the Universe is characterized by "great emptiness"~\cite{CWGE}. It is a vacuum, for which only expectation values and fluctuations matter. There are essentially no propagating particles. Any particle excitation propagates similar to photons, being massless with an ultrarelativistic dispersion relation. The beginning state can extend to the infinite past. The evolution near the infinite past is very slow. The infinite past is a fixed point with associated quantum scale symmetry. As compared to the more general setting discussed in ref.~\cite{CWGE}, the scaling solution characterizing the beginning in our model of pregeometry has the additional property that the vacuum expectation values for $\tilde{e}_{\mu}{}^{m}$, $A_{\mu mn}$ and $\chi$ all vanish. 

We may extend the discussion of the beginning state to small expectation values for $\tilde{e}_{\mu}{}^{m}$, $A_{\mu mn}$ and $\chi$ . As long as they are sufficiently small, they have not much impact on the properties of the correlation functions. Since the expectation values increase towards later time, their influence on the correlation functions also increases. For the early attractor solution the influence on the vierbein correlation becomes important only once $k\eta$ approaches $-1$, as we will see below. 


\vspace{2\baselineskip}
\zwisch{Vierbein correlation function}

We next establish a close connection between the correlation function or propagator and the inhomogeneous solutions in the linear approximation. This will permit us to extrapolate relations of the type \eqref{B17} beyond the range of validity of the linear approximation.

The vierbein correlator is given by the connected two-point function
\be
\label{B24}
\tilde{G}_{\mu\nu}{}^{mn}=\langle e_{\mu}'{}^{m}e_{\nu}'{}^{n}\rangle_{c}=\langle e_{\mu}'{}^{m}e_{\nu}'{}^{n}\rangle -\tilde{e}_{\mu}{}^{m}\tilde{e}_{\nu}{}^{n}\ .
\ee
Using the general form \eqref{G7},
\be
\label{B25}
e_{\mu}'{}^{m}=\tilde{e}_{\mu}{}^{m}+\frac{1}{2}\tilde{H}_{\mu\nu}'\tilde{e}^{\: m\nu}=\tilde{e}_{\mu}{}^{m}+\frac{1}{2}\tilde{H}_{\mu}'{}^{m}\ ,
\ee
one has
\be
\label{B26}
\tilde{G}_{\mu\nu}{}^{mn}(x,y)=\frac{1}{4}\langle \tilde{H}_{\mu}'{}^{m}(x)\tilde{H}_{\nu}'{}^{n}(y)\rangle \ ,
\ee
where we employ $\langle \tilde{H}_{\mu}'{}^{m}(x)\rangle=0$ .

The vierbein correlator is part of a larger system of correlation functions which includes all fields in a model. Let us collectively denote the fields by $\varphi_{i}'(\eta)$ , where $i$ is a multi-index which specifies position $\vec{x}$ or momentum $\vec{k}$ , as well as the species of fields and indices as $\mu$ or $m$ for $\tilde{e}_{\mu}{}^{m}$ or $\mu$ and $(m,n)$ for $A_{\mu mn}$ . The general correlation function,
\be
\label{B27}
\tilde{G}_{ij}(\eta , \eta ')=\langle\varphi_{i}'(\eta)\varphi_{j}'(\eta')\rangle_{c}\ ,
\ee
can be considered as a symmetric matrix,
\be
\label{B28}
\tilde{G}_{ji}(\eta' , \eta)=\tilde{G}_{ij}(\eta , \eta ')\ .
\ee

The connected two-point functions obey an exact functional identity, 
\be
\label{B29}
\int_{\eta}\Gamma_{\: ki}^{(2)}(\eta'' , \eta)\tilde{G}_{ij}(\eta , \eta')=\delta_{kj}\: \delta(\eta'' -\eta')\ .
\ee
Here $\Gamma^{(2)}$ is the matrix of second functional derivatives of the effective action, as defined by the quadratic expansion in infinitesimal $\tilde{h}$ , 
\ba 
\label{B30}
\Gamma \big{[}\tilde{\varphi}+\tilde{h}\big{]}=&\Gamma_{0}\big{[}\tilde{\varphi}\big{]}+\int_{\eta}\Gamma_{i}^{(1)}\big{[}\eta ;\tilde{\varphi}\big{]}\tilde{h}_{i}(\eta)
\\
&+\frac{1}{2}\int_{\eta''\eta}\tilde{h}_{k}(\eta'')\Gamma_{ki}^{(2)}\big{[}\eta'' , \eta ; \tilde{\varphi}\big{]}\tilde{h}_{i}(\eta)+\dots\nn
\end{align}
The relation~\eqref{B29} is an exact identity. In contrast to the evolution equation for small fluctuations it does not assume the validity of a linear expansion for a particular solution. The expansion~\eqref{B30} uses only the definition of the functional derivative, and $\tilde{h}_{i}(\eta)$ can be taken infinitesimal.

In the absence of additional sources the exact quantum field equations are given by the vanishing of the first functional derivative of $\Gamma$,
\be
\label{B31}
\Gamma_{i}^{(1)}\big{[}\eta ; \tilde{\varphi}\big{]}=0\ .
\ee
Consider now a homogeneous solution of the field equations $\tilde{\varphi}_{0}$ , with $\Gamma_{i}^{(1)}\big{[}\eta ; \:\tilde{\varphi}_{0}\big{]}=0$. The field equations for neighboring inhomogeneous fields,  $\tilde{\varphi}_{i}|\eta |= \tilde{\varphi}_{0,i}(\eta )+\tilde{H}_{i}(\eta )$, can be linearized for small enough $\tilde{H}_{i}(\eta )$. Expanding
\ba 
\label{B32}
\Gamma_{i}^{(1)}\big{[}\eta ; \tilde{\varphi}_{0}+\tilde{H}&\big{]}=\Gamma_{i}^{(1)}\big{[}\eta ; \tilde{\varphi}_{0}\big{]}\\
&+\int_{\eta '}\Gamma_{ij}^{(2)}\big{[}\eta ,  \eta' ; \varphi_{0}\big{]}\tilde{H}_{j}(\eta ') + \dots \: =0\ ,\nn
\end{align}
we observe that the same operator $\Gamma^{(2)}\big{[}\tilde{\varphi}_{0}\big{]}$ appears in the quadratic expansion~\eqref{B30} of $\Gamma$ around $\tilde{\varphi}_{0}$ , and in the linearized field equation~\eqref{B32}.
This will relate solutions of linearized field equations to the exact propagator which no longer involves any linearization.

Let us denote by $\tilde{H}_{i}^{\alpha}(\eta)$ different solutions of the linearized field equations
\be
\label{B33}
\int_{\eta'}\Gamma_{ij}^{(2)}\big{[}\eta ,  \eta' ;\tilde{\varphi}_{0}\big{]}\tilde{H}_{j}^{\alpha}(\eta')=D_{ij}\big{[}  \eta ,\tilde{\varphi}_{0}\big{]}\tilde{H}_{j}^{\alpha}(\eta)=0\ .
\ee
Here we have expressed $\Gamma^{(2)}(\eta , \eta' )=\delta (\eta -\eta')D(\eta)$ as a differential operator $D(\eta)$.
Possible solutions of the exact propagator equation~\eqref{B29} are then given by the ansatz
\be
\label{B34}
\tilde{G}_{ij}(\eta ,  \eta')=c_{\alpha\beta}\tilde{H}_{i}^{\alpha}(\eta)\tilde{H}_{j}^{\beta}(\eta')\ ,\quad  c_{\beta\alpha}=c_{\alpha\beta}\ .
\ee
This guaranties for eq.~\eqref{B29}
\be
\label{B35}
\int_{\eta}\Gamma_{ki}^{(2)}(\eta'' ,  \eta)\tilde{G}_{ij}(\eta ,  \eta')=0 \quad \text{for} \; \eta''\neq \eta'\ .
\ee
The $\delta$-distribution for $\eta''=\eta'$ in eq.~\eqref{B29} places constraints on the coefficients $c_{\alpha\beta}$, partially normalizing the propagator~\cite{Wetterich_2015, CWMF}.

We concentrate on the vierbein components $\tilde{H}_{\mu}{}^{m}$ corresponding to the graviton fluctuations around the early scaling solutions
\be
\label{B36}
\tilde{H}_{i}{}^{j}=t_{ik}\tilde{e}^{kj}\ .
\ee
For the scaling solution reflecting the Bunch-Davies vacuum the propagator in the graviton sector reads in Fourier space
\begin{align}
\label{B37}
&\tilde{G}_{ik}{}^{jl}\big{(}\eta , \vec{k} ; \eta' ,\vec{k}'\big{)}=\\
&\frac{1}{4}A(\eta)A(\eta')\tilde{G}_{\text{grav}}(k,\eta, \eta')
P_{\dub i\ k}^{(\gamma)j\ l}(\vec{k})\delta(\vec{k}-\vec{k}')\ ,\nn 
\end{align}
in close analogy to eq.~\eqref{292C}. This yields the equal-time propagator in position space
\be
\label{B38}
\tilde{G}_{ik}{}^{jl}(\eta , \vec{x} , \eta , \vec{y})=\int_{k} A^{2}(\eta)|\tilde{w}_{k}^{-}(\eta)|^{2}e^{i\vec{k}(\vec{x}-\vec{y})}P_{\dub i\ k}^{(\gamma)j\ l}(\vec{k})\ .
\ee
Omitting indices and the projector insertion, the frame-invariant version of eq.~\eqref{G49} yields
\be
\label{B39}
\tilde{G}(\eta ,\:  \vec{x}-\vec{y})=\int_{k}\frac{1}{2k}\Big{(}1+\frac{1}{k^{2}\eta^{2}}\Big{)}e^{i\vec{k}(\vec{x}-\vec{y})}\ .
\ee
In the limit $k\eta\to -\infty$ this equals the propagator in flat space for a free massless bosonic particle. It is the same as for Einstein gravity in flat space. There is no singularity in the vierbein propagator if we extrapolate to the infinite past $\eta\to -\infty$. The influence of the non-vanishing expectation values for the homogeneous solution is reflected in term $\sim (k\eta )^{-2}$ in eq.~\eqref{B39}. It becomes important only near "horizon crossing" at $k\eta =-1$ .

If the correlators for other field components behave similarly for $\eta \to -\infty$ , the infinite past corresponds indeed to a fixed point in the evolution of the system of correlation functions. It is unstable, however, with respect to the evolution of small homogeneous expectation values as $\tilde{e}_{\mu}{}^{m}$ or $\chi$. This instability explains why the Universe does not stay forever at this fixed point. In this more general view very early cosmology is characterized by a crossover in the extended space of all correlation functions.
This crossover is no longer restricted to a crossover between the two types of de Sitter solutions. In general, there are two quantities characterizing the crossover trajectories. They can be taken as $A(\overline{\eta})$ and $\chi(\overline{\eta})$ at some particular time $\overline{\eta}$. One can be absorbed in the definition of $\eta$. In case of an additional crossover in the space of homogeneous solutions, as discussed for the two types of de Sitter solutions, there will be one (or more) additional parameters characterizing the crossover solution, as we have discussed earlier.

\zwisch{Inhomogeneous cosmologies beyond the linear approximation}

We have seen that the computation of the propagator does not involve a linear approximation. The relation between the propagator and inhomogeneous cosmological solutions allow us to draw conclusions for the behavior of inhomogeneous solutions beyond the validity of the linear approximation, e.g. for $\eta\to -\infty$. The central conclusion is that inhomogeneous field configurations at some initial time $\eta_{0}$ arbitrarily far in the past cannot all be damped to zero at a given finite $\eta$, even in the limit $\eta - \eta_{0}\to \infty$ . This follows from the  simple observation that the propagator~\eqref{B39} obtains from a probability distribution over corresponding inhomogeneous vierbein configurations. Since the propagator at finite $\eta$ does not vanish, a vanishing of all corresponding inhomogeneous vierbein configurations at $\eta$ is excluded. At least some of these configurations must remain different from zero for an arbitrary long time difference  $\eta - \eta_{0}$. For these configurations eq.~\eqref{B17} is qualitatively correct even for $\eta < \eta_{\text{nl}}$ when the linear approximation breaks down and interactions have to be taken into account. 

This statement does not imply that the backwards extrapolation of some given inhomogeneous cosmological solution remains quantitatively correct when extrapolated backwards beyond $\eta_{\text{nl}}$ . The propagator only specifies a probabilistic average over many inhomogeneous solutions. It also does not imply that arbitrary inhomogeneous configurations can be extrapolated backwards without encountering a singularity. The predictive power of crossover trajectories characterized by a few parameters suggest the contrary. At finite $\eta$ many properties of inhomogeneous solutions are fixed as a prediction of the model. Starting with a configuration not obeying these properties necessarily leads to a singularity when extrapolated backwards. Otherwise, there would be no prediction, since initial conditions leading to a violation of this property would be possible. These simple observations have been demonstrated explicitly for the crossover between the two types of de Sitter solutions in the beginning of this section. The same concepts and general properties are valid for the crossover in the much larger space of evolving correlation functions. 

Our observed Universe is described by a particular inhomogeneous solution. It is useful to discuss how such a particular "realization" is described in our general probabilistic framework. The overall probabilistic description specifies only a probability for a given inhomogeneous solution. What we do in practice, however, is to use conditional probabilities. Typical questions are: 
For a given observed distribution of anisotropies in the cosmic microwave background at some time after its emission, what will be the probability for observing some particular properties at some later time. Conditional probabilities are efficiently described by a ``reduction of the probability distribution", in complete correspondence to the reduction of the wave function in quantum mechanics~\cite{CWPW}. This reduced probability distribution may then be well described by a particular inhomogeneous cosmological solution. The properties of the evolution of particular inhomogeneous solutions have to be compatible with the evolution of the propagator. For $\eta \gg \eta_{\text{nl}}$ , where the linear approximation for inhomogeneous solutions can be trusted, this is manifestly the case. 

We finally observe that for $\eta > \eta_{\text{nl}}$ the homogeneous expectation values start to dominate the full two-point function $\langle \tilde{e}_{\mu}'{}^{m}\tilde{e}_{\nu}'{}^{n}\rangle $. 
As time increases, the relative importance of inhomogeneities becomes less and less important. The homogeneous early attractor becomes a better and better approximation for any given particular inhomogeneous realization. This reflects the role of the early attractor solution as an attractive partial fixed point with respect to relative inhomogeneities.

\zwisch{Scaling frame}

Our frame-invariant formulation allows us to translate all results into arbitrary metric frames related by a Weyl scaling. These frames may be classified by the behavior of $M^{2}(\chi)$. A first ``scale invariant frame" is adapted to the properties of scaling solutions for functional flow equations and the associated fundamental scale invariance~\cite{CWFSI}. This is the frame underlying our discussion of crossover models in sect.~\ref{section:CM}. The Planck mass obeys $M^{2}(\chi)=2w_{0}k^{2}+\xi\chi^{2}$. 
For the beginning epoch the term $\xi\chi^{2}$ can be neglected, such that $M^{2}=2w_{0}k^{2}$ is a constant. Up to a proportionality factor the scale factor $a(\eta)=A(\eta)/(\sqrt{2w_{0}}\: k ) $ shows the same evolution as $A(\eta )$. Cosmic time $t$ in this frame obeys for the beginning epoch 
\be
\label{B40}
\frac{\partial t}{\partial \hat{t}}=\frac{\partial t}{\partial \eta}\frac{\partial \eta}{\partial \hat{t}}=\frac{a(\eta)}{A(\eta)}=\frac{1}{\sqrt{2w_{0}} \: k}\ ,
\ee
such that
\be
\label{B41}
t=\frac{1}{\sqrt{2w_{0}} \: k}\:  \hat{t}\; , \quad H=\sqrt{2w_{0}}\: k \widehat{H}\ .
\ee

A second metric frame is the Einstein frame with constant $M(\chi )=\overline{M} $ , and
\be
\label{B42}
t=\frac{\hat{t}}{\overline{M}}\; , \quad H=\overline{M}\widehat{H}\; , \quad a= \frac{A}{\overline{M}}\ .
\ee
For the beginning epoch this looks rather similar to the scale invariant frame. There is a large difference in the overall scale, however. While $\overline{M}=2.44\cdot 10^{18}$ GeV, one has $\sqrt{2w_{0}} \: k$ in the order of magnitude of $10^{-3} $ eV, with details depending on $u_{0}$ and $w_{0}$ according to
\be
\label{B43}
k \approx 2 u_{0}^{-\frac{1}{4}} \: 10^{-3}\:  \text{eV} \ .
\ee
The latter estimate equates the value of the effective potential $V=u_{0}k^{4}$ with the present dark energy density. The huge ratio of scales $\overline{M}/k$ arises from the fact that today's value $M^{2}(\chi_{0})$ is set equal to the fixed Planck mass
\be
\label{B43a}
M^{2}\big{(}\chi(t_{0})\big{)}=\overline{M}^{2}=\xi \chi^{2}(t_{0})\ .
\ee
Since at present $\chi (t_{0})$ has grown many orders of magnitude large than $k$ this explains the huge factor.

As a third metric frame we take the scaling frame
\be
\label{B44}
M^{2} (\chi)=\chi^{2}\ .
\ee
Up to a trivial rescaling of $\chi$ this equals the scale invariant frame for large values of $\chi$ . For the beginning epoch it differs 
substantially from both the scale invariant and the Einstein frame. The choice of the scaling frame can be motivated by the standard realization of quantum scale symmetry at the $UV$-fixed point for $\chi\to 0$, as discussed in sect.~\ref{section:QS}. 

In the scaling frame the geometry for the beginning epoch is different from the two other frames. With 
\be
\label{B45}
A(\eta )=\chi (\eta ) a (\eta )\ ,
\ee
the increase of $A(\eta )$ can be due essentially to the increase of $\chi (\eta )$ , rather than being associated with the variation of geometry encoded in $a(\eta )$. In this frame the Hubble parameter obeys
\be
\label{B46}
H=\widehat{H}\chi -\frac{d \chi}{d \hat{t}}=\chi\Big{(}\widehat{H}+\frac{1}{2\big{(}\hat{t}-\hat{t}_{s}\big{)}}\Big{)}\ ,
\ee
where the second identity uses eq.~\eqref{B14}. With constant $\widehat{H}$ towards the infinite past $\widehat{t}\to -\infty$ , the Hubble parameter in the scaling frame vanishes due to the vanishing of $\chi$. The geometry is no longer a de Sitter space. 

The scaling frame is not unique. One may use $M^{2}=\tilde{\chi}^{2}$, with $\tilde{\chi}$ a function of $\chi$. For a suitable choice of $\tilde{\chi}(\chi ) $ one obtains a ``primordial flat frame"~\cite{CWPFF} for which geometry approaches flat Minkowski space in the infinite past. This frame is useful in order to understand that there are no pure geometrical problems. Geodesics are complete for Minkowski space and there cannot be any singularities in the metric. Many features of great emptiness are particularly simple to understand in the primordial flat frame. This concerns, in particular, the property that all particles are effectively massless. For masses $\sim k$ in the scale invariant frame the mass vanishes $\sim \chi$ in the scaling frames, including the primordial flat frame.

A disadvantage of the scaling frame for the infinite past is the divergence of the inhomogeneous vierbein fluctuations.
For $a(\eta\to -\infty )=a_{0}$ the frame-invariant property that the relative fluctuations diverge has to translate into divergent fluctuations. 
While the scale invariant frame provides for a natural description of the beginning without any singularity, the singular Weyl transformation leads to singular inhomogeneities. The origin of this singularity is easy to understand. The vanishing expectation value of $\tilde{e}_{\mu}{}^{m}$ for $\eta\to - \infty$ also results in vanishing $e_{\mu}{}^{m}$ for the scale invariant frame. 
For constant inhomogeneities this is reflected by divergent relative inhomogeneities. If one translates this by a Weyl transformation to a constant vierbein in the primordial flat frame, the frame-invariant divergence of the relative inhomogeneities is translated to diverging inhomogeneous fluctuations. The regular frame invariant formulation shows that this divergence is an artifact of the choice of ``field coordinates". The singular propagator and inhomogeneous fluctuations in the primordial flat frame are a field singularity rather than a physical singularity. 

There exist other potentially useful choices for the metric frame. For the "standard potential frame" one chooses $M^{2}(\chi )=V(\chi )/\mu^{2} $ . In this frame the Hubble parameter is typically a constant during long epochs, with $H^{2}\sim V/ M^{2}=\mu^{2}$ . In this version of a scaling frame the Universe typically shrinks during the matter dominated epoch~\cite{CWUWE}. 

Field relativity \cite{CWUWE}, \cite{CWIQM} states that all these different pictures or frames yield the same predictions for observations. Geometry looses its absolute meaning - it becomes an issue of the choice of the metric field. While our model of pregeometry predicts the emergence of geometry, the precise geometry is a matter of the choice of fields. Only statements based on the frame invariant metric $\tilde{g}_{\mu\nu}$ are the same in all geometric pictures related by Weyl scalings. 

\zwisch{Fluctuation metric}

The choice of a frame invariant metric is not unique either. So far we have defined it as
\be
\label{B47}
\tilde{g}_{\mu\nu}=\tilde{e}_{\mu}{}^{m}\tilde{e}_{\nu m}\ .
\ee
This does not seem very appropriate for a situation where the expectation value $\tilde{e}_{\mu}{}^{m}$ vanishes, as for the infinite past of the homogeneous solutions. A more natural choice is the ``fluctuation metric", as defined by the two-point function 
\be
\label{B48}
\overline{G}_{\mu\nu}=\langle e_{\mu}'{}^{m}e_{\nu m}'\rangle\ .
 \ee
Such a choice is in line with the observation that distances can be defined by correlation functions~\cite{CWGGS}.

With eq.~\eqref{B24} we can express the fluctuation metric in terms of the contracted vierbein propagator
\be
\label{B49}
\overline{G}_{\mu\nu}=\eta_{mn}\tilde{G}_{\mu\nu}{}^{mn}+\tilde{g}_{\mu\nu}\ .
\ee
Except for very early cosmology the fluctuation metric is dominated by $\tilde{g}_{\mu\nu}$. This typically holds for $\eta >\eta_{nl}$ . For the beginning, however, the metric is dominated by the vierbein correlator. In view of the above discussion it typically approaches a constant for $\eta\to -\infty$. It seems not unlikely that in this limit the fluctuation metric becomes Lorentz-invariant, $\overline{G}_{\mu\nu}\sim\eta_{\mu\nu}$. Geometries defined by the fluctuation metric have a much smoother beginning as compared to $\tilde{g}_{\mu\nu}$. They reflect well the fixed point behavior.


\section{Conclusions}
\label{section:C}

Our model of pregeometry leads to interesting and rather realistic cosmological solutions. Variable gravity and general relativity emerge naturally for late times. The early epoch describes inflation. For the late epoch it predicts dynamical dark energy which vanishes in the infinite future. The model also contains a candidate for dark matter.

These statements are perhaps surprising. The non-compact character of the gauge group could lead to kinetic terms for some of the gauge bosons with the "wrong" sign, which could destabilize any solution with vanishing gauge fields in Minkowski space. This is indeed what happens if one takes into account only the standard gauge invariant kinetic term for gauge fields, e.g. $\Ztil=Z>0$. The presence of the additional vector field corresponding to the vierbein permits us to construct further gauge invariant kinetic terms for the gauge bosons. For an appropriate range of the couplings, $\Ztil<0$, these additional invariants stabilize the flat space solution, which can now be approached asymptotically for increasing time. This property is requested for any realistic cosmology, since Minkowski space is a very good approximation for particle physics and gravity at length scales that are small as compared to the size of the observable universe and away from objects with very large gravitational fields.

For large parameter ranges our model contains no tachyonic instabilities in flat space. We have numerically verified that for these parameter ranges the cosmological solutions indeed approach Minkowski space for increasing time.
As an interesting observation we note that ghost instabilities do not matter for the solutions we have investigated, both homogeneous and inhomogeneous in linear approximation. The spin-two sector fluctuations around flat space is free of ghosts for $Z<Z_c$, $Z_c=y/(1-y)$, $y=\M/\m$, while a ghost pole appears for $Z>Z_c$.
For our figures we use $y=0.2$, $Z_c=0.25$, with $Z=1$ for Figs. \ref{fig:A}-\ref{fig:3}, \ref{fig:5A}, \ref{fig:5}, while $Z=0.1$ for Figs. \ref{fig:4}, \ref{fig:B}, \ref{fig:C}. The results are very similar with and without a ghost pole. In contrast to tachyonic instabilities which appear in linear order and have to be avoided, potential instabilities due to ghosts are non-linear and have to be established explicitly for a given model and class of solutions. We also have seen that the energy density is not at the absolute minimum for flat space, while stability of flat space with respect to neighboring cosmological solutions is found nevertheless.

A geometric description becomes possible in terms of a composite metric which is a bilinear in the vierbein. In this sense geometry emerges from pregeometry. We observe that rather arbitrary field configurations evolve fast towards a situation where only the composite metric and the scalar field are needed for a description of the evolution of the universe and local gravitational physics. General relativity, and more generally variable gravity, which is a rather moderate and conservative version of modified gravity, emerge as effective theories at late times. For homogeneous solutions the deviations from variable gravity are encoded in the functions $f$ and $c$. They vanish rapidly, as seen in Figs. \ref{fig:1}, \ref{fig:2} or \ref{fig:5}.

We have formulated our model and the corresponding field equations in terms of frame invariant fields. For example, the frame invariant vierbein $\etil_{\mu}^{\ m}$ is related to the vierbein by a function involving the scalar field $\chi$, $\etil_{\mu}^{\ m}=M(\chi)\e$. Frame invariant fields and coupling functions are the same in all metric frames related by a field-dependent conformal transformation or Weyl scaling. This approach allows for a rather simple formulation and understanding of models with fundamental scale invariance. We have investigated a simple family of models of this type. They exhibit an ultraviolet fixed point for $\chi\to 0$ and an infrared fixed point for $\chi\to\infty$. As for all fixed points, quantum scale symmetry becomes an exact symmetry precisely at these fixed points. In the vicinity of the fixed points, for small non-zero $\chi$ or for large finite $\chi$, quantum scale symmetry is an approximate symmetry.

The overall cosmology of this model is characterized by a crossover between the two fixed points. The solution of the field equations leads to an evolution of the scalar field from the ultraviolet (UV) fixed point for $\chi\to 0$ to the infrared (IR) fixed point for $\chi\to\infty$. The UV-fixed point characterizes the infinite past, and the IR-fixed point will be reached in the infinite future. This crossover permits for a non-trivial evolution of the universe. The (approximately) stationary state at (near) a fixed point only characterizes the infinite past and future.

In early cosmology rather arbitrary initial conditions are attracted towards an early attractor solution. This is given by an approximate de Sitter solution for which variable gravity is valid. Due to the very slow evolution of the scalar field an exact de Sitter solution is a very good approximation. The approximate de Sitter solution can be identified with the inflationary epoch in cosmology. Inflation arises very naturally in our model of pregeometry, without any tuning of parameters. Realistic inflation requires an amplitude and spectrum of the primordial fluctuations consistent with observation. This can be achieved for a suitable choice of coupling functions.

Late cosmology is characterized by a late attractor solution which differs from the early attractor solution. It describes the approach towards flat space in the infinite future. Our model of pregeometry predicts for the present cosmological epoch the presence of some form of dynamical dark energy, related to the potential and kinetic energy of the slowly evolving scalar field. This very light scalar field arises naturally as the pseudo-Goldstone boson of spontaneously broken scale symmetry. In the infinite future scale symmetry becomes exact at the infrared fixed point, and the scalar field becomes massless. A typical present value of the varying mass is of the order of the Hubble parameter. The same scalar field is responsible for inflation and dynamical dark energy or quintessence, such that our model realizes "quintessential inflation" \cite{PEVI, BRMA} or "cosmon inflation" \cite{CWCI, CWIQM, RUCW, HMSS, HMSS2}. Our model also leads to a candidate for dark matter. This is constituted by the potential and kinetic energy of very rapidly oscillating gauge fields. They behave as cold dark matter, similar to the axion. The mass of the associated particle is, however, of the order of the Planck mass.

The crossover from the early attractor solution to the late attractor solution can be associated to the end of inflation. The precise timing of the end of inflation depends on the properties of the coupling functions. In turn, this determines the amplitude and spectral properties of the observed primordial fluctuations.

Our model of pregeometry contains important ingredients for a realistic cosmology: general relativity, inflation, dark energy and dark matter. This is quite remarkable for such a simple model. While this model may be a valid approximation for early cosmology including inflation, it needs to be extended for late cosmology after inflation. One has to add particle physics, which will influence the heating of the universe after inflation and provide for the main ingredients of the following radiation dominated epoch. The addition of the fields for particle physics is straightforward. This can again be implemented in a form respecting fundamental scale invariance. We give a short account in appendix~\ref{app:A*} and leave a more detailed investigation of this issue for further work.

As long as the coupling functions in the quantum effective action can be chosen freely it seems rather likely that a realistic cosmology can be found. This situation will change once a quantum field theory computation restricts the properties of the effective action. Our model of pregeometry can then become highly predictive! For a given particle content realistic cosmology will be no longer guaranteed.

Finally, our model of pregeometry also entails a picture for the beginning of the universe. The universe emerges in the infinite past from a scaling solution characterizing an ultraviolet fixed point. For this fixed point scaling solution the expectation values of all fields vanish. This vacuum state is characterized by non-vanishing static correlation functions for the fluctuations of all fields. These correlation functions are finite and no singularity occurs. The fixed point scaling solution is unstable with respect to the slow increase of small expectation values of fields. This instability triggers first a crossover to the early scaling solution and later further crossovers towards the scaling solution for the infinite future.

In summary, the evolution of the universe undergoes several crossovers between different approximate scaling solutions that are each valid for long cosmological epochs. The driving agents for these crossovers are slowly evolving expectation values of fields, in particular the scalar field. This evolution is dictated by the properties of solutions of the field equations which are derived from a rather simple quantum effective action for our model of pregeometry. Many ingredients for realistic cosmology, that are often "put in by hand", arise naturally from pregeometry.


\appendix

\begin{appendices}

\section{Pregeometric standard model}
\label{app:A*}

In this appendix we extend our model of pregeometry by the inclusion of additional fields for the standard model of particle physics or extensions thereof. The three generations of fermions can all be be taken as left-handed Weyl spinors. These include left-handed positrons or antiquarks, as familiar from representations of GUT-symmetries as $SO(10)$. The corresponding Grassmann fields $\psi_{\alpha a}(x)$ carry a Lorentz-index $\alpha=1,2$ and a species index $a$ which accounts for the different charges and generations of fermions. Denoting the generators of the $SO(1,3)$-gauge group by hermitian $2\times2$-matrices $\gl \sigma^{mn}\gr_{\alpha\beta}$, and the generators of the $SU(3)\times SU(2)\times U(1)$-gauge symmetry (or grand unified extensions thereof) by $\gl T^z\gr_{ab}$, the covariant derivative reads
\bel{AA1}
D_\mu\psi=\partial_\mu\psi-\frac i2\sigma^{mn}A_{\mu mn}\psi-iT^zB_\mu^z\psi\ ,
\ee
with $B_\mu^z$ the gauge bosons of the standard model or beyond. Fermions are singlets with respect to diffeomorphisms such that the geometric connection does not appear in the covariant derivative. They only  couple to the $SO(1,3)$-gauge bosons $A_{\mu mn}$ and the additional gauge bosons $B_\mu^z$. The generators $T^z$ are block-diagonal for the different irreducible representations of the gauge group and they are unit matrices in generation space.

The gauge invariant kinetic term for the fermions is given by
\bel{AA2}
\Gamma_\psi=\int_xiee_m^{\ \mu}\overline\psi\gamma^mD_\mu\psi\ ,
\ee
where $ee_m^{\ \mu}\sim\eps^{\mu\nu\rho\sigma}\eps_{mnpq}e_\nu^{\ n}e_\rho^{\ p}e_\sigma^{\ q}$. For the low-energy effective theory the gauge fields $A_{\mu mn}$ equal the spin connection $\omega_{\mu mn}$. In this limit the fermions have the standard couplings to gravity as for Cartan's geometry. Possible modifications of the gravitational couplings of the fermions can only arise beyond the effective low energy theory by the coupling to $U_{\mu\nu}^{\ \ m}$ according to
\bel{AA3}
A_{\mu mn}=\omega_{\mu mn}-U_{\mu mn}\ .
\ee

The invariant kinetic term for the gauge bosons takes the standard form
\bel{AA4}
\Gamma_B=\frac14\int_xeZ_{yz}^{(B)}B^{y\mu\nu}B_{\mu\nu}^z\ ,
\ee
with $B_{\mu\nu}^z$ the (non-abelian) field strength. The diagonal matrix $Z^{(B)}=\text{diag}\gl Z^{(1)},Z^{(2)},Z^{(3)}\gr$, $Z^{(i)})=g_i^{-2}$, involves the gauge couplings $g_i$ for $U(1)$, $SU(2)$ and $SU(3)$ of the standard model. (Here $Z^{(2)}$ and $Z^{(3)}$ correspond to $3$- and $8$-dimensional unit matrices. The structure can easily be extended to other gauge groups.) For the low-energy effective theory we can replace $e$ by $\sqrt{g}$ such that no anomalous gravitational couplings are present.

The standard model contains further the Higgs doublet. Extensions thereof, as grand unified theories, involve additional scalar fields. The interactions of these scalars with the fermions are of the standard form, involving the dimensionless Yukawa couplings. Also the interactions with the gauge bosons $B_\mu^z$ are given by the standard covariant derivative. These additional scalars are singlets with respect to $SO(1,3)$, such that the covariant derivative involves neither $A_{\mu mn}$ nor the geometric connection.

Possible mass scales can appear in the effective potential for the Higgs-doublet or further scalars. For large values of the singlet field, $\chi^2\gg k^2$, we assume scale symmetry as required by the infrared fixed point. In this limit all quadratic mass terms $\mu_i^2$ are proportional to $\chi^2$, involving dimensionless couplings $\eps_i$ according to $\mu_i^2=\eps_i\chi^2$. In particular the Fermi scale, given by the expectation value of the Higgs doublet, is proportional to $\chi$. As a consequence, all fermion masses induced by the electroweak symmetry breaking are proportional to $\chi$, and mass ratios as electron mass over effective Planck mass become independent of $\chi$.

Furthermore, for the scale invariant standard model~\cite{Wetterich_1988, SZE} the confinement scale $\Lambda_{\text{QCD}}$ is proportional to $\chi$, such that the ratio of electron mass over proton mass becomes $\chi$-independent. Scale symmetry also requires the $\chi$-independence of the dimensionless couplings, as gauge couplings, Yukawa couplings or quartic scalar couplings. As a result, there will be no observation of a time-variation of couplings or apparent violations of the equivalence principle in the limit of exact quantum scale symmetry~\cite{CWQS}. This holds despite the fact that $\chi$ depends on time for the cosmologies discussed in this work. Possible very small observable effects can only arise by small deviations from quantum scale symmetry since $\chi\to\infty$ is not yet reached for the present cosmological epoch. The quantum scale invariant setting can easily be extended beyond the standard model, with possible cubic scalar couplings proportional to $\chi$.

\section{Stability of Minkowski space}
\label{app:A}

Lorentz symmetry is extremely well tested in fundamental particle physics. The small violations of Lorentz symmetry due to local gravitational or electromagnetic fields on Earth or in galaxies are well understood.
For any realistic cosmology the (averaged) homogeneous isotropic metric should therefore approach the metric of flat Minkowski space during its evolution to the present time, with very small deviations due to the small size of the present Hubble parameter $H_{0}$.
A prerequisite for realistic cosmologies is the stability of Minkowski space, or neighboring evolving cosmologies with small $H_{0}$. More precisely, tachyonic fluctuation modes with mass terms $|\mu |$ much larger than $H_{0}$ should be absent. Otherwise fluctuations would grow $\sim \exp\big{(}|\mu | \, t\big{)}$ and obstruct any approach to flat space. 

In Einstein gravity with vanishing cosmological constant Minkowski space is stable. This extends to the Friedman universe in the presence of matter and radiation, and to the inclusion of a small cosmological constant. For our model of pregeometry Minkowski space with vanishing gauge fields is a solution of the field equations for $V=0$. The stability of this flat space solution is not guaranteed, however for arbitrary values of the couplings or coupling functions. We deal with a non-compact gauge symmetry $SO(1, 3) $ such that negative signs of the kinetic terms for some of the gauge fields may lead to tachyonic instabilities. Also the mode mixing due to a generalized Higgs mechanism~\cite{CWPG, wetterich2021pregeometry} can induce tachyonic instabilities.

Before a detailed analysis of possible homogeneous isotropic solutions of the field equations we should therefore clarify the stability properties of the flat space solution. This will restrict the range of allowed parameters for which realistic late cosmology is possible. In this appendix we concentrate on constant couplings, i.e. $\chi$-independent coupling functions. It will become clear in sect. \ref{section:CM} that this analysis extends to $\chi$-dependent coupling functions of a crossover model. We find that stability of Minkowski space can be realized for a negative value of the coefficient of the term $\sim F^{2}$, i.e
\be
\label{M0}
C<-\frac{1}{12}\, (Z+4B)\ .
\ee

\zwisch{Field expansion around flat space}

For the stability analysis we take $V=0$ and consider constant coupling functions. We first omit the scalar field $\chi$ which will be included at the end. We expand around the flat space solution with vanishing gauge fields, $\e=\delta_{\mu}^{m}$, $g_{\mu\nu}=\eta_{\mu\nu}\ $. The fluctuations of the vierbein are parametrized by
\be
\label{M1}
\e=\delta_{\mu}^{m}+\frac{1}{2}H_{\mu\nu}\, \eta^{\nu m}\ .
\ee
For a decomposition in terms of irreducible representations of the Lorentz group we employ for the vierbein the analogue of ref~ \cite{CWMF}, 
\be
\label{M2}
H_{\mu\nu}=t_{\mu\nu}+\frac{1}{3}P_{\mu\nu}\sigma +b_{\mu\nu}+\partial_{\mu}\gamma_{\nu}-\partial_{\nu}\gamma_{\mu}\ ,
\ee
with
\be
\label{M3}
P_{\mu\nu}=\eta_{\mu\nu}-\frac{\partial_{\mu}\partial_{\nu}}{\partial^{2}\phantom{\Big{|}}}\  .
\ee
The fluctuations obeys the conditions
\begin{align} 
\label{M4}
&t_{\nu\mu}=t_{\mu\nu}\; ,\quad \eta^{\mu\nu} t_{\mu\nu}=0 \; , \quad \partial^{\mu}t_{\mu\nu}=0\ , \nn\\
&b_{\nu\mu}=-b_{\mu\nu}\; , \quad \partial^{\mu}b_{\mu\nu}=0\; ,\quad \partial^{\mu}\gamma_{\mu}=0\ .
\end{align}
For the gauge fields we expand \cite{CWPG, wetterich2021pregeometry}
\ba
\label{M5}
A_{\mu\nu\rho}&=\frac{1}{2}(\partial_{\nu}E_{\mu\rho}-\partial_{\rho}E_{\mu\nu})+C_{\mu\nu\rho}\nn\\
+&\frac{1}{3}(P_{\mu\nu}w_{\rho}-P_{\mu\rho}w_{\nu})+\frac{1}{4}\varepsilon_{\nu\rho}{}^{\sigma\tau}(P_{\mu\sigma}v_{\tau}-P_{\mu\tau}v_{\sigma})\ ,
\end{align}
where the fluctuations obey
\begin{align}
\label{M6}
&E_{\nu\mu}=E_{\mu\nu}\; ,\quad \eta^{\mu\nu}E_{\mu\nu}=0\; , \quad\partial^{\mu}E_{\mu\nu}=0\ , \nn\\
&C_{\mu\nu\rho}=-C_{\mu\rho\nu}\; , \quad \partial^{\mu}C_{\mu\nu\rho}=0\; , \quad \partial^{\rho}C_{\mu\nu\rho}=0\; ,\nn\\
&\varepsilon^{\sigma\mu\nu\rho}C_{\mu\nu\rho}=0\ ,\quad \eta^{\mu\nu}C_{\mu\nu\rho}=\eta^{\mu\rho}C_{\mu\nu\rho}=0\ , \nn \\&\tilde{P}_{\sigma\tau}{}^{\nu\rho}C_{\mu\nu\rho}=0\ ,
\end{align}
with projector
\be
\label{M7}
\tilde{P}_{\sigma\tau}{}^{\nu\rho}=\frac{1}{2\partial^{2}}\big{(}\partial_{\sigma}\partial^{\nu}\delta_{\tau}^{\rho}-\partial_{\tau}\partial^{\nu}\delta_{\sigma}^{\rho}-\partial_{\sigma}\partial^{\rho}\delta_{\tau}^{\nu}+\partial_{\tau}\partial^{\rho}\delta_{\sigma}^{\nu}\big{)} \ .
\ee
The four-vectors $v_{\mu}$ and $w_{\mu} $ decompose into transversal and scalar parts
\begin{align}
\label{M8}
&v_{\mu} =v_{\mu}^{(t)}+\partial_{\mu}\tilde{v}\; , \quad w_{\mu} =w_{\mu} ^{(t)}+\partial_{\mu} \tilde{w}\  , \nn\\
&\partial^{\mu}v_{\mu} ^{(t)}=0\;,\quad \partial^{\mu}w_{\mu} ^{(t)}=0\ .
\end{align}
We have left out the gauge modes, which may be removed by gauge fixing~\cite{CWPG}.

In linear order in the gauge field fluctuations one has
\begin{align}
\label{M9}
F&_{\alpha\mu\nu\rho}=\half \big{(}\partial_{\alpha}\partial_{\nu}E_{\mu\rho}-\partial_{\alpha}\partial_{\rho}E_{\mu\nu}-\partial_{\mu}\partial_{\nu}E_{\alpha\rho}+\partial_{\mu}\partial_{\rho}E_{\alpha\nu}\big{)} \nn\\
&+\partial_{\alpha}C_{\mu\nu\rho}-\partial_{\mu}C_{\alpha\nu\rho}\nn\\
&+\frac{1}{3}\big{(}P_{\mu\nu}\partial_{\alpha}w_{\rho}-P_{\mu\rho}\partial_{\alpha}w_{\nu}-P_{\alpha\nu}\partial_{\mu}w_{\rho}+P_{\alpha\rho}\partial_{\mu}w_{\nu}\big{)} \nn\\
&+\frac{1}{4}\varepsilon_{\nu\rho}{}^{\sigma\tau}\big{(}P_{\mu\sigma}\partial_{\alpha}v_{\tau}-P_{\mu\tau}\partial_{\alpha}v_{\sigma}-P_{\alpha\sigma}\partial_{\mu}v_{\tau}+P_{\alpha\tau}\partial_{\mu}v_{\sigma}\big{)} \ .
\end{align}
For the contractions one obtains
\ba
\label{M10}
F_{\mu\rho}=&\eta^{\alpha\nu}F_{\alpha\mu\nu\rho}=\half \partial^{2}E_{\mu\rho}\nn\\
&-\frac{2}{3}\partial_{\mu}w_{\rho}^{(t)}-\frac{1}{3}\big{(}\eta_{\mu\rho}\partial^{2}+2\partial_{\mu}\partial_{\rho}\big{)}\tilde{w}\ ,
\end{align}
and
\be
\label{M11}
F=\eta^{\mu\rho}F_{\mu\rho}=-2\partial^{2}\tilde{w}\ .
\ee

\zwisch{Quadratic expansion of effective action}

Expanding in quadratic order in the fluctuations one finds in momentum space, with $q^{2}=q_{\mu}q^{\mu}=-q_{0}^{2}+\vec{q}^{\, 2}\; $,
\ba
\label{M12}
\frac{1}{8}\int_{x}& eZF_{\mu\nu\rho\sigma}F^{\mu\nu\rho\sigma}=\frac{Z}{4}\int_{q}\bigg{\lbrace}\frac{q^{4}}{2}E_{\mu\nu}^{*}E^{\mu\nu} \nn\\
+&q^{2}C_{\mu\nu\rho}^{*}C^{\mu\nu\rho}+q^{2}v_{\mu}^{(t)*}v^{(t)\mu}+\frac{3q^{4}}{2}v^{*}v\nn\\
+&\frac{4 q^{2}}{9}w_{\mu}^{(t)*}w^{(t)\mu}+\frac{2 q^{4}}{3}\tilde{w}^{*}\tilde{w}\bigg{\rbrace}\  .
\end{align}
Here $\tilde{w}^{*}\tilde{w}$ stands for $\tilde{w}^{*}(q)\tilde{w}(q)=w(-q)w(q)$ etc, and $q^{2}\tilde{w}^{*}\tilde{w}$ corresponds in position space to $\partial_{\mu}\tilde{w}^{*}\partial^{\mu}w$.
Similarly, one obtains
\ba 
\label{M13}
\half\int_{x}e\Big{\lbrace}BF_{\mu\nu}F^{\mu\nu}+CF^{2}\Big{\rbrace}=\int_{q}\bigg{\lbrace}\frac{Bq^{4}}{8}E_{\mu\nu}^{*}E^{\mu\nu}\nn\\
+\frac{2Bq^{2}}{9}w_{\mu}^{(t)*}w^{(t)\mu}+\bigg{(}\frac{2B}{3}+2C\bigg{)}q^{4}\tilde{w}^{*}\tilde{w}\bigg{\rbrace}\ .
\end{align}

The linear expansion for $U_{\mu\nu\rho}$ reads
\ba
\label{M14}
&U_{\mu\nu\rho}=\half\bigg{\lbrace}\partial_{\rho}\big{(}t_{\mu\nu}+E_{\mu\nu}\big{)}-\partial_{\nu}\big{(}t_{\mu\rho}+E_{\mu\rho}\big{)} \\
&+\frac{1}{3}\Big{[}P_{\mu\nu}\partial_{\rho}(\sigma-2\tilde{w})-P_{\mu\rho}\partial_{\nu}(\sigma-2\tilde{w})\Big{]}\nn\\
&+\partial_{\mu}b_{\nu\rho}+\partial_{\mu}\big{(}\partial_{\nu}\gamma_{\rho}-\partial_{\rho}\gamma_{\nu}\big{)}-2C_{\mu\nu\rho}\nn\\
&-\frac{2}{3}\Big{(}P_{\mu\nu}w_{\rho}^{(t)}-P_{\mu\rho}w_{\nu}^{(t)}\Big{)}-\half\varepsilon_{\nu\rho}{}^{\sigma\tau}\big{(}P_{\mu\sigma}\nu_{\tau}-P_{\mu\tau}\nu_{\sigma}\big{)}\bigg{\rbrace}\ ,\nn
\end{align}
and the contracted part is given by
\be
\label{M15}
U_{\mu\ \rho}^{\ \mu}=\eta^{\mu\nu}U_{\mu\nu\rho}=\half\partial_{\rho}(\sigma-2\tilde{w})+\half\partial^{2}\gamma_{\rho}-\frac{2}{3}w_{\rho}^{(t)}\ .
\ee
This yields in quadratic order
\ba
\label{M16}
&\int_{x}e\, \bigg{\lbrace}\frac{m^{2}}{4}U_{\mu\nu\rho}U^{\mu\nu\rho}+\frac{n^{2}}{2}U_{\mu}{}^{\mu\rho}U_{\nu\ \rho}^{\ \nu}\bigg{\rbrace}\\
&=\int_{q}\bigg{\lbrace}\frac{m^{2}}{4}\Big{[}\frac{q^{2}}{2}\big{(}t_{\mu\nu}^{*}+E_{\mu\nu}^{*}\big{)}\big{(}t^{\mu\nu}+E^{\mu\nu}\big{)}+\frac{q^{2}}{4}b_{\mu\nu}^{*}b^{\mu\nu }\nn\\
&\quad +C_{\mu\nu\rho}^{*}C^{\mu\nu\rho}+v_{\ \mu}^{(t)*} v^{(t)\mu}+\frac{3}{2}q^{2}\tilde{v}^{*}\tilde{v}\Big{]}\nn\\
&+\frac{q^{2}}{24}\big{(}m^{2}+3n^{2}\big{)}\big{(}\sigma^{*}-2\tilde{w}^{*}\big{)}\big{(}\sigma-2\tilde{w}\big{)}+\frac{q^{4}}{8}\big{(}m^{2}+n^{2}\big{)}\gamma_{\mu}^{*}\gamma^{\mu}\nn\\
&+\frac{2}{9}\big{(}2m^{2}+n^{2}\big{)}w_{\ \mu}^{(t)*}w^{(t)\mu}+\frac{q^{2}n^{2}}{6}\big{(}w_{\ \mu}^{(t)*}\gamma^{\mu}+\gamma_{\mu}^{*}w^{(t)\mu}\big{)}\bigg{\rbrace}\ .\nn
\end{align}
The quadratic approximation to the term linear in $F$ yields
\ba
\label{M17}
&-\half \int_{x}e M^{2}F=-\int_{q}\frac{M^{2}}{4}\bigg{\{}\frac{q^{2}}{2}\big{(}E_{\mu\nu}^{*}E^{\mu\nu}+E_{\mu\nu}^{*}t^{\mu\nu}+t_{\mu\nu}^{*}E^{\mu\nu}\big{)}\nn\\
&+\frac{2q^{2}}{3}\big{(}\tilde{w}^{*}\sigma+\sigma^{*}\tilde{w}-2\tilde{w}^{*}\tilde{w}\big{)}-i\frac{q_{\nu}}{2}\varepsilon^{\mu\nu\rho\sigma}\big{(}v_{\mu}^{(t)*}b_{\rho\sigma}-b_{\rho\sigma}^{*}v_{\ \mu}^{(t)}\big{)}\nn\\
&-v_{\ \mu}^{(t)*}v^{(t)\mu}-\frac{2q^{2}}{3}\big{(}w_{\ \mu}^{(t)*}\gamma^{\mu}+\gamma_{\mu}^{*}w^{(t)\mu}\big{)}-\frac{4}{9}w_{\ \mu}^{(t)*}w^{(t)\mu}\nn\\
&-2C_{\mu\nu\rho}^{*}C^{\mu\nu\rho}-3q^{2}\tilde{v}^{*}\tilde{v}\bigg{\}}\ .
\end{align}

\zwisch{Absence of tachyons}

Stability of Minkowski space depends on the poles of the propagators. Let us start with the fluctuations $\tilde{v}$ and $C_{\mu\nu\rho}$ that do not mix with other fluctuations. The relevant quadratic part of the effective action reads
\be
\label{M18}
\Gamma_{v C}=\half \int_{q} \Big{\lbrace}\tilde{v}^{*}(q)P_{\tilde{v}}\big{(}q^{2}\big{)}v(q)+C_{\mu\nu\rho}^{*}(q)P_{C}\big{(}q^{2}\big{)}C^{\mu\nu\rho}(q)\Big{\rbrace} \ .
\ee
The propagators for these fluctuations are given by $P_{\tilde{v}}^{-1} $ and $P_{C}^{-1}$, multiplied by appropriate projection operators and normalization factors that do not play a role for this discussion. The inverse propagators are given in our case by
\ba
\label{M19}
P_{\tilde{v}}&=\frac{3Zq^{2}}{4}\big{(}q^{2}+\mu_{\tilde{v}}^{2}\big{)}\ ,\nn\\
P_{C}&=\frac{Z}{2}\big{(}q^{2}+\mu_{C}^{2}\big{)}\ ,
\end{align}
with
\be
\label{M20}
\mu_{\tilde{v}}^{2}=\mu_{C}^{2}=\frac{m^{2}+2M^{2}}{Z}\ .
\ee
For positive $\tilde{\mu}^{2}$ these formulations can be associated with stable particles.

Indeed, the field equations $\big{(}q^{2}+\mu^{2}\big{)}\psi=0$ read, with $-q_{0}^{2}=\partial_{\eta}^{2}$ , $\;\vec{q}^{\, 2}=k^{2}$ , 
\be
\label{M21}
\big{(}\partial_{\eta}^{2}+k^{2}+\mu^{2}\big{)}\psi=0\ .
\ee
The general solution for the Fourier modes,
\be
\label{M22}
\psi(\vec{q},\eta)= \psi_{+}\big{(}\vec{q}\big{)} e^{i\lambda_{k}\eta}+\psi_{-}\big{(}\vec{q}\big{)} e^{-i\lambda_{k}\eta}\ ,
\ee
with
\be
\label{M23}
\lambda_{k}^{2}=k^{2}+\mu^{2}\ ,
\ee
results in a stable oscillatory behavior for $\mu^{2}>0$.
(For an expanding Universe Hubble damping induces a decreasing fluctuation~ amplitude). In contrast, for $\mu^{2}<0$ the momentum modes with $k^{2}<{-\mu}^{2}$ lead to imaginary $\lambda_{k}$. In this case one has solutions whose amplitude grows exponentially, $\psi\sim\exp\Big{(}\sqrt{-\big{(}\mu^{2}+k^{2}}\big{)}\,\eta\Big{)}$.
The presence of tachyonic modes with $\mu^{2}<0$ leads to an instability of Minkowski space. Parameter ranges leading to tachyons cannot result in realistic cosmologies and have to be excluded. We consider
\be
\label{M24}
Z>0\;,\quad m^{2}>0\;,\quad M^{2}>0\ ,
\ee
such that the modes $\tilde{w}$ and $C$ are stable. (The prefactor $\sim q^{2}$ for $P_{\tilde{v}}$ is due to normalization and not associated to a pole.)

\zwisch{Traceless transversal tensors and scalars}

The traceless transverse tensor modes $t_{\mu\nu}$ and $E_{\mu\nu}$ mix. The inverse propagator is now given by a $2\times 2$ - matrix $P_{Et}\big{(}q^{2}\big{)}$ 
\be
\label{M25}
P_{Et}=\frac{q^{2}}{4}\renewcommand*{\arraystretch}{1.2}\begin{pmatrix}Z_{t}q^{2}+\m-\M &\m-\M \\ \m-\M &\m\end{pmatrix}\ ,
\ee
with 
\be
\label{M26}
Z_{t}=Z+B .
\ee
Poles of the propagators correspond to zero eigenvalues of $P_{Et}$. The pole at $q^{2}=0$ corresponds to the massless graviton. The other zero of $P_{Et}$ is found by a vanishing determinant of the $2\times 2 $ matrix in eq.~\eqref{M25}
\be
\label{M27}
Z_{t}\m q^{2}+\M\big{(}\m-\M\big{)}=0 \ ,
\ee
leading to
\be
\label{M28}
\mu^{2}_{Et}=\frac{\M}{Z_{t}}\Big{(}1-\frac{\M}{\m}\Big{)}\ .
\ee
We require the stability conditions \cite{CWFSI, CWPG}
\be
\label{M29}
Z+B>0\;,\quad 0<\M <\m \ .
\ee
The properties of the graviton propagator obtained from eq.~\eqref{M25} are discussed in detail in refs.~\cite{CWPG, wetterich2021pregeometry}.

We next turn to the sector of scalars $\tilde{w}$ and $\sigma$.
We write the inverse propagator matrix in the form
\be
\label{M30}
P_{\tilde{w}\sigma}=\frac{q^{2}}{12}\renewcommand*{\arraystretch}{1.2}\begin{pmatrix}4\big{(}\tilde{Z}q^{2}+\tilde{m}^{2}+2\M\big{)} &-2\big{(}\tilde{m}^{2}+2\M\big{)}\\ -2\big{(}\mtil+2\M\big{)} &\tilde{m}^{2}\end{pmatrix}\ ,
\ee
with
\be
\label{M31}
\tilde{Z}=Z+4B+12C\;,\quad \tilde{m}^{2}=\m+3n^{2}\ .
\ee
One pole occurs for $q^{2}=0$, and the other is given by the condition 
\be
\label{M32}
\tilde{Z}q^{2}-2\M(1+2\tilde{y})=0\;,\quad \tilde{y}=\frac{\M}{\tilde{m}^{2}}\ .
\ee
With 
\be
\label{M33}
\mu_{\tilde{w}\sigma}^{2}=-\frac{2\M}{\tilde{Z}}(1+2\tilde{y})\ ,
\ee
the stability of Minkowski space requires either
\be
\label{M34}
\tilde{Z}>0\;,\quad \tilde{y}<-\half\ ,
\ee
or
\be
\label{M35}
\tilde{Z}<0\;,\quad\tilde{y}>-\half \ .
\ee

\zwisch{Transverse vectors}

For the coupled sector of the transverse vectors $\gamma_{\mu}$ and $w^{(t)}_{\ \mu} $ one has
\be
\label{M36}
\! \! P_{w\gamma}\!=\! \frac{2}{9}\!  \renewcommand*{\arraystretch}{1.2}\begin{pmatrix}\! \! (Z+2B)q^{2}+4\m +2n^{2}+\M & \frac{3}{2}q^{2}\big{(}n^{2}+\M\big{)}\\ \frac{3}{2}q^{2}\big{(}n^{2}+\M\big{)}&\frac{9}{8}q^{4}\big{(}\m +n^{2}\big{)}\! \! \! 
\end{pmatrix}
\ee
Besides the pole at $q^{2.}=0$ the other pole is determined by
\ba
\label{M37}
(Z+2B)\big{(}\m +&\n \big{)}q^{2}+\big{(}4\m +2\n+\M\big{)}\big{(}\m +\n \big{)}\nn\\
&-2\big{(}\n+\M\big{)}^{2}=0\ .
\end{align}
The corresponding mass term, 
\ba
\label{M38}
\mu_{w\gamma}^{2}&=(Z+2B)^{-1}\big{(}\m +\n \big{)}^{-1} A_{w\gamma}\ , \\
A_{w\gamma}&=\big{(}4 \m +2\n +\M\big{)}\big{(}\m +\n \big{)}-2\big{(}\n +\M \big{)}^{2}\ ,\nn
\end{align}
varies with $M^{2}/m^{2}$, with limiting values
\ba
\label{M39}
A_{w\gamma}&\big{(}\M =0 \big{)} = 4\m \big{(}\m +\frac{3}{2} \n \big{)} \ ,\nn\\
A_{w\gamma}&\big{(}\M =\m \big{)} = 3\m \big{(}\m+ \n \big{)} \ ,\nn\\
A_{w\gamma}&\big{(}\n =0 \big{)} = \big{(}4\m +\M \big{)} \m -2M^{4} \ .
\end{align}
For
\be
\label{M39A}
Z+2B>0
\ee
no tachyonic instability occurs for a wide range of $n^{2}/m^{2}$.

Finally, in the sector of $v^{(t)}_{\ \mu}$ and $b_{\nu\rho}$ the quadratic approximation to the effective action reads
\ba
\label{M40}
\Gamma_{2}^{(vb)}=&\int_{x}\bigg{\lbrace}\frac{1}{4}v^{(t)\mu}\big{(}-Z\partial^{2}+\m +\M \big{)}v^{(t)\mu}\nn\\
&+\frac{\m}{16}\partial^{\nu}b^{\rho\sigma}\partial_{\nu}b_{\rho\sigma}+\frac{\M}{4}v_{\mu}^{(t)}\varepsilon^{\mu\nu\rho\sigma}\partial_{\nu} b_{\rho\sigma}\bigg{\rbrace} \ .
\end{align}
We define the vector
\be
\label{M41}
\tilde{b}^{ \mu}=\varepsilon^{\mu\nu\rho\sigma}\partial_{\nu} b_{\rho\sigma}\; , \quad \tilde{b}^{\mu}\tilde{b}_{\mu}=2\partial^{\nu}b^{\rho\sigma}\partial_{\nu}b_{\rho\sigma}\ ,
\ee
and extract the inverse propagator matrix
\be
\label{M42}
P_{vb}=\frac{1}{2}  \renewcommand*{\arraystretch}{1.2}\begin{pmatrix}Zq^{2}+\m+\M & \frac{\M}{2}\\  \frac{\M}{2}& \frac{\m}{2}
\end{pmatrix}
\ee
The vanishing of the eigenvalue of $P_{vb}$ , determined by $q^{2}+\mu_{vb}^{2}=0$, corresponds to a stable particle since
\be
\label{M43}
\mu_{vb}^{2}=\frac{1}{Z}\Big{(}\m+\M-\frac{M^{4}}{2\m}\Big{)}
\ee
is positive for the conditions~\eqref{M24},~\eqref{M29}.

\zwisch{Saddle point of effective action}

We conclude that Minkowski space is indeed stable for a suitable range of parameters in the effective action. A rather natural choice takes $Z>0$, $Z+2B>0$ and negative $C$ according to eq.~\eqref{M0}. This corresponds to $\tilde{Z}<0$ in eq.~\eqref{M35}. A negative sign of $C$ resembles the situation in higher derivative gravity where the coefficient of the term $\sim R^{2}$ has to be negative for a stable asymptotic Minkowski space. Decomposing $F_{\mu\nu\rho\sigma}$ into irreducible pieces,
\be
\label{M44}
F_{\mu\nu\rho\sigma}=\Ctil_{\mu\nu\rho\sigma}+\Etil_{\mu\nu\rho\sigma}+\Stil_{\mu\nu\rho\sigma} \ ,
\ee
with
\ba
\label{M45}
&\Etil_{\mu\nu\rho\sigma}=\half\big{(}g_{\mu\rho}\tilde{F}_{\nu\sigma}+g_{\nu\sigma}\tilde{F}_{\mu\rho}-g_{\mu\sigma}\tilde{F}_{\nu\rho}-g_{\nu\rho}\tilde{F}_{\mu\sigma}{)}\ ,\nn\\
&\tilde{F}_{\mu\nu}=F_{\mu\nu}-\frac{1}{4}Fg_{\mu\nu}\ , \nn\\
&\Stil_{\mu\nu\rho\sigma}=\frac{1}{12}F(g_{\mu\rho}g_{\nu\sigma}-g_{\mu\sigma}g_{\nu\rho})\ ,
\end{align}
one has
\ba
\label{M46}
\frac{Z}{8}&F_{\mu\nu\rho\sigma}F^{\mu\nu\rho\sigma}+\frac{B}{2}F_{\mu\nu}F^{\mu\nu}+\frac{C}{2}F^{2}\nn\\
=&\frac{Z}{8}\Ctil_{\mu\nu\rho\sigma}\Ctil^{\mu\nu\rho\sigma}+\frac{1}{8}(Z+2B)\Etil_{\mu\nu\rho\sigma}\Etil^{\mu\nu\rho\sigma}\nn\\
&+\dfrac{1}{48}\big{(}2\tilde{Z}-(Z+2B)\big{)}F^{2}\ .
\end{align}
With positive $Z$ and $Z+2B$ and negative $\tilde{Z}$ the coefficient of $F^{2}$ in eq.~\eqref{M46} is negative. A negative value of $\tilde{Z}$ does therefore not correspond to a minimum for the combined field strength terms, but rather to a saddle point.

\zwisch{Mixing of scalar modes}

We finally include the fluctuations of the scalar field $\chi$ around some constant value $\chi_{0}$. For $Y\chi_{0}=0$ the fluctuations of $\chi$ decouple from the other fluctuations. Stability of Minkowski space requires $K>0$ and $\chi_{0}$ to be a minimum of $V$.
For $Y\chi_{0}\neq 0$ we take the approximation of a flat potential, $V=0$.
We observe now a mixing of $\chi$ with the scalar metric fluctuations $\sigma$ and $\wtil$ according to
\ba
\label{M47}
&\int_{x} e \bigg{\lbrace}\frac{K}{2}\partial^{\mu}\chi\partial_{\mu}\chi+YU_{\mu}{}^{\mu\nu}\chi\partial_{\nu}\chi\bigg{\rbrace} \\
&\!\!=\int_{q}  \bigg{\lbrace}\!\frac{K}{2}q^{2}\delta\chi^{*}\delta\chi+\frac{Y\chi_{0}}{4}q^{2}\Big{[}\big{(}\sigma^{*}-2\tilde{w}^{*})\delta\chi+\delta\chi^{*}(\sigma -2\tilde{w})\Big{]}\!\bigg{\rbrace} \ .\nn
\end{align}
This extends the inverse propagator matrix~\eqref{M30} to a $3\times 3$ matrix
\ba
\label{M48}
&P_{\tilde{w}\sigma\chi}=\\
&=\frac{q^{2}}{12}  \renewcommand*{\arraystretch}{1.2}\begin{pmatrix}4\big{(}\tilde{Z}q^{2}+\tilde{m}^{2}+2\M\big{)} & -2\big{(}\tilde{m}^{2}+2\M\big{)} & -12Y\chi_{0}\\  
-2\big{(}\tilde{m}^{2}+2\M\big{)}& \tilde{m}^{2}& 6Y\chi_{0} \\
-12Y\chi_{0}&6Y\chi_{0} & 12K
\end{pmatrix}\ .\nn
\end{align}
The mass term is shifted to
\be
\label{M49}
\mu^{2}=-\frac{2\M}{\tilde{Z}}\bigg{(}1+\frac{2K\M}{K\tilde{m}^{2}-3Y^{2}\chi_{0}^{2}}\bigg{)} \ .
\ee
We recover eq.~\eqref{M33} for $Y\chi_{0}\to 0$. For $K>0$, $M^{2}>0$ the conditions of stability depend now on the ratio $r$,
\be
\label{M50}
r=\frac{3Y^{2}\chi_{0}^{2}}{K\tilde{m}^{2}}\;,\quad \mu^{2}=-\frac{2\M}{\tilde{Z}}\bigg{(}1+\frac{2\tilde{y}}{1-r}\bigg{)}\ .
\ee
They are given for $\tilde{y}>0$ by
\ba
\label{M51}
&\tilde{Z}>0  &&\text{for } 1<r<1+2\tilde{y}\nn\\
&\tilde{Z}<0  &&\text{for  }  r<1  \quad \text{or }r>1+2\tilde{y}\ .
\end{align}
For $\tilde{y}<0\,, \;  r<0$ the range of stability for $\tilde{Z}<0$ becomes $r<1+2\tilde{y}\ $ or $\tilde{y}>-\half (1-r)$ .
This amounts to the condition
\be
\label{M52}
\tilde{m}^{2}<-2\M + \frac{3Y^{2}\chi_{0}^{2}}{K} .
\ee


\vspace{\baselineskip}
\zwisch{Stability of Minkowski space}

For the crossover models discussed in sect.~\ref{section:CM} the late Universe is characterized by a ``runaway cosmology" where $\chi$ increases to infinity in the infinite future. For large $\chi$ one has $M^{2}\sim \chi^{2} $, and similarly for other parameters. The question of stability of Minkowski space concerns the region of large $\chi$ and the approximation of constant coupling functions cannot be used. For the relevant range of large $\chi$ we can use the frame-invariant coupling functions. This amounts to a stability discussion which can be performed in complete correspondence to the discussion above, with the replacements
\ba
\label{377A}
&\chi\to s \; , \quad M^{2}\to 1\; , \quad m^{2} \to \frac{1}{y}\; , \quad \tilde{m}^{2}\to\frac{1}{\tilde{y}}\ , \nn\\
&K\to \overline{K}\; ,\quad Y\chi_{0}\to \overline{Y} \ .
\end{align}
The potential~$\widehat{V}(s)$ given by eq.~\eqref{eq:QS21} is very small for large $s$, such that the approximation $V=0$ holds very accurately. The main modification from a non-zero potential is a small scalar mass term 
\be 
\label{377B}
\tilde{\mu}_{s}^{2}=\frac{\partial^{2}\widehat{V}}{\partial s^{2}}\ ,
\ee
which replaces in the propagator matrix \eqref{M48}
\be 
\label{377C}
K\to\overline{K}+\frac{\tilde{\mu}_{s}^{2}}{q^{2}}\ .
\ee
For typical runaway solutions of the field equations one finds a $\tilde{\mu}_{s}^{2}\approx \widehat{H}^{2}$ , such that for all distances which are small as compared to the horizon we can neglect $\tilde{\mu}_{s}^{2}$ .

We concentrate in the main text on $B=0$, $M^{2}>0$ , and
\be 
\label{377Ca}
Z>0\; , \quad \tilde{Z}<0\; , \quad y>0\; .
\ee
The stability condition in this scalar sector is then given by eqs. \eqref{M51}, \eqref{M52} with
\bel{A58A}
r=\frac{3\Ybar^{2}y}{\Kbar}\ .
\ee
In particular, for $\ytil<0$ one has
\be 
\label{377D}
1+2\tilde{y}>\frac{3\overline{Y}^{2}\tilde{y}}{\overline{K}}\; , \quad \frac{\widehat{k}^{2}(1+2\tilde{y})}{\overline{K}}>0\ .
\ee
We require a positive kinetic term for the scalar in the effective variable gravity model
\be 
\label{377E}
\widehat{k}^{2}>0\;,\quad \overline{K}(1+2\tilde{y})>0\ .
\ee
The stability condition for the traceless tensor modes reads 
\be 
\label{377F}
y<1  \ ,
\ee
which also guarantees the positivity of $\mu^{2}_{vb}\sim\big{(}1+y-\frac{y^{2}}{2}\big{)}$.
Finally, the positivity of $\mu_{w\gamma}^{2}$ in eq.~\eqref{M38} requires
\be 
\label{377G}
Z\tilde{\mu}_{w\gamma}^{2}=\frac{6}{y(y+2\tilde{y})}\Big{(}y+\tilde{y}+y\tilde{y}-\frac{1}{2}y^{2}-y^{2}\tilde{y}\Big{)}\geqslant 0 \ .
\ee
For the choice $\tilde{y}=-y$ this condition is obeyed, $Z\tilde{\mu}_{w\gamma}^{2}=9-6y>0$ .


\section{Stability of de Sitter solutions with vanishing scalar field}
\label{app:B}

In this appendix we discuss the stability properties of the de Sitter solutions \eqref{FS9}, \eqref{FS10} for $\chi=0$, $c=0$. For this purpose we consider the linear expansion
\bel{ST1}
\btil=\btil_{0}+v\ ,\quad H=H_{0}+w\ ,
\ee
with $\btil_{0}$ and $H_{0}$ the values according to the de Sitter solution. The linearized eq. \eqref{MS8} reads
\bel{ST2}
\dt^{2}v+3H_{0}\dt v+2H_{0}^{2}v+\btil_{0}\dt w+4H_{0}\btil_{0}w=\frac{\mub}{2\ytil}(v-w)+6\btil_{0}^{2}v\ .
\ee
For the expansion of the Hubble parameter we employ eq. \eqref{FS12}
\bel{ST3}
H_{0}w=(1+2\ytil)\btil_{0} v+\frac{\Ztil}{\mtil}\big{(}H_{0}\btil_{0}\dt v+H_{0}^{2}\btil_{0}v+H_{0}\btil_{0}^{2}w-2\btil_{0}^{3}v\big{)}\ ,
\ee
or
\bel{ST4}
w=\bigg{(}1-\frac{\Ztil\btil_{0}^{2}}{\mtil}\bigg{)}^{-1}\bigg{[}(1+2\ytil)\frac{\btil_{0}}{H_{0}}v+\frac{\Ztil}{\mtil}\big{(}\btil_{0}\dt v+H_{0}\btil_{0}v-\frac{2\btil_{0}^{3}}{H_{0}}v\big{)}\bigg{]}\ .
\ee
Insertion into eq. \eqref{ST2} yields a closed linear second order differential equation for $v$ which allows a standard stability analysis.

Consider first the de Sitter solutions of type 1 with $\btil_{0}=H_{0}$. We denote
\bel{ST5}
g=\frac{\Ztil\btil_{0}^{2}}{\mtil}
\ee
such that
\bel{ST6}
w=(1-g)^{-1}\big{[}(1+2\ytil-g)v+\frac{g}{H_{0}}\dt v\big{]}\ .
\ee
We eliminate $H_{0}$ by $z=H_{0}t$ and use eq. \eqref{MS7} for $\mubtil$, resulting for eq. \eqref{ST2} in
\bel{ST7}
\dz^{2}v+3\dz v+\dz w=\bigg{(}4-\frac{1+2\ytil}{g}\bigg{)}(v-w)\ .
\ee
The derivative of eq. \eqref{ST6} yields
\bel{ST8}
\dz w=\frac{g}{1-g}\dz^{2}v+\dz v+\frac{2\ytil}{1-g}\dz v\ .
\ee
Insertion of eqs. \eqref{ST8} and \eqref{ST6} into eq. \eqref{ST7} results in the simple linear evolution equation
\bel{ST9}
\dz^{2}v+3\dz v+\muvtil v=0\ ,
\ee
with
\begin{align}
\label{ST10}
\muvtil&=-2\ytil\bigg{(}\frac{1+2\ytil}{g}-4\bigg{)}=\frac{\mub}{H_{0}^{2}}+8\ytil\nn\\
&=-\frac{6(1+2\ytil)M^{4}}{\Ztil V}+8\ytil\ .
\end{align}

With $$v=v_{+}\exp\big{(}\lambda_{+}H_{0}t\big{)}+v_{-}\exp\big{(}\lambda_{-}H_{0}t\big{)}$$ the stability properties depend on the eigenvalues $\lambda_{\pm}$ which are solutions of the quadratic equation
\bel{ST11}
\lambda^{2}+3\lambda+\muvtil=0
\ee
or
\bel{ST12}
\lambda_{\pm}=\half\Big{(}-3\pm\sqrt{9-4\muvtil}\Big{)}\ .
\ee
For positive $\muvtil$ in the range $0<\muvtil<9/4$ one finds both eigenvalues $\lambda_{\pm}$ real and positive, while for $\muvtil>9/4$ one observes damped oscillations. In both cases the de Sitter solution of type 1 is a stable attractor. The rate of attraction is given by the real part of $\lambda_{+}$. In particular, for $\muvtil>9/4$ the difference from the early attractor solution decreases by a factor $\exp(-3/2)$ for each $e$-folding. In contrast, for $\muvtil<0$ the de Sitter solution of type 1 becomes unstable, characterized by an exponential growth of $v$ with rate given by $\lambda_{+}$.

The range of stability is given by eq. \eqref{ST10}. For $\Ztil<0$, $1+2\ytil>0$ one finds stability unless $V$ is very large. For the parameters chosen for Figs. \ref{fig:1}, \ref{fig:2} one has $\muvtil=3.6/V-1.6=10.4>9/4$, corresponding to the visible damped oscillations.

A similar analysis can be done for the de Sitter solutions of type 2. It is algebraically a bit more lengthy and not presented here. As visible from fig. \ref{fig:3} the de Sitter solutions of type 2 are unstable for the parameters used.

Finally, one can perform a similar stability investigation for the gauge fields $\ctil$. For the de Sitter solutions of type 1 the linearized field equation \eqref{MS9} reads
\bel{ST13}
\big{(}\dz^{2}+3\dz+\muctil\big{)}c=0\ ,
\ee
with
\bel{ST14}
\muctil=\frac{\muc}{H_{0}^{2}}-\frac{4\Ztil}{Z}\ .
\ee
Stability is guaranteed for $\muc>0$, $\Ztil/Z<0$. If initial conditions are set for $t_{0}\to -\infty$ both $v$ and $c$ are zero for any finite $t$.

\section{Variable gravity approximation for \\homogeneous field equations}
\label{app:C}

In this appendix we discuss the embedding of variable gravity into our model of pregeometry on the level of the effective action and field equations for homogeneous isotropic configurations. While the results are, in principle, already contained in the more general discussion of sect. \ref{section:EVG}, the present treatment shows in a concrete way the approximations made when one uses the field equations of variable gravity.

\zwisch{Effective action for homogeneous configurations}

We may investigate the issue first by looking at the effective action for homogeneous isotropic cosmologies. For the homogeneous isotropic configurations \eqref{eq:8} one finds for $\Delta$ in eq. \eqref{26}
\begin{align}
\Delta &= -\frac{6}{a^{2}}\big{[}\deta(b-\hub) + (b+\hub)(b-\hub) - c^{2}\big{]} \nonumber \\
&= -\frac{6}{a^{2}}\big{[}\deta f + (f+2\hub)f - c^{2}\big{]}\ .
\end{align}
We observe from eq. \eqref{eq:9} that $\Gamma_{2}$ in eq. \eqref{25} contains a piece $\Gamma_{f}$ that vanishes for $f=0$, $c=0$, and a "higher derivative piece" $\Gamma_{HD}$,
\be
\label{27A}
\Gamma_{2} = \Gamma_{f} + \Gamma_{HD}\ ,
\ee
where (for $B=0$)
\begin{align}
\label{27B}
\Gamma_{f} &= \Omega_{3}\int_{\eta}\bigg{\{}\frac{3a^{2}}{2}\big{(}\m c^{2}-\mtil f^{2}\big{)}\nn \\
&- 3M^{2}a^{2}\big{[}\deta f + (2\hub + f)f -c^{2}\big{]}+3Ya^{2}\chi\deta\chi f \nonumber \\
&+\frac{3Z}{2}\bigg{[}4\hub^{3}f + 6\hub^{2}f^{2} + 4\hub f^{3} + f^{4} + c^{4}\nonumber \\
&\trip\trip\trip- 6(\hub^{2}+2\hub f+f^{2})c^{2} \nonumber \\
&\trip\trip\trip + 2\deta\hub\deta f + (\deta f)^{2} - (\deta c)^{2}\bigg{]}\nn \\
&+18C\bigg{[}4\hub^{3}f+6\hub^{2}f^{2}+4\hub f^{3}+f^{4}+c^{4}\nn\\
&-2\big{(}2\hub f+f^{2}\big{)}c^{2}+2\big{(}\hub^{2}+\deta\hub+2\hub f+f^{2}\big{)}\deta f\nn\\
&\trip\trip\trip+2\deta\hub\big{(}2\hub f+f^{2}\big{)}+(\deta f)^{2}\bigg{]}\bigg{\}} \ ,
\end{align}
and
\be
\label{27C}
\Gamma_{HD} = \Omega_{3}\int_{\eta}\bigg{\{}\frac{3Z}{2}\big{(}\hub^{4} + (\deta\hub)^{2}\big{)}+18C\big{(}\hub^{2}+\deta\hub\big{)}^{2}\bigg{\}}\ .
\ee
The higher derivative term \eqref{26} corresponds to
\be
\label{27D}
\Gamma_{HD} = \int_{x}\sqrt{g}\bigg{\{}\frac{Z}{8}R_{\mu\nu\rho\sigma}R^{\mu\nu\rho\sigma}+\frac{C}{2}R^{2}\bigg{\}}\ ,
\ee
with $R_{\mu\nu\rho\sigma}$ the Riemann curvature tensor formed from the metric $g_{\mu\nu}$ and $R$ the corresponding curvature scalar. It obtains by inserting the homogeneous isotropic ansatz \eqref{eq:8}, $g_{\mu\nu} = a^{2}\eta_{\mu\nu}$. This term can be interpreted as a higher derivative term within variable gravity.

For $f=c=0$ one recovers indeed the higher derivative extension of variable gravity whose cosmology has been discussed in ref. \cite{CWVG, CWIQM}. For late cosmology higher derivative terms play no role if their coefficients are not huge. In many circumstances the effects of the higher derivative term are negligible even for early stages of inflation or the "beginning" of the universe, and cosmology can be well described by the simple form \eqref{23}. The effective action $\Gamma_{1} + \Gamma_{HD}$ only involves the functions $a(\eta)$ and $\chi(\eta)$. Any new effects of pregeometry are therefore connected to non-zero $f$ or $c$.

Since $c$ appears only quadratically in $\Gamma_{f}$ we will find consistent solutions with $c=0$. We will pursue this type of solutions here. In contrast, $\Gamma_{f}$ contains terms that are linear in $f$,
\begin{align}
\label{28}
\Gamma_{f,1} &= \Omega_{3}\int_{\eta}\bigg{\{}-3M^{2}\deta (a^{2}f) + 3\Ztil (2\hub^{3}f+\deta\hub\deta f)\nn \\
&\trip\trip\trip\trip+3(\Ztil-Z)\deta(\hub^{2}f)+3Ya^{2}\chi\deta\chi f\bigg{\}} \nonumber \\
&= \Omega_{3}\int_{\eta}\bigg{\{}3a^{2}\bigg{(}\frac{\partial M^{2}}{\partial \chi}+Y\chi\bigg{)}\deta\chi f\nn \\
&\trip\trip\trip\trip+3\Ztil\Big{(}2\hub^{3} - \deta^{2}\hub - \frac{\partial \ln \Ztil}{\partial \chi}\deta\chi\deta\hub\Big{)}f\nn\\
&\trip\trip\trip\trip-3\hub^{2}f\partial_{\chi}(\Ztil-Z)\deta\chi\bigg{\}}\ ,
\end{align}
where the second line obtains by partial integration. For constant $\Ztil$, $Z$ the derivative of this expression with respect to $f$ produces the driving term in the field equation \eqref{eq:21} for $f$. Due to this driving term the cosmological solutions of variable gravity will be modified whenever $f$ plays a role. If we omit for a moment $\partial_{\chi}Z$, $\partial_{\chi}\Ztil$ and powers of $\hub$ or $\deta\hub$, a typical size of $f$ is given by
\be
\label{29}
f \sim \bigg{(}\frac{\partial M^{2}}{\partial \chi}+Y\chi\bigg{)}\deta\chi(\mtil+2M^{2})^{-1}\ .
\ee
This quantity vanishes if $Y=0$ and $M^{2}$ is independent of $\chi$, as well as for $\mtil\to\infty$. For the particular case $Y=0$, $M^{2}\sim\chi^{2}$ and constant $\ytil=\M/\mtil$ one has
\be
\label{30}
f\sim\frac{2\ytil}{1+2\ytil}\deta\ln\chi\ .
\ee

For $\tilde{K}=K$ and $f=0$ the scalar field equation \eqref{33} agrees with eq. \eqref{eq:20}. On the other hand, for $Y=0$ and $f=0$ eq. \eqref{31} agrees with eq. \eqref{eq:19} provided that $C_{R}=0$. These findings are consistent for $\partial\M/\partial\chi = 0$. In this case $f$ can be neglected, and also $C_{R}$ vanishes. For non-zero $\partial\M/\partial\chi$, as typically relevant for late cosmology where $\chi^{2}$ is much larger than $k^{2}$, or nonzero $Y$, one needs to account both for non-zero $f$ and non-zero $C_{R}$. These effects of non-vanishing $f$ and $C_{R}$ can be summarized by the shift in the kinetial \eqref{23A}.

\zwisch{Leading effects}

In the leading order we can approximate
\be
\label{VG10}
\Gamma_{f} = \Gamma_{f,1}  + \Gamma_{f,2}\ ,
\ee
with
\be
\label{VG11}
\Gamma_{f,2} = \Omega_{3}\int_{\eta}-\frac{a^{2}}{2}(3\mtil + 6\M)f^{2}\ ,
\ee
and $\Gamma_{f,1}$ given by eq. \eqref{28}. Insertion of the approximate solution,
\be
\label{VG12}
f = \frac{1}{\mtil+2\M}\bigg{\{}\bigg{(}\frac{\partial\M}{\partial\chi}+Y\chi\bigg{)}\deta\chi+ \frac{\Ztil}{a^{2}}(2\hub^{3}-\deta^{2}\hub)\bigg{\}}\ ,
\ee
yields
\begin{align}
\label{VG13}
\Gamma_{f} &= \frac{3}{2}\int_{x}\frac{1}{\mtil+2\M}\bigg{\{}a^{2}\bigg{(}\frac{\partial\M}{\partial\chi}+Y\chi\bigg{)}^{2}(\deta\chi)^{2}+\dots\bigg{\}} \nonumber \\
&= -\frac{3}{2}\int_{x}e\frac{1}{\mtil+2\M}\bigg{\{}\bigg{(}\frac{\partial\M}{\partial\chi}+Y\chi\bigg{)}^{2}\partial^{\mu}\chi\partial_{\mu}\chi +\dots\bigg{\}}\ ,
\end{align}
where we employ $e=a^{4}$, $(\deta\chi)^{2}=-a^{2}\partial^{\mu}\chi\partial_{\mu}\chi$. Eq. \eqref{VG13} agrees indeed with the first term in eq. \eqref{VG9}. We note for $U_{\tau\ m}^{\ \tau}$ the relations $U_{\tau\ 0}^{\ \tau}=-3f/a$, $U_{\tau\ k}^{\ \tau}=0$.

\vspace{2\baselineskip}
\zwisch{Equivalence of homogeneous field equations}

Up to omitted higher derivative contributions we can verify that the scalar field equation \eqref{eq:20} in pregeometry indeed agrees with the scalar field equation \eqref{33} of variable gravity by use of the identity
\begin{align}
\label{VG14}
3&\bigg{(}\frac{\partial\M}{\partial\chi}+Y\chi\bigg{)}\big{(}\deta f + 2\hub f + f^{2}\big{)} + \frac{3}{2}\frac{\partial\mtil}{\partial\chi}f^{2} \\
&= (K-\Ktil)\big{(}\deta^{2}\chi + 2\hub\deta\chi\big{)} +\half\frac{\partial}{\partial\chi}(K-\Ktil)(\deta\chi)^{2}\nn\ .
\end{align}
Similarly, the gravitational field equation \eqref{eq:19} in pregeometry agrees with \eqref{31} in variable gravity, using
\begin{align}
\label{VG15}
(&K-\Ktil)(\deta\chi)^{2} -a^{2}C_{R} \\
&=-3(\mtil+2\M)\big{(}\deta f + 2\hub f + f^{2}\big{)} -3\frac{\partial\mtil}{\partial\chi}\deta\chi f\nn\ .
\end{align}

The field equations of variable gravity are considerably simpler than the ones in pregeometry. Only the metric and the scalar field remain as relevant degrees of freedom. The homogeneous isotropic field equations only depend on $a$ and $\chi$, while $f$ and $c$ are eliminated since the gauge fields become fixed functions of the vierbein in the low energy effective theory.
When the conditions of validity of the low energy effective theory are met we will directly use the field equations of variable gravity. This has the additional advantage that the substantial body of cosmological investigations for variable gravity can be taken over to our model of pregeometry. Already in the late phases of the inflationary epoch the Hubble parameter is substantially smaller than the effective Planck mass, $H^{2}/\M \ll 1$. This is the range of applicability of the low energy effective theory. The predictions of pregeometry for the properties of the observable primordial cosmic fluctuations are precisely those of the corresponding model of variable gravity.

We will discuss in sect. \ref{section:DMDE} that beyond the leading order \eqref{VG12} for $f$ and $c=0$ there can be small oscillations. They contribute in variable gravity as an effective energy momentum tensor for cold dark matter.

\section{Traceless transversal tensor fluctuations for de Sitter solutions of type 2}
\label{app:D}

In this appendix we discuss the transverse traceless tensor fluctuations in the background of the de Sitter solutions of type 2.
For the second family of de Sitter solutions \eqref{FS10} we substitute the constants $V$ and $Z$ by the constants $\btil$ and $H$. With
\be
\label{G57}
b=\alpha\hub\ ,\quad \alpha=\frac{\btil}{H}\ ,
\ee
eq. \eqref{FS8} determines $Va^{2}$ and $Z/a^{2}$ as functions of $\hub$ and $\alpha$, namely
\be
\label{G58}
\frac{Va^{2}}{\m}=c_{V}\hub^{2}\ ,\quad c_{V}=\frac{3\mtil}{4\m}\big{[}\alpha(1+\alpha)(1+2\ytil)-2\big{]}\ ,
\ee
and
\be
\label{G59}
\frac{\m a^{2}}{Z}=c_{Z}\hub^{2}\ ,\quad c_{Z}=\frac{2\alpha(1+\alpha)\m\Ztil}{(1+2\ytil)\mtil Z}\ .
\ee
For solutions with $\deta\chi=0$ this yields for the elements in eq. \eqref{G28} the operators
\be
\label{G60}
A=\deta^{2}-2(2-\alpha)\hub\deta+k^{2}+\big{[}2-2\alpha+(1-y)c_{Z}\big{]}\hub^{2}\ ,
\ee
and
\be
\label{G61}
B=\big{[}2\alpha^{2}+(1-y)c_{Z}\big{]}\hub^{2}\ .
\ee
For the elements $C$ and $D$ eqs. \eqref{G26}, \eqref{G27} imply
\begin{align}
\label{G62}
C&=\m\bigg{\{}\big{(}1-y+\frac{2\alpha}{c_{Z}}\big{)}\deta^{2}+\big{(}1-y+\frac{2\alpha^{2}}{c_{Z}}\big{)}k^{2}\nn\\
&\quad -\Big{(}2+y(\alpha-3)+\frac{2\alpha(3-\alpha^{2})}{c_{Z}}\Big{)}\hub\deta+\gamma_{C}\hub^{2}\bigg{\}}\ .
\end{align}
and
\bel{G63}
D=\m\Big{[}\deta^{2}-(1+\alpha)\hub\deta+k^{2}+\gamma_{D}\hub^{2}\Big{]}\ .
\ee

We observe that for $\alpha\neq 1$ the operator $C$ is not Lorentz invariant even for the modes with very high $k^{2}$. The coefficients of $\deta^{2}$ and $k^{2}$ in this operator are no longer the same due to the difference between $\deta b$ and $b^{2}$. This contrasts with the de Sitter solution of type 1. The origin of the breaking of Lorentz symmetry are the non-zero gauge fields $b$.

For early enough epochs or $k^{2}\gg\hub^{2}$ we can neglect the terms $\sim\hub^{2}$ in the operators $A$, $B$, $C$ and $D$. With $B=0$ one has in this limit the equations
\bel{299}
\big{[}\deta^{2}-2(2-\alpha)\hub\deta+k^{2}\big{]}E_{ij}=0\ ,
\ee
and
\bel{300}
\big{[}\deta^{2}-(1+\alpha)\hub\deta+k^{2}\big{]}t_{ij}=-\frac{C}{\m}E_{ij}\ .
\ee
One of the eigenmodes corresponds to $E_{ij}=0$, $t_{ij}\neq 0$ and may be called the $t$-mode. For the other eigenmode, the $E$-mode, one has $E_{ij}\neq 0$, obeying eq. \eqref{299}. The corresponding fluctuation of $t_{ij}$ is then given by eq. \eqref{300}. We observe that the violation of Lorentz symmetry for $k^{2}\to\infty$ only concerns the relation between $t_{ij}$ and $E_{ij}$ for the $E$-mode, where $C$ appears explicitly. The oscillations of $E_{ij}$ for the $E$-mode and $t_{ij}$ for the $t$-mode both approach the plane waves in Minkowski space.

The damping term $\sim\hub\deta$ depends on $\alpha$. For $\alpha<1$ one has $1+\alpha<2$ and $4-2\alpha>2$. In this case the $E$-mode becomes dominant, increasing faster than the $t$-mode. For a general evolution equation
\bel{301}
\big{[}\deta^{2}-2\delta\hub\deta+k^{2}+\gamma\hub^{2}\big{]}\varphi=0\ ,
\ee
one finds for the asymptotic solution for $\hub^{2}\ll k^{2}$
\bel{302}
\varphi=a^{\delta}\tilde{\varphi}\ ,
\ee
where $\tilde{\varphi}$ oscillates with constant amplitude similar to $\tilde{t}_{ij}$ in eq. \eqref{G43}. Starting with similar amplitudes and evolving for many $e$-foldings only the mode with larger $\delta$ survives effectively. We can therefore always concentrate on the dominant mode. In later stages when the universe evolves away from the de Sitter solutions the two modes will mix, such that also the $t$-mode will get an amplitude comparable to the $E$-mode. For $\alpha<1$ the leading $E$-mode increases with $\delta=2-\alpha>1$, and therefore faster than the graviton modes for the type 1 de Sitter solutions which both increase with $\delta=1$. The relative strength of the vierbein-fluctuations increases or decreases $\sim a^{\delta-2}$
, cf. eqs. \eqref{G44}, \eqref{G45}. For $\delta=2$ or $\alpha=0$ the relative strength remains constant. For $\alpha<0$ one finds positive $\delta-2$ such that the relative strength of inhomogeneities increases with increasing time. An approximately homogeneous initial state becomes rapidly strongly inhomogeneous. Finally, for $\alpha>1$ the dominant mode is the $t$-mode, with $\delta=(1+\alpha)/2$. It is now this mode which increases faster than the graviton modes for the type 1 de Sitter solution. Again, for $\alpha>3$ the relative strength of the inhomogeneities increases. We conclude that for $\alpha<0$ or $\alpha>3$ the de Sitter solutions exhibit a further instability beyond the one in the scalar $\chi$. The transversal traceless tensor fluctuations $t_{ij}$, $E_{ij}$ are unstable.

We can compute the parameter $\alpha$ from the model parameters by inserting the explicit de Sitter solution. From eq. \eqref{FS8} one infers the relation
\bel{303}
\frac{4\alpha}{1+\alpha}=F\pm\sqrt{F^{2}-16}\ ,
\ee
where
\bel{304}
F=5+2\ytil-\frac{8V\Ztil}{3(1+2\ytil)\tilde{m}^{4}}=5+2\ytil-\frac{8x}{1+2\ytil}\ .
\ee
Using equation \eqref{FS12A} we infer for the two possible de Sitter solutions of type 2 the relation
\bel{305}
\alpha=-\half\pm\half\sqrt{\frac{x_{+}-x}{x_{c}-x}}\ ,
\ee
where we recall $0<x_{c}<x_{+}$ and $x<0$ for $V\Ztil<0$. In particular, for $V=0$, $x=0$ one has
\bel{306}
\alpha=-\half\pm\half\sqrt{\frac{x_{+}}{x_{c}}}=-\half\pm\half\sqrt{\frac{9+2\ytil}{1+2\ytil}}\ .
\ee
In the limit $\ytil\to 0$ this yields $\alpha_{-}=-2$, $\alpha_{+}=1$. We observe that the de Sitter solution with the minus sign of the root always leads to negative $\alpha$ and is therefore unstable with respect to the $E$-mode. For the positive root we can write
\bel{307}
\alpha=\half\Big{[}3\big{(}1+\frac{r_{\alpha}}{9}\big{)}^{\half}\big{(}1+r_{\alpha}\big{)}^{-\half}-1\Big{]}\ ,
\ee
with
\bel{308}
r_{\alpha}=2\ytil-\frac{8x}{1+2\ytil}\ .
\ee
For negative $r_{\alpha}$, as characteristic for $\ytil<0$ and small $\abs{x}$, one finds $\alpha>1$. For the corresponding de Sitter solution the $t$-mode dominates, with $\delta>1$.

For both de Sitter solutions of type 2 the inhomogeneous part of the vierbein and the metric vanish for the leading mode for $a\to 0$ proportional to $a^{\delta-1}$. This contrasts with the de Sitter solutions of type 1 for which the inhomogeneous part of the vierbein and the metric remain constant for $a\to 0$. For the type 2 solutions the dominant inhomogeneous part for $a\to 0$ is given by the non-leading or decaying mode. While this mode becomes unimportant as time progresses, it dominates for the backward extrapolation.

\end{appendices}

\nocite{*}
\bibliography{refs}

\end{document}